\documentclass[titlepage,twoside,11pt]{report}
 \usepackage{amssymb}
 \usepackage[english]{babel}
 \usepackage{epsfig}
 \usepackage{pstricks}
 \usepackage{latexsym}

 \input{def.cmd}
 \author{Michael Pl\"umacher}

\begin{document}
    \begin{titlepage}

\begin{center}

\vspace*{0.5cm}

\huge\bf 

Baryon Asymmetry, Neutrino Mixing\\
and\\ 
Supersymmetric SO(10) Unification

\rm

\vfill

\large

\vfill

 Dissertation \\
 zur Erlangung des Doktorgrades\\
 des Fachbereichs Physik\\
 der Universit\"at Hamburg\\

\vfill

 vorgelegt von \\
 Michael Pl\"umacher \\
 aus Antananarivo\\

\vfill

 Hamburg\\
  1998

\vspace*{1cm}

\end{center}
\end{titlepage}
\mbox{ }\\
\vfill
\pagestyle{empty}
\begin{tabular}{ll}
\begin{minipage}[t]{6.0cm}
Gutachter der Dissertation:
\end{minipage} 
& 
\begin{minipage}[t]{5.0cm}
Prof.~Dr.~W.~Buchm\"uller\\
Prof.~Dr.~G.~Kramer\\
\end{minipage} \\
\bigskip

\begin{minipage}[t]{6.0cm}
Gutachter der Disputation:
\end{minipage} 
& 
\begin{minipage}[t]{5.0cm}
Prof.~Dr.~W.~Buchm\"uller\\
Prof.~Dr.~J.~Bartels\\
\end{minipage} \\
\bigskip

\begin{minipage}[t]{6.0cm}
Datum der Disputation:
\end{minipage} 
& 
\begin{minipage}[t]{5.0cm}
19.~Mai 1998
\end{minipage} \\
\bigskip

\begin{minipage}[t]{6.0cm}
Sprecher des Fachbereichs Physik \\
und Vorsitzender des \\
Promotionsausschusses:
\end{minipage} 
& 
\begin{minipage}[t]{5.0cm}
${}$ \\
${}$ \\
Prof.~Dr.~B.~Kramer\\
\end{minipage} 
\end{tabular}

\newpage

\begin{center}
\begin{minipage}[t]{12cm}
{\begin{center}\bf{Abstract}\end{center}}
  The baryon asymmetry of the universe can be explained by the
  out-of-equilibrium decays of heavy right-handed neutrinos. We
  analyse this mechanism in the framework of a supersymmetric
  extension of the Standard Model and show that lepton number
  violating scatterings are indispensable for baryogenesis, even
  though they may wash-out a generated asymmetry. By assuming a
  similar pattern of mixings and masses for neutrinos and up-type
  quarks, as suggested by SO(10) unification, we can generate the
  observed baryon asymmetry without any fine tuning, if $(B-L)$ is
  broken at the unification scale $\Lambda_{\mbox{\tiny GUT}}\sim 
  10^{16}\;$GeV and, if $m_{\n_\m} \sim 3\cdot 10^{-3}\;$eV as 
  preferred by the MSW solution to the solar neutrino deficit.\\[10ex]
{\begin{center}\bf{Zusammenfassung}\end{center}}
  Die Baryonasymmetrie des Universums kann durch den Zerfall schwerer
  rechtsh\"andiger Neutrinos au\ss{}erhalb des thermischen
  Gleichgewichts erkl\"art werden.  Wir untersuchen dies im Rahmen
  einer supersym\-metrischen Erweiterung des Standard-Modells und
  zeigen, da\ss{} lepton\-zahlverletzende Streuprozesse, die eine
  erzeugte Asym\-metrie wieder vernichten k\"onnen, f\"ur die
  Baryogenese unverzichtbar sind. Nimmt man f\"ur Quarks und Leptonen
  \"ahnliche Massen und Mischungswinkel an --- wie von
  SO(10)-vereinheitlichten Modellen nahegelegt --- so kann man die
  beobachtete Baryonasymmetrie erzeugen. Dazu wird $(B-L)$ an der
  Vereinheitlichungsskala $\Lambda_{\mbox{\tiny GUT}}\sim
  10^{16}\;$GeV gebrochen, und $m_{\n_\m}\sim 3\cdot 10^{-3}\;$eV
  angenommen, wie es die MSW-L\"osung des solaren Neutrino-Problems
  nahelegt.
\end{minipage}
\end{center}

\pagebreak
\pagestyle{empty}
\mbox{}

\newpage
\pagestyle{headings}
\pagenumbering{roman}
\setcounter{page}{1}
\tableofcontents

    \clearpage
\pagenumbering{arabic}
\setcounter{page}{1}
\noindent
\begin{minipage}[t]{16.5cm}
\hspace{\fill}
\minipage{3.0cm} {\small \it
seeker of truth\\[0.7ex]
follow no path\\[-0.8ex]
all paths lead where\\[0.7ex]
truth is here\\[1ex]
{\rm\footnotesize E.~E.~Cummings \cite{eecummings}}  }
\endminipage
\vspace{2cm}
\noindent
\chapter*{Introduction}
\addcontentsline{toc}{chapter}{Introduction}
\end{minipage}
\markboth{INTRODUCTION}{INTRODUCTION}\\
  The observed baryon asymmetry of the universe is one of the most
  intriguing problems of particle physics and cosmology. This
  asymmetry, which is usually expressed as ratio of the baryon density
  $n_B$ to the entropy density $s$ of the universe,
  \beqx
    Y_B={n_B\over s}=(0.6-1)\cdot 10^{-10}\;,
  \eeqx
  could in principle be an initial condition of the cosmological
  evolution. However, this is not compatible with an inflationary
  phase which seems to be required in a consistent cosmological model
  \cite{kt1}. Hence, the baryon asymmetry has to be generated
  dynamically during the evolution of the universe. This is possible
  if baryon number is not conserved, if $C$ and $CP$ are violated, and
  if the universe is not in thermal equilibrium \cite{sakharov}.
  
  Although the Standard Model (SM) contains all the necessary
  ingredients, it is not possible to explain the baryon asymmetry
  within the SM, i.e.\ one has to envisage extended theories.
  Grand unified theories (GUTs) are attractive for various reasons and
  there have been many attempts to generate $Y_B$ at the GUT scale
  \cite{kt1}. However, these mechanisms are difficult to reconcile
  with inflationary scenarios which require reheating temperatures
  well below the GUT scale.
  
  Preheating, i.e.\ the non-thermal decay of the oscillating inflaton
  at the end of inflation via parametric resonance \cite{preheating},
  may re-open the window for GUT baryogenesis, since it enables the
  coherent decay of the inflaton condensate into particles that are
  more massive than the inflaton itself.  However, recent calculations
  indicate that parametric resonance may be ineffective in most
  inflationary models, if the back reaction of the produced particles
  onto the condensate, the rescattering of the decay products, and the
  expansion of the universe are taken into account \cite{nope}.

  In supersymmetric theories, the influence of baryon number carrying
  scalar condensates along flat directions of the scalar potential,
  i.e.\ Affleck-Dine baryogenesis \cite{AD}, requires further studies,
  since it is not clear under which conditions this mechanism can
  generate a baryon asymmetry of the requested magnitude
  \cite{ADnope}.

  During the evolution of the early universe, the electroweak phase
  transition is the last opportunity to generate a baryon asymmetry
  without being in conflict with the strong experimental bounds on
  baryon number violation at low energies \cite{phase}. However, the
  thermodynamics of this transition indicates that such scenarios are
  rather unlikely \cite{jansen}.
  
  Therefore, the baryon asymmetry has to be generated between the
  reheating scale and the electroweak scale, where baryon plus lepton
  number $(B+L)$ violating anomalous processes are in thermal
  equilibrium \cite{sphal2}, thereby making a $(B-L)$ violation
  necessary for baryogenesis. Hence, no asymmetry can be generated
  within GUT scenarios based on the gauge group SU(5), where $(B-L)$ is
  a conserved quantity.
  
  Gauge groups containing SO(10) predict the existence of right-handed
  neutrinos. In such theories $(B-L)$ is spontaneously broken, one
  consequence being that the right-handed neutrinos can acquire large
  Majorana masses, thereby explaining the smallness of the light
  neutrino masses via the see-saw mechanism \cite{seesaw}.  Heavy
  right-handed Majorana neutrinos violate lepton number in their
  decays, thus implementing the required $(B-L)$ breaking as lepton
  number violation. This leptogenesis mechanism was first suggested by
  Fukugita and Yanagida \cite{fy} and has subsequently been studied by
  several authors (see, e.g., refs.~\citer{luty,pil}). As detailed
  studies have shown, the observed baryon asymmetry can be generated
  in non-supersymmetric \cite{luty,pluemi,bp} and supersymmetric
  theories \cite{camp,pluemi2}.
  
  If one assumes a similar pattern of mass ratios and mixings for
  leptons and quarks and, if $m_{\n_{\m}}\sim3\cdot10^{-3}\;$eV as
  preferred by the MSW solution to the solar neutrino problem,
  leptogenesis implies that $(B-L)$ is broken at the unification scale
  \cite{bp}.  This suggests a grand unified theory based on the group
  SO(10), or one of its extensions, which is directly broken into the
  standard model gauge group at the unification scale
  $\sim10^{16}\;$GeV. However, for a successful gauge coupling
  unification, such a GUT scenario requires low-energy supersymmetry.
  
  Supersymmetric leptogenesis has already been considered in
  refs.~\cite{camp,covi} in the approxi\-mation that there are no
  lepton number violating scatterings which can inhibit the generation
  of a lepton number. Another usually neglected problem of
  leptogenesis scenarios is the necessary production of the
  right-handed neutrinos after reheating. In non-supersymmetric
  scenarios one has to assume additional interactions of the
  right-handed neutrinos for successful leptogenesis \cite{pluemi}.
  
  In this thesis, we investigate supersymmetric leptogenesis within
  the framework of the mini\-mal supersymmetric standard model (MSSM),
  to which we add right-handed Majorana neutrinos, as suggested by
  SO(10) unification \cite{pluemi2}. Since $CP$ asymmetries in the
  decays of these neutrinos are one of the principal ingredients of
  this model, we start by considering possible sources of $CP$
  violation in decays of Majorana neutrinos in the next chapter. In
  addition to the usually considered one-loop vertex corrections
  \cite{kt1}, we show how self-energy contributions to the $CP$
  asymmetry, which have previously been considered in
  refs.~\citer{covi,pil}, can be consistently taken into
  account \cite{bp2}.  For simplicity we only consider the
  non-supersymmetric leptogenesis scenario. However, our results are
  easily generalized to the supersymmetric case.
  
  In chapter \ref{SuperPert} we present superfield techniques, which
  simplify calculations in theories with exact supersymmetry These
  techniques are used in chapter \ref{theory} where we introduce
  supersymmetric leptogenesis. In particular, we discuss the neutrino
  decays and scattering processes that one has to take into account to
  be consistent \cite{pluemi2}. In chapter \ref{results} we develop
  the full network of Boltzmann equations necessary to get a reliable
  relation between the input parameters and the final baryon
  asymmetry.  We work out the parameter dependence of the generated
  baryon asymmetry, and show that by neglecting the lepton number
  violating scatterings one largely overestimates the generated
  asymmetry, and that in our scenario the Yukawa interactions are
  strong enough to produce a thermal population of right-handed
  neutrinos at high temperatures.  Finally, we show in chapter
  \ref{Yuk} that by assuming a similar pattern of masses and mixings
  for leptons and quarks one gets the required value for the baryon
  asymmetry without any fine tuning, provided $(B-L)$ is broken at the
  GUT scale, and the Dirac mass scale for the neutrinos is of order of
  the top quark mass, as suggested by SO(10) unification
  \cite{bp,pluemi2}.
  
  In appendix \ref{integrals} we summarize some standard formulae for
  one-loop integrals. In appendix \ref{SpinorConv} we introduce our
  spinor notation and compile formulae which are needed for the
  superfield calculations of chapters \ref{SuperPert} and
  \ref{theory}, while the Feynman rules for component field
  calculations are presented in appendix \ref{AppFeynm}. After a brief
  review of thermodynamics in an expanding universe in appendix
  \ref{appB}, we present the cross sections for the scattering
  processes discussed in chapter \ref{theory} in appendix \ref{appC}.
  Finally, in appendix \ref{appD} we discuss some limiting cases in
  which the corresponding reaction densities can be calculated
  analytically.

    \clearpage\chapter{CP Asymmetry in Majorana Neutrino Decays 
  \label{decaychapter}}
  In this chapter we study how self-energy diagrams can be
  consistently taken into account when computing $CP$ asymmetries in
  heavy particle decays. This is not obvious, since the naive
  prescription leads to a well-defined result for the $CP$ asymmetry,
  whereas the individual partial decay widths are infinite.
  
  We investigate this problem in the case of heavy Majorana neutrinos,
  which are obtained as mass eigenstates if right-handed neutrinos are
  added to the standard model. Since they are unstable, they cannot
  appear as in- or out-states of S-matrix elements. Rather, their
  properties are defined by S-matrix elements for scatterings of
  stable particles mediated by the unstable neutrino \cite{veltman}.
  By using a resummed propagator for the intermediate neutrino, we can
  separate two-body scattering processes in resonance contributions
  and remainder. While the $CP$ asymmetries of two-body processes vanish
  \cite{rcv}, the resonance contributions yield a finite $CP$ asymmetry
  which can be assigned to the intermediate neutrino

\section{Self-energy and vertex corrections}
  We consider the standard model with three additional right-handed
  neutrinos. The corresponding Lagrangian for Yukawa couplings and
  masses of charged leptons and neutrinos reads
  \beq
    \cl_Y = \overline{l_{\mbox{\tiny L}}}\,H\,\l^*_l\,e_{\mbox{\tiny R}}
      +\overline{l_{\mbox{\tiny L}}}\e H^{\dg}\,\l_{\n}^*\,\n_{\mbox{\tiny R}}
      -{1\over2}\,\overline{\n_{\mbox{\tiny R}}^c}\,M\,\n_{\mbox{\tiny R}}
      +\mbox{ h.c.}\;,
  \eeq    
  where $l_{\mbox{\tiny L}}=(\n_{\mbox{\tiny L}},e_{\mbox{\tiny L}})$
  is the left-handed lepton doublet and $H=(H^+,H^0)$ is the standard
  model Higgs doublet. $\l_{l}$, $\l_{\n}$ and $M$ are $3\times3$
  complex matrices in the case of three generations.  One can always
  choose a basis for the fields $\n_{\mbox{\tiny R}}$ such that the
  mass matrix $M$ is diagonal and real with eigenvalues $M_i$. The
  corresponding physical mass eigenstates are then the three Majorana
  neutrinos $N_i=\n_{\mbox{\tiny R}_{\scriptstyle i}} +
  \n_{\mbox{\tiny R}_{\scriptstyle i}}^c$.  At tree level the
  propagator matrix of these Majorana neutrinos reads
  \beq
    i\,S_0(q)={i\over\slash{q}-M+i\e}\;,\label{S0}
  \eeq  
  This propagator has poles at $q^2=M_i^2$ corresponding to stable
  particles, whereas the physical Majorana neutrinos are unstable.
  This is taken into account by summing self-energy diagrams in the
  usual way, which leads to the resummed propagator
  \beq
    i\,S(q)={i\over\slash{q}-M-\S(q)}\;.\label{S}
  \eeq

\begin{figure}[t]
    \input{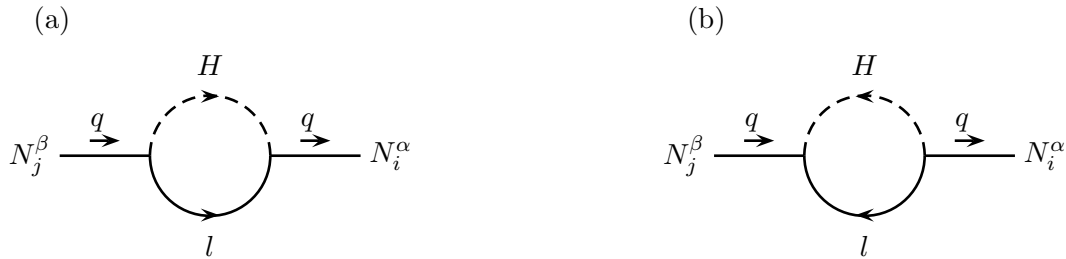}
    \caption{\it Leading order contributions to the self-energy of
      the heavy Majorana neutrinos. The fermion flow has been chosen
      parallel to the external momentum $q$.\label{chap1_fig01}}
  \end{figure}

  At one-loop level the two diagrams in fig.~\ref{chap1_fig01} yield the
  self energy\footnote{For the calculations we use the
    non-supersymmetric subset of the Feynman rules in
    App.~\ref{AppFeynm}, by identifying the SM Higgs doublet $H$ with
    the supersymmetric scalar Higgs doublet $H_2$.}
  \beq
    \S_{\a\b}^{ij}(q)=\left(\slash{q}\,P_{\mbox{\tiny R}}\right)_{\a\b}
    \,\S_{\mbox{\tiny R}}^{ij}(q^2)+\left(\slash{q}\,
    P_{\mbox{\tiny L}}\right)_{\a\b}\,
    \S_{\mbox{\tiny L}}^{ij}(q^2)\;,\label{sigma2}
  \eeq
  where $P_{\mbox{\tiny R,L}}={1\over2}(1\pm\g_5)$ are the projectors
  on right- and left-handed chiral states.
  $\S_{\mbox{\tiny R}}$ and $\S_{\mbox{\tiny L}}$ are the
  contributions of the diagrams figs.~(\ref{chap1_fig01}a) and
  (\ref{chap1_fig01}b), respectively.  They can be written as products
  of a complex function $a(q^2)$ and a hermitian matrix $K$,
  \beq
    \S_{\mbox{\tiny L}}(q^2)=\left(\S_{\mbox{\tiny R}}(q^2)\right)^T
     =a(q^2)\,K\; ,\quad K=\l_{\n}^{\dg}\l_{\n}\; .\label{def_aK}
  \eeq  
  $a(q^2)$ is given by the usual form factor $B_0(q^2,0,0)$ defined in
  eq.~(\ref{B0def}), whose finite part reads in the $\Bar{MS}$-scheme,
  \beq
    a(q^2) = {1\over 16\p^2}\left(\ln{{|q^2|\over \m^2}} - 
    2 - i\p\Theta(q^2)\right)\;.\label{aq2}
  \eeq
  For simplicity we will often omit the argument of $a$ in the following,
  however one should keep in mind that $a$ depends on $q^2$.

  According to eqs.~(\ref{S}) and (\ref{sigma2}) the resummed
  propagator $S(q)$ satisfies
  \beq
    \left[\,\slash{q}\,\Big(\,(1-\S_{\mbox{\tiny R}}(q^2))
    P_{\mbox{\tiny R}}+(1-\S_{\mbox{\tiny L}}(q^2))P_{\mbox{\tiny L}}
    \,\Big)\,-M\,\right]\,S(q)\;=1\;.\label{propre}
  \eeq
  The fermion propagator $S(q)$ consists of four chiral parts
  \beq
    S(q) = P_{\mbox{\tiny R}}\,S^{\mbox{\tiny RR}}(q^2)
      +P_{\mbox{\tiny L}}\,S^{\mbox{\tiny LL}}(q^2)
      + P_{\mbox{\tiny L}}\,\slash{q}\,S^{\mbox{\tiny LR}}(q^2)
      + P_{\mbox{\tiny R}}\,\slash{q}\,S^{\mbox{\tiny RL}}(q^2)\;.
  \eeq  
  Inserting this decomposition into eq.~(\ref{propre}), and
  multiplying the resulting equation from the left and the right
  with chiral projectors $P_{R,L}$, yields a system of four coupled
  linear equations for the four parts of the propagator. The
  solution reads
  \beqa
    S^{\mbox{\tiny RR}}(q^2)&=&\left[(1-\S_{\mbox{\tiny L}}(q^2))
      {\textstyle q^2\over\textstyle M}
      (1-\S_{\mbox{\tiny R}}(q^2))-M\right]^{-1}\;,\label{SRR}\\[1ex]
    S^{\mbox{\tiny LR}}(q^2)&=&{1\over M}(1-\S_{\mbox{\tiny R}}(q^2))
      S^{\mbox{\tiny RR}}(q^2)\;,\label{SLR}\\[1ex]
    S^{\mbox{\tiny LL}}(q^2)&=&\left[(1-\S_{\mbox{\tiny R}}(q^2))
      {\textstyle q^2\over\textstyle M}
      (1-\S_{\mbox{\tiny L}}(q^2))-M\right]^{-1}\;,\label{SLL}\\[1ex]
    S^{\mbox{\tiny RL}}(q^2)&=&{1\over M}(1-\S_{\mbox{\tiny L}}(q^2))
      S^{\mbox{\tiny LL}}(q^2)\;.\label{SRL}
  \eeqa
  As discussed below, the diagonal elements of $S(q)$ have approximately
  the usual Breit-Wigner form.
  \begin{figure}[t]
    \input{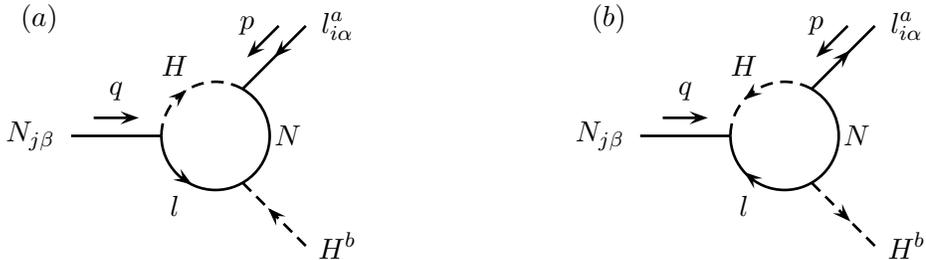}
    \caption{\it One-loop corrections to the couplings of heavy Majorana
        neutrinos $N_j$ to anti-lepton Higgs states (a) and lepton
        Higgs states (b). The fermion flow has been chosen according
        to the external lepton lines. \label{vert}}
  \end{figure}
  
  In addition to the self-energy we need the one-loop vertex function.
  The two expressions for the coupling of $N$ to $\bar{l},{H^{\dg}}$
  (fig.~\ref{vert}a) and $N$ to $l,H$ (fig.~\ref{vert}b) can be 
  written as
  \beqa
    \Bar{\G}^{ji}_{\b\a,ab}(q,p)&=&+i\e_{ab}
      \left[\left(K M b(q,p) \l_\n^T\,\right)_{ji} q_\m + 
      \left(K M c(q,p) \l_\n^T\,\right)_{ji}p_\m\right] 
      (P_{\mbox{\tiny R}}\g^{\m})_{\b\a}\;,\label{gam}\\[1ex]
    \G^{ij}_{\a\b,ab}(q,p)&=&-i\e_{ab}\left[
      \left(\l_\n^*Mb(q,p)K\right)_{ij}q_\m+
      \left(\l_\n^* M c(q,p) K\right)_{ij} p_\m\right]
      (P_{\mbox{\tiny R}} \g^\m)_{\a\b}\; .\label{gamb}
  \eeqa
  Here $b(q,p)$ and $c(q,p)$ are diagonal matrices whose elements are
  given by the standard three point form factors $C_0$ and $C_{12}$
  defined in appendix \ref{A3},
  \beqa
    b_k(q,p)&=&{1\over 16\p^2}\left[C_0(-p-q,q,M_k,0,0)
           +C_{12}(-p-q,q,M_k,0,0)\right]\;,\label{bk}\\[1ex]
    c_k(q,p)&=&{1\over 16\p^2}\left[C_0(-p-q,q,M_k,0,0)
           + 2 C_{12}(-p-q,q,M_k,0,0)\right]\;.
  \eeqa
  Since we shall only consider amplitudes with massless on-shell
  leptons, the terms proportional to $c_k$ will not contribute. We
  shall only need the imaginary part of $b_k$ which is given by
  \beq
    \mbox{Im}\{b_k(q^2)\} = {1\over 16\p \sqrt{q^2}M_k} 
    f\left({M_k^2\over q^2}\right) \Theta(q^2)\;,\label{imbk}
  \eeq
  where the function $f$ is defined as
  \beq
    f(x) = \sqrt{x}\left(1-(1+x)\ln\left({1+x\over x}\right)\right)
    \;.\label{fx}
  \eeq 

\section{Transition matrix elements}
\begin{figure}[t]
    \input{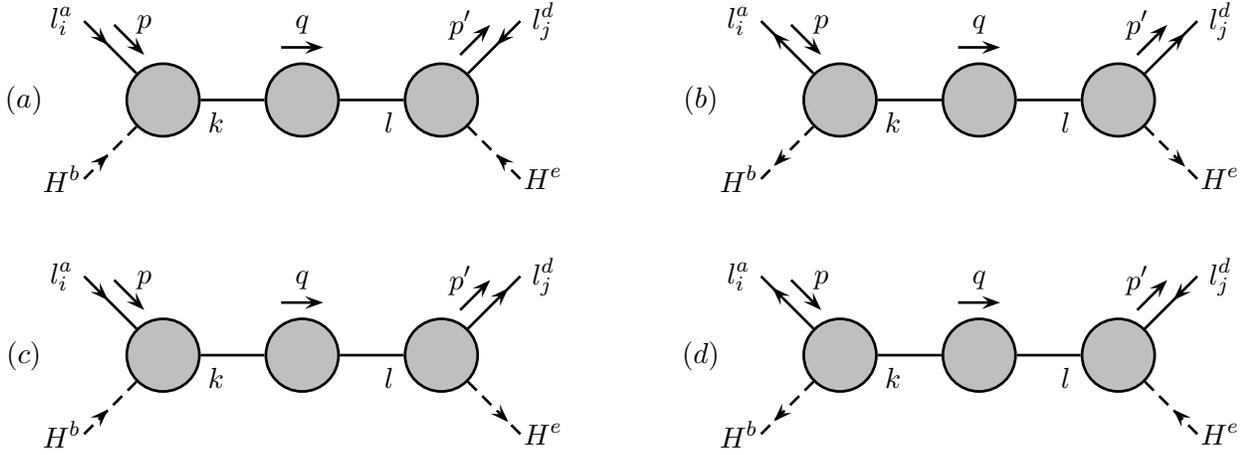}
    \caption{\it s-channel contributions to lepton-Higgs scattering, 
     including full propagators and vertices. \label{schannel}}
  \end{figure}

  The two lepton-number violating and the two lepton-number conserving
  processes are shown in figs.~\ref{schannel}a-\ref{schannel}d.
  Consider first the contributions of the full propagator, where the
  full vertices are replaced by tree couplings. The four scattering
  amplitudes read
  \beqa
    \langle\bar{l}^d_j(p')H^{e\dg}(q-p')|l^a_i(p)H^b(q-p)\rangle&=&
      +i\e_{ab}\e_{de} (\l^T_\n)_{lj}(\l^T_\n)_{ki}\NO\\
    &&\qquad (C P_{\mbox{\tiny L}} v(p'))^T\,S^{\mbox{\tiny LL}}_{lk}(q) 
      \,(P_{\mbox{\tiny L}} u(p))\;,\\[1ex]
    \langle l^d_j(p')H^e(q-p')|\bar{l}^a_i(p)H^{b\dg}(q-p)\rangle&=&
      +i\e_{ab}\e_{de} (\l^{\dg}_\n)_{lj}(\l^{\dg}_\n)_{ki}\NO\\
    &&\qquad (\bar{u}(p') P_{\mbox{\tiny R}})\,S^{\mbox{\tiny RR}}_{lk}(q)
      \,(\bar{v}(p)P_{\mbox{\tiny R}}C)^T\;,\\[1ex]
    \langle l^d_j(p')H^e(q-p')|l^a_i(p) H^b(q-p)\rangle
    &=& -i \e_{ab}\e_{de} (\l^{\dg}_\n)_{lj}(\l^T_\n)_{ki}\NO\\
    &&\qquad (\bar{u}(p') P_{\mbox{\tiny R}})\,S^{\mbox{\tiny RL}}_{lk}(q)
      \,\slash{q}\,(P_{\mbox{\tiny L}} u(p))\;,\\[1ex]
    \langle \bar{l}^d_j(p')H^{e\dg}(q-p')|\bar{l}^a_i(p)H^{b\dg}(q-p)
    \rangle&=& -i \e_{ab}\e_{de} (\l^T_\n)_{lj}(\l^{\dg}_\n)_{ki}\NO\\
    &&\qquad(C P_{\mbox{\tiny L}}v(p'))^T \,S^{\mbox{\tiny LR}}_{lk}(q)
      \,\slash{q}\,(\bar{v}(p) P_{\mbox{\tiny R}}C)^T\,.
  \eeqa
  Here $a,b,d,e$ denote the SU(2) indices of lepton and Higgs fields and
  $i,j,k,l$ are generation indices. The relative signs follow from Fermi 
  statistics.
 
  We are particularly interested in the contributions of a single
  heavy neutrino to the scattering amplitudes. In order to determine
  these contributions we have to find the poles and the residues of
  the propagator matrix. Here an unfamiliar complication arises due to
  the fact that the self-energy matrix is different for left- and
  right-handed states.  Hence, the different chiral projections of the
  propagator matrix are diagonalized by different matrices.
  
  $S^{\mbox{\tiny LL}}$ and $S^{\mbox{\tiny RR}}$ are symmetric
  complex matrices, since $\S_{\mbox{\tiny L}}(q^2)=
  \left(\S_{\mbox{\tiny R}}(q^2)\right)^T$.  Hence, $S^{\mbox{\tiny LL}}$
  and $S^{\mbox{\tiny RR}}$ can be diagonalized by complex orthogonal
  matrices $V$ and $U$, respectively,
  \beq
    S^{\mbox{\tiny LL}}(q^2) = V^T(q^2) M D(q^2) V(q^2)\;,\quad
    S^{\mbox{\tiny RR}}(q^2) = U^T(q^2) M D(q^2) U(q^2)\;.\label{sdiag}
  \eeq
  Splitting the self-energy into a diagonal and an off-diagonal part, 
  \beq
    \S_{\mbox{\tiny L}}(q^2) = \S_{\mbox{\tiny D}}(q^2) 
    + \S_{\mbox{\tiny N}}(q^2)\;,
  \eeq
  one finds
  \beq
    D^{-1}(q^2) = q^2(1-\S_{\mbox{\tiny D}}(q^2))^2 - M^2 + 
      \co(\S_{\mbox{\tiny N}}^2)\;.\label{dresum}
  \eeq
  One can easily identify real and imaginary parts of the propagator
  poles. The pole masses are given by
  \beq
    \Bar{M}_i^2 = Z_{\scr M_{\scriptstyle i}} M_i^2\; ,\quad 
    Z_{\scr M_{\scriptstyle i}}=1 + {K_{ii}\over 8\p^2}\left(
    \ln{M_i^2\over \m^2} - 2\right)\;,\label{polem}
  \eeq
  and the widths are $\G_i = K_{ii}M_i/(8\p)$. In the vicinity of the
  poles the propagator has the familiar Breit-Wigner form
  \beq
    D_i(q^2) \simeq {Z_i\over q^2 - \Bar{M}_i^2 + i \Bar{M}_i \G_i}
    \;,\quad Z_i = 1 + {K_{ii}\over 8\p^2}
    \left(\ln{M_i^2\over\m^2}-1\right)\;.
  \eeq 
 
  We can now easily write down the contribution of a single resonance
  $N_l$ with spin $s$ to the lepton-Higgs scattering amplitudes.
  Suppressing spin indices for massless fermions, one has
  \beqa
    \langle\bar{l}^d_j(p')H^{e\dg}(q-p')|l^a_i(p)H^b(q-p)\rangle_l&=&
      \langle \bar{l}^d_j(p') {H^{\dg}}^e(q-p')|N_l(q,s)\rangle \NO\\
    &&\quad i D_l(q^2)\ \langle N_l(q,s)|l^a_i(p) H^b(q-p)\rangle
      \;,\label{lbl}\\[1ex]
    \langle l^d_j(p')H^e(q-p')|\bar{l}^a_i(p)H^{b\dg}(q-p)\rangle_l&=&
      \langle l^d_j(p') H^e(q-p')|N_l(q,s)\rangle \NO\\
    &&\quad i D_l(q^2)\ \langle N_l(q,s)|\bar{l}^a_i(p)H^{b\dg}(q-p)
      \rangle\;,\\[1ex]
    \langle l^d_j(p') H^e(q-p')|l^a_i(p) H^b(q-p)\rangle_l&=&
      \langle l^d_j(p') H^e(q-p')|N_l(q,s)\rangle_{LC} \NO\\
    &&\quad i D_l(q^2)\ \langle N_l(q,s)|l^a_i(p) H^b(q-p)\rangle
      \;,\\[1ex]
    \langle\bar{l}^d_j(p')H^{e\dg}(q-p')|\bar{l}^a_i(p)H^{b\dg}(q-p)
      \rangle_l&=&\langle\bar{l}^d_j(p')H^{e\dg}(q-p')|N_l(q,s)
      \rangle_{LC} \NO\\
    &&\quad i D_l(q^2)\ \langle N_l(q,s)|\bar{l}^a_i(p)H^{b\dg}(q-p)
      \rangle\;.
  \eeqa
  Here the subscript $LC$ distinguishes an amplitude defined by a
  lepton-number conserving process from the same amplitude defined by
  a lepton-number violating process. From
  eqs.~(\ref{SRR})-(\ref{SRL}) and (\ref{sdiag}) one finds
  \beqa
    \langle N_l(q,s)| l^a_i(p) H^b(q-p)\rangle&=&
      +i \e_{ab}\left(V(q^2)\l_\n^T\right)_{li}\ \bar{u}_s(q,M_l)
      P_{\mbox{\tiny L}} u(p)\;,\label{ln}\\[1ex]
    \langle \bar{l}^d_j(p')H^{e\dg}(q-p')|N_l(q,s)\rangle &=&
      -i\e_{de}\left(\l_\n V^T(q^2)\right)_{jl}\bar{v}_s(q,M_l)
      P_{\mbox{\tiny L}} v(p')\;,\label{nlc}\\[1ex]
    \langle N_l(q,s)|\bar{l}^a_i(p)H^{b\dg}(q-p)\rangle &=&
      -i \e_{ab}\left(U(q^2)\l_\n^{\dg}\right)_{li}\ \bar{v}(p)
      P_{\mbox{\tiny R}} v_s(q,M_l)\;,\label{lcn}\\[1ex]
    \langle l^d_j(p') H^e(q-p')|N_l(q,s)\rangle &=&
      +i\e_{de}\left(\l_\n^* U^T(q^2)\right)_{jl}\bar{u}(p')
      P_{\mbox{\tiny R}} u_s(q,M_l)\;,\label{nl}\\[1ex]
    \langle l^d_j(p') H^e(q-p')|N_l(q,s)\rangle_{LC} &=&
      +i\e_{de}\left(\l_\n^*{1\over M}(1 - \S_{\mbox{\tiny L}}(q^2))
      V^T(q^2)M\right)_{jl}\NO\\ 
    &&\hspace{1.2cm}\bar{u}(p')P_{\mbox{\tiny R}} u_s(q,M_l)
      \;,\label{nlLC}\\[1ex]
    \langle\bar{l}^d_j(p')H^{e\dg}(q-p')|N_l(q,s)\rangle_{LC}&=&
      -i\e_{de}\left(\l_\n {1\over M} (1 - \S_{\mbox{\tiny R}}(q^2)) 
      U^T(q^2) M\right)_{jl}\NO\\
    &&\hspace{1.2cm}\bar{v}_s(q,M_l)P_{\mbox{\tiny L}} v(p')\;,\label{nlcLC}
  \eeqa
  where we have used the identity $C \bar{v}^T_s(p) = u_s(p)$.
  
  Eqs.~(\ref{ln}) and (\ref{nlc}) describe the coupling of the
  Majorana field $N$ to the lepton fields $l_i$ and the Higgs field
  $H$, and eqs.~(\ref{lcn}) and (\ref{nl}) give the couplings of $N$
  to the charge conjugated fields $\bar{l}_i$ and $H^{\dg}$. In the
  case of $CP$ conservation, one has $\l_{\n_{\scriptstyle ij}}
  = \l_{\n_{\scriptstyle ij}}^*$, which implies $K=K^T$ and therefore
  \beq
    \S_{\mbox{\tiny L}}(q^2) = \S_{\mbox{\tiny R}}(q^2)\;, 
    \quad V(q^2) = U(q^2)\; .
  \eeq
  This yields
  \beq
    \langle N_l(q,s)| l^a_i(p) H^b(q-p)\rangle 
    = \langle N_l(\tilde{q},s)|\bar{l}^a_i(\tilde{p}) 
      H^{b\dg}(\tilde{q}-\tilde{p})\rangle\; ,
  \eeq
  with $\tilde{q}=(q_0,-\vec{q}\,)$, $\tilde{p}=(p_0,-\vec{p}\,)$,
  as required by $CP$ invariance.
  
  The amplitudes given in eqs.~(\ref{ln}) - (\ref{nl}) have been
  obtained from the lepton-number violating processes
  figs.~\ref{schannel}a and \ref{schannel}b. The lepton-number
  conserving processes figs.~\ref{schannel}c and \ref{schannel}d
  yield the amplitudes given in eqs.~(\ref{nlLC}) and (\ref{nlcLC}).
  The consistent definition of an on-shell contribution of a single
  heavy Majorana neutrino to the two-body scattering amplitudes
  requires that the transition amplitudes extracted from lepton-number
  conserving and lepton-number violating processes are consistent.
  This implies
  \beqa  
    \langle l^d_j(p'\,) H^e(q-p'\,)|N_l(q,s)\rangle&=&
      \langle l^d_j(p'\,) H^e(q-p'\,)|N_l(q,s)\rangle_{LC}\; ,\\ [1ex]
    \langle\bar{l}^d_j(p'\,)H^{e\dg}(q-p'\,)|N_l(q,s)\rangle&=&
    \langle\bar{l}^d_j(p'\,)H^{e\dg}(q-p'\,)|N_l(q,s)\rangle_{LC}\; .
  \eeqa  
  From eqs.~(\ref{nlc}), (\ref{nl}), (\ref{nlLC}) and (\ref{nlcLC}) it
  is clear that these relations are fulfilled if the mixing matrices
  $V(q^2)$ and $U(q^2)$ satisfy certain consistency relations.
  Assuming that the matrix $\l_\n$ has an inverse, one reads off
  \beqa
    U_{ij}(M_i^2) &=& \left(M V(M_i^2)
      \left(1-\S_{\mbox{\tiny R}}(M_i^2)\right)
      {1\over M}\right)_{ij}\;,\label{consU}\\[1ex]
    V_{ij}(M_i^2) &=& \left(M U(M_i^2)
      \left(1-\S_{\mbox{\tiny L}}(M_i^2)\right)
      {1\over M}\right)_{ij}\label{consV}\;.
  \eeqa

  The matrices $V$ and $U$ are determined by the requirement that the
  expressions (cf.~eqs.~(\ref{sdiag}))
  \beqa
    V(q^2)\left(S^{\mbox{\tiny LL}}(q^2)\right)^{-1}V^T(q^2)= 
      V(q^2)\left(\left(1-\S_{\mbox{\tiny R}}(q^2)\right){q^2\over M}
      \left(1-\S_{\mbox{\tiny L}}(q^2)\right)-M\right)V^T(q^2)
      \;,\\[1ex]
    U(q^2)\left(S^{\mbox{\tiny RR}}(q^2)\right)^{-1}U^T(q^2) = 
      U(q^2)\left(\left(1-\S_{\mbox{\tiny L}}(q^2)\right){q^2\over M}
      \left(1-\S_{\mbox{\tiny R}}(q^2)\right)-M\right)U^T(q^2)\;,
  \eeqa
  are diagonal on-shell, i.e., at $q^2=M_i^2$. Using $\S_{\mbox{\tiny L}}
  = \S_{\mbox{\tiny D}} + \S_{\mbox{\tiny N}}$, and writing
  \beqa
    V(q^2) &=& 1 + v(q^2)\; ,\quad v(q^2)=-v^T(q^2)\; ,\label{Vv}\\[1ex]
    U(q^2) &=& 1 + u(q^2)\; ,\quad u(q^2)=-u^T(q^2)\; ,\label{Uu}
  \eeqa
  a straightforward calculation yields
  \beqa
    v_{ij}(q^2)&=&w_{ij}(q^2)\left(
      M_i \S_{\mbox{\tiny N}_{\scriptstyle ji}}(q^2)
      + M_j  \S_{\mbox{\tiny N}_{\scriptstyle ij}}(q^2)
      \right)\;,\label{vw}\\[1ex]
    u_{ij}(q^2)&=&w_{ij}(q^2)\left(
      M_i \S_{\mbox{\tiny N}_{\scriptstyle ij}}(q^2)
      + M_j  \S_{\mbox{\tiny N}_{\scriptstyle ji}}(q^2)
      \right)\;,\label{uw}
  \eeqa
  where
  \beq
    w_{ij}(q^2)^{-1} = (M_i-M_j)\left(1 + {M_i M_j\over q^2}\right)
    - 2 a(q^2) \left(M_i K_{jj} - M_j K_{ii}\right)\;.\label{ww}
  \eeq
  These equations give the matrices $V$ and $U$ to leading order in
  $\S_{\mbox{\tiny N}}$. They are meaningful as long as the matrix
  elements of $\S_{\mbox{\tiny N}}$ are small compared to those of
  $w^{-1}$.
  
  Inserting eqs.~(\ref{vw}) and (\ref{uw}) in eqs.~(\ref{consU}) and
  (\ref{consV}), one finds that the consistency conditions for the
  mixing matrices $V$ and $U$ are fulfilled to leading order in
  $\S_{\mbox{\tiny N}}$. We conclude that the contribution of a single
  heavy neutrino to two-body scattering processes can indeed be 
  consistently defined. The pole masses are given by eq.~(\ref{polem})
  and the couplings to lepton-Higgs initial and final states are
  given by eqs.~(\ref{ln})-(\ref{nl}).
 
\section{CP asymmetry in heavy neutrino decays}

  It is now straightforward to evaluate the $CP$ asymmetry in the decay
  of a heavy Majorana neutrino,
  \beq
    \ve_i = {\G(N_i\rightarrow lH) - \G(N_i\rightarrow \bar{l}{H^{\dg}}) 
    \over\G(N_i\rightarrow lH) + \G(N_i\rightarrow \bar{l}{H^{\dg}})}
    \;.\label{cpasym}
  \eeq 
  From eqs.~(\ref{nlc}) and (\ref{nl}) one obtains for the partial
  decay widths, including mixing effects,
  \beqa
    \G_{\mbox{\tiny M}}(N_i\rightarrow \bar{l}{H^{\dg}}) &\propto& 
      \sum_j |(\l_\n V^T(M_i^2))_{ji}|^2 \;,\label{gnbarl}\\
    \G_{\mbox{\tiny M}}(N_i\rightarrow lH) &\propto& 
      \sum_j |(\l_\n^* U^T(M_i^2))_{ji}|^2\;.\label{gnl}
  \eeqa
  To leading order in $\l_\n^2$ this yields the asymmetry 
  (cf.~eqs.~(\ref{Vv}), (\ref{Uu})),
  \beq
    \ve_i^{\mbox{\tiny M}}={1\over K_{ii}}\mbox{Re}
    \left\{(u(M_i^2)K)_{ii}-(v(M_i^2)K^T)_{ii}\right\}\;.
  \eeq
  Using eqs.~(\ref{vw}) - (\ref{ww}) and (\ref{aq2}), one finally
  obtains
  \beq
    \ve_i^{\mbox{\tiny M}}=-{1\over 8\p}\sum_j|w_{ij}(M_i^2)|^2
    (M_i^2-M_j^2){M_j\over M_i}{\mbox{Im}
    \{K_{\mbox{\tiny N}_{\scriptstyle ij}}^2\}\over K_{ii}}
    \;.\label{cpan}
  \eeq

  Consider first the case where differences between heavy neutrino masses
  are large, i.e., $|M_i - M_j|\gg |\G_i - \G_j|$. Eq.~(\ref{cpan}) then
  simplifies to
  \beq
    \ve_i^{\mbox{\tiny M}}=-{1\over8\p}\sum_j{M_iM_j\over M_i^2-M_j^2}
    {\mbox{Im}\{K_{\mbox{\tiny N}_{\scriptstyle ij}}^2\}\over K_{ii}}
    \;.\label{cpwave}
  \eeq
  This is the familiar $CP$ asymmetry due to flavour mixing
  \cite{covi}. It has previously been obtained by considering directly
  the self-energy correction to the Majorana neutrino decay, without
  any resummation. The $CP$ asymmetry $\ve_i$ reaches its maximum for
  $|M_i - M_j|\sim |\G_i - \G_j|$, where the perturbative expansion
  breaks down.
  
  Interesting is also the limiting case where the heavy neutrinos
  become mass degenerate. From eq.~(\ref{cpan}) it is obvious that the
  $CP$ asymmetry vanishes in this limit. The vanishing of the $CP$
  asymmetry for mass degenerate heavy neutrinos is expected on general
  grounds, since in this case the $CP$ violating phases of the matrix
  $K$ can be eliminated by a change of basis.
 
  The $CP$ asymmetry due to the vertex corrections is easily obtained
  using eqs.~(\ref{gam}), (\ref{gamb}), (\ref{imbk}) and (\ref{fx}).
  The partial decay widths corresponding to the full vertex read
  \beqa
    \G_{\mbox{\tiny V}}(N_i\rightarrow \bar{l}{H^{\dg}}) &\propto& 
      \sum_j|(\l_\n(1-MbK^TM))_{ji}|^2 \;,\label{gvbarl}\\[1ex]
    \G_{\mbox{\tiny V}}(N_i\rightarrow lH)&\propto&
      \sum_j|(\l_\n^*(1-MbKM))_{ji}|^2 \;.\label{gvl}
  \eeqa
  For the corresponding $CP$ asymmetry (\ref{cpasym}) one obtains the
  familiar result
  \beq
    \ve_i^{\mbox{\tiny V}} =-{1\over 8\p}\sum_j {\mbox{Im}\{
      K_{\mbox{\tiny N}_{\scriptstyle ij}}^2\}\over K_{ii}} 
      f\left({M_j^2\over M_i^2}\right) \; ,
  \eeq
  where the function $f(x)$ has been defined in eq.~(\ref{fx}).

\section{CP asymmetries in two-body processes}
\begin{figure}[t]
    \input{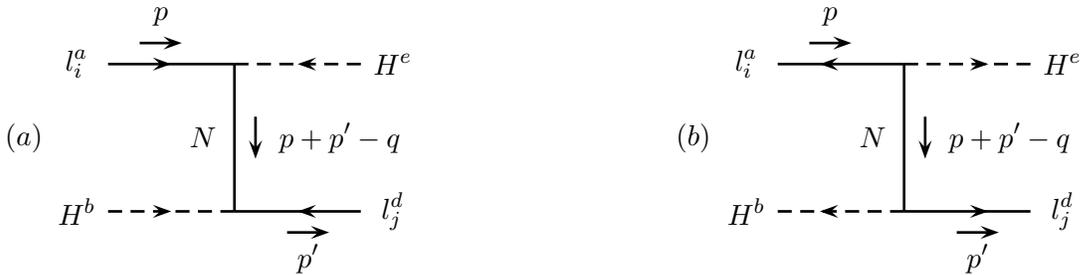}
    \caption{\it u-channel contributions to lepton-Higgs scattering. 
     \label{uchannel}}
 \end{figure}

  Let us now consider the $CP$ asymmetries in two-body processes. Here
  we have to take into account the s-channel amplitudes shown in
  figs.~\ref{schannel}a and \ref{schannel}b, with vertex functions up
  to one-loop, and the two u-channel amplitudes depicted in
  figs.~\ref{uchannel} and \ref{uchannel}b. For the u-channel
  amplitudes vertex and self-energy corrections can be omitted to
  leading order since the absorptive parts vanish.
 
  In the following we shall evaluate various contributions to the $CP$
  asymmetry
  \beq
    \ve \equiv {\Delta |\cm|^2\over 2 |\cm |^2}
    \equiv {|\cm (\bar{l}{H^{\dg}} \rightarrow lH)|^2 -
      |\cm (lH \rightarrow \bar{l}{H^{\dg}})|^2 \over
      |\cm (\bar{l}{H^{\dg}} \rightarrow lH)|^2 +
      |\cm (lH \rightarrow \bar{l}{H^{\dg}})|^2} \;,
  \eeq  
  where we always sum over generations in initial and final states.
  There are contributions from the full s-channel propagator, $\Delta
  |\cm|^2_s$, from the interference between s-channel amplitudes at
  tree-level and with one-loop vertex corrections, $\Delta
  |\cm|^2_{s,\G}$, the interference between tree-level s-channel and
  u-channel amplitudes, $\Delta |\cm|^2_{s,u}$, and the interference
  between s-channel with one-loop vertex corrections and u-channel
  amplitudes, $\Delta |\cm|^2_{u,\G}$.

  Consider first the $CP$ asymmetry $\ve_s$ due to the full
  propagator. The contribution of a single intermediate neutrino
  $N_i$ is (cf.~(\ref{lbl}), (\ref{ln}), (\ref{nlc})) 
  \beq
    |\cm_i(lH \rightarrow \bar{l}{H^{\dg}})|^2_s \propto |D_i(q^2)|^2
    \sum_j|(V(q^2)\l_\n^T\,)_{ij}|^2\  \sum_k|(\l_\n V^T(q^2))_{ki}|^2\; .
  \eeq
  Comparison with eq.~(\ref{gnbarl}) yields immediately
  \beq
    |\cm_i(lH \rightarrow \bar{l}{H^{\dg}})|^2_s \propto |D_i(q^2)|^2
    \G_{\mbox{\tiny M}}(N_i \rightarrow \bar{l}{H^{\dg}})^2\; .
  \eeq
  Similarly, one has for the charge conjugated process 
  \beq
    |\cm_i(\bar{l}{H^{\dg}} \rightarrow lH)|^2_s \propto |D_i(q^2)|^2
    \G_{\mbox{\tiny M}}(N_i\rightarrow lH)^2\; .
  \eeq
  The corresponding $CP$ asymmetry is, as expected, twice the asymmetry
  in the decay due to mixing,
  \beq
    \ve_s^{(i)}={\Delta|\cm_i|^2_s\over2|\cm_i|^2_s}
    \simeq2\ve^{\mbox{\tiny M}}_i\; .
  \eeq
  
  It is very instructive to compare the contribution of a single
  resonance with the $CP$ asymmetry $\ve_s$ for the full propagator.
  Due to the structure of the propagators $S^{\mbox{\tiny LL}}$ and
  $S^{\mbox{\tiny RR}}$ it is difficult to evaluate $\ve_s$ exactly.
  However, one may easily calculate $\ve_s$ perturbatively in powers of
  $\S_{\mbox{\tiny N}}$, like the mixing matrices $V(q^2)$ and
  $U(q^2)$ in the previous section.

  The full propagator (cf.~(\ref{propre})) reads to first order in
  $\S_{\mbox{\tiny N}}$,
  \beq
    S(q)=S_{\mbox{\tiny D}}(q) 
    + S_{\mbox{\tiny D}}(q) \slash{q}\left[\S_{\mbox{\tiny N}}^T(q^2)
    P_{\mbox{\tiny R}}+\S_{\mbox{\tiny N}}(q^2)P_{\mbox{\tiny L}}\right]
    S_{\mbox{\tiny D}}(q)+\ldots\;,
  \eeq
  where (cf.~(\ref{dresum}))
  \beq
    S_{\mbox{\tiny D}}(q)=
    \left[\slash{q}(1-\S_{\mbox{\tiny D}}(q^2))+M\right]D(q^2)\;. 
  \eeq

  It is now straightforward to calculate the matrix elements of the
  two-body processes, summed over generations in initial and final states,
  \beqa
    |\cm (lH \rightarrow \bar{l}H^{\dg})|^2_s&=&16\,p\cdot p'\, q^2 
      \Bigg({1\over 2 q^2}\mbox{Tr}\left[KMD(q^2)K^TMD^*(q^2)\right]
      +\NO\\[1ex]
    &&\mbox{Re}\Big\{\mbox{Tr}\Big[KMD(q^2)\S_{\mbox{\tiny N}}^T(q^2)
      (1-\S_{\mbox{\tiny D}}(q^2))D(q^2)K^TMD^*(q^2)+\NO\\[1ex]
    &&KD(q^2)(1-\S_{\mbox{\tiny D}}(q^2))
      \S_{\mbox{\tiny N}}(q^2)MD(q^2)K^TMD^*(q^2)\Big]\Big\} 
      + \ldots\Bigg)\, .
  \eeqa
  This yields for the sum and the difference of the $CP$ conjugated matrix 
  elements,
  \beqa
    2 |\cm|^2_s &=& 16\, p\cdot p'\sum_{i,j} A_{ij}+\ldots\;,\\
    \Delta|\cm|^2_s&=&-16\,p\cdot p'\sum_{i,j}(B_{ij}+C_{ij})+\ldots\;,
  \eeqa
  where
  \beqa
    A_{ij}&=&\mbox{Re}\left\{K_{ij}^2 M_i M_j D_j(q^2) D_i^*(q^2)\right\}
      \;,\label{aij}\\[1ex]
    B_{ij}&=& i \mbox{Im}\{K_{\mbox{\tiny N}_{\scriptstyle ij}}\}^2 
      M_i M_j D_j(q^2) D_i^*(q^2)\;,\label{bij}\\[1ex]
    C_{ij}&=& 4 q^2 \mbox{Re}\left\{ia(q^2) \mbox{Im}\{
      K_{\mbox{\tiny N}_{\scriptstyle ij}}^2\} M_i M_j
      \left(1-\S_{\mbox{\tiny D}}(q^2)_i\right)K_{ii}D_j(q^2)
      |D_i(q^2)|^2\right\}\;.\label{cij}
  \eeqa

  For $q^2 \simeq M_i^2$ the expressions $A_{ij}$ and $C_{ij}$ are
  dominated by the contribution of a single resonance $N_i$,
  \beqa
    A_{ii} &\simeq& K_{ii}^2 M_i^2 |D_i(q^2)|^2 \;,\label{Apole}\\[1ex]
    C_{ij} &\simeq& {1\over 4\pi} \mbox{Im}\{
      K_{\mbox{\tiny N}_{\scriptstyle ij}}^2\} {M_i^3 M_j\over 
      M_i^2-M_j^2} K_{ii} |D_i(q^2)|^2\;. \label{Cpole}
  \eeqa  
  From eqs.~(\ref{cpwave}), (\ref{Apole}) and (\ref{Cpole}) one reads
  off that the sum over $C_{ij}$ yields precisely the contribution of
  the resonance $N_i$ to the $CP$ asymmetry,
  \beq
    -{\sum_j C_{ij}\over A_{ii}} = 
    -{1\over 4\p}\sum_j {M_i M_j \over M_i^2 - M_j^2}
    {\mbox{Im}\{K_{\mbox{\tiny N}_{\scriptstyle ij}}^2\}
    \over K_{ii}} =2 \ve_i^{\mbox{\tiny M}}\;.
  \eeq
  
  The second contribution to the $CP$ asymmetry $\ve_s$ is due to the
  sum over $B_{ij}$ (cf.~eq.~(\ref{bij})). $B_{ij}$ involves two
  different propagators ($i\neq j$) and corresponds to an interference
  term. Using $D_j^{*-1}(q^2)= q^2-M_j^2-2a^*(q^2) q^2 K_{jj}$ and $2
  q^2 \mbox{Im}\{a(q^2)\} K_{ii}= -\mbox{Im}\{D_i^{-1}(q^2)\}$, one
  can rewrite $C_{ij}$ as follows,
  \beq
    C_{ij} = -i\,\mbox{Im}\,\{K_{\mbox{\tiny N}_{\scriptstyle ij}}^2\} 
    M_i M_j D_j^{*-1}(q^2) D_i^{-1}(q^2) |D_i(q^2)|^2
    |D_j(q^2)|^2\;.\label{cijre}
  \eeq
  Comparing eqs.~(\ref{bij}) and (\ref{cijre}) it is obvious that the
  sum of both terms, i.e., the $CP$ asymmetry $\ve_s$ corresponding to
  the full propagator, is identically zero! The pole contribution is
  cancelled by the interference of the pole term with an off-shell
  propagator.
  
  The contribution to the $CP$ asymmetry $\Delta |\cm|^2_{s,\G}$ can be
  computed in a similar manner. The diagrams fig.~\ref{schannel}a and
  \ref{schannel}b yield two contributions for the two vertices. After
  some algebra one obtains the result (cf.~(\ref{bk}))
  \beqa
    \Delta |\cm|^2_{s,\G} &=& - 64\ p\cdot p'\ q^2 \sum_{i,j,k} D_{ijk} 
      + \ldots\;,\label{dmgs} \\
    D_{ijk} &=& \mbox{Im}\{K_{ik}K_{jk}K_{ij}\}
      \mbox{Im}\{b_k(q^2)\} M_k M_j
      D_i(q^2) D_j^*(q^2)\;.\label{dijk}
  \eeqa
  For $q^2\simeq M_i^2$, one reads off that the sum over $D_{ijk}$
  yields, as expected, twice the vertex $CP$ asymmetry,
  \beq
    \ve_{s,\G}(M_i^2)\simeq{\sum_kD_{iik}\over A_{ii}}=-{1\over 4\p}
    \sum_k{\mbox{Im}\{K_{\mbox{\tiny N}_{\scriptstyle ik}}^2\}
    \over K_{ii}}f\left({M_k^2\over M_i^2}\right) = 
    2 \ve_i^{\mbox{\tiny V}}\; .
  \eeq
  A result very similar to eqs.~(\ref{dmgs}), (\ref{dijk}) is obtained
  for the asymmetry $\Delta|\cm |^2_{s,u}$, the interference between
  tree-level s-channel and u-channel amplitudes. One finds 
  ($u=(q-p-p')^2$), 
  \beqa
    \Delta |\cm|^2_{s,u} &=& - 32\ p\cdot p'\ q^2 \sum_{i,j,k} E_{ijk} + 
      \ldots\;,\label{emgs} \\
    E_{ijk} &=& \mbox{Im}\{K_{ik}K_{jk}K_{ij}\}\mbox{Im}\{a(q^2)\}M_kM_j
      D_i(q^2) D_j^*(q^2) D_k^*(u)\;.\label{eijk}
  \eeqa
  Integrating the expressions over phase space and using
  \beq
    \int_{-q^2}^0 du {2 p\cdot p'\over u - M_k^2}
    = {q^2 \sqrt{q^2}\over M_k} f\left({M_k^2\over q^2}\right)\; ,
  \eeq
  one finds the cancellation
  \beq
    \int_{-q^2}^0 du (\Delta |\cm|^2_{s,\G} + \Delta |\cm|^2_{s,u}) = 0\; .
  \eeq

  Finally, we have to consider the $CP$ asymmetry $\Delta|\cm|^2_{u,\G}$.
  A straightforward calculation yields
  \beqa
    \Delta |\cm|^2_{u,\G} &=& - 32\ p\cdot p'\ q^2 \sum_{i,j,k} F_{ijk} 
      + \ldots\;,\label{fmgs} \\
    F_{ijk} &=& \mbox{Im}\{K_{ik}K_{jk}K_{ji}\}
      \mbox{Im}\{b_k(q^2)\} M_k M_j D_i(q^2) D_j^*(u) \;.\label{fijk}
  \eeqa  
  After integration over $u$ the resulting matrix $\Bar{F}_{ijk}$ is
  antisymmetric in the indices $j$ and $k$. As a consequence, the
  asymmetry $\Delta |\cm|^2_{u,\G}$ is identically zero.
  
  As we have seen, the total $CP$ asymmetry vanishes to leading order
  in $\l_\n^2$. This result has previously been obtained in
  \cite{rcv}. It follows from unitarity and $CPT$ invariance. The
  considered T-matrix elements satisfy the unitarity relation
  \beq
    2\,\mbox{Im}\langle lH|T|lH\rangle = 
    \langle lH|T^\dagger T|lH\rangle\;.\label{imt}
  \eeq
  If, in perturbation theory, the leading contribution to the
  right-hand side is given by two-particle intermediate states, one has
  \beq
    \sum_l \langle lH|T^\dagger T|lH\rangle = \sum_{l,l'}\left(
    |\langle l'H|T|lH\rangle|^2 +|\langle \bar{l'}{H^{\dg}}|T|lH\rangle|^2 
    \right) + \ldots\;.\label{twpst}
  \eeq
  $CPT$ invariance implies
  \beq
    \langle l'H|T|lH\rangle = 
    \langle \bar{l}{H^{\dg}}|T|\bar{l'}{H^{\dg}}\rangle\;.\label{cpt}
  \eeq
  From eqs.~(\ref{imt})~-~(\ref{cpt}) one then obtains
  \beq
    \sum_{l,l'}\left(
    |\langle \bar{l'}{H^{\dg}}|T|lH\rangle|^2 -
    |\langle l'H|T|\bar{l}{H^{\dg}}\rangle|^2 \right) + \ldots = 0 \;.
  \label{cpaunit}
  \eeq
  In \cite{rcv} it was concluded that away from resonance poles, where
  ordinary perturbation theory holds, the $CP$ asymmetry
  (\ref{cpaunit}) vanishes to order $\l_\n^6$. Corrections due to
  four-particle intermediate states are $\co (\l_\n^8)$. In this paper
  we have developed a resummed perturbative expansion in powers of
  $\S_{\mbox{\tiny N}}$ which is also valid for $s\simeq M_i^2$. The
  same argument then implies that in this case the $CP$ asymmetry
  (\ref{cpaunit}) vanishes to order $\l_\n^2$ whith corrections $\co
  (\l_\n^4)$.
  
  The nature of the cancelation is different for different
  subprocesses. For the full propagator, the $CP$ asymmetry vanishes
  identically for fixed external momenta. Interference contributions
  between various s-channel and u-channel amplitudes cancel after
  phase space integration. In applications at finite temperature the
  standard practice \cite{kt1} is to treat in the Boltzmann equations
  resonance contributions and the remaining two-body cross sections
  differently. This procedure yields for the $CP$ asymmetry of the
  decaying heavy neutrino $N_i$ the sum of mixing and vertex
  contribution, $\ve_i = \ve_i^{\mbox{\tiny M}} 
  + \ve_i^{\mbox{\tiny V}}$.

    \clearpage\chapter{Perturbation Theory in Superspace
  \label{SuperPert}}
  In this chapter we give a short review of the superspace formulation
  of theories with global supersymmetry. Since in the following
  chapters we will only encounter chiral superfields, we restrict
  ourselves to chiral superfields and leave aside the superspace
  formulation of gauge theories, which can be found e.g.\ in
  ref.~\cite{gates}. In particular, we will present the powerful
  calculational tool of perturbation theory in superspace, which
  drastically simplifies calculations of $S$-matrix elements in
  theories with exact supersymmetry.

\section{Superspace}
  Supersymmetry transformations are generated by operators $Q$ which
  transform bosons into fermions, i.e.\ these generators have
  fermionic character \cite{nilles,sohnius}. The operators $Q$ and
  their Hermitian adjoints $\Bar{Q}$ can be chosen to be Weyl 
  spinors\footnote{For our notations and conventions see 
    app.~\ref{SpinorConv}.}, which obey anticommutation relations
  \beqa
    &&\{Q_{\a},Q_{\b}\}=\{\,\Bar{Q}_{\dt{\a}},
      \Bar{Q}_{\dt{\b}}\}=0\;,\\[1ex]
    &&\{Q_{\a},\Bar{Q}_{\dt{\a}}\}=
      2{\s_{\a\dt{\a}}}^{\m}P_{\m}\;,\\[1ex]
    &&[Q_{\a},P_{\m}]=
      [\,\Bar{Q}_{\dt{\a}},P_{\m}]=0\;,
  \eeqa
  where $P_{\m}$ is the energy-momentum operator. Together with the
  familiar commutation relations for the generators $P_{\m}$ and
  $M_{\m\n}$ of the Poincar\'e group, $Q$ and $\Bar{Q}$ form a closed
  algebra, the so-called super-Poincar\'e algebra.

  A compact technique for working out representations of the
  supersymmetry algebra was proposed by Salam and Strathdee
  \cite{salam}. They introduced Grassmann variables $\q^{\a}$
  ($\a=1,2$) and $\Bar{\q}_{\dt{\a}}$ ($\dt{\a}=1,2$), 
  \beq
    \{\q^{\a},\q^{\b}\}=\{\,\Bar{\q}^{\,\dt{\a}},\Bar{\q}^{\,\dt{\b}}\,\}
    =\{\q^{\a},\Bar{\q}^{\,\dt{\b}}\,\}=0\;.
  \eeq
  With these anticommuting parameters the supersymmetry algebra can be
  expressed in terms of commutators,
  \beqa
    &&[\q Q,\q Q]=[\,\Bar{Q}\,\Bar{\q},\Bar{Q}\,\Bar{\q}\,]=0\;,\\[1ex]
    &&[\q Q,\Bar{Q}\,\Bar{\q}\,]=2\q\s^{\m}\Bar{\q}P_{\m}\;,\\[1ex]
    &&[\q Q,P_{\m}]=[\,\Bar{Q}\,\Bar{\q},P_{\m}]=0\;.
  \eeqa
  Now we are able to exponentiate the super-Poincar\'e algebra into a
  group in such a way that the product of two group elements is again
  a group element. An element of this super-Poincar\'e group is given
  by
  \beq
    G(x,\q,\Bar{\q}\,)=\mbox{e}^{i(x\cdot P+\q Q
    +\Bar{\q}\,\Bar{Q}\,)}\;.\label{superp}
  \eeq
  These group elements generate transformations in the
  eight-dimensional superspace parametrized by the coordinates
  $(x^{\m}, \q^{\a}, \Bar{\q}_{\dt{\a}})$. In the following we use
  $z=(x^{\m}, \q^{\a}, \Bar{\q}_{\dt{\a}})$ to denote a point in
  superspace.

  Left action of the group element $G(a,\h,\Bar{\h}\,)$ induces a
  motion in superspace,
  \beq
    (x^{\m},\q,\Bar{\q}\,)\to(x^{\m}+a^{\m}+i\h\s^{\m}\Bar{\q}-
    i\q\s^{\m}\Bar{\h},\q+\h,\Bar{\q}+\Bar{\h}\,)\;.
  \eeq
  This transformation is realized by the differential
  operator representation of the algebra
  \beqa
    &&P_{\m}=i\partial_{\m}\;,\\[1ex]
    &&Q_{\a}={\partial\over\partial\q^{\a}}+i{\s_{\a\dt{\a}}}^{\m}
      \Bar{\q}^{\,\dt{\a}}\partial_{\m}\;,\\[1ex]
    &&\Bar{Q}_{\dt{\a}}=-{\partial\over\partial\Bar{\q}^{\,\dt{\a}}}-
      i\q^{\a}{\s_{\a\dt{\a}}}^{\m}\partial_{\m}\;,
  \eeqa
  where differentiation with respect to spinor parameters $\q$ and
  $\Bar{\q}$ is defined by
  \beq
    {\partial\over\partial\q^{\a}}\q^{\b}\equiv\d_{\a}^{\b}\;,\qquad
    {\partial\over\partial\Bar{\q}^{\,\dt{\a}}}\Bar{\q}^{\,\dt{\b}}
    \equiv\d_{\dt{\a}}^{\dt{\b}}\;.
  \eeq
  The usual rules for raising and lowering spinor indices
  (cf.~app.~\ref{SpinorConv}) therefore give an additional sign, if
  the index position in the differentiations is changed,
  \beq
    \ve^{\a\b}{\partial\over\partial\q^{\b}}=
    -{\partial\over\partial\q_{\a}}\;,\qquad
    \ve_{\dt{\a}\dt{\b}}{\partial\over\partial\Bar{\q}_{\dt{\b}}}=
    -{\partial\over\partial\Bar{\q}^{\,\dt{\a}}}\;.
  \eeq

  Correspondingly, right action of the group elements induces an
  anti-realization of the super-Poincar\'e group generated by
  covariant derivatives $D$ and $\Bar{D}$,
  \beqa
    D_{\a}&=&{\partial\over\partial\q^{\a}}-i{\s_{\a\dt{\a}}}^{\m}
      \Bar{\q}^{\,\dt{\a}}\partial_{\m}\;,\\[1ex]
    \Bar{D}_{\dt{\a}}&=&-{\partial\over\partial\Bar{\q}^{\,\dt{\a}}}+
      i\q^{\a}{\s_{\a\dt{\a}}}^{\m}\partial_{\m}\;,
  \eeqa
  which obey the anticommutation rules
  \beqa 
    &&\{D_{\a},D_{\b}\}=
      \{\,\Bar{D}_{\dt{\a}},\Bar{D}_{\dt{\b}}\}=0\\[1ex]
    &&\{D_{\a},\Bar{D}_{\dt{\a}}\}=
      2{\s_{\a\dt{\a}}}^{\m}P_{\m}\;.
  \eeqa
  Furthermore, one can derive the following useful identities
  \beqa
    D^{\a}D^{\b}&=&-{1\over2}\,\ve^{\a\b}D^2\;,\label{DD1}\\[1ex]
    \Bar{D}^{\,\dt{\a}}\,\Bar{D}^{\,\dt{\b}}&=&
      {1\over2}\,\ve^{\dt{\a}\dt{\b}}\,\Bar{D}^{\,2}\;,\label{DD2}\\[1ex]
    D^{\a}\,\Bar{D}^{\,2}D_{\a}&=&\Bar{D}_{\dt{\a}}D^2\,
      \Bar{D}^{\,\dt{\a}}\;,\label{DD3}\\[1ex]
    \Bar{D}^{\,2}\,D^2\,\Bar{D}^{\,2}&=&-16\,\bo\Bar{D}^{\,2}
      \;,\label{DD4}\\[1ex]
    D^2\,\Bar{D}^{\,2}D^2&=&-16\,\bo D^2\;,\label{DD5}
  \eeqa
  where the squared covariant derivatives are given by
  \beqa
    D^2\equiv D^{\a}D_{\a}&=&-\ve^{\a\b}{\partial\over\partial\q^{\a}}
      {\partial\over\partial\q^{\b}}+2i\left(\,\Bar{\q}\,\Bar{\s}^{\,\m}
      {\partial\over\partial\q}\right)\partial_{\m}+\Bar{\q}^{\,2}\bo
      \;,\\[1ex]
    \Bar{D}^{\,2}\equiv \Bar{D}_{\dt{\a}}\Bar{D}^{\,\dt{\a}}&=&
      \ve^{\dt{\a}\dt{\b}}{\partial\over\partial\Bar{\q}^{\,\dt{\a}}}
      {\partial\over\partial\Bar{\q}^{\,\dt{\b}}}+2i\left(\q\s^{\m}
      {\partial\over\partial\Bar{\q}}\right)\partial_{\m}+\q^2\bo\;.
  \eeqa

\section{Superfields \label{superfields}}
  Supersymmetric theories are most easily formulated in terms of
  superfields in superspace. In order to get a feeling for how to
  define a superfield, let us first consider an ordinary quantum field
  $\f(x)$ which depends only on the coordinates $x^{\m}$ of
  Minkowski space. Translations of these coordinates are generated by
  the operator $P_{\m}$, and we can consider $\f(x)$ to have been
  translated from $x^{\m}=0$,
  \beq
    \f(x)=\mbox{e}^{ix\cdot P}\f(0)\mbox{e}^{-ix\cdot P}\;.
  \eeq
  In complete analogy, a superfield $F(x,\q,\Bar{\q}\,)$ can be defined
  as \cite{sohnius}
  \beq
    F(x,\q,\Bar{\q}\,)=
      G(x,\q,\Bar{\q}\,)F(0,0,0)G^{-1}(x,\q,\Bar{\q}\,)\;,
  \eeq 
  where $G(x,\q,\Bar{\q}\,)$ is an element of the super-Poincar\'e
  group given by eq.~(\ref{superp}).  This means that a
  superfield is defined as a Taylor expansion in $\q$ and $\Bar{\q}$
  with coefficients which are themselves local fields in Minkowski
  space. Due to the Grassmann nature of $\q$ and $\Bar{\q}$, this
  expansion breaks off, and the most general superfield
  reads
  \beqa
    F(x,\q,\Bar{\q}\,)&=&f(x)+\q\f(x)+\Bar{\q}\,\Bar{\c}(x)
      +\q^2m(x)+\Bar{\q}^{\,2}n(x)+\q\s^{\m}\Bar{\q}v_{\m}(x)\NO\\[1ex]
    &&+\q^2\,\Bar{\q}\,\Bar{\l}(x)+\Bar{\q}^{\,2}\q\j(x)
      +\q^2\Bar{\q}^{\,2}d(x)\;.
  \eeqa
  This superfield contains as Taylor coefficients four complex scalar
  fields $f$, $m$, $n$ and $d$, one complex vector $v_{\m}$, two
  spinors $\f$ and $\j$ in the $({1\over2},0)$ representation and two
  spinors $\Bar{\c}$ and $\Bar{\l}$ in the $(0,{1\over2})$
  representation of the Lorentz group, altogether 16 fermionic and 16
  bosonic field components.
  
  Consequently, superfields form linear representations of the
  supersymmetry algebra which are, in general, highly reducible.
  Irreducible representations can be constructed by imposing
  constraints on the superfields. Like all covariant derivatives, $D$
  and $\Bar{D}$ can be used to impose covariant conditions. Chiral
  superfields $\F$ are characterized by the condition
  \beq
    \Bar{D}_{\dt{\a}}\F=0\;.\label{chirality}
  \eeq
  This first order differential equation is most easily solved in
  terms of the variables $y^{\m}=x^{\m}-i\q\s^{\m}\Bar{\q}$ and $\q$
  since
  \beq
    \Bar{D}_{\dt{\a}}\left(x^{\m}-i\q\s^{\m}\Bar{\q}\,\right)=0\quad
    \mbox{and}\quad\Bar{D}_{\dt{\a}}\q=0\;.
  \eeq
  Then an arbitrary function of $y$ and $\q$ is a chiral superfield,
  \beq
    \F(y,\q)=A(y)+\sqrt{2}\q\j(y)+\q^2 F(y)\;.\label{chiral1}
  \eeq
  This is the most general solution to eq.~(\ref{chirality}), as may
  be seen by expressing the covariant derivatives in terms of $y$, 
  $\q$ and $\Bar{\q}$,
  \beqa
    D_{\a}&=&{\partial\over\partial\q^{\a}}-2i{\s_{\a\dt{\a}}}^{\m}
      \Bar{\q}^{\,\dt{\a}}{\partial\over\partial y^{\m}}\;,\\[1ex]
    \Bar{D}_{\dt{\a}}&=&-{\partial\over\partial\Bar{\q}^{\,\dt{\a}}}\;.
  \eeqa

  By Taylor expansion in $\q$ and $\Bar{\q}$, we can write a chiral
  superfield as a function of the original superspace coordinates
  $x^{\m}$, $\q$ and $\Bar{\q}$,
  \beqa
    \F(x,\q,\Bar{\q}\,)&=&A(x)-i\q\s^{\m}\Bar{\q}\partial_{\m}A(x)
       -{1\over4}\q^2\Bar{\q}^{\,2}\bo A(x)\label{chiral2}\\[1ex]
    &&+\sqrt{2}\q\j(x)
       +{i\over\sqrt{2}}\q^2\partial_{\m}\j(x)\sigma^{\m}\Bar{\q}
       +\q^2 F(x)\;.\NO
  \eeqa

  Conjugation gives an antichiral superfield $\Bar{\F}$ which
  satisfies the constraint 
  \beq
    D_{\a}\Bar{\F}=0\;.\label{antichirality}
  \eeq
  It is a natural function of $\Bar{y}^{\m}=x^{\m}+i\q\s^{\m}\Bar{\q}$
  and $\Bar{\q}$, and its power series expansio is obtained from 
  eqs.~(\ref{chiral1}) and (\ref{chiral2}) by conjugation.

  Supersymmetry invariant actions can be constructed from chiral
  superfields and their products. It is clear from the expansion
  (\ref{chiral1}) that a product of chiral superfields is again a
  chiral superfield (cf.~app.~\ref{sfproducts}), whereas a product of
  a chiral and an antichiral superfield will satisfy neither
  eq.~(\ref{chirality}) nor eq.~(\ref{antichirality}). However, not
  every component of these product superfields can be used to
  construct supersymmetric actions. To be able to formulate
  supersymmetric theories we have to isolate the components which are
  invariant under supersymmetry transformations, up to total derivatives.

\section{Superspace Invariants}
  The general method by which a translation invariant action is
  derived from fields is to integrate a Lagrange density $\cl(x)$
  over $d^4x$. The result is translationally invariant if surface
  terms vanish. Similarly, SUSY invariant actions can be constructed
  by integration over superspace, once we have defined an integral
  over the Grassmann variables $\q$ and $\Bar{\q}$. This Berezin
  integral is determined by imposing linearity and translation
  invariance, except for the normalization which is fixed by the
  definitions \cite{srivastava}
  \beq
    \int d^2\q\,\q^2=1\qquad\mbox{and}\qquad
    \int d^2\Bar{\q}\;\Bar{\q}^{\,2}=1\;,
  \eeq
  with all other integrals vanishing. The two-dimensional volume
  elements are defined by 
  \beqa
    d^2\q&=&-{1\over4}d\q^{\a}d\q_{\a}\;,\\[1ex]
    d^2\Bar{\q}&=&-{1\over4}d\Bar{\q}_{\dt{\a}}d\Bar{\q}^{\,\dt{\a}}\;.
  \eeqa
  For integration over superspace we introduce the following
  integration measures
  \beqa
    &d^6s\equiv d^4x\,d^2\q\;,\qquad d^6\Bar{s}\equiv 
      d^4x\,d^2\Bar{\q}\;,&\\[1ex]
    &d^8z\equiv d^4x\,d^2\q\,d^2\Bar{\q}\;.&
  \eeqa

  If we adopt the convention of dropping total divergences, i.e.\ 
  surface integrals, the differential operators $-{1\over4}D^2$ and
  $-{1\over4}\Bar{D}^{\,2}$ are equivalent to $d^2\q$ and
  $d^2\Bar{\q}$ under a space-time volume integral,
  \beqa
    \int d^8z\,F(x,\q,\Bar{\q}\,)&=&\int d^6s\,
      \left(-{1\over4}\Bar{D}^{\,2}\right)\,F(x,\q,\Bar{\q}\,)
      \label{d2qa}\\[1ex]
    &=&\int d^6\Bar{s}\,\left(-{1\over4}D^2\right)\,
      F(x,\q,\Bar{\q}\,)\\[1ex]
    &=&\int d^4x\,{D^2\Bar{D}^{\,2}\over16}\,F(x,\q,\Bar{\q}\,)\\[1ex]
    &=&\int d^4x\,{\Bar{D}^{\,2}D^2\over16}\,F(x,\q,\Bar{\q}\,)
      \;,\label{d2qb}
  \eeqa
  where $F$ is an arbitrary function of $x$, $\q$ and $\Bar{\q}$,
  i.e.\ a superfield. It follows that
  \beq
    \int d^8z\,D_{\a}\,F(x,\q,\Bar{\q}\,)=
    \int d^8z\,\Bar{D}_{\dt{\a}}\,F(x,\q,\Bar{\q}\,)=0\;.
  \eeq
  Hence, we immediately get the following rules of integration by
  parts
  \beqa
    &&\int d^8z\,F_1\,D_{\a}\,F_2=
    \mp\int d^8z\,(D_{\a}\,F_1)\,F_2\;,\label{partint1}\\[1ex]
    &&\int d^8z\,F_1\,\Bar{D}_{\a}\,F_2=
    \mp\int d^8z\,(\,\Bar{D}_{\a}\,F_1)\,F_2\;,\label{partint2}
  \eeqa
  where the upper (lower) sign is valid if $F_1$ is an even (odd)
  Grassmann function. Similarly, higher powers of covariant
  derivatives can be partially integrated by means of the following
  formulae 
  \beqa
    \int d^8z\,F_1\left(D^2\,F_2\right)&=&\int d^8z\,\left(D^2\,F_1
      \right)\,F_2\;,\label{partint3}\\[1ex]
    \int d^8z\,F_1\left(\,\Bar{D}^{\,2}\,F_2\right)&=&\int d^8z\,
      \left(\,\Bar{D}^{\,2}\,F_1\right)\,F_2\;,
      \label{partint4}\\[1ex]
    \int d^8z\,F_1\left(D^2\,\Bar{D}^{\,2}\,F_2\right)&=&\int d^8z\,
      \left(\,\Bar{D}^{\,2}\,D^2\,F_1\right)\,F_2\;,
      \label{partint5}\\[1ex]
    \int d^8z\,F_1\left(D^{\a}\,\Bar{D}^{\,2}\,D_{\a}F_2\right)&=&
      \mp\int d^8z\,\left(\,\Bar{D}^{\,2}\,D^{\a}\,F_1\right)\,
      D_{\a}\,F_2\;,\label{partint6}\\[1ex]
    &=&\int d^8z\,\left(D^{\a}\,\Bar{D}^{\,2}\,D_{\a}\,F_1\right)\,F_2
      \;.\label{partint7}
  \eeqa

  We may also define superspace delta distributions
  \beqa
    &\d^2(\q-\q')\equiv(\q-\q')^2\;,\qquad\d^2(\,\Bar{\q}-\Bar{\q'}\,)
     \equiv(\,\Bar{\q}-\Bar{\q'}\,)^2\;,&\\[1ex]
    &\d^8(z-z')\equiv\d^4(x-x')\d^2(\q-\q')
     \d^2(\,\Bar{\q}-\Bar{\q'}\,)\;.&
  \eeqa
  Applying covariant derivatives to these delta functions yields
  \beqa
    D_1^2\d^2(\q_1-\q_2)&=&-4\exp\left[
      i(\q_1-\q_2)\s^{\m}\Bar{\q_1}\partial_{1,\m}\right]\;,\\[1ex]
    \Bar{D_1}^{\,2}\d^2(\,\Bar{\q_1}-\Bar{\q_2}\,)&=&-4\exp\left[
      -i\q_1\s^{\m}(\,\Bar{\q_1}-\Bar{\q_2}\,)\partial_{1,\m}\right]
      \;,\\[1ex]
    \Bar{D_1}^{\,2}D_1^2\d^2(\q_1-\q_2)\d^2(\,\Bar{\q_1}-\Bar{\q_2}\,)&=&
      16\exp\left[-i\left(\q_1\s^{\m}\Bar{\q_1}+\q_2\s^{\m}\Bar{\q_2}
      -2\q_1\s^{\m}\Bar{\q_2}\,\right)\partial_{1,\m}\right]
      \;,\\[1ex]
    D_1^2\Bar{D_1}^{\,2}\d^2(\q_1-\q_2)\d^2(\,\Bar{\q_1}-\Bar{\q_2}\,)&=&
      16\exp\left[i\left(\q_1\s^{\m}\Bar{\q_1}+\q_2\s^{\m}\Bar{\q_2}
      -2\q_2\s^{\m}\Bar{\q_1}\,\right)\partial_{1,\m}\right]\;,
  \eeqa
  where $D_1$ and $\Bar{D_1}$ act on $x_1$, $\q_1$ and $\Bar{\q_1}$.
  The argument of the derivatives can be changed through the following
  transfer rules
  \beqa
    D_1^{\a}\d^8(z_1-z_2)&=&-D_2^{\a}\d^8(z_1-z_2)\;,\label{Ddelta1}\\[1ex]
    \Bar{D_1}^{\,\dt{\a}}\d^8(z_1-z_2)&=&
      -\Bar{D_2}^{\,\dt{\a}}\d^8(z_1-z_2)\;,\label{Ddelta2}\\[1ex]
    D_1^2\d^8(z_1-z_2)&=&D_2^2\d^8(z_1-z_2)\;,\label{Ddelta3}\\[1ex]
    \Bar{D_1}^{\,2}\d^8(z_1-z_2)&=&
      \Bar{D_2}^{\,2}\d^8(z_1-z_2)\;,\label{Ddelta4}
  \eeqa
  and since covariant derivatives with different arguments anticommute
  we also have
  \beqa
    \Bar{D_1}^{\,2}D_1^2\d^8(z_1-z_2)&=& 
      \Bar{D_1}^{\,2}D_2^2\d^8(z_1-z_2)\label{Ddelta5}\\[1ex]
    &=&D_2^2\,\Bar{D_1}^{\,2}\d^8(z_1-z_2)\NO\\[1ex]
    &=&D_2^2\,\Bar{D_2}^{\,2}\d^8(z_1-z_2)\;.\NO
  \eeqa

  Superspace integration can be used to construct invariant
  actions. Consider first the integral over a chiral superfield. Due
  to the constraint (\ref{chirality}), a chiral superfield is
  independent of $\Bar{\q}$, i.e.\ $\int d^2\Bar{\q}$, and hence the
  full superspace integral gives zero.
  
  Since for chiral superfields the supersymmetry algebra can be
  realized as coordinate transformations of the chiral subspace of
  superspace alone, which has coordinates $y^{\m}$ and $\q^{\a}$ but
  not $\Bar{\q}_{\dt{\a}}$, the $\int d^4x\,d^2\q$ integral, without
  the $d^2\Bar{\q}$, is already an invariant integral for chiral
  superfields.  Therefore, the most general supersymmetric
  renormalizable Lagrange density involving only one chiral
  superfield reads
  \beq
    \cl = \int d^2\q\,d^2\Bar{\q}\;\Bar{\F}\,\F+\left[\int d^2\q\,\left(
    {1\over2}m\F^2+{1\over3}\l\F^3+g\F\right)
    +\mbox{h.c.}\right]\;.
  \eeq
  We will omit the tadpole term $g\F$ in the following, since it can
  always be eliminated by field redefinitions.

\section{Superfield Propagator}
  In close analogy to the usual perturbation theory one can develop a
  perturbation theory in superspace \cite{wb,gates}. Our goal is to
  compute Green functions for superfields,
  \beq
    G^{(N)}\left(z^1,\ldots,z^r;z^{r+1},\dots,z^N\right)=
    \left\langle0\left|\mbox{T}\left\{\F(z^1)\cdots\F(z^r)
    \Bar{\F}(z^{r+1})\cdots\Bar{\F}(z^N)\right\}\right|0\right\rangle\;,
  \eeq
  where $z^i=({x^i}^{\m},{\q^i}^{\a},\Bar{\q^i}_{\dt{\a}})$ denotes a
  point in configuration superspace. Let us start by computing the
  propagator for a chiral superfield, which is constructed from the
  free Lagrangian
  \beq
    L_0 = \int d^8z\,\Bar{\F}\F+\left(\int d^6s\,
    {1\over2}m\F^2+\mbox{h.c.}\right)\;.
  \eeq
  Since the operator $-\Bar{D}^{\,2}D^2/(16\,\bo)$ projects on chiral
  fields,
  \beq
    -{1\over16}{\Bar{D}^{\,2}D^2\over\bo}\F=\F\quad\mbox{if}\quad
    \Bar{D}\F=0\;,\label{chiral_eigen}
  \eeq
  we can use eqs.~(\ref{d2qa})-(\ref{d2qb}), and rewrite the $d^6s$
  integration in the mass term of the free Lagrangian into an
  integration over the whole superspace,
  \beqa
    L_0&=&\int d^8z\,\left\{\Bar{\F}\F+{1\over8}m\left(
      \F{D^2\over\bo}\F+\Bar{\F}{\Bar{D}^{\,2}\over\bo}\Bar{\F}
      \right)\right\}\NO\\[1ex]
    &=&\int d^8z\,{1\over2}\left(\F,\;\Bar{\F}\,\right)\cm
      {\F\choose \Bar{\F}}\;,\label{free1}
  \eeqa
  with the matrix
  \beq
    \cm=\left(
    \begin{array}{cc} 
      \displaystyle{1\over4}{m\over\bo}D^2 & \displaystyle 1\\[2ex]
      \displaystyle 1 & \displaystyle {1\over4}{m\over\bo}\,\Bar{D}^{\,2}
    \end{array}\right)\;,
  \eeq
  where we have assumed a real mass $m$.
  
  To derive equations of motion we have to define a functional
  derivative in superspace, where we have to take into account the
  chirality constraint $\Bar{D}\,\F=0$. This constraint is 
  automatically respected by varying in the $y$ basis,
  \beq
    {\d\over\d\F(y,\q)}\,\F(y',\q')=\d^4(y-y')\d^2(\q-\q')\;.
  \eeq
  Going back to the variable $x$, the variation under superspace
  integration reads
  \beq
    {\d\over\d\F(x,\q,\Bar{\q}\,)}\,\int d^8z'\,
    \F(x',\q',\Bar{\q'}\,)\,F(x',\q',\Bar{\q'}\,)=-{1\over4}\Bar{D}^{\,2}\,
    F(x,\q,\Bar{\q}\,)\;.
  \eeq  
  This leads to the formal definition
  \beq
    {\d\over\d\F(x,\q,\Bar{\q}\,)}\,\F(x',\q',\Bar{\q'}\,)=
    -{1\over4}\Bar{D}^{\,2}\,\d^8(z-z')\;.\label{funct_der}
  \eeq

  Variation of the free Lagrangian (\ref{free1}) then gives equations
  of motion,
  \beq
    {1\over4}\left(\begin{array}{cc} 
          \displaystyle\Bar{D}^{\,2} & \displaystyle 0\\
          \displaystyle 0 & \displaystyle D^2
     \end{array}\right)\cm{\F\choose \Bar{\F}}=0\;.
  \eeq
  With eq.~(\ref{chiral_eigen}) this leads to
  \beqa
    m\F-{1\over4}\Bar{D}^{\,2}\,\Bar{\F}&=&0\;,\\[1ex]
    m\Bar{\F}-{1\over4}D^2\F&=&0\;.
  \eeqa

  The propagator is defined as Green function of the
  operator
  \beq
    {1\over4}\left(\begin{array}{cc} 
          \displaystyle\Bar{D}^{\,2} & \displaystyle 0\\
          \displaystyle 0 & \displaystyle D^2
     \end{array}\right)\cm\;,
  \eeq
  i.e.\ the differential equation defining the two point function $\D$
  reads
  \beq
    {1\over4}\left(\begin{array}{cc} 
          \displaystyle\Bar{D}^{\,2} & \displaystyle 0\\
          \displaystyle 0 & \displaystyle D^2
    \end{array}\right)\cm\D=\left(\begin{array}{cc} 
          \displaystyle{1\over4}{\Bar{D}^{\,2}\over\bo} & 
          \displaystyle 0\\[1ex]
          \displaystyle 0& \displaystyle {1\over4}{D^2\over\bo}
        \end{array}\right)\d(z-z')\;,
    \label{prop_eq}
  \eeq
  where the differential operator on the right-hand side implements
  the chirality constraint (\ref{chirality}). Solving this equation,
  one gets the propagator for a chiral superfield
  \beq
    \D(z,z')={-1\over\bo+m^2}\left(
        \begin{array}{cc} 
          \displaystyle{m\over4}\Bar{D}^{\,2} & 
          \displaystyle {1\over16}\Bar{D}^{\,2}D^2\\[2ex]
          \displaystyle {1\over16}D^2\Bar{D}^{\,2} & 
          \displaystyle {m\over4}D^2
        \end{array}\right)\d(z-z')\;,\label{prop}
  \eeq
  where $-1/(\bo+m^2)$ is a symbolic notation for the Green
  function of the Klein-Gordon operator $\bo+m^2$.

\section{The Generating Functional \label{generating}}
  The generating functional for free Green functions is given by the
  vacuum-to-vacuum amplitude in the presence of an external classical
  chiral source $J$ coupled to a free chiral field $\F$,
  \beqa
    Z_0[J,\Bar{J}\,]&=&{\Big\langle}\,0\,{\Big|}\;
    T\exp\left[i\int d^8z\,\left(J,\;\Bar{J}\,\right)
    \left(\begin{array}{cc} 
        \displaystyle{1\over4}{D^2\over\bo} & \displaystyle 0\\[1ex]
        \displaystyle 0& \displaystyle {1\over4}{\Bar{D}^{\,2}\over\bo}
    \end{array}\right){\F\choose \Bar{\F}}\right]\;
    {\Big|}\,0\,{\Big\rangle}\\[1ex]
   &=&N\,\int\cd\F\,\cd\Bar{\F}\,
      \exp\left\{iL_0+i\int d^8z\,\left(J,\;\Bar{J}\,\right)
    \left(\begin{array}{cc} 
        \displaystyle{1\over4}{D^2\over\bo} & \displaystyle 0\\[1ex]
        \displaystyle 0& \displaystyle {1\over4}{\Bar{D}^{\,2}\over\bo}
    \end{array}\right){\F\choose \Bar{\F}}\right\}\;,\label{path}
  \eeqa
  where $L_0$ is the free Lagrangian from eq.~(\ref{free1}), and $N$ is
  a normalization factor which can be chosen such that $Z_0[0,0]=1$.
  In non-supersymmetric quantum field theories the role of this
  normalization factor is to take out disconnected vacuum bubbles,
  which would otherwise contribute to Green functions. Although vacuum
  diagrams vanish in supersymmetric theories because of the
  non-renormalization theorems (cf.~section \ref{feynman}), the 
  normalization factor does not equal unity when the volume of the
  system tends to infinity \cite{zumino}.
  
  The oscillatory path integral (\ref{path}) is not well defined, and
  has to be Wick rotated to Euclidean space to be evaluated
  unambiguously. The Green functions calculated in Euclidean space
  then yield Green functions in Minkowski space by analytic
  continuation. We will write all quantities in Minkowski space with
  the understanding that they can be justified in Euclidean space.

  Performing the functional integral with standard techniques, one gets
  \beqa
    Z_0[J,\Bar{J}\,]=\exp\left\{-{i\over2}\int d^8z\,d^8z'\,
      \left(J(z),\;\Bar{J}(z)\right)\left(\begin{array}{cc} 
        \displaystyle{1\over4}{D^2\over\bo} & \displaystyle 0\\[1ex]
        \displaystyle 0& \displaystyle {1\over4}{\Bar{D}^{\,2}\over\bo}
      \end{array}\right)\D(z,z')\;\times\right.\NO\\[2ex]
    \left.\times\;\left(\begin{array}{cc} 
        \displaystyle{1\over4}{{D'}^2\over\bo} & \displaystyle 0\\[1ex]
        \displaystyle 0& \displaystyle {1\over4}{\Bar{D'}^{\,2}\over\bo}
    \end{array}\right){J(z')\choose \Bar{J}(z')}\right\}\;,
  \eeqa
  where $\D(z,z')$ is the chiral superfield propagator (\ref{prop}). 
  With eq.~(\ref{chiral_eigen}) this can be brought to a familiar form,
  \beq
   Z_0[J,\Bar{J}\,]=\exp\left\{-{i\over2}\int d^4z\,d^4z'\,
    \left(J(z),\;\Bar{J}(z)\right)\D_{\mbox{\tiny GRS}}(z,z')
    {J(z')\choose \Bar{J}(z')}\right\}\;,\label{Z0}
  \eeq
  where $\D_{\mbox{\tiny GRS}}(z,z')$ is the superfield propagator of
  Grisaru, Ro\v{c}ek and Siegel \cite{grisaru},
  \beq
    \D_{\mbox{\tiny GRS}}(z,z')={-1\over\bo+m^2}\left(
        \begin{array}{cc} 
          \displaystyle-{m\over4}{D^2\over\bo} & 
          \displaystyle 1 \\[2ex]
          \displaystyle 1 & 
          \displaystyle -{m\over4}{\Bar{D}^{\,2}\over\bo}
        \end{array}\right)\d(z-z')\;.\label{GRS}
  \eeq

  \pagebreak
  Since $J$ is a chiral source, its functional derivative is defined
  like in eq.~(\ref{funct_der}). The functional derivative of $Z_0$
  then reads
  \beq
    \left(\begin{array}{l}\displaystyle {1\over i}{\d\over\d J(z)}\\[2ex]
    \displaystyle {1\over i}{\d\over\d \Bar{J}(z)}\end{array}\right)\,Z_0=
    -\int d^4z'\,\D(x,x')\,{1\over4\bo}\left(\begin{array}{l}
    \displaystyle D^2J(z)\\[2ex]\displaystyle\Bar{D}^{\,2}\,\Bar{J}(z)
    \end{array}\right)\,Z_0\;.
  \eeq
  With eqs.~(\ref{prop_eq}) and (\ref{chiral_eigen}) this yields a
  functional equation for $Z_0$,
  \beq
    {1\over4}\left(\begin{array}{cc} 
          \displaystyle\Bar{D}^{\,2} & \displaystyle 0\\
          \displaystyle 0 & \displaystyle D^2
    \end{array}\right)\cm\left(\begin{array}{l}
    \displaystyle {1\over i}{\d\over\d J(z)}\\[2ex]
    \displaystyle {1\over i}{\d\over\d \Bar{J}(z)}\end{array}\right)\,Z_0=
    \left(\begin{array}{l}
    \displaystyle J(z)\\[2ex]\displaystyle \Bar{J}(z)
    \end{array}\right)\,Z_0\;.\label{Z0_eq}
  \eeq
  This equation can easily be generalized to the interacting case. For
  the $\F^3$ theory coupled to an external source the equations of
  motion read
  \beq
    {1\over4}\left(\begin{array}{cc} 
          \displaystyle\Bar{D}^{\,2} & \displaystyle 0\\
          \displaystyle 0 & \displaystyle D^2
     \end{array}\right)\cm{\F\choose \Bar{\F}}
    -\l{\F^2\choose {\Bar{\F}}^2}={J\choose \Bar{J}}\;.
  \eeq 
  By comparison with the functional equation for $Z_0$ (\ref{Z0_eq})
  we can write down the defining equation for the full generating
  functional $Z[J,\Bar{J}]$,
  \beq
    {1\over4}\left(\begin{array}{cc} 
          \displaystyle\Bar{D}^{\,2} & \displaystyle 0\\
          \displaystyle 0 & \displaystyle D^2
    \end{array}\right)\cm\left(\begin{array}{l}
    \displaystyle {1\over i}{\d\over\d J(z)}\\[2ex]
    \displaystyle {1\over i}{\d\over\d \Bar{J}(z)}\end{array}\right)\,Z=
    \left\{\left(\begin{array}{l}
    \displaystyle J(z)\\[2ex]\displaystyle \Bar{J}(z)
    \end{array}\right)+\l\left(\begin{array}{l}\displaystyle
    \left({1\over i}{\d\over\d J(z)}\right)^2\\[2ex]
    \displaystyle\left({1\over i}{\d\over\d \Bar{J}(z)}\right)^2
    \end{array}\right)\right\}\,Z\;.\label{Z_eq}
  \eeq
  Using the interaction Lagrangian $\cl_{\mbox{\tiny INT}}$ 
  \beq
    \cl_{\mbox{\tiny INT}} = \int d^2\q\,{1\over3}\l\F^3 +
    \int d^2\Bar{\q}\,{1\over3}\l\Bar{\F}^{\,3}\;,
  \eeq
  the operator on the right-hand side of eq.~(\ref{Z_eq}) can be
  rewritten as
  \beqa
    \lefteqn{\left(\begin{array}{l}
      \displaystyle J(z)\\[2ex]\displaystyle \Bar{J}(z)
      \end{array}\right)+\l\left(\begin{array}{l}\displaystyle
      \left({1\over i}
      {\d\over\d J(z)}\right)^2\\[2ex]
      \displaystyle\left({1\over i}
      {\d\over\d \Bar{J}(z)}\right)^2\end{array}\right)=}\\[1ex]
    &&=\mbox{e}^{\raisebox{1ex}{$i\int d^4x'\,\cl_{\mbox{\tiny INT}}
      \left({1\over i}{\d\over\d J},\;{1\over i}
      {\d\over\d \Bar{J}}\right)$}}\left(\begin{array}{l}
      \displaystyle J(z)\\[2ex]\displaystyle 
      \Bar{J}(z)\end{array}\right)\mbox{e}^{\raisebox{1ex}{
      $-i\int d^4x'\,\cl_{\mbox{\tiny INT}}\left({1\over i}{\d\over\d J},
      {1\over i}{\d\over\d \Bar{J}}\right)$}}\NO\;.
  \eeqa
  Hence, eq.~(\ref{Z_eq}) yields
  \beqa
    {1\over4}\left(\begin{array}{cc} 
          \displaystyle\Bar{D}^{\,2} & \displaystyle 0\\
          \displaystyle 0 & \displaystyle D^2
    \end{array}\right)\cm\left(\begin{array}{l}
    \displaystyle {1\over i}{\d\over\d J(z)}\\[2ex]
    \displaystyle {1\over i}{\d\over\d \Bar{J}(z)}\end{array}\right)
    \mbox{e}^{\raisebox{1ex}{$-i\int d^4x'\,\cl_{\mbox{\tiny INT}}
    \left({1\over i}{\d\over\d J},\;{1\over i}
    {\d\over\d \Bar{J}}\right)$}}\,Z=\\[2ex]
    =\left(\begin{array}{l}\displaystyle J(z)\\[2ex]\displaystyle 
    \Bar{J}(z)\end{array}\right)\mbox{e}^{\raisebox{1ex}{
    $-i\int d^4x'\,\cl_{\mbox{\tiny INT}}\left({1\over i}{\d\over\d J},
    {1\over i}{\d\over\d \Bar{J}}\right)$}}\,Z\NO\;.
  \eeqa
  By comparison with the functional equation for $Z_0$ (\ref{Z0_eq}),
  the generating functional $Z$ can be related to the free-field
  generating functional
  \beq
    Z[J,\;\Bar{J}\,]=\mbox{e}^{\raisebox{1ex}{$i\int d^4x'\,
    \cl_{\mbox{\tiny INT}}\left({1\over i}{\d\over\d J},{1\over i}
    {\d\over\d \Bar{J}}\right)$}}\,Z_0[J,\;\Bar{J}\,]\;.
    \label{Z_pert}
  \eeq
  This relation, familiar from ordinary quantum field theory, is the
  starting point of perturbation theory in superspace.

\section{Feynman Rules in Superspace \label{feynman}}
  $N$-point Green functions are obtained from the generating
  functional by functional derivation,
  \beqa
    \lefteqn{G^{(N)}\left(z^1,\ldots,z^r;z^{r+1},\dots,z^N\right)\equiv}
      \NO\\[1ex]
    &\equiv&\left.(-i)^N{\d\over\d J(z^1)}\cdots{\d\over\d J(z^r)}
      {\d\over\d \Bar{J}(z^{r+1})}\cdots{\d\over\d \Bar{J}(z^N)}
      Z[J,\Bar{J}\,]\right|_{J=\Bar{J}=0}\\[1ex]
    &=&(-i)^N{\d\over\d J(z^1)}\cdots{\d\over\d J(z^r)}
      {\d\over\d \Bar{J}(z^{r+1})}\cdots{\d\over\d \Bar{J}(z^N)}
      \;\times\NO\\[1ex]
    &&\left.\times\sum\limits_{n=0}^{\infty}{(i)^n\over n!}
      \prod\limits_{j=1}^n\int d^4{x'}^j\,\cl_{\mbox{\tiny INT}}
      \left({1\over i}{\d\over\d J},{1\over i}{\d\over\d \Bar{J}}
      \right)\,Z_0[J,\;\Bar{J}\,]\right|_{J=\Bar{J}=0}\;,
  \eeqa
  where we have used eq.~(\ref{Z_pert}) in the last step. The factors
  \beq
    \int d^4{x'}^j\,\cl_{\mbox{\tiny INT}}\left({1\over i}
    {\d\over\d J},{1\over i}{\d\over\d \Bar{J}}\right)
  \eeq
  generate vertices at the superspace points ${z'}^j$, and the
  functional derivatives ${\d\over\d J}$ in $\cl_{\mbox{\tiny INT}}$,
  when acting on $Z_0$, generate propagators connecting different
  vertices. The operators ${\d\over\d J(z^i)}$ not in
  $\cl_{\mbox{\tiny INT}}$ generate propagators on external lines,
  which have to be amputated and replaced by superfields $\F(z^i)$ in
  order to get an effective contribution to the Lagrangian.
  This leads to the following Feynman rules in configuration
  superspace
  \begin{enumerate}
  \item Provide external lines with factors $\F(z)$ or $\Bar{\F}(z)$.
  \item Due to the chirality of the functional derivative
    (cf.~eq.~(\ref{funct_der})), one has to include one (two) factors 
    $-{1\over4}\Bar{D}^{\,2}$ acting on internal propagators to each
    $\F^3$ vertex with two (three) internal lines. The same applies to
    $\Bar{\F}^{\,3}$ vertices and factors $-{1\over4}D^2$.
  \item Include a coupling constant ${i\over3}\l$ and a superspace
    integration $\int d^8z$ for each vertex.
  \item Use the GRS-propagators $i\D_{\mbox{\tiny GRS}}$ (\ref{GRS})
     for internal lines.
  \item Include the usual symmetry factors.
  \end{enumerate} 
  In the next chapter we will illustrate these Feynman rules by
  performing several sample calculations. Let us first investigate the
  general structure of diagrams calculated with these rules.
  
  The $d^2\q^i\,d^2\Bar{\q}^{\,i}$ integrals at each vertex can be
  done, leaving us with one overall $d^2\q\,d^2\Bar{\q}$ integral and
  the usual Minkowski space integrals over $d^4x^i$ for each diagram.
  To see how this comes about let us follow the $\q$-integrations
  around an arbitrary closed loop\footnote{We will assume that loop
    divergences have been properly regularized.}.  It consists of
  propagators, including factors $\d^8(z_i-z_{i+1})$ and covariant
  derivatives acting on them, external superfield factors, and
  $d^2\q^i\,d^2\Bar{\q}^{\,i}$ integrals.  Higher powers of covariant
  derivatives can be reduced by using the identities
  (\ref{DD1})-(\ref{DD5}).
  
  Consider now the propagator from one vertex $z^i$ to another one
  $z^j$, and integrate by parts using
  eqs.~(\ref{partint1})-(\ref{partint7}) to remove all the covariant
  derivatives from its $\d$-function. The original contribution
  becomes a sum of terms. If there are other propagators connecting
  $z^i$ and $z^j$ we can use the relations
  \beqa
    && \d_{ij}\d_{ij}=0\;,\label{partial3}\\[1ex]
    && \d_{ij}D^{\a}\d_{ij}=0\;,\label{partial4}\\[1ex]
    && \d_{ij}D^2\d_{ij}=0\;,\label{partial5}\\[1ex]
    && \d_{ij}D^{\a}\Bar{D}^{\,\dt{\a}}\d_{ij}=0
       \;,\label{partial6}\\[1ex]
    && \d_{ij}D^{\a}\Bar{D}^{\,2}\d_{ij}=0\;,\label{partial7}\\[1ex]
    && \d_{ij}D^2\Bar{D}^{\,2}\d_{ij}=\d_{ij}\Bar{D}^{\,2}D^2\d_{ij}
       =\d_{ij}D^{\a}\Bar{D}^{\,2}D_{\a}\d_{ij}
       =\d_{ij}\,\Bar{D}_{\dt{\a}}D^2\,\Bar{D}^{\,\dt{\a}}\,\d_{ij}
       =16\d_{ij}\,\d^4(x_i-x_j)\;,\hspace{1cm}\mbox{ }\label{partial8}\\[1ex]
    && \d_{ij}D^{\a}\Bar{D}^{\,2}D^{\b}\d_{ij}=-8\ve^{\a\b}\d_{ij}
       \,\d^4(x_i-x_j)\;,\label{partial8a}\\[1ex]
    && \d_{ij}\Bar{D}^{\,\dt{\a}}D^2\Bar{D}^{\,\dt{\b}}\d_{ij}=
       8\ve^{\dt{\a}\dt{\b}}\d_{ij}\,\d^4(x_i-x_j)\;,\label{partial9}
  \eeqa
  where $\d_{ij}\equiv\d^8(z_i-z_j)$. Hence, the terms generated by
  the partial integration vanish, unless each of the other
  $\d$-functions has exactly two $D$'s and two $\Bar{D}$'s acting on
  it. Now the free $\d$-function can be used to perform the
  $d^2\q^j\,d^2\Bar{\q}^{\,j}$ integral and shrink all the propagators
  between $z^i$ and $z^j$ to a point in $\q$-space. This procedure can
  be repeated, until we have removed all $\d$-functions and performed
  all $\q$-integrals except the original one at $z^i$. We are left
  with a sum of terms, all with a single $d^2\q\,d^2\Bar{\q}$
  integral, various d'Alembert operators from eqs.~(\ref{DD4}) and
  (\ref{DD5}), as well as covariant derivatives acting on the external
  superfields.
  
  Hence, we have ended up with a $d^2\q\,d^2\Bar{\q}$ integral, even
  though in the original Lagrangian we may have had chiral $d^2\q$
  integrals. This is the perturbative no-renormalization theorem for
  chiral superfields \cite{iliopoulos}: radiative corrections do not
  induce renormalizations of F-terms, i.e.\ purely chiral mass or
  interaction terms. Furthermore, all vacuum diagrams vanish, since
  the $d^2\q\,d^2\Bar{\q}$ integral without any external superfield
  vanishes.

    \clearpage\chapter{Supersymmetric Leptogenesis \label{theory}}
  In this chapter, which is based on ref.~\cite{pluemi2}, we present
  the supersymmetric generalization of the leptogenesis scenario
  suggested by Fukugita and Yanagida \cite{fy}. After having
  introduced the superpotential, we compute all the relevant decay
  widths, $CP$ asymmetries and scattering cross sections. In order to
  check the results we have performed two independent calculations.
  First by using the component field Feynman rules from appendix 
  \ref{AppFeynm}, and then by using the superfield techniques
  introduced in the last chapter.
\section{The Superpotential \label{superpot}}
  In supersymmetric unification scenarios based on SO(10), the
  effective theory below the $(B-L)$ breaking scale is the MSSM
  supplemented by right-handed Majorana neutrinos.  Neglecting soft
  breaking terms, the masses and Yukawa couplings relevant for
  leptogenesis are given by the superpotential
  \beq
    \cw = {1\over2}N^cMN^c + \m H_1\e H_2 
    + H_1 \e Q \l_d D^c + H_1 \e L \l_l E^c 
    + H_2 \e Q \l_u U^c + H_2 \e L \l_{\n} N^c
    \;,\label{spotential}
  \eeq
  where we have chosen a basis in which the Majorana mass matrix
  $M$ and the Yukawa coupling matrices $\l_d$ and $\l_l$ for the
  down-type quarks and the charged leptons are diagonal with real
  and positive eigenvalues. The corresponding Lagrange density reads
  \beq
    \cl = \int d^2\q\,\cw+\int d^2\Bar{\q}\,\Bar{\cw}\;.
  \eeq

  The chiral superfields in the superpotential (\ref{spotential})
  are most conveniently parametrized in the $y$-basis
  (cf.~section~\ref{superfields}).
  \beqa
    H_i(y,\q)&=&H_i(y)+\sqrt{2}\,\q\wt{H}_i(y)
      +\q^2\,F_{\scr H_i}(y)\;,\label{comp1}\\[1ex]
    Q(y,\q)&=&\wt{q}(y)+\sqrt{2}\,\q q_{\mbox{\tiny L}}(y)
      +\q^2\,F_{\scr Q}(y)\;,\\[1ex]
    L(y,\q)&=&\wt{l}(y)+\sqrt{2}\,\q l_{\mbox{\tiny L}}(y)
      +\q^2\,F_{l}(y)\;,\\[1ex]
    U^c(y,\q)&=&\wt{U^c}(y)+\sqrt{2}\,\q{u_{\mbox{\tiny R}}}^c(y)
      +\q^2\,F_{\scr U^c}(y)\;,\\[1ex]
    D^c(y,\q)&=&\wt{D^c}(y)+\sqrt{2}\,\q{d_{\mbox{\tiny R}}}^c(y)
      +\q^2\,F_{\scr D^c}(y)\;,\\[1ex]
    E^c(y,\q)&=&\wt{E^c}(y)+\sqrt{2}\,\q{e_{\mbox{\tiny R}}}^c(y)
      +\q^2\,F_{\scr E^c}(y)\;,\\[1ex]
    N^c(y,\q)&=&\wt{N^c}(y)+\sqrt{2}\,\q{\n_{\mbox{\tiny R}}}^c(y)
      +\q^2\,F_{\scr N^c}(y)\;.\label{comp2}
  \eeqa
  $Q$ and $L$ stand for the left-handed quark and lepton doublets,
  $U^c$, $D^c$, $E^c$ and $N^c$ are the right-handed singlet fields,
  and $H_i$ denotes the two Higgs-doublets,
  \beq
    H_1=\left(\begin{array}{c}H_1^0\\-H_1^-\end{array}\right)
    \qquad\mbox{and}\qquad
    H_2=\left(\begin{array}{c}H_2^+\\H_2^0\end{array}\right)\;.
  \eeq
  
  Besides the usual bispinors for quarks and charged leptons we can
  introduce Majorana-spinors for the right- and left-handed neutrinos,
  \beq
    N=\left(\begin{array}{c}{{\n_{\mbox{\tiny R}}}^c}_{\a}\\[1ex]
    {\Bar{{\n_{\mbox{\tiny R}}}^c}}^{\;\dt{\a}}\end{array}\right)
    \qquad\mbox{and}\qquad
    \n=\left(\begin{array}{c}{\n_{\mbox{\tiny L}}}_{\a}\\[1ex]
    {\Bar{\n_{\mbox{\tiny L}}}}^{\;\dt{\a}}\end{array}\right)\;.
  \eeq
  In the symmetric phase of the MSSM no mixing occurs between the
  fermionic partners of gauge and Higgs bosons. Therefore,
  we have two Dirac higgsinos 
  \beq
    \wt{h^0}=\left(\begin{array}{c}{\wt{H_1^0}}_{\a}\\[1ex]
    {\Bar{\wt{H_2^0}}}^{\;\dt{\a}}\end{array}\right)
    \qquad\mbox{and}\qquad
    \wt{h^-}=\left(\begin{array}{c}{\wt{H_1^-}}_{\a}\\[1ex]
    {\Bar{\wt{H_2^+}}}^{\;\dt{\a}}\end{array}\right)\;,
  \eeq
  which again form an isospin doublet,
  \beq
    \wt{h}=\left(\begin{array}{c}\wt{h^0}\\[1ex]
     -\wt{h^-}\end{array}\right)\;.
  \eeq

  The auxiliary component fields of the chiral superfields 
  (\ref{comp1})--(\ref{comp2}) are obtained from the Lagrange density 
  \cite{witten},
  \beq
    F_i=-\left({\partial \cl\over\partial F_i}\right)^{\dg}\;.
  \eeq
  The superpotential (\ref{spotential}) then yields the following
  contributions to the auxiliary fields
  \beqa
    F_{\scr H_1}&=&\left[\m H_2^{\dg}
      +\left(\wt{q}\,\l_d\,\wt{D^c}\right)^{\dg}
      +\left(\wt{l}\,\l_l\,\wt{E^c}\right)^{\dg}\right]\e\;,
      \label{aux1}\\[1ex]
    F_{\scr H_2}&=&\left[-\m H_1^{\dg}
      +\left(\wt{q}\,\l_u\,\wt{U^c}\right)^{\dg}
      +\left(\wt{l}\,\l_{\n}\,\wt{N^c}\right)^{\dg}\right]\e\;,
      \label{aux2}\\[1ex]
    F_{\scr Q_i}&=&\e\left[
      H_2^{\dg}\,\wt{U^c_j}^{\dg}\,\big(\l_u^{\dg}\big)_{ji}+
      H_1^{\dg}\,\wt{D^c_j}^{\dg}\,(\l_d^{\dg})_{ji}
      \right]\;,\label{auxQ}\\[1ex]
    F_{\scr L_i}&=&\e\left[
      H_2^{\dg}\,\wt{N^c_j}^{\dg}\,(\l_{\n}^{\dg})_{ji}+
      H_1^{\dg}\,\wt{E^c_j}^{\dg}\,(\l_l^{\dg})_{ji}
      \right]\;,\label{auxL}\\[1ex]
    F_{\scr U^c_i}&=&(\l_u^{\dg})_{ij}\,
      \wt{q_j}^{\dg}\,\e\,H_2^{\dg}\;,\label{auxU}\\[1ex]
    F_{\scr D^c_i}&=&(\l_d^{\dg})_{ij}\,
      \wt{q_j}^{\dg}\,\e\,H_1^{\dg}\;,\label{auxD}\\[1ex]
    F_{\scr E^c_i}&=&(\l_l^{\dg})_{ij}\,
      \wt{l_j}^{\dg}\,\e\,H_1^{\dg}\;,\label{auxE}\\[1ex]
    F_{\scr N^c_i}&=&-M_i\,\wt{N^c_i}^{\dg}+
      (\l_{\n}^{\dg})_{ij}\,\wt{l_j}^{\dg}\,\e\,H_2^{\dg}\;.
      \label{auxN}
  \eeqa

  The vacuum expectation values of the neutral Higgs fields generate
  Dirac masses for the down-type quarks and the charged leptons
  \beq
    v_1=\left\langle H_1^0\right\rangle\ne0\qquad\Rightarrow\qquad
    m_d=\l_d\;v_1\quad\mbox{and}\quad m_l=\l_l\;v_1\;,
  \eeq
  and for the up-type quarks and the neutrinos
  \beq
    v_2=\left\langle H_2^0\right\rangle\ne0\qquad\Rightarrow\qquad
    m_u=\l_u\;v_2\quad\mbox{and}\quad m_{\scr D}=\l_{\n}\;v_2\;.
  \eeq
  The Majorana masses $M$ for the right-handed neutrinos, which have
  to be much larger than the Dirac masses $m_{\scr D}$, offer a natural 
  explanation for the smallness of the light neutrino masses via the
  see-saw mechanism \cite{seesaw}. 
  
  To generate a non-vanishing baryon asymmetry, one needs a hierarchy
  in the Majorana mass matrix $M$. Then the scale at which the
  asymmetry is generated is given by the mass $M_1$ of the lightest
  right-handed neutrino. Hence, it is convenient to write all the
  masses and energies in units of $M_1$,
  \beq
    a_j = \left({M_j\over M_1}\right)^2\;,\qquad
    x={s\over M_1^2}\quad\mbox{and}\quad
    z={M_1\over T}\;,
  \eeq
  where $M_j$ are the masses of the heavier right-handed neutrinos,
  $s$ is the squared centre of mass energy of a scattering process
  and $T$ is the temperature.

\section{The Decay Channels of Heavy Neutrinos \label{DecaySection}}
  \begin{figure}[t]
    \centerline{\input{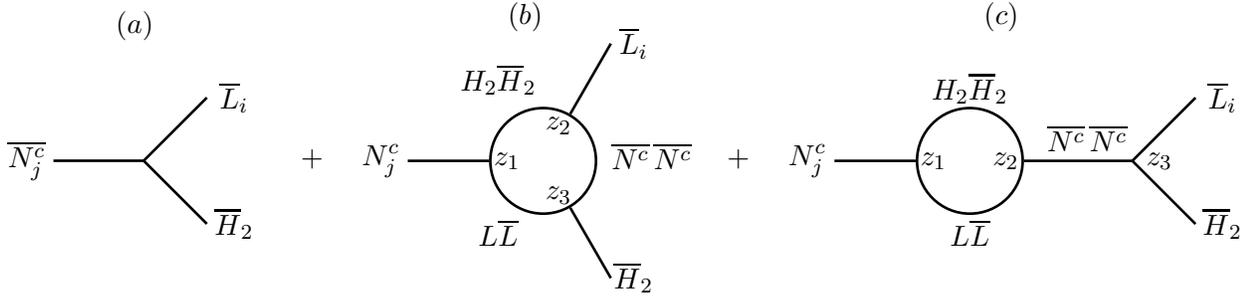}}
    \caption{\it Decay modes of the right-handed neutrino
      superfield. \label{chap3_fig01a}}
  \end{figure}
\begin{figure}[ht]
    \centerline{\input{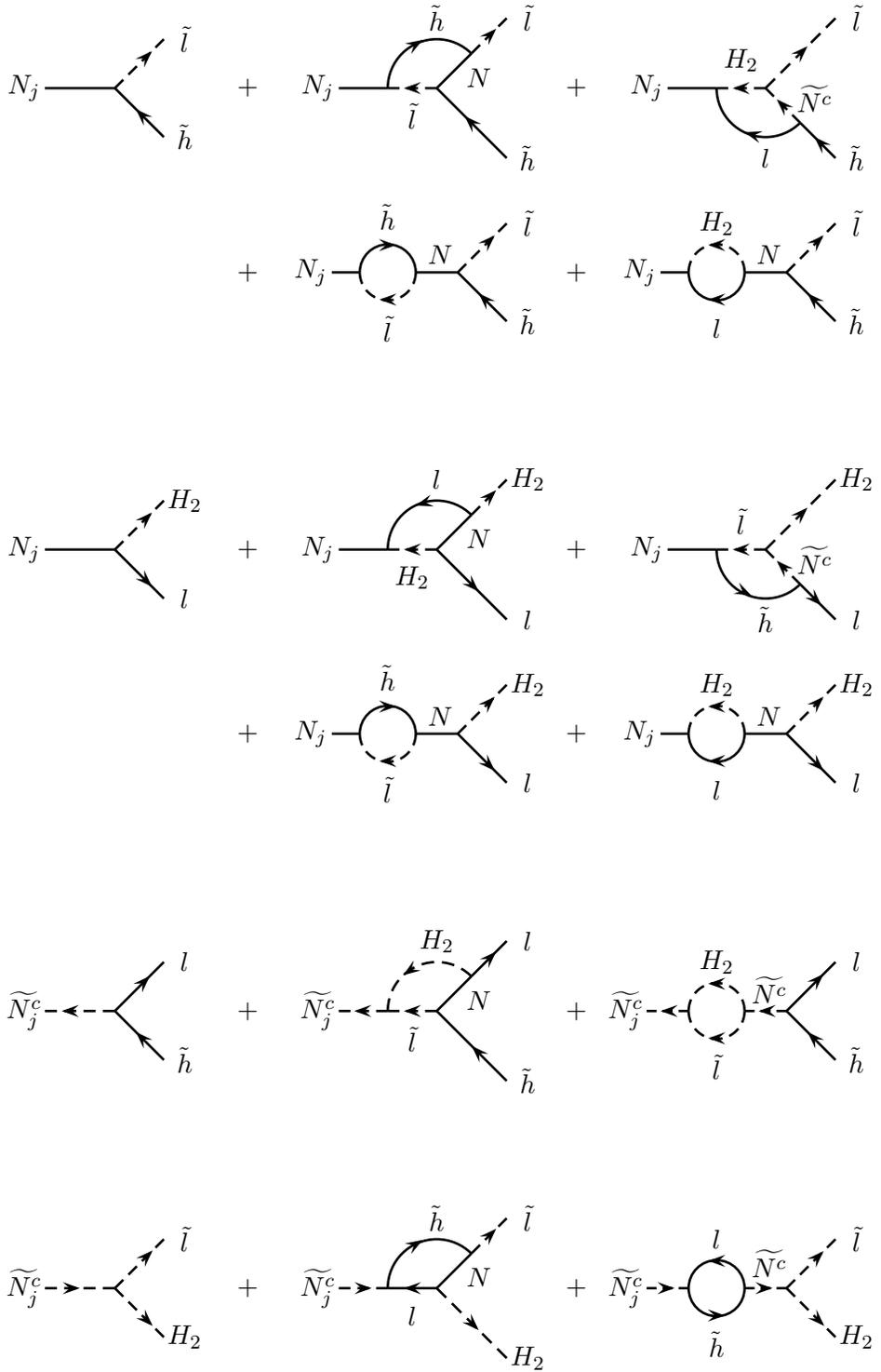}}
    \caption{\it Decay modes of the right-handed Majorana neutrinos
      and their scalar partners. \label{chap3_fig01b}}
  \end{figure}

  Since at these energy scales supersymmetry breaking terms can be
  safely neglected, we are working in a theory with exact
  supersymmetry, i.e.\ we can use the superfield techniques exposed in
  the last chapter to compute decay and scattering amplitudes. To
  leading order, the decay modes of the right-handed neutrino and its
  scalar partner with positive lepton number in the final state are
  all contained in the superfield diagram fig.~\ref{chap3_fig01a}a,
  corresponding to the superpotential term
  $-\Bar{N^c}\l_{\n}^{\dg}\,\Bar{L}\e\Bar{H}_2$.  Choosing the usual
  decomposition of the $S$-matrix
  \beq
    S=1+iT\;,
  \eeq
  we get the following tree level contribution to the $T$-matrix
  \beq
    T^{(a)}_{\scr N}=
    -\int d^4x\,\int d^2\Bar{\q}\,\Bar{N^c}
    \l_{\n}^{\dg}\,\Bar{L}\e\Bar{H}_2\;.
  \eeq  
  The decay amplitudes we are interested in are then given by the
  matrix elements of $iT^{(a)}_{\scr N}$. Let us first
  concentrate on two-particle final states. According to the component
  field decompositions (\ref{comp1})--(\ref{comp2}) of the chiral
  superfields, the right-handed Majorana neutrinos $N_j$ can decay
  into a lepton and a Higgs-boson or into a slepton and a higgsino,
  while their scalar partners $\snj$ can decay into a lepton and a
  higgsino or into a slepton and a Higgs boson
  (cf.~fig.~\ref{chap3_fig01b}). The decay widths at tree level read
  \cite{covi}
  \beqa
    {1\over4}\Gnj&:=&\G\Big(N_j\to\wt{l}+\Bar{\wt{h}}\;\Big)
      =\G\Big(N_j\to{\wt{l}{ }}^{\,\dg}+\wt{h}\;\Big)\NO\\
    &=&\G\Big(N_j\to l+H_2\Big)=\G\Big(N_j\to 
      \Bar{l}+H_2^{\dg}\Big)
      ={M_j\over16\p}\;{\mmjj\over v_2^2}\;,
      \label{decay1}\\[1ex]
    {1\over2}\Gsnj&:=&\G\Big(\snj\to\wt{l}+H_2\Big)
      =\G\Big(\snj\to\Bar{l}+\wt{h}\;\Big)\NO\\
    &=&\G\Big(\snj^{\,\dg}\to{\wt{l}{ }}^{\,\dg}+H_2^{\dg}\Big)
      =\G\Big(\snj^{\,\dg}\to l+\Bar{\wt{h}}\;\Big)=
      {M_j\over8\p}\;{\mmjj\over v_2^2}\;.
      \label{decay2}
  \eeqa
  According to eq.~(\ref{decay}), the reaction densities for 
  these decays are then given by
  \beq
    \gnj=2\,\gsnj = {M_1^4\over4\p^3}\;{\mmjj\over v_2^2}\;
      {a_j\sqrt{a_j}\over z}\mbox{K}_1(z\sqrt{a_j})\;.
  \eeq
  \pagebreak
  \clearpage

  All these decay modes are $CP$ violating, the dominant
  contribution to $CP$ violation coming about through interference
  between the tree level and the one-loop diagrams shown in
  fig.~\ref{chap3_fig01b}. These one-loop diagrams can be summarized
  by the one-loop superfield diagrams in figs.~\ref{chap3_fig01a}b and
  \ref{chap3_fig01a}c. Note that the non-renormalization theorems do
  not apply to these diagrams, since they are contributions to the
  D-term $N^c_j\Bar{L}_i\e \Bar{H}_2$. Using the superspace Feynman
  rules discussed in section \ref{feynman} the diagram in
  fig.~\ref{chap3_fig01a}b yields the following contribution to the
  $T$-matrix
  \beqa
    \lefteqn{iT^{(b)}_{\scr N}=\sum\limits_n\llnj
      \left(\l_{\n}^{\dg}\right)_{ni}\int d^8z_1\,d^8z_2\,d^8z_3\,
      N^c_j(z_1)\Bar{L}_i(z_2)\e \Bar{H}_2(z_3)\times}\\[1ex]
    &&\times\left[{\Bar{D}^{\,2}(z_1)\over4}{\d^8(z_1-z_2)\over\bo_1-i\ve}
      \right]\left[{D^2(z_3)\over4}{\d^8(z_1-z_3)\over\bo_1-i\ve}
      \right]\left[-{M_n\over16}{D^2(z_2)\Bar{D}^{\,2}(z_2)\over\bo_2}
      {\d^8(z_2-z_3)\over\bo_2+M_n^2-i\ve}\right]\;.\NO
  \eeqa
  After partial integration of the $D_{\a}(z_3)$ derivatives, the only
  non-vanishing contribution reads
  \beqa
    iT^{(b)}_{\scr N}&=&\sum\limits_n\llnj
      \left(\l_{\n}^{\dg}\right)_{ni}\int d^8z_1\,d^8z_2\,d^8z_3\,
      N^c_j(z_1)\Bar{L}_i(z_2)\e \Bar{H}_2(z_3)
      {1\over4}{\d(z_1-z_3)\over\bo_1-i\ve}\times\NO\\[1ex]
    &&\times\left[{\Bar{D}^{\,2}(z_1)\over4}{\d(z_1-z_2)\over\bo_1-i\ve}
      \right]\left[-{M_n\over16}{D^2(z_2)\Bar{D}^{\,2}(z_2)\over\bo_2}
      D^2(z_3){\d(z_2-z_3)\over\bo_2+M_n^2-i\ve}\right]\\[1ex]
    &=&\sum\limits_n\llnj\left(\l_{\n}^{\dg}\right)_{ni}\int 
      d^8z_1\,d^8z_2\,d^8z_3\,N^c_j(z_1)\Bar{L}_i(z_2)\e \Bar{H}_2(z_3)
      {1\over4}{\d(z_1-z_3)\over\bo_1-i\ve}\times\NO\\[1ex]
    &&\times\left[{\Bar{D}^{\,2}(z_2)\over4}{\d(z_1-z_2)\over\bo_1-i\ve}
      \right]\left[M_n D^2(z_2){\d(z_2-z_3)\over\bo_2+M_n^2-i\ve}\right]\;,
  \eeqa  
  where we have used eqs.~(\ref{Ddelta3}), (\ref{Ddelta4}) and
  (\ref{DD5}) in the last step. Now we can perform the 
  $d^2\q_3\,d^2\Bar{\q_3}$ integration,
  \beqa
    iT^{(b)}_{\scr N}&=&\sum\limits_n\llnj
      \left(\l_{\n}^{\dg}\right)_{ni}\int d^8z_1\,d^8z_2\,d^4x_3\,
      N^c_j(z_1)\Bar{L}_i(z_2)\e \Bar{H}_2(z_1)
      {1\over4}{\d^4(x_1-x_3)\over\bo_1-i\ve}\times\NO\\[1ex]
    &&\times\left[{\Bar{D}^{\,2}(z_2)\over4}
      {\d^8(z_1-z_2)\over\bo_1-i\ve}\right]\left[
      M_n D^2(z_2){\d^8(z_2-z_1)\over\bo_2+M_n^2-i\ve}\right]\;.
  \eeqa  
  Similarly, we can partially integrate the $\Bar{D}^{\,2}(z_2)$
  derivatives, and perform the $d^2\q_2\,d^2\Bar{\q_2}$ integrations
  after having used eq.~(\ref{partial8}) to remove all covariant
  derivatives. Then $iT^{(b)}$ reads
  \beqa
    iT^{(b)}_{\scr N}&=&\sum\limits_nM_n\llnj
      \left(\l_{\n}^{\dg}\right)_{ni}\int d^4x_1\,d^4x_2\,d^4x_3\,
      {\d^4(x_1-x_3)\over\bo_1-i\ve}{\d^4(x_1-x_2)\over\bo_1-i\ve}
      {\d^4(x_2-x_3)\over \bo_2+M_n^2-i\ve}\times\NO\\[1ex]
    &&\times\int d^2\q\,d^2\Bar{\q}\,N^c_j(x_1,\q,\Bar{\q}\,)
      \Bar{L}_i(x_2,\q,\Bar{\q}\,)\e \Bar{H}_2(x_3,\q,\Bar{\q}\,)\;.
  \eeqa
  By Fourier transforming the loop propagators, we get
  \beqa
    iT^{(b)}_{\scr N}&=&-\sum\limits_nM_n\llnj
      \left(\l_{\n}^{\dg}\right)_{ni}\times\NO\\[1ex]
   &&\times\int d^4x_1\,d^4x_2\,d^4x_3\,\int{d^4k_1\over(2\p)^4}\,
      {d^4k_2\over(2\p)^4}\,{d^4k_3\over(2\p)^4}\,
      {\mbox{e}^{ix_1(k_1+k_2)}\mbox{e}^{ix_2(k_3-k_2)}
      \mbox{e}^{-ix_3(k_1+k_3)}\over(k_1^2+i\ve)(k_2^2+i\ve)
      (k_3^2-M_n^2+i\ve)}\NO\\[1ex]
    &&\times\int d^2\q\,d^2\Bar{\q}\,N^c_j(x_1,\q,\Bar{\q}\,)
      \Bar{L}_i(x_2,\q,\Bar{\q}\,)\e \Bar{H}_2(x_3,\q,\Bar{\q}\,)\;.
  \eeqa

  Analogously, we can compute the contribution of the diagram in
  fig.~\ref{chap3_fig01a}c to the $T$-matrix,
  \beqa
    iT^{(c)}_{\scr N}&=&2\sum\limits_nM_n\llnj
      \left(\l_{\n}^{\dg}\right)_{ni}\int d^4x_1\,d^4x_2\,d^4x_3\,
      \left({\d^4(x_1-x_2)\over\bo_1-i\ve}\right)^2
      {\d^4(x_2-x_3)\over \bo_3+M_n^2-i\ve}\times\NO\\[1ex]
    &&\times\int d^2\q\,d^2\Bar{\q}\,N^c_j(x_1,\q,\Bar{\q}\,)
      \Bar{L}_i(x_3,\q,\Bar{\q}\,)\e \Bar{H}_2(x_3,\q,\Bar{\q}\,)\;.
  \eeqa
  Again using the Fourier representation for the propagators, we find
  \beqa
    iT^{(c)}_{\scr N}&=&{-i\over8\p^2}
      \sum\limits_nM_n\llnj\left(\l_{\n}^{\dg}\right)_{ni}
      \int{d^4k\over(2\p)^4}\,{B_0(-k,0,0)\over k^2-M_n^2+i\ve}
      \int d^4x_1\,d^4x_3\,\mbox{e}^{ik(x_1-x_3)}\times\NO\\[1ex]
    &&\times\int d^2\q\,d^2\Bar{\q}\,N^c_j(x_1,\q,\Bar{\q}\,)
      \Bar{L}_i(x_3,\q,\Bar{\q}\,)\e \Bar{H}_2(x_3,\q,\Bar{\q}\,)\;,
  \eeqa
  where $B_0(-k,0,0)$ is the massless two-point scalar integral
  defined in eq.~(\ref{B0def}). One-loop corrections to the decay
  amplitudes are given by the matrix elements of 
  $iT^{(b)}_{\scr N}+iT^{(c)}_{\scr N}$.

  Interference between these one-loop diagrams and the tree-level
  amplitudes gives rise to $CP$ asymmetries in the different decay
  channels of $N_j$ and $\snj$, which can all be expressed by the same
  $CP$ violation parameter $\ve_j$,
  \beqa
    \lefteqn{\ve_j:={\G\Big(N_j\to\wt{l}+\Bar{\wt{h}}\,\Big)
       -\G\Big(N_j\to{\wt{l}{ }}^{\;\dg}+\wt{h}\,\Big)\over
       \G\Big(N_j\to\wt{l}+\Bar{\wt{h}}\,\Big)
       +\G\Big(N_j\to{\wt{l}{ }}^{\;\dg}+\wt{h}\,\Big)}
    ={\G\Big(N_j\to l+H_2\Big)-\G\Big(N_j\to\Bar{l}
       +H_2^{\dg}\Big)\over\G\Big(N_j\to l+H_2\Big)
       +\G\Big(N_j\to\Bar{l}+H_2^{\dg}\Big)}}\NO\\[1ex]
    &=&{\G\Big(\snj^{\dg}\to l+\Bar{\wt{h}}\,\Big)-
       \G\Big(\snj\to \Bar{l}+\wt{h}\,\Big)\over
       \G\Big(\snj^{\dg}\to l+\Bar{\wt{h}}\,\Big)+
       \G\Big(\snj\to \Bar{l}+\wt{h}\,\Big)}
    ={\G\Big(\snj\to\wt{l}+H_2\Big)-\G\Big(\snj^{\dg}
       \to{\wt{l}{ }}^{\;\dg}+H_2^{\dg}\Big)\over\G\Big(\snj
       \to\wt{l}+H_2\Big)+\G\Big(\snj^{\dg}\to
       {\wt{l}{ }}^{\;\dg}+H_2^{\dg}\Big)}\NO\\[1ex]
    &=& -{1\over8\p v_2^2}\;{1\over\mmjj}\sum\limits_{n\ne j}
       \mbox{Im}\left[(m_{\scr D}^{\dg}m_{\scr D})^2_{nj}\right]
       \;g\Big({a_n\over a_j}\Big)\;,\label{SUSY_CP}\\[1ex]
    &&\mbox{with}\quad g(x)=\sqrt{x}\left[\mbox{ln}\left({1+x\over x}
       \right)+{2\over x-1}\right]\;\approx {3\over\sqrt{x}}\quad
       \mbox{for}\quad x\gg1\;.\NO
  \eeqa
  Here $n$ is the flavour index of the heavy (s)neutrino in the
  loop. This result agrees with the one in ref.~\cite{covi} and is
  of the same order as the $CP$ asymmetry in ref.~\cite{camp}. 

  With $\ve_j$ we can parametrize the reaction densities for the
  decays and inverse decays in the following way
  \beqa
    {1\over4}(1+\ve_j)\gnj&=&
      \g\Big(N_j\to\wt{l}+\Bar{\wt{h}}\;\Big)
      =\g\Big(N_j\to l+H_2\Big)\\[1ex]
      &=&\g\Big(\,{\wt{l}{ }}^{\;\dg}+\wt{h}\to N_j\Big)
      =\g\Big(\Bar{l}+H_2^{\dg}\to N_j\Big)\;,\NO\\[1ex]
    {1\over4}(1-\ve_j)\gnj&=&
      \g\Big(N_j\to{\wt{l}{ }}^{\;\dg}+\wt{h}\;\Big)
      =\g\Big(N_j\to\Bar{l}+H_2^{\dg}\Big)\\[1ex]
    &=&\g\Big(\wt{l}+\Bar{\wt{h}}\to N_j\;\Big)
      =\g\Big(l+H_2\to N_j\Big)\;,\NO\\[1ex]
    {1\over2}(1+\ve_j)\gsnj&=&
      \g\Big(\snj\to\wt{l}+H_2\Big)
      =\g\Big(\snj^{\dg}\to l+\Bar{\wt{h}}\;\Big)\\[1ex]
    &=&\g\Big(\,{\wt{l}{ }}^{\;\dg}+H_2^{\dg}\to\snj^{\dg}\Big)
      =\g\Big(\Bar{l}+\wt{h}\to\snj\Big)\;,\NO\\[1ex]
    {1\over2}(1-\ve_j)\gsnj&=&
      \g\Big(\snj^{\dg}\to{\wt{l}{ }}^{\;\dg}+H_2^{\dg}\Big)
      =\g\Big(\snj\to \Bar{l}+\wt{h}\;\Big)\\[1ex]
    &=&\g\Big(\wt{l}+H_2\to\snj\Big)
      =\g\Big(l+\Bar{\wt{h}}\to\snj^{\dg}\Big)\;.\NO
  \eeqa

  \begin{figure}[t]
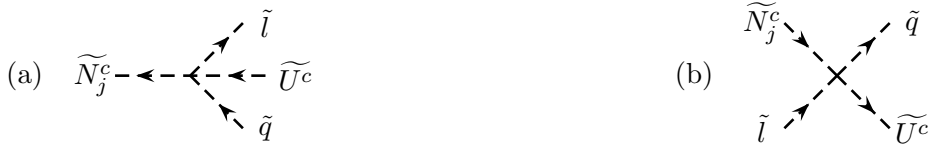

    \begin{center}
\parbox[c]{12cm}{
\pspicture(-0.6,0)(3,2)
\psline[linewidth=1pt,linestyle=dashed](0.6,1)(2.6,1)
\psline[linewidth=1pt,linestyle=dashed](1.6,1)(2.3,1.7)
\psline[linewidth=1pt,linestyle=dashed](1.6,1)(2.3,0.3)
\psline[linewidth=1pt]{<-}(0.88,1)(0.98,1)
\psline[linewidth=1pt]{<-}(2.1,1)(2.2,1)
\psline[linewidth=1pt]{->}(2.02,1.42)(2.12,1.52)
\psline[linewidth=1pt]{<-}(1.98,0.62)(2.08,0.52)
\rput[cc]{0}(-0.6,1){(a)}
\rput[cc]{0}(0.3,1){$\sn$}
\rput[cc]{0}(2.6,1.7){$\tilde{l}$}
\rput[cc]{0}(3.0,1){$\sur$}
\rput[cc]{0}(2.6,0.3){$\tilde{q}$}
\endpspicture\hspace{\fill}
\pspicture(-0.6,0)(2.5,2)
\psline[linewidth=1pt,linestyle=dashed](0.6,0.3)(1.3,1)
\psline[linewidth=1pt,linestyle=dashed](0.6,1.7)(1.3,1)
\psline[linewidth=1pt,linestyle=dashed](1.3,1)(2,1.7)
\psline[linewidth=1pt,linestyle=dashed](1.3,1)(2,0.3)
\psline[linewidth=1pt]{->}(0.83,0.53)(0.93,0.63)
\psline[linewidth=1pt]{->}(0.83,1.47)(0.93,1.37)
\psline[linewidth=1pt]{->}(1.7,1.4)(1.8,1.5)
\psline[linewidth=1pt]{->}(1.7,0.6)(1.8,0.5)
\rput[cc]{0}(-0.6,1){(b)}
\rput[cc]{0}(0.3,1.7){$\sn$}
\rput[cc]{0}(0.3,0.3){$\tilde{l}$}
\rput[cc]{0}(2.3,0.3){$\sur$}
\rput[cc]{0}(2.3,1.7){$\tilde{q}$}
\endpspicture
}
\end{center}
    \caption{\it Contributions of the scalar potential to the decay
      width and the  interactions of a scalar neutrino.\label{chap2_fig02}}
  \end{figure}
  Additionally, the scalar potential contains quartic scalar
  couplings, which enable the decay of $\snj$ into three particles via
  the diagram shown in fig.~\ref{chap2_fig02}a. This is just the
  contribution of the auxiliary field $F_{\scr H_2}$
  given in eq.~(\ref{aux2}) to the superfield decay amplitude in
  fig.~\ref{chap3_fig01a}a. The partial width for this decay is given
  by 
  \beq
    \Gtr:=\G\Big(\snj^{\;\dg}\to \wt{l}+\sur^{\dg}
      +\wt{q}^{\dg}\;\Big)={3\,\a_uM_j\over64\p^2}\;
      {\mmjj\over v_2^2}\quad\mbox{with}\quad
      \a_u={\mbox{Tr}\Big(\l_u^{\dg}\l_u\Big)\over4\p}\;,
  \eeq
  and the corresponding reaction density reads
  \beq
    \gtr={3\,\a_uM_1^4\over128\p^4}\;{\mmjj\over v_2^2}\;
      {a_j\sqrt{a_j}\over z}\mbox{K}_1(z\sqrt{a_j})
      ={3\,\a_u\over16\p}\,\gsnj\;.
  \eeq
  Since the Yukawa coupling of the top quark and its scalar partner is
  large, $\a_u$ can be of order one. But even then $\gtr$ is much
  smaller than $\gsnj$. Hence, the three particle decays give only a
  small correction, which we have taken into account for completeness.
  However, we have neglected the $CP$ asymmetry in this decay which
  comes about through the one-loop diagrams in fig.~\ref{chap3_fig01a}.

  Furthermore, we have neglected the leptonic auxiliary field 
  $F_{\scr L_i}$ given in eq.~(\ref{auxL}), since its 
  contribution to the sneutrino decay width will be of order 
  $(\l_{\n}^{\dg}\l_l^2\l_{\n})_{jj}$, i.e. much smaller
  than the other partial decay widths, at least for the lightest
  sneutrino ($j=1$).

  \begin{figure}[t]
    \centerline{\input{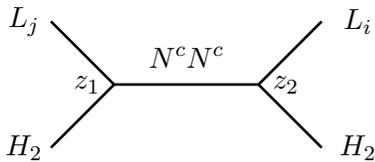}}
    \caption{\it Lepton number violating scatterings mediated by a
        neutrino superfield.\label{chap2_fig03super}}
  \end{figure}
  The dimensionless squared total decay widths of $N_j$ and $\snj$
  are then finally given by
  \beqa
    c_j&:=&\left({\Gnj\over M_1}\right)^2
      ={a_j\over16\p^2}\;{\mmjj^2\over v_2^4}\;,\\[1ex]
    \wt{c_j}&:=&\left({\Gsnj+\Gtr\over M_1}\right)^2
      ={a_j\over16\p^2}\;{\mmjj^2\over v_2^4}\;
      \left[1+{3\,\a_u\over16\p}\right]^2\;.
  \eeqa

  The vertex in fig.~\ref{chap2_fig02}a also gives $2\to2$ scattering
  processes involving one scalar neutrino, like
  $\snj+\wt{l}\to\wt{q}+\sur$ (cf.~fig.~\ref{chap2_fig02}b). The
  reduced cross section for this process reads
  \beq
    \hat{\s}_{22_j}(x)=3\a_u\;{\mmjj\over v_2^2}\;{x-a_j\over x}\;.
  \eeq
  For the processes $\snj+\wt{q}^{\dg}\to{\wt{l}{ }}^{\,\dg}+\sur$
  and $\snj+\sur^{\dg}\to{\wt{l}{ }}^{\,\dg}+\wt{q}$, the
  corresponding back reactions and the $CP$ conjugated processes we
  find the same result. The corresponding reaction density can then
  be calculated according to eq.~(\ref{22scatt}). One finds
  \beq
    \g_{22_j}(z)={3\,\a_uM_1^4\over16\p^4}\;{\mmjj\over v_2^2}\;
    {\sqrt{a_j}\over z^3}
    \,\mbox{K}_1(z\sqrt{a_j})={3\,\a_u\over4\p\,a_jz^2}\;\gnj(z)\;.
  \eeq
  Hence, $\g_{22_j}$ will be much larger than $\gnj$ and $\gsnj$ for
  small $a_jz^2$, i.e.\ for high temperatures $T\gg M_j$. Together
  with similar scatterings, which we are going to discuss in section
  \ref{Ntopsection}, these processes will therefore be very effective
  in bringing the heavy (s)neutrinos into thermal equilibrium at high
  temperatures where decays and inverse decays are suppressed by a
  time dilatation factor.

\section[Lepton Number Violating Scatterings]{Lepton Number Violating
         Scatterings Mediated by Right-Handed Neutrinos}
  \begin{figure}[t]
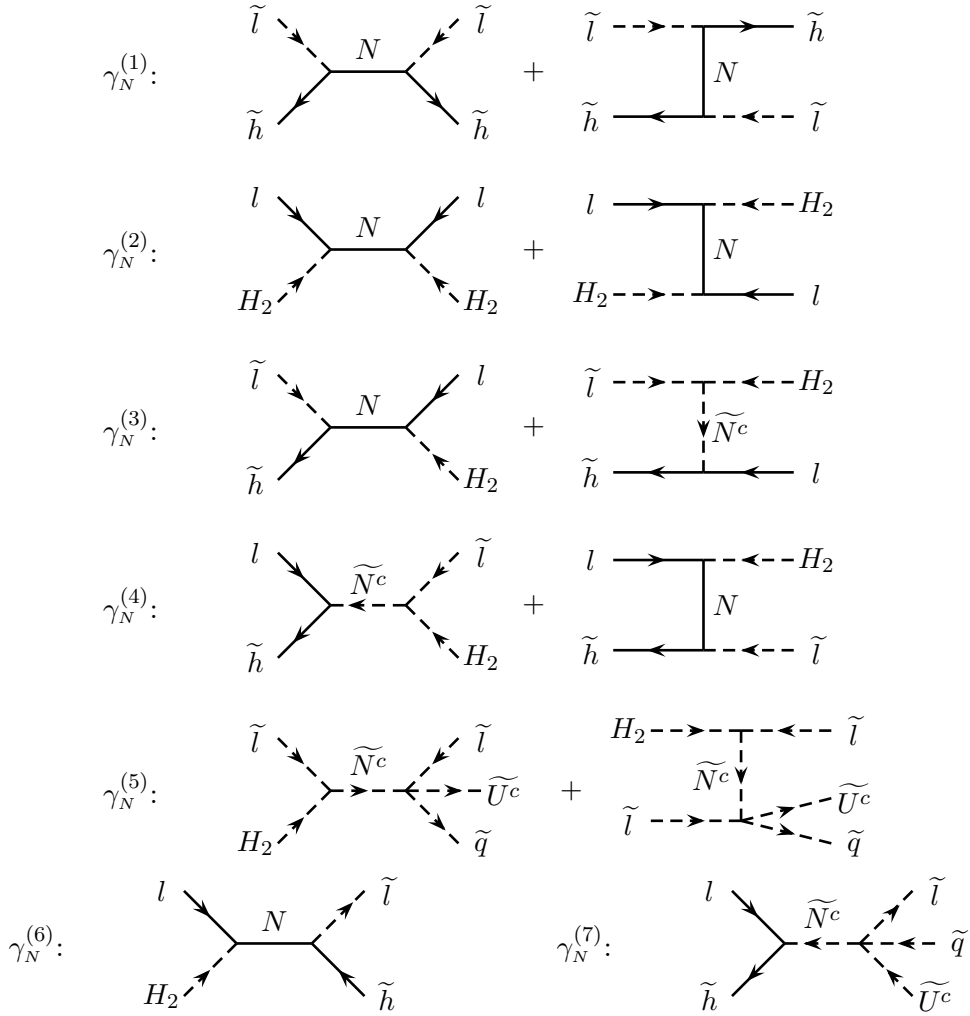

    \begin{center}
\parbox[c]{11.5cm}{
\pspicture(-2,0)(3.5,2)
\rput[rc]{0}(-1,1){$\g_{\scr N}^{(1)}$:}
\psline[linewidth=1pt](0.6,0.3)(1.3,1)
\psline[linewidth=1pt,linestyle=dashed](0.6,1.7)(1.3,1)
\psline[linewidth=1pt](1.3,1)(2.3,1)
\psline[linewidth=1pt,linestyle=dashed](2.3,1)(3,1.7)
\psline[linewidth=1pt](2.3,1)(3,0.3)
\psline[linewidth=1pt]{<-}(0.8,0.5)(0.9,0.6)
\psline[linewidth=1pt]{->}(0.85,1.45)(0.95,1.35)
\psline[linewidth=1pt]{<-}(2.65,1.35)(2.75,1.45)
\psline[linewidth=1pt]{->}(2.7,0.6)(2.8,0.5)
\rput[cc]{0}(0.3,1.7){$\widetilde{l}$}
\rput[cc]{0}(0.3,0.3){$\widetilde{h}$}
\rput[cc]{0}(1.8,1.3){$N$}
\rput[cc]{0}(3.3,0.3){$\widetilde{h}$}
\rput[cc]{0}(3.3,1.7){$\widetilde{l}$}
\rput[cc]{0}(4.0,1.0){$+$}
\endpspicture\hspace{5ex}
\pspicture(0,-0.1)(3.5,1.8)
\psline[linewidth=1pt,linestyle=dashed](0.6,1.5)(1.8,1.5)
\psline[linewidth=1pt](1.8,1.5)(3,1.5)
\psline[linewidth=1pt](1.8,1.5)(1.8,0.3)
\psline[linewidth=1pt](0.6,0.3)(1.8,0.3)
\psline[linewidth=1pt,linestyle=dashed](1.8,0.3)(3,0.3)
\psline[linewidth=1pt]{->}(1.2,1.5)(1.3,1.5)
\psline[linewidth=1pt]{->}(2.42,1.5)(2.52,1.5)
\psline[linewidth=1pt]{<-}(1.05,0.3)(1.15,0.3)
\psline[linewidth=1pt]{<-}(2.3,0.3)(2.4,0.3)
\rput[cc]{0}(0.3,1.5){$\widetilde{l}$}
\rput[cc]{0}(0.3,0.3){$\widetilde{h}$}
\rput[cc]{0}(2.1,0.9){$N$}
\rput[cc]{0}(3.3,1.5){$\widetilde{h}$}
\rput[cc]{0}(3.3,0.3){$\widetilde{l}$}
\endpspicture\\[2ex]
\pspicture(-2,0)(3.5,2)
\rput[rc]{0}(-1,1){$\g_{\scr N}^{(2)}$:}
\psline[linewidth=1pt,linestyle=dashed](0.6,0.3)(1.3,1)
\psline[linewidth=1pt](0.6,1.7)(1.3,1)
\psline[linewidth=1pt](1.3,1)(2.3,1)
\psline[linewidth=1pt](2.3,1)(3,1.7)
\psline[linewidth=1pt,linestyle=dashed](2.3,1)(3,0.3)
\psline[linewidth=1pt]{->}(0.85,0.55)(0.95,0.65)
\psline[linewidth=1pt]{->}(0.85,1.45)(0.95,1.35)
\psline[linewidth=1pt]{<-}(2.65,1.35)(2.75,1.45)
\psline[linewidth=1pt]{<-}(2.65,0.65)(2.75,0.55)
\rput[cc]{0}(0.3,1.7){$l$}
\rput[cc]{0}(0.3,0.3){$H_2$}
\rput[cc]{0}(1.8,1.3){$N$}
\rput[cc]{0}(3.3,0.3){$H_2$}
\rput[cc]{0}(3.3,1.7){$l$}
\rput[cc]{0}(4.0,1.0){$+$}
\endpspicture
\hspace{5ex}
\pspicture(0,-0.1)(3.5,1.8)
\psline[linewidth=1pt](0.6,1.5)(1.8,1.5)
\psline[linewidth=1pt,linestyle=dashed](1.8,1.5)(3,1.5)
\psline[linewidth=1pt](1.8,1.5)(1.8,0.3)
\psline[linewidth=1pt,linestyle=dashed](0.6,0.3)(1.8,0.3)
\psline[linewidth=1pt](1.8,0.3)(3,0.3)
\psline[linewidth=1pt]{->}(1.2,1.5)(1.3,1.5)
\psline[linewidth=1pt]{<-}(2.3,1.5)(2.4,1.5)
\psline[linewidth=1pt]{->}(1.2,0.3)(1.3,0.3)
\psline[linewidth=1pt]{<-}(2.3,0.3)(2.4,0.3)
\rput[cc]{0}(0.3,1.5){$l$}
\rput[cc]{0}(0.3,0.3){$H_2$}
\rput[cc]{0}(2.1,0.9){$N$}
\rput[cc]{0}(3.3,1.5){$H_2$}
\rput[cc]{0}(3.3,0.3){$l$}
\endpspicture\\[2ex]
\pspicture(-2,0)(3.5,2)
\rput[rc]{0}(-1,1){$\g_{\scr N}^{(3)}$:}
\psline[linewidth=1pt](0.6,0.3)(1.3,1)
\psline[linewidth=1pt,linestyle=dashed](0.6,1.7)(1.3,1)
\psline[linewidth=1pt](1.3,1)(2.3,1)
\psline[linewidth=1pt](2.3,1)(3,1.7)
\psline[linewidth=1pt,linestyle=dashed](2.3,1)(3,0.3)
\psline[linewidth=1pt]{<-}(0.8,0.5)(0.9,0.6)
\psline[linewidth=1pt]{->}(0.85,1.45)(0.95,1.35)
\psline[linewidth=1pt]{<-}(2.65,1.35)(2.75,1.45)
\psline[linewidth=1pt]{<-}(2.65,0.65)(2.75,0.55)
\rput[cc]{0}(0.3,1.7){$\widetilde{l}$}
\rput[cc]{0}(0.3,0.3){$\widetilde{h}$}
\rput[cc]{0}(1.8,1.3){$N$}
\rput[cc]{0}(3.3,0.3){$H_2$}
\rput[cc]{0}(3.3,1.7){$l$}
\rput[cc]{0}(4.0,1.0){$+$}
\endpspicture\hspace{5ex}
\pspicture(0,-0.1)(3.5,1.8)
\psline[linewidth=1pt,linestyle=dashed](0.6,1.5)(1.8,1.5)
\psline[linewidth=1pt,linestyle=dashed](1.8,1.5)(3,1.5)
\psline[linewidth=1pt,linestyle=dashed](1.8,1.5)(1.8,0.3)
\psline[linewidth=1pt](0.6,0.3)(1.8,0.3)
\psline[linewidth=1pt](1.8,0.3)(3,0.3)
\psline[linewidth=1pt]{->}(1.2,1.5)(1.3,1.5)
\psline[linewidth=1pt]{<-}(2.3,1.5)(2.4,1.5)
\psline[linewidth=1pt]{->}(1.8,0.86)(1.8,0.76)
\psline[linewidth=1pt]{<-}(1.05,0.3)(1.15,0.3)
\psline[linewidth=1pt]{<-}(2.3,0.3)(2.4,0.3)
\rput[cc]{0}(0.3,1.5){$\widetilde{l}$}
\rput[cc]{0}(0.3,0.3){$\widetilde{h}$}
\rput[cc]{0}(2.15,0.9){$\widetilde{N^c}$}
\rput[cc]{0}(3.3,1.5){$H_2$}
\rput[cc]{0}(3.3,0.3){$l$}
\endpspicture\\[2ex]
\pspicture(-2,0)(3.5,2)
\rput[rc]{0}(-1,1){$\g_{\scr N}^{(4)}$:}
\psline[linewidth=1pt](0.6,0.3)(1.3,1)
\psline[linewidth=1pt](0.6,1.7)(1.3,1)
\psline[linewidth=1pt,linestyle=dashed](1.3,1)(2.3,1)
\psline[linewidth=1pt,linestyle=dashed](2.3,1)(3,1.7)
\psline[linewidth=1pt,linestyle=dashed](2.3,1)(3,0.3)
\psline[linewidth=1pt]{<-}(0.8,0.5)(0.9,0.6)
\psline[linewidth=1pt]{->}(0.85,1.45)(0.95,1.35)
\psline[linewidth=1pt]{<-}(1.53,1)(1.63,1)
\psline[linewidth=1pt]{<-}(2.65,1.35)(2.75,1.45)
\psline[linewidth=1pt]{<-}(2.65,0.65)(2.75,0.55)
\rput[cc]{0}(0.3,1.7){$l$}
\rput[cc]{0}(0.3,0.3){$\widetilde{h}$}
\rput[cc]{0}(1.8,1.35){$\widetilde{N^c}$}
\rput[cc]{0}(3.3,0.3){$H_2$}
\rput[cc]{0}(3.3,1.7){$\widetilde{l}$}
\rput[cc]{0}(4.0,1.0){$+$}
\endpspicture\hspace{5ex}
\pspicture(0,-0.1)(3.5,1.8)
\psline[linewidth=1pt](0.6,1.5)(1.8,1.5)
\psline[linewidth=1pt,linestyle=dashed](1.8,1.5)(3,1.5)
\psline[linewidth=1pt](1.8,1.5)(1.8,0.3)
\psline[linewidth=1pt](0.6,0.3)(1.8,0.3)
\psline[linewidth=1pt,linestyle=dashed](1.8,0.3)(3,0.3)
\psline[linewidth=1pt]{->}(1.2,1.5)(1.3,1.5)
\psline[linewidth=1pt]{<-}(2.3,1.5)(2.4,1.5)
\psline[linewidth=1pt]{<-}(1.05,0.3)(1.15,0.3)
\psline[linewidth=1pt]{<-}(2.3,0.3)(2.4,0.3)
\rput[cc]{0}(0.3,1.5){$l$}
\rput[cc]{0}(0.3,0.3){$\widetilde{h}$}
\rput[cc]{0}(2.1,0.9){$N$}
\rput[cc]{0}(3.3,1.5){$H_2$}
\rput[cc]{0}(3.3,0.3){$\widetilde{l}$}
\endpspicture\\[2ex]
\pspicture(-2,0)(4.0,2)
\rput[rc]{0}(-1,1){$\g_{\scr N}^{(5)}$:}
\psline[linewidth=1pt,linestyle=dashed](0.6,0.3)(1.3,1)
\psline[linewidth=1pt,linestyle=dashed](0.6,1.7)(1.3,1)
\psline[linewidth=1pt,linestyle=dashed](1.3,1)(2.3,1)
\psline[linewidth=1pt,linestyle=dashed](2.3,1)(3.3,1)
\psline[linewidth=1pt,linestyle=dashed](2.3,1)(3,1.7)
\psline[linewidth=1pt,linestyle=dashed](2.3,1)(3,0.3)
\psline[linewidth=1pt]{->}(0.85,0.55)(0.95,0.65)
\psline[linewidth=1pt]{->}(0.85,1.45)(0.95,1.35)
\psline[linewidth=1pt]{->}(1.7,1)(1.8,1)
\psline[linewidth=1pt]{<-}(2.65,1.35)(2.75,1.45)
\psline[linewidth=1pt]{->}(2.95,1)(3.05,1)
\psline[linewidth=1pt]{->}(2.73,0.57)(2.83,0.47)
\rput[cc]{0}(0.3,1.7){$\widetilde{l}$}
\rput[cc]{0}(0.3,0.3){$H_2$}
\rput[cc]{0}(1.8,1.4){$\widetilde{N^c}$}
\rput[cc]{0}(3.3,0.3){$\widetilde{q}$}
\rput[cc]{0}(3.3,1.7){$\widetilde{l}$}
\rput[cc]{0}(3.6,1){$\widetilde{U^c}$}
\rput[cc]{0}(4.5,1.0){$+$}
\endpspicture\hspace{5ex}
\pspicture(0,0)(3.5,2.1)
\psline[linewidth=1pt,linestyle=dashed](0.6,1.8)(1.8,1.8)
\psline[linewidth=1pt,linestyle=dashed](1.8,1.8)(3,1.8)
\psline[linewidth=1pt,linestyle=dashed](1.8,1.8)(1.8,0.6)
\psline[linewidth=1pt,linestyle=dashed](0.6,0.6)(1.8,0.6)
\psline[linewidth=1pt,linestyle=dashed](1.8,0.6)(3,0.9)
\psline[linewidth=1pt,linestyle=dashed](1.8,0.6)(3,0.3)
\psline[linewidth=1pt]{->}(1.22,1.8)(1.32,1.8)
\psline[linewidth=1pt]{<-}(2.28,1.8)(2.38,1.8)
\psline[linewidth=1pt]{<-}(1.8,1.08)(1.8,1.18)
\psline[linewidth=1pt]{->}(1.22,0.6)(1.32,0.6)
\psline[linewidth=1pt]{->}(2.42,0.45)(2.52,0.42)
\psline[linewidth=1pt]{->}(2.42,0.75)(2.52,0.78)
\rput[cc]{0}(0.3,1.8){$H_2$}
\rput[cc]{0}(0.3,0.6){$\widetilde{l}$}
\rput[cc]{0}(1.4,1.2){$\widetilde{N^c}$}
\rput[cc]{0}(3.3,1.8){$\widetilde{l}$}
\rput[cc]{0}(3.3,0.9){$\widetilde{U^c}$}
\rput[cc]{0}(3.3,0.3){$\widetilde{q}$}
\endpspicture
}
\parbox[c]{14cm}{
\pspicture(-2,0)(3.5,2)
\rput[rc]{0}(-1,1){$\g_{\scr N}^{(6)}$:}
\psline[linewidth=1pt,linestyle=dashed](0.6,0.3)(1.3,1)
\psline[linewidth=1pt](0.6,1.7)(1.3,1)
\psline[linewidth=1pt](1.3,1)(2.3,1)
\psline[linewidth=1pt,linestyle=dashed](2.3,1)(3,1.7)
\psline[linewidth=1pt](2.3,1)(3,0.3)
\psline[linewidth=1pt]{->}(0.85,0.55)(0.95,0.65)
\psline[linewidth=1pt]{->}(0.85,1.45)(0.95,1.35)
\psline[linewidth=1pt]{->}(2.75,1.45)(2.85,1.55)
\psline[linewidth=1pt]{<-}(2.65,0.65)(2.75,0.55)
\rput[cc]{0}(0.3,1.7){$l$}
\rput[cc]{0}(0.3,0.3){$H_2$}
\rput[cc]{0}(1.8,1.3){$N$}
\rput[cc]{0}(3.3,0.3){$\widetilde{h}$}
\rput[cc]{0}(3.3,1.7){$\widetilde{l}$}
\endpspicture\hspace{10ex}
\pspicture(-2,0)(4.0,2)
\rput[rc]{0}(-1,1){$\g_{\scr N}^{(7)}$:}
\psline[linewidth=1pt](0.6,0.3)(1.3,1)
\psline[linewidth=1pt](0.6,1.7)(1.3,1)
\psline[linewidth=1pt,linestyle=dashed](1.3,1)(2.3,1)
\psline[linewidth=1pt,linestyle=dashed](2.3,1)(3.3,1)
\psline[linewidth=1pt,linestyle=dashed](2.3,1)(3,1.7)
\psline[linewidth=1pt,linestyle=dashed](2.3,1)(3,0.3)
\psline[linewidth=1pt]{<-}(0.8,0.5)(0.9,0.6)
\psline[linewidth=1pt]{->}(0.85,1.45)(0.95,1.35)
\psline[linewidth=1pt]{<-}(1.55,1)(1.65,1)
\psline[linewidth=1pt]{<-}(2.8,1)(2.9,1)
\psline[linewidth=1pt]{->}(2.73,1.41)(2.83,1.51)
\psline[linewidth=1pt]{<-}(2.65,0.66)(2.75,0.56)
\rput[cc]{0}(0.3,1.7){$l$}
\rput[cc]{0}(0.3,0.3){$\widetilde{h}$}
\rput[cc]{0}(1.8,1.4){$\widetilde{N^c}$}
\rput[cc]{0}(3.3,0.3){$\widetilde{U^c}$}
\rput[cc]{0}(3.3,1.7){$\widetilde{l}$}
\rput[cc]{0}(3.6,1){$\widetilde{q}$}
\endpspicture
}
\end{center}
    \caption{\it $L$ violating processes mediated by a virtual
      Majorana neutrino or its scalar partner. \label{chap2_fig03}}
  \end{figure}    
  \begin{figure}[t]
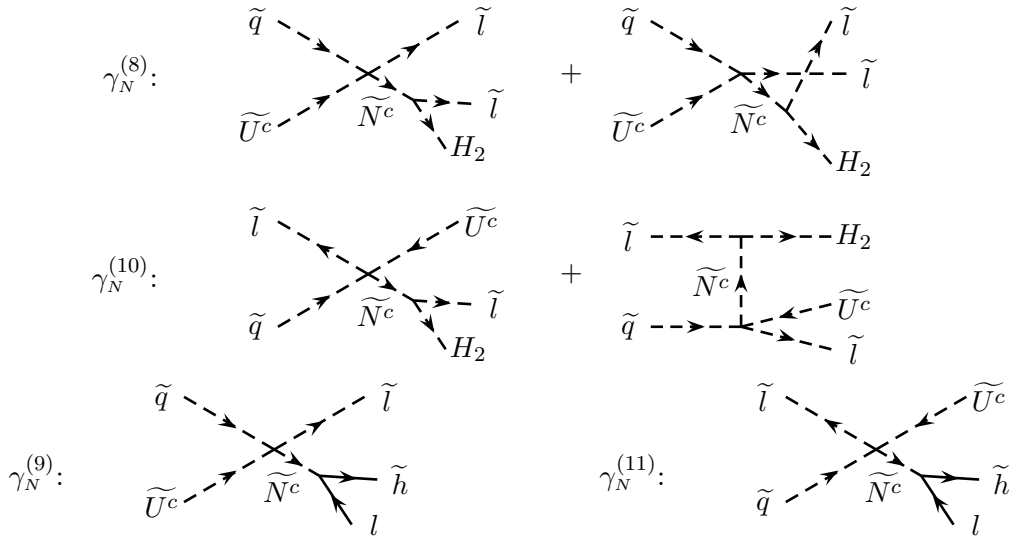

    \begin{center}
\parbox[c]{11.5cm}{
\pspicture(-2,0)(4.0,2.3)
\rput[rc]{0}(-1,1.3){$\g_{\scr N}^{(8)}$:}
\psline[linewidth=1pt,linestyle=dashed](0.6,0.6)(1.8,1.3)
\psline[linewidth=1pt,linestyle=dashed](0.6,2)(1.8,1.3)
\psline[linewidth=1pt,linestyle=dashed](1.8,1.3)(3,2)
\psline[linewidth=1pt,linestyle=dashed](1.8,1.3)(2.4,0.95)
\psline[linewidth=1pt,linestyle=dashed](2.4,0.95)(3.17,0.9)
\psline[linewidth=1pt,linestyle=dashed](2.4,0.95)(2.83,0.3)
\psline[linewidth=1pt]{->}(1.2,0.96)(1.3,1.02)
\psline[linewidth=1pt]{->}(1.2,1.64)(1.3,1.58)
\psline[linewidth=1pt]{->}(2.38,1.63)(2.48,1.69)
\psline[linewidth=1pt]{->}(2.08,1.14)(2.18,1.08)
\psline[linewidth=1pt]{->}(2.8,0.92)(2.9,0.91)
\psline[linewidth=1pt]{->}(2.65,0.57)(2.68,0.52)
\rput[cc]{0}(0.3,2){$\widetilde{q}$}
\rput[cc]{0}(0.3,0.6){$\widetilde{U^c}$}
\rput[cc]{0}(3.3,2){$\widetilde{l}$}
\rput[cc]{0}(1.9,0.8){$\widetilde{N^c}$}
\rput[cc]{0}(3.47,0.9){$\widetilde{l}$}
\rput[cc]{0}(3.13,0.3){$H_2$}
\rput[cc]{0}(4.5,1.3){$+$}
\endpspicture\hspace{5ex}
\pspicture(0,0)(4.0,2.3)
\psline[linewidth=1pt,linestyle=dashed](0.6,0.6)(1.8,1.3)
\psline[linewidth=1pt,linestyle=dashed](0.6,2)(1.8,1.3)
\psline[linewidth=1pt,linestyle=dashed](1.8,1.3)(2.4,0.8)
\psline[linewidth=1pt,linestyle=dashed](2.4,0.8)(3.0,0.1)
\psline[linewidth=1pt,linestyle=dashed](2.4,0.8)(3.0,2.0)
\psline[linewidth=1pt,linestyle=dashed](1.8,1.3)(3.2,1.3)
\psline[linewidth=1pt]{->}(1.2,0.96)(1.3,1.02)
\psline[linewidth=1pt]{->}(1.2,1.64)(1.3,1.58)
\psline[linewidth=1pt]{->}(2.22,1.3)(2.32,1.3)
\psline[linewidth=1pt]{->}(2.82,1.66)(2.87,1.76)
\psline[linewidth=1pt]{->}(2.09,1.06)(2.19,0.98)
\psline[linewidth=1pt]{->}(2.75,0.38)(2.85,0.28)
\rput[cc]{0}(0.3,2){$\widetilde{q}$}
\rput[cc]{0}(0.3,0.6){$\widetilde{U^c}$}
\rput[cc]{0}(3.2,2){$\widetilde{l}$}
\rput[cc]{0}(1.9,0.7){$\widetilde{N^c}$}
\rput[cc]{0}(3.47,1.3){$\widetilde{l}$}
\rput[cc]{0}(3.3,0.1){$H_2$}
\endpspicture\\[2ex]
\pspicture(-2,0)(4.0,2.3)
\rput[rc]{0}(-1,1.3){$\g_{\scr N}^{(10)}$:}
\psline[linewidth=1pt,linestyle=dashed](0.6,0.6)(1.8,1.3)
\psline[linewidth=1pt,linestyle=dashed](0.6,2)(1.8,1.3)
\psline[linewidth=1pt,linestyle=dashed](1.8,1.3)(3,2)
\psline[linewidth=1pt,linestyle=dashed](1.8,1.3)(2.4,0.95)
\psline[linewidth=1pt,linestyle=dashed](2.4,0.95)(3.17,0.9)
\psline[linewidth=1pt,linestyle=dashed](2.4,0.95)(2.83,0.3)
\psline[linewidth=1pt]{->}(1.2,0.96)(1.3,1.02)
\psline[linewidth=1pt]{<-}(1.11,1.7)(1.21,1.64)
\psline[linewidth=1pt]{<-}(2.33,1.6)(2.43,1.66)
\psline[linewidth=1pt]{->}(2.08,1.14)(2.18,1.08)
\psline[linewidth=1pt]{->}(2.8,0.92)(2.9,0.91)
\psline[linewidth=1pt]{->}(2.65,0.57)(2.68,0.52)
\rput[cc]{0}(0.3,2){$\widetilde{l}$}
\rput[cc]{0}(0.3,0.6){$\widetilde{q}$}
\rput[cc]{0}(3.3,2){$\widetilde{U^c}$}
\rput[cc]{0}(1.9,0.8){$\widetilde{N^c}$}
\rput[cc]{0}(3.47,0.9){$\widetilde{l}$}
\rput[cc]{0}(3.13,0.3){$H_2$}
\rput[cc]{0}(4.5,1.3){$+$}
\endpspicture\hspace{5ex}
\pspicture(0,0)(3.5,2.1)
\psline[linewidth=1pt,linestyle=dashed](0.6,1.8)(1.8,1.8)
\psline[linewidth=1pt,linestyle=dashed](1.8,1.8)(3,1.8)
\psline[linewidth=1pt,linestyle=dashed](1.8,1.8)(1.8,0.6)
\psline[linewidth=1pt,linestyle=dashed](0.6,0.6)(1.8,0.6)
\psline[linewidth=1pt,linestyle=dashed](1.8,0.6)(3,0.9)
\psline[linewidth=1pt,linestyle=dashed](1.8,0.6)(3,0.3)
\psline[linewidth=1pt]{<-}(1.06,1.8)(1.16,1.8)
\psline[linewidth=1pt]{->}(2.44,1.8)(2.54,1.8)
\psline[linewidth=1pt]{->}(1.8,1.2)(1.8,1.3)
\psline[linewidth=1pt]{->}(1.22,0.6)(1.32,0.6)
\psline[linewidth=1pt]{->}(2.42,0.45)(2.52,0.42)
\psline[linewidth=1pt]{<-}(2.3,0.72)(2.4,0.75)
\rput[cc]{0}(0.3,1.8){$\widetilde{l}$}
\rput[cc]{0}(0.3,0.6){$\widetilde{q}$}
\rput[cc]{0}(1.4,1.2){$\widetilde{N^c}$}
\rput[cc]{0}(3.3,1.8){$H_2$}
\rput[cc]{0}(3.3,0.9){$\widetilde{U^c}$}
\rput[cc]{0}(3.3,0.3){$\widetilde{l}$}
\endpspicture
}
\parbox[c]{14cm}{
\pspicture(-2,0)(3.5,2.3)
\rput[rc]{0}(-1,1){$\g_{\scr N}^{(9)}$:}
\psline[linewidth=1pt,linestyle=dashed](0.6,0.6)(1.8,1.3)
\psline[linewidth=1pt,linestyle=dashed](0.6,2)(1.8,1.3)
\psline[linewidth=1pt,linestyle=dashed](1.8,1.3)(3,2)
\psline[linewidth=1pt,linestyle=dashed](1.8,1.3)(2.4,0.95)
\psline[linewidth=1pt](2.4,0.95)(3.17,0.9)
\psline[linewidth=1pt](2.4,0.95)(2.83,0.3)
\psline[linewidth=1pt]{->}(1.2,0.96)(1.3,1.02)
\psline[linewidth=1pt]{->}(1.2,1.64)(1.3,1.58)
\psline[linewidth=1pt]{->}(2.36,1.62)(2.46,1.68)
\psline[linewidth=1pt]{->}(2.08,1.14)(2.18,1.08)
\psline[linewidth=1pt]{->}(2.8,0.92)(2.9,0.91)
\psline[linewidth=1pt]{<-}(2.55,0.7)(2.58,0.66)
\rput[cc]{0}(0.3,2){$\widetilde{q}$}
\rput[cc]{0}(0.3,0.6){$\widetilde{U^c}$}
\rput[cc]{0}(3.3,2){$\widetilde{l}$}
\rput[cc]{0}(1.9,0.8){$\widetilde{N^c}$}
\rput[cc]{0}(3.47,0.9){$\widetilde{h}$}
\rput[cc]{0}(3.13,0.3){$l$}
\endpspicture\hspace{\fill}
\pspicture(-2,0)(4.0,2.3)
\rput[rc]{0}(-1,1){$\g_{\scr N}^{(11)}$:}
\psline[linewidth=1pt,linestyle=dashed](0.6,0.6)(1.8,1.3)
\psline[linewidth=1pt,linestyle=dashed](0.6,2)(1.8,1.3)
\psline[linewidth=1pt,linestyle=dashed](1.8,1.3)(3,2)
\psline[linewidth=1pt,linestyle=dashed](1.8,1.3)(2.4,0.95)
\psline[linewidth=1pt](2.4,0.95)(3.17,0.9)
\psline[linewidth=1pt](2.4,0.95)(2.83,0.3)
\psline[linewidth=1pt]{->}(1.2,0.96)(1.3,1.02)
\psline[linewidth=1pt]{<-}(1.11,1.7)(1.21,1.64)
\psline[linewidth=1pt]{<-}(2.33,1.6)(2.43,1.66)
\psline[linewidth=1pt]{->}(2.08,1.14)(2.18,1.08)
\psline[linewidth=1pt]{->}(2.8,0.92)(2.9,0.91)
\psline[linewidth=1pt]{<-}(2.55,0.7)(2.58,0.66)
\rput[cc]{0}(0.3,2){$\widetilde{l}$}
\rput[cc]{0}(0.3,0.6){$\widetilde{q}$}
\rput[cc]{0}(3.3,2){$\widetilde{U^c}$}
\rput[cc]{0}(1.9,0.8){$\widetilde{N^c}$}
\rput[cc]{0}(3.47,0.9){$\widetilde{h}$}
\rput[cc]{0}(3.13,0.3){$l$}
\endpspicture
}
\end{center}
    \caption{\it Diagrams contributing to the lepton number
      violating scatterings via heavy sneutrino exchange.
      \label{chap2_fig04}}
  \end{figure}
  \begin{figure}[ht]
    \input{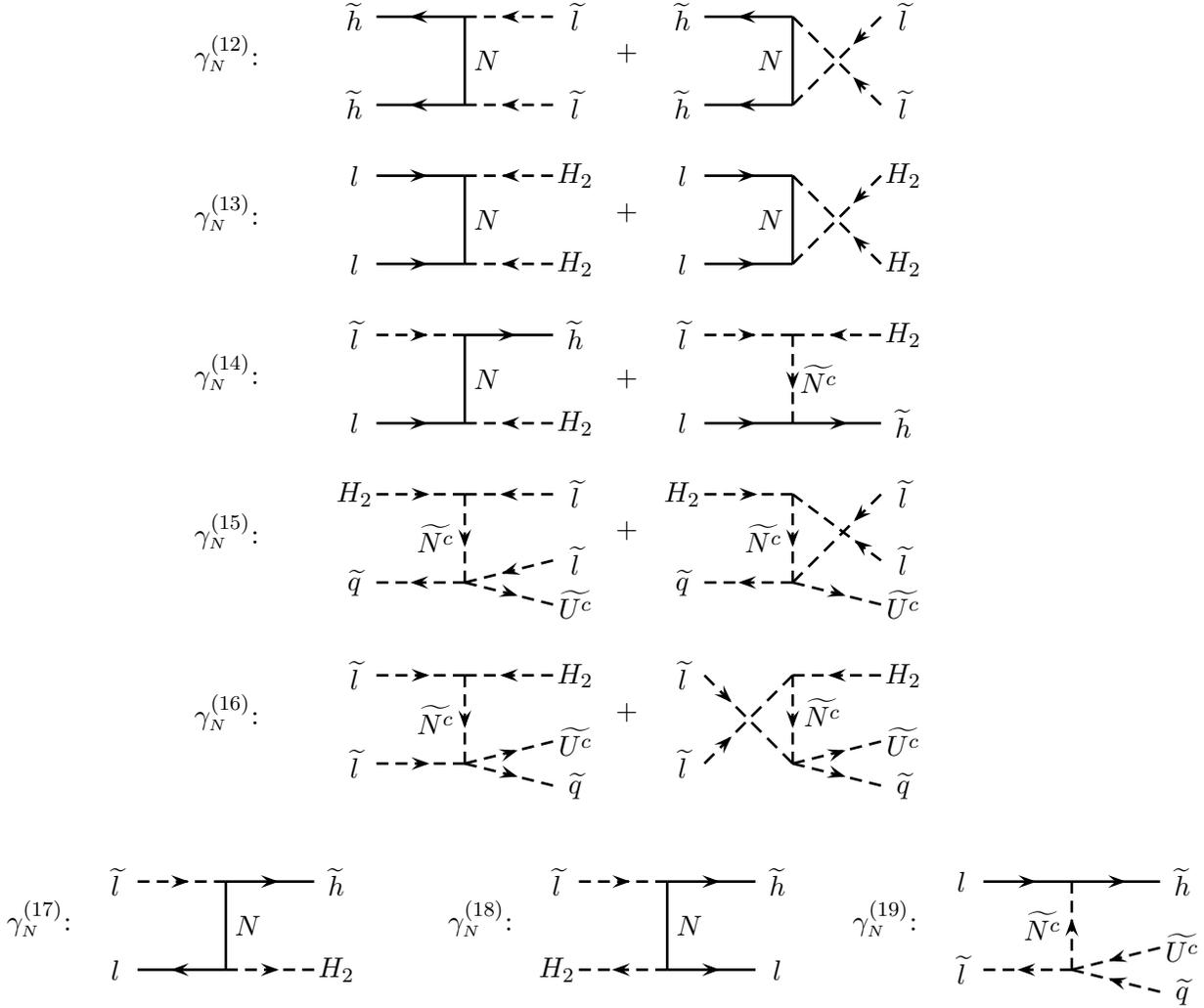}
    \caption{\it $L$ violating processes mediated by a right-handed
      (s)neutrino in the $t$-channel. \label{chap2_fig05}}
  \end{figure}
  Using the tree level vertices from figs.~\ref{chap3_fig01b} and
  \ref{chap2_fig02} as building blocks we can construct lepton number
  violating scatterings mediated by a heavy (s)neutrino. Although of
  higher order than the tree level decays, these diagrams have to be
  taken into consideration to avoid the generation of an asymmetry in
  thermal equilibrium, which is forbidden by $CPT$ invariance 
  \cite{susskind,kw}.
  
  All these processes can be summarized in the configuration space
  superfield diagram shown in fig.~\ref{chap2_fig03super}. It yields
  the following contribution to the $T$-matrix 
  \beqa
    iT_{\scr \D L}&=&{i\over2}\sum\limits_k
      \left(\l_{\n}\right)_{ik}\left(\l_{\n}\right)_{jk}
      \int d^4x_1\,d^4x_2\int{d^4k\over(2\p)^4}\,
      {\mbox{e}^{ik(x_1-x_2)}\over k^2-M_k^2+i\ve}\times\\[1ex]
    &&\times\int d^2\q\,d^2\Bar{\q}\,
      L_j(x_1,\q,\Bar{\q}\,)\e H_2(x_1,\q,\Bar{\q}\,)\,
      {M_k\over4}{D_2^2\over\bo_2}\,
      L_i(x_2,\q,\Bar{\q}\,)\e H_2(x_2,\q,\Bar{\q}\,)\;.\NO
  \eeqa
  The superfield product can be evaluated with eq.~(\ref{phi4prod}).
  The matrix elements of $iT_{\scr \D L}$ then correspond to the
  component field processes that we are going to discuss in the
  following.  In this section we will only mention the different
  processes which have to be considered.  The corresponding reduced
  cross sections can be found in appendix~\ref{appC} and the reaction
  densities are discussed in appendix~\ref{appD}.
  
  By combining two of the decay vertices (cf.~fig.~\ref{chap3_fig01b}
  and fig.~\ref{chap2_fig02}a) one gets the processes that we have
  shown in fig.~\ref{chap2_fig03} and the corresponding $CP$
  conjugated processes.  We will use the following abbreviations for
  the reaction densities
  \beqx
    \begin{array}{l@{\qquad\qquad}l}
      \g_{\scr N}^{(1)}=\g\Big(\wt{l}+\Bar{\wt{h}}\leftrightarrow 
        {\wt{l}{}}^{\;\dg}+\wt{h}\Big)\;, &
      \g_{\scr N}^{(2)}=\g\Big(l+H_2\leftrightarrow
        \Bar{l}+H_2^{\dg}\Big)\;,\\[1ex]
      \g_{\scr N}^{(3)}=\g\Big(\wt{l}+\Bar{\wt{h}}\leftrightarrow
        \Bar{l}+H_2^{\dg}\Big)\;, &
      \g_{\scr N}^{(4)}=\g\Big(l+\Bar{\wt{h}}\leftrightarrow
        {\wt{l}{}}^{\;\dg}+H_2^{\dg}\Big)\;,\\[1ex]
      \g_{\scr N}^{(5)}=\g\Big(\wt{l}+H_2\leftrightarrow{\wt{l}{}}^{\;\dg}
        +\sur+\wt{q}\Big)\;, &
      \g_{\scr N}^{(6)}=\g\Big(l+H_2\leftrightarrow\wt{l}+\Bar{\wt{h}}\;
        \Big)\;,\\[1ex]
      \g_{\scr N}^{(7)}=\g\Big(l+\Bar{\wt{h}}\leftrightarrow
        \wt{l}+\wt{q}^{\,\dg}+\sur^{\dg}\Big)\;. &
    \end{array}
  \eeqx 
  The contributions from on-shell (s)neutrinos contained in these
  reactions have already been taken into account as inverse decay
  followed by a decay. Hence, one has to subtract the contributions
  from real intermediate states to avoid a double counting of
  reactions \cite{kw}.

  {}From the scattering vertex in fig.~\ref{chap2_fig02}b and the decay
  vertices we can construct the following processes
  \beqa
     \g_{\scr N}^{(8)}&=&\g\Big(\sur+\wt{q}\leftrightarrow\wt{l}+\wt{l}
       +H_2\Big)\;,\qquad\qquad
       \g_{\scr N}^{(9)}=\g\Big(\wt{q}+\sur\leftrightarrow\wt{l}+
       \Bar{l}+\wt{h}\Big)\;,\NO\\[1ex]
     \g_{\scr N}^{(10)}&=&\g\Big({\wt{l}{}}^{\;\dg}+\wt{q}\leftrightarrow
       \wt{l}+\sur^{\dg}+H_2\Big)
       =\g\Big({\wt{l}{}}^{\;\dg}+\sur\leftrightarrow
       \wt{l}+\wt{q}^{\,\dg}+H_2\Big)\;,\NO\\[1ex]
     \g_{\scr N}^{(11)}&=&\g\Big({\wt{l}{}}^{\;\dg}+\wt{q}\leftrightarrow
       \Bar{l}+\wt{h}+\sur^{\dg}\Big)
       =\g\Big({\wt{l}{}}^{\;\dg}+\sur\leftrightarrow
       \Bar{l}+\wt{h}+\wt{q}^{\,\dg}\Big)\;.\NO
   \eeqa   
   In fig.~\ref{chap2_fig04} we have shown one typical diagram for
   each of these reaction densities. Again, these diagrams have
   on-shell contributions which have to be subtracted, since they can
   be described as decay of a sneutrino which has been produced in a
   scattering process.
   
   Up to now we have only considered processes with a neutrino or its
   scalar partner in the $s$-channel. In fig.~\ref{chap2_fig05} we
   have shown a selection of diagrams without on-shell contributions.
   The corresponding reaction densities will be denoted by
  \beqa
    \g_{\scr N}^{(12)}&=&\g\Big(\Bar{\wt{h}}+\Bar{\wt{h}}\leftrightarrow
       {\wt{l}{}}^{\;\dg}+{\wt{l}{}}^{\;\dg}\;\Big)\;,\qquad\qquad
      \g_{\scr N}^{(13)}=\g\Big(l+l\leftrightarrow H_2^{\dg}+H_2^{\dg}
      \Big)\;,\NO\\[1ex]
    \g_{\scr N}^{(14)}&=&\g\Big(\wt{l}+l\leftrightarrow\wt{h}
      +H_2^{\dg}\Big)\;,\qquad\qquad
      \g_{\scr N}^{(16)}=\g\Big(\wt{l}+\wt{l}\leftrightarrow\sur+\wt{q}
      +H_2^{\dg}\Big)\;,\NO\\[1ex]
    \g_{\scr N}^{(15)}&=&\g\Big(H_2+\wt{q}^{\,\dg}\leftrightarrow
      {\wt{l}{}}^{\;\dg}+{\wt{l}{}}^{\;\dg}+\sur\Big)
      =\g\Big(H_2+\sur^{\dg}\leftrightarrow{\wt{l}{}}^{\;\dg}
      +{\wt{l}{}}^{\;\dg}+\wt{q}\Big)\;,\NO\\[1ex]
    \g_{\scr N}^{(17)}&=&\g\Big(\wt{l}+\Bar{l}\leftrightarrow
      \wt{h}+H_2\Big)\;,\qquad\qquad
    \g_{\scr N}^{(18)}=\g\Big(\wt{l}+H_2^{\dg}\leftrightarrow
      \wt{h}+l\Big)\;,\NO\\[1ex]
    \g_{\scr N}^{(19)}&=&\g\Big(l+{\wt{l}{}}^{\;\dg}\leftrightarrow
      \wt{h}+\wt{q}^{\,\dg}+\sur^{\dg}\Big)
    =\g\Big(l+\wt{q}\leftrightarrow\wt{l}+\sur^{\dg}+\wt{h}
      \Big)\NO\\[1ex]
    &=&\g\Big(l+\sur\leftrightarrow\wt{l}+\wt{q}^{\,\dg}+\wt{h}
      \Big)
    =\g\Big({\wt{l}{}}^{\;\dg}+\Bar{\wt{h}}\leftrightarrow
      \Bar{l}+\wt{q}^{\,\dg}+\sur^{\dg}\Big)\NO\\[1ex]
    &=&\g\Big(\wt{q}+\Bar{\wt{h}}\leftrightarrow\Bar{l}+\wt{l}+
      \sur^{\dg}\Big)
    =\g\Big(\sur+\Bar{\wt{h}}\leftrightarrow\Bar{l}+\wt{l}+
      \wt{q}^{\,\dg}\Big)\;.\NO
  \eeqa  
  At first sight one may think that these diagrams could be neglected,
  since they are suppressed at intermediate temperatures, i.e.\ 
  intermediate energies $x\approx a_j$. However, they give an
  important contribution to the effective lepton number violating
  interactions at low energies and therefore have to be taken into
  consideration.

\section{Interactions with a Top or a Stop \label{Ntopsection}}
  \begin{figure}[b]
    \centerline{\input{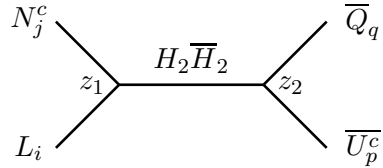}}
    \caption{\it Neutrino-(s)top scattering in configuration 
      superspace. $i$, $j$, $q$ and $p$ are flavour indices.
      \label{Ntopsuper}}
  \end{figure}
  \begin{figure}[ht]
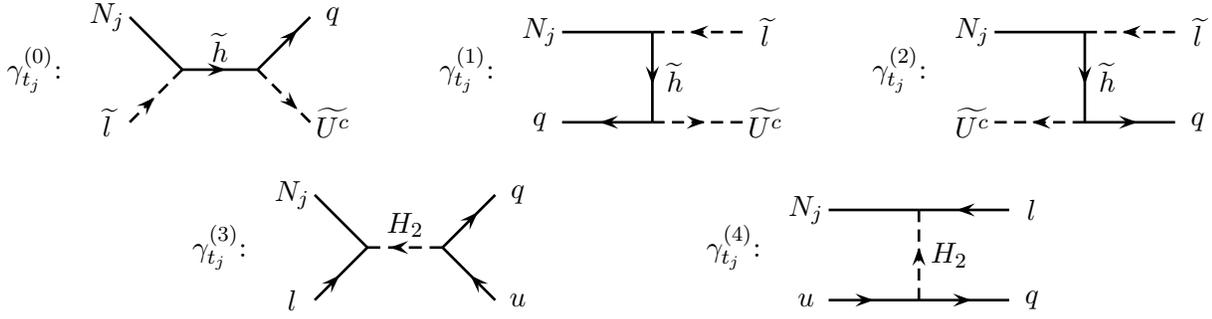

    \begin{center}
\parbox[c]{16cm}{
\pspicture(-1,0)(3.5,2)
\rput[rc]{0}(-0.3,1){$\g_{t_j}^{(0)}$:}
\psline[linewidth=1pt,linestyle=dashed](0.6,0.3)(1.3,1)
\psline[linewidth=1pt](0.6,1.7)(1.3,1)
\psline[linewidth=1pt](1.3,1)(2.3,1)
\psline[linewidth=1pt](2.3,1)(3,1.7)
\psline[linewidth=1pt,linestyle=dashed](2.3,1)(3,0.3)
\psline[linewidth=1pt]{->}(0.83,0.53)(0.93,0.63)
\psline[linewidth=1pt]{->}(1.78,1)(1.88,1)
\psline[linewidth=1pt]{->}(2.73,1.43)(2.83,1.53)
\psline[linewidth=1pt]{->}(2.73,0.57)(2.83,0.47)
\rput[cc]{0}(0.3,1.7){$N_j$}
\rput[cc]{0}(0.3,0.3){$\widetilde{l}$}
\rput[cc]{0}(1.8,1.3){$\widetilde{h}$}
\rput[cc]{0}(3.3,0.3){$\widetilde{U^c}$}
\rput[cc]{0}(3.3,1.7){$q$}
\endpspicture\hspace{\fill}
\pspicture(-1,0)(3.5,1.8)
\rput[rc]{0}(-0.3,1){$\g_{t_j}^{(1)}$:}
\psline[linewidth=1pt](0.6,1.5)(1.8,1.5)
\psline[linewidth=1pt,linestyle=dashed](1.8,1.5)(3,1.5)
\psline[linewidth=1pt](1.8,1.5)(1.8,0.3)
\psline[linewidth=1pt](0.6,0.3)(1.8,0.3)
\psline[linewidth=1pt,linestyle=dashed](1.8,0.3)(3,0.3)
\psline[linewidth=1pt]{<-}(2.3,1.5)(2.4,1.5)
\psline[linewidth=1pt]{->}(1.8,0.9)(1.8,0.8)
\psline[linewidth=1pt]{->}(1.2,0.3)(1.1,0.3)
\psline[linewidth=1pt]{<-}(2.5,0.3)(2.4,0.3)
\rput[cc]{0}(0.3,1.5){$N_j$}
\rput[cc]{0}(0.3,0.3){$q$}
\rput[cc]{0}(2.1,0.9){$\widetilde{h}$}
\rput[cc]{0}(3.3,1.5){$\widetilde{l}$}
\rput[cc]{0}(3.3,0.3){$\widetilde{U^c}$}
\endpspicture\hspace{\fill}
\pspicture(-1,0)(3.5,1.8)
\rput[rc]{0}(-0.3,1){$\g_{t_j}^{(2)}$:}
\psline[linewidth=1pt](0.6,1.5)(1.8,1.5)
\psline[linewidth=1pt,linestyle=dashed](1.8,1.5)(3,1.5)
\psline[linewidth=1pt](1.8,1.5)(1.8,0.3)
\psline[linewidth=1pt,linestyle=dashed](0.6,0.3)(1.8,0.3)
\psline[linewidth=1pt](1.8,0.3)(3,0.3)
\psline[linewidth=1pt]{<-}(2.3,1.5)(2.4,1.5)
\psline[linewidth=1pt]{->}(1.8,0.9)(1.8,0.8)
\psline[linewidth=1pt]{->}(1.2,0.3)(1.1,0.3)
\psline[linewidth=1pt]{<-}(2.5,0.3)(2.4,0.3)
\rput[cc]{0}(0.3,1.5){$N_j$}
\rput[cc]{0}(0.3,0.3){$\widetilde{U^c}$}
\rput[cc]{0}(2.1,0.9){$\widetilde{h}$}
\rput[cc]{0}(3.3,1.5){$\widetilde{l}$}
\rput[cc]{0}(3.3,0.3){$q$}
\endpspicture\\[2ex]
\mbox{ }\hspace{\fill}
\pspicture(-1,0)(3.5,2)
\rput[rc]{0}(-0.3,1){$\g_{t_j}^{(3)}$:}
\psline[linewidth=1pt](0.6,0.3)(1.3,1)
\psline[linewidth=1pt](0.6,1.7)(1.3,1)
\psline[linewidth=1pt,linestyle=dashed](1.3,1)(2.3,1)
\psline[linewidth=1pt](2.3,1)(3,1.7)
\psline[linewidth=1pt](2.3,1)(3,0.3)
\psline[linewidth=1pt]{->}(0.83,0.53)(0.93,0.63)
\psline[linewidth=1pt]{->}(1.7,1)(1.6,1)
\psline[linewidth=1pt]{->}(2.67,1.37)(2.77,1.47)
\psline[linewidth=1pt]{->}(2.77,0.53)(2.67,0.63)
\rput[cc]{0}(0.3,1.7){$N_j$}
\rput[cc]{0}(0.3,0.3){$l$}
\rput[cc]{0}(1.8,1.3){$H_2$}
\rput[cc]{0}(3.3,0.3){$u$}
\rput[cc]{0}(3.3,1.7){$q$}
\endpspicture\hspace{\fill}
\pspicture(-1,0)(3.5,1.8)
\rput[rc]{0}(-0.3,1){$\g_{t_j}^{(4)}$:}
\psline[linewidth=1pt](0.6,1.5)(1.8,1.5)
\psline[linewidth=1pt](1.8,1.5)(3,1.5)
\psline[linewidth=1pt,linestyle=dashed](1.8,1.5)(1.8,0.3)
\psline[linewidth=1pt](0.6,0.3)(1.8,0.3)
\psline[linewidth=1pt](1.8,0.3)(3,0.3)
\psline[linewidth=1pt]{<-}(2.3,1.5)(2.4,1.5)
\psline[linewidth=1pt]{->}(1.8,0.9)(1.8,1)
\psline[linewidth=1pt]{->}(1.1,0.3)(1.2,0.3)
\psline[linewidth=1pt]{<-}(2.5,0.3)(2.4,0.3)
\rput[cc]{0}(0.3,1.5){$N_j$}
\rput[cc]{0}(0.3,0.3){$u$}
\rput[cc]{0}(2.2,0.9){$H_2$}
\rput[cc]{0}(3.3,1.5){$l$}
\rput[cc]{0}(3.3,0.3){$q$}
\endpspicture\hspace{\fill}
}
\end{center}
    \caption{\it Neutrino-(s)top scattering. \label{Ntop}}
  \end{figure}  
  The Yukawa coupling of the top quark is large. Thus we have to take
  into account lepton number violating interactions of a right-handed
  neutrino with a top quark or its scalar partner. In addition to the
  processes already considered in section \ref{DecaySection}
  (cf.~fig.~\ref{chap2_fig02}b), we have the superfield diagram shown
  in fig.\ref{Ntopsuper}, which gives the following contribution to
  the $T$-matrix
  \beqa
    iT_{t_j}&=&-i(\l_{\n})_{ij}(\l_u^{\dg})_{pq}
      \int d^4x_1\,d^4x_2\int{d^4k\over(2\p)^4}\,
      {\mbox{e}^{ik(x_1-x_2)}\over k^2+i\ve}\times\\[1ex]
    &&\times\int d^2\q\,d^2\Bar{\q}\,N_j^c(x_1,\q,\Bar{\q}\,)
      \left(L_i(x_1,\q,\Bar{\q}\,)\,\Bar{Q}_q(x_2,\q,\Bar{\q}\,)\right)
      \Bar{U_p^c}(x_2,\q,\Bar{\q}\,)\;.\NO
  \eeqa
  In component fields, we have the following processes with a Majorana
  neutrino $N_j$ as external line (cf.~fig.~\ref{Ntop})
  \beqa
    \g_{t_j}^{(0)}&=&\g\Big(N_j+\wt{l}\leftrightarrow q+\sur\Big)
      =\g\Big(N_j+\wt{l}\leftrightarrow\wt{q}+\Bar{u}\Big)
      \;,\NO\\[1ex]
    \g_{t_j}^{(1)}&=&\g\Big(N_j+\Bar{q}\leftrightarrow{\wt{l}{}}^{\;\dg}
      +\sur\Big)=\g\Big(N_j+u\leftrightarrow{\wt{l}{}}^{\;\dg}
      +\wt{q}\Big)\;,\NO\\[1ex]
    \g_{t_j}^{(2)}&=&\g\Big(N_j+\sur^{\dg}\leftrightarrow
      {\wt{l}{}}^{\;\dg}+q\Big)=\g\Big(N_j+\wt{q}^{\,\dg}
      \leftrightarrow{\wt{l}{}}^{\;\dg}+\Bar{u}\Big)\;,\NO \\[1ex]
    \g_{t_j}^{(3)}&=&\g\Big(N_j+l\leftrightarrow q+\Bar{u}\Big)
      \;,\NO\\[1ex]
    \g_{t_j}^{(4)}&=&\g\Big(N_j+u\leftrightarrow\Bar{l}+q\Big)=
      \g\Big(N_j+\Bar{q}\leftrightarrow\Bar{l}+\Bar{u}\Big)\;.\NO
  \eeqa
  At this order of perturbation theory these processes are $CP$
  invariant. Hence, we have the same reaction densities for the $CP$
  conjugated processes.

  \begin{figure}[t]
    \begin{center}
\parbox[c]{16cm}{
\pspicture(-1,0)(3.5,2)
\rput[rc]{0}(-0.3,1){$\g_{t_j}^{(5)}$:}
\psline[linewidth=1pt](0.6,0.3)(1.3,1)
\psline[linewidth=1pt,linestyle=dashed](0.6,1.7)(1.3,1)
\psline[linewidth=1pt](1.3,1)(2.3,1)
\psline[linewidth=1pt,linestyle=dashed](2.3,1)(3,1.7)
\psline[linewidth=1pt](2.3,1)(3,0.3)
\psline[linewidth=1pt]{->}(0.83,0.53)(0.93,0.63)
\psline[linewidth=1pt]{->}(0.83,1.47)(0.93,1.37)
\psline[linewidth=1pt]{->}(1.78,1)(1.88,1)
\psline[linewidth=1pt]{->}(2.73,1.43)(2.83,1.53)
\psline[linewidth=1pt]{->}(2.73,0.57)(2.83,0.47)
\rput[cc]{0}(0.3,1.7){$\snj$}
\rput[cc]{0}(0.3,0.3){$l$}
\rput[cc]{0}(1.8,1.3){$\widetilde{h}$}
\rput[cc]{0}(3.3,0.3){$q$}
\rput[cc]{0}(3.3,1.7){$\widetilde{U^c}$}
\endpspicture\hspace{\fill}
\pspicture(-1,0)(3.5,1.8)
\rput[rc]{0}(-0.3,1){$\g_{t_j}^{(6)}$:}
\psline[linewidth=1pt,linestyle=dashed](0.6,1.5)(1.8,1.5)
\psline[linewidth=1pt](1.8,1.5)(3,1.5)
\psline[linewidth=1pt](1.8,1.5)(1.8,0.3)
\psline[linewidth=1pt,linestyle=dashed](0.6,0.3)(1.8,0.3)
\psline[linewidth=1pt](1.8,0.3)(3,0.3)
\psline[linewidth=1pt]{->}(1.2,1.5)(1.3,1.5)
\psline[linewidth=1pt]{<-}(2.3,1.5)(2.4,1.5)
\psline[linewidth=1pt]{->}(1.8,0.9)(1.8,0.8)
\psline[linewidth=1pt]{->}(1.2,0.3)(1.1,0.3)
\psline[linewidth=1pt]{<-}(2.5,0.3)(2.4,0.3)
\rput[cc]{0}(0.3,1.5){$\snj$}
\rput[cc]{0}(0.3,0.3){$\widetilde{U^c}$}
\rput[cc]{0}(2.1,0.9){$\widetilde{h}$}
\rput[cc]{0}(3.3,1.5){$l$}
\rput[cc]{0}(3.3,0.3){$q$}
\endpspicture\hspace{\fill}
\pspicture(-1,0)(3.5,1.8)
\rput[rc]{0}(-0.3,1){$\g_{t_j}^{(7)}$:}
\psline[linewidth=1pt,linestyle=dashed](0.6,1.5)(1.8,1.5)
\psline[linewidth=1pt](1.8,1.5)(3,1.5)
\psline[linewidth=1pt](1.8,1.5)(1.8,0.3)
\psline[linewidth=1pt](0.6,0.3)(1.8,0.3)
\psline[linewidth=1pt,linestyle=dashed](1.8,0.3)(3,0.3)
\psline[linewidth=1pt]{->}(1.2,1.5)(1.3,1.5)
\psline[linewidth=1pt]{<-}(2.3,1.5)(2.4,1.5)
\psline[linewidth=1pt]{->}(1.8,0.9)(1.8,0.8)
\psline[linewidth=1pt]{->}(1.2,0.3)(1.1,0.3)
\psline[linewidth=1pt]{<-}(2.5,0.3)(2.4,0.3)
\rput[cc]{0}(0.3,1.5){$\snj$}
\rput[cc]{0}(0.3,0.3){$q$}
\rput[cc]{0}(2.1,0.9){$\widetilde{h}$}
\rput[cc]{0}(3.3,1.5){$l$}
\rput[cc]{0}(3.3,0.3){$\widetilde{U^c}$}
\endpspicture\\[2ex]
\mbox{ }\hspace{\fill}
\pspicture(-1,0)(3.5,2)
\rput[rc]{0}(-0.3,1){$\g_{t_j}^{(8)}$:}
\psline[linewidth=1pt,linestyle=dashed](0.6,0.3)(1.3,1)
\psline[linewidth=1pt,linestyle=dashed](0.6,1.7)(1.3,1)
\psline[linewidth=1pt,linestyle=dashed](1.3,1)(2.3,1)
\psline[linewidth=1pt](2.3,1)(3,1.7)
\psline[linewidth=1pt](2.3,1)(3,0.3)
\psline[linewidth=1pt]{->}(0.9,0.6)(0.8,0.5)
\psline[linewidth=1pt]{->}(0.83,1.47)(0.93,1.37)
\psline[linewidth=1pt]{->}(1.7,1)(1.8,1)
\psline[linewidth=1pt]{->}(2.67,1.37)(2.77,1.47)
\psline[linewidth=1pt]{->}(2.77,0.53)(2.67,0.63)
\rput[cc]{0}(0.3,1.7){$\snj$}
\rput[cc]{0}(0.3,0.3){$\widetilde{l}$}
\rput[cc]{0}(1.8,1.3){$H_2$}
\rput[cc]{0}(3.3,0.3){$q$}
\rput[cc]{0}(3.3,1.7){$u$}
\endpspicture\hspace{\fill}
\pspicture(-2,0)(3.5,1.8)
\rput[rc]{0}(-1,1){$\g_{t_j}^{(9)}$:}
\psline[linewidth=1pt,linestyle=dashed](0.6,1.5)(1.8,1.5)
\psline[linewidth=1pt,linestyle=dashed](1.8,1.5)(3,1.5)
\psline[linewidth=1pt,linestyle=dashed](1.8,1.5)(1.8,0.3)
\psline[linewidth=1pt](0.6,0.3)(1.8,0.3)
\psline[linewidth=1pt](1.8,0.3)(3,0.3)
\psline[linewidth=1pt]{->}(1.2,1.5)(1.3,1.5)
\psline[linewidth=1pt]{<-}(2.5,1.5)(2.4,1.5)
\psline[linewidth=1pt]{->}(1.8,0.9)(1.8,0.8)
\psline[linewidth=1pt]{->}(1.1,0.3)(1.2,0.3)
\psline[linewidth=1pt]{<-}(2.5,0.3)(2.4,0.3)
\rput[cc]{0}(0.3,1.5){$\snj$}
\rput[cc]{0}(0.3,0.3){$q$}
\rput[cc]{0}(2.2,0.9){$H_2$}
\rput[cc]{0}(3.3,1.5){$\widetilde{l}$}
\rput[cc]{0}(3.3,0.3){$u$}
\endpspicture\hspace{\fill}
}
\end{center}
    \caption{\it Sneutrino-(s)top scattering. \label{Nttop}}
  \end{figure}
  For the scalar neutrinos we have similarly (cf.~fig.~\ref{Nttop})
  \beqa
    \g_{t_j}^{(5)}&=&\g\Big(\snj+l\leftrightarrow q+\sur\Big)
      =\g\Big(\snj+l\leftrightarrow\wt{q}+\Bar{u}\Big)
      \;,\NO\\[1ex]
    \g_{t_j}^{(6)}&=&\g\Big(\snj+\sur^{\dg}\leftrightarrow\Bar{l}+q
      \Big)=\g\Big(\snj+\wt{q}^{\,\dg}\leftrightarrow\Bar{l}
      +\Bar{u}\Big)\;,\NO\\[1ex]
    \g_{t_j}^{(7)}&=&\g\Big(\snj+\Bar{q}\leftrightarrow\Bar{l}+\sur\Big)
      =\g\Big(\snj+u\leftrightarrow\Bar{l}+\wt{q}\Big)\;,\NO \\[1ex]
    \g_{t_j}^{(8)}&=&\g\Big(\snj+{\wt{l}{}}^{\;\dg}\leftrightarrow
      \Bar{q}+u\Big)\;,\NO\\[1ex]
    \g_{t_j}^{(9)}&=&\g\Big(\snj+q\leftrightarrow\wt{l}+u\Big)=
      \g\Big(\snj+\Bar{u}\leftrightarrow\wt{l}+\Bar{q}\Big)\;.\NO
  \eeqa
  The quartic scalar couplings from the scalar potential give
  additional $2\to3$, $3\to3$ and $2\to4$ processes, which can be
  neglected since they are phase space suppressed.

\section{Neutrino Pair Creation and Annihilation}
  \begin{figure}[b]
    \centerline{\input{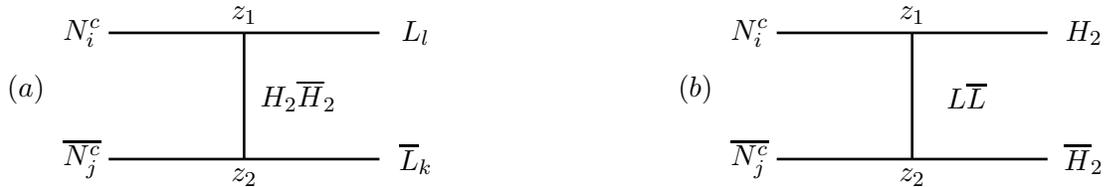}}
    \caption{\it Pair annihilation of singlet neutrino superfields in
      configuration superspace. \label{NNsuper}}
  \end{figure}
  \begin{figure}[t]
    \begin{center}
\parbox[c]{10.7cm}{
\pspicture(-2,0)(3.5,1.8)
\rput[rc]{0}(-1,1){$\g_{\scr N_iN_j}^{(1)}$:}
\psline[linewidth=1pt](0.6,1.5)(1.8,1.5)
\psline[linewidth=1pt,linestyle=dashed](1.8,1.5)(3,1.5)
\psline[linewidth=1pt](1.8,1.5)(1.8,0.3)
\psline[linewidth=1pt](0.6,0.3)(1.8,0.3)
\psline[linewidth=1pt,linestyle=dashed](1.8,0.3)(3,0.3)
\psline[linewidth=1pt]{<-}(2.3,1.5)(2.4,1.5)
\psline[linewidth=1pt]{->}(1.8,0.9)(1.8,0.8)
\psline[linewidth=1pt]{<-}(2.5,0.3)(2.4,0.3)
\rput[cc]{0}(0.3,1.5){$N_i$}
\rput[cc]{0}(0.3,0.3){$N_j$}
\rput[cc]{0}(2.1,0.9){$\widetilde{h}$}
\rput[cc]{0}(3.3,1.5){$\widetilde{l}$}
\rput[cc]{0}(3.3,0.3){$\widetilde{l}$}
\rput[cc]{0}(4.0,0.9){$+$}
\endpspicture\hspace{5ex}
\pspicture(0,0)(3.5,1.8)
\psline[linewidth=1pt](0.6,1.5)(1.8,0.3)
\psline[linewidth=1pt](0.6,0.3)(1.8,1.5)
\psline[linewidth=1pt](1.8,1.5)(1.8,0.3)
\psline[linewidth=1pt,linestyle=dashed](1.8,0.3)(3,0.3)
\psline[linewidth=1pt,linestyle=dashed](1.8,1.5)(3,1.5)
\psline[linewidth=1pt]{<-}(2.3,1.5)(2.4,1.5)
\psline[linewidth=1pt]{->}(1.8,0.9)(1.8,0.8)
\psline[linewidth=1pt]{<-}(2.5,0.3)(2.4,0.3)
\rput[cc]{0}(0.3,1.5){$N_i$}
\rput[cc]{0}(0.3,0.3){$N_j$}
\rput[cc]{0}(2.1,0.9){$\widetilde{h}$}
\rput[cc]{0}(3.3,1.5){$\widetilde{l}$}
\rput[cc]{0}(3.3,0.3){$\widetilde{l}$}
\endpspicture\\[2ex]
\pspicture(-2,0)(3.5,1.8)
\rput[rc]{0}(-1,1){$\g_{\scr N_iN_j}^{(2)}$:}
\psline[linewidth=1pt](0.6,1.5)(1.8,1.5)
\psline[linewidth=1pt](1.8,1.5)(3,1.5)
\psline[linewidth=1pt,linestyle=dashed](1.8,1.5)(1.8,0.3)
\psline[linewidth=1pt](0.6,0.3)(1.8,0.3)
\psline[linewidth=1pt](1.8,0.3)(3,0.3)
\psline[linewidth=1pt]{<-}(2.3,1.5)(2.4,1.5)
\psline[linewidth=1pt]{->}(1.8,0.9)(1.8,1)
\psline[linewidth=1pt]{<-}(2.5,0.3)(2.4,0.3)
\rput[cc]{0}(0.3,1.5){$N_i$}
\rput[cc]{0}(0.3,0.3){$N_j$}
\rput[cc]{0}(2.2,0.9){$H_2$}
\rput[cc]{0}(3.3,1.5){$l$}
\rput[cc]{0}(3.3,0.3){$l$}
\rput[cc]{0}(4.0,0.9){$+$}
\endpspicture\hspace{5ex}
\pspicture(0,0)(3.5,1.8)
\psline[linewidth=1pt](0.6,1.5)(1.8,0.3)
\psline[linewidth=1pt](1.8,1.5)(3,1.5)
\psline[linewidth=1pt,linestyle=dashed](1.8,1.5)(1.8,0.3)
\psline[linewidth=1pt](0.6,0.3)(1.8,1.5)
\psline[linewidth=1pt](1.8,0.3)(3,0.3)
\psline[linewidth=1pt]{<-}(2.3,1.5)(2.4,1.5)
\psline[linewidth=1pt]{->}(1.8,0.9)(1.8,1)
\psline[linewidth=1pt]{<-}(2.5,0.3)(2.4,0.3)
\rput[cc]{0}(0.3,1.5){$N_i$}
\rput[cc]{0}(0.3,0.3){$N_j$}
\rput[cc]{0}(2.2,0.9){$H_2$}
\rput[cc]{0}(3.3,1.5){$l$}
\rput[cc]{0}(3.3,0.3){$l$}
\endpspicture\\[2ex]
\pspicture(-2,0)(3.5,1.8)
\rput[rc]{0}(-1,1){$\g_{\scr N_iN_j}^{(3)}$:}
\psline[linewidth=1pt](0.6,1.5)(1.8,1.5)
\psline[linewidth=1pt,linestyle=dashed](1.8,1.5)(3,1.5)
\psline[linewidth=1pt](1.8,1.5)(1.8,0.3)
\psline[linewidth=1pt](0.6,0.3)(1.8,0.3)
\psline[linewidth=1pt,linestyle=dashed](1.8,0.3)(3,0.3)
\psline[linewidth=1pt]{<-}(2.3,1.5)(2.4,1.5)
\psline[linewidth=1pt]{<-}(1.8,1.0)(1.8,0.9)
\psline[linewidth=1pt]{<-}(2.5,0.3)(2.4,0.3)
\rput[cc]{0}(0.3,1.5){$N_i$}
\rput[cc]{0}(0.3,0.3){$N_j$}
\rput[cc]{0}(2.1,0.9){$l$}
\rput[cc]{0}(3.3,1.5){$H_2$}
\rput[cc]{0}(3.3,0.3){$H_2$}
\rput[cc]{0}(4.0,0.9){$+$}
\endpspicture\hspace{5ex}
\pspicture(0,0)(3.5,1.8)
\psline[linewidth=1pt](0.6,1.5)(1.8,0.3)
\psline[linewidth=1pt,linestyle=dashed](1.8,1.5)(3,1.5)
\psline[linewidth=1pt](1.8,1.5)(1.8,0.3)
\psline[linewidth=1pt](0.6,0.3)(1.8,1.5)
\psline[linewidth=1pt,linestyle=dashed](1.8,0.3)(3,0.3)
\psline[linewidth=1pt]{<-}(2.3,1.5)(2.4,1.5)
\psline[linewidth=1pt]{<-}(1.8,1.0)(1.8,0.9)
\psline[linewidth=1pt]{<-}(2.5,0.3)(2.4,0.3)
\rput[cc]{0}(0.3,1.5){$N_i$}
\rput[cc]{0}(0.3,0.3){$N_j$}
\rput[cc]{0}(2.1,0.9){$l$}
\rput[cc]{0}(3.3,1.5){$H_2$}
\rput[cc]{0}(3.3,0.3){$H_2$}
\endpspicture\\[2ex]
\pspicture(-2,0)(3.5,1.8)
\rput[rc]{0}(-1,1){$\g_{\scr N_iN_j}^{(4)}$:}
\psline[linewidth=1pt](0.6,1.5)(1.8,1.5)
\psline[linewidth=1pt](1.8,1.5)(3,1.5)
\psline[linewidth=1pt,linestyle=dashed](1.8,1.5)(1.8,0.3)
\psline[linewidth=1pt](0.6,0.3)(1.8,0.3)
\psline[linewidth=1pt](1.8,0.3)(3,0.3)
\psline[linewidth=1pt]{<-}(2.3,1.5)(2.4,1.5)
\psline[linewidth=1pt]{<-}(1.8,0.8)(1.8,0.9)
\psline[linewidth=1pt]{<-}(2.5,0.3)(2.4,0.3)
\rput[cc]{0}(0.3,1.5){$N_i$}
\rput[cc]{0}(0.3,0.3){$N_j$}
\rput[cc]{0}(2.2,0.9){$\widetilde{l}$}
\rput[cc]{0}(3.3,1.5){$\widetilde{h}$}
\rput[cc]{0}(3.3,0.3){$\widetilde{h}$}
\rput[cc]{0}(4.0,0.9){$+$}
\endpspicture\hspace{5ex}
\pspicture(0,0)(3.5,1.8)
\psline[linewidth=1pt](0.6,1.5)(1.8,0.3)
\psline[linewidth=1pt](1.8,1.5)(3,1.5)
\psline[linewidth=1pt,linestyle=dashed](1.8,1.5)(1.8,0.3)
\psline[linewidth=1pt](0.6,0.3)(1.8,1.5)
\psline[linewidth=1pt](1.8,0.3)(3,0.3)
\psline[linewidth=1pt]{<-}(2.3,1.5)(2.4,1.5)
\psline[linewidth=1pt]{<-}(1.8,0.8)(1.8,0.9)
\psline[linewidth=1pt]{<-}(2.5,0.3)(2.4,0.3)
\rput[cc]{0}(0.3,1.5){$N_i$}
\rput[cc]{0}(0.3,0.3){$N_j$}
\rput[cc]{0}(2.2,0.9){$\widetilde{l}$}
\rput[cc]{0}(3.3,1.5){$\widetilde{h}$}
\rput[cc]{0}(3.3,0.3){$\widetilde{h}$}
\endpspicture
}
\end{center}
    \caption{\it Neutrino pair annihilation. \label{NN}}
  \end{figure}
  \begin{figure}[t]
    \begin{center}
\parbox[c]{13cm}{
\pspicture(-2,0)(3.5,1.8)
\rput[rc]{0}(-1,1){$\g_{\scr \sni\snj}^{(1)}$:}
\psline[linewidth=1pt,linestyle=dashed](0.6,1.5)(1.8,1.5)
\psline[linewidth=1pt](1.8,1.5)(3,1.5)
\psline[linewidth=1pt](1.8,1.5)(1.8,0.3)
\psline[linewidth=1pt,linestyle=dashed](0.6,0.3)(1.8,0.3)
\psline[linewidth=1pt](1.8,0.3)(3,0.3)
\psline[linewidth=1pt]{->}(1.2,1.5)(1.3,1.5)
\psline[linewidth=1pt]{<-}(2.3,1.5)(2.4,1.5)
\psline[linewidth=1pt]{->}(1.8,0.9)(1.8,0.8)
\psline[linewidth=1pt]{->}(1.2,0.3)(1.1,0.3)
\psline[linewidth=1pt]{<-}(2.5,0.3)(2.4,0.3)
\rput[cc]{0}(0.3,1.5){$\widetilde{N^c_i}$}
\rput[cc]{0}(0.3,0.3){$\widetilde{N^c_j}$}
\rput[cc]{0}(2.1,0.9){$\widetilde{h}$}
\rput[cc]{0}(3.3,1.5){$l$}
\rput[cc]{0}(3.3,0.3){$l$}
\endpspicture\hspace{\fill}
\pspicture(-2,0)(3.5,1.8)
\rput[rc]{0}(-1,1){$\g_{\scr \sni\snj}^{(3)}$:}
\psline[linewidth=1pt,linestyle=dashed](0.6,0.3)(1.8,0.3)
\psline[linewidth=1pt](1.8,0.3)(3,0.3)
\psline[linewidth=1pt](1.8,0.3)(1.8,1.5)
\psline[linewidth=1pt,linestyle=dashed](0.6,1.5)(1.8,1.5)
\psline[linewidth=1pt](1.8,1.5)(3,1.5)
\psline[linewidth=1pt]{->}(1.2,1.5)(1.3,1.5)
\psline[linewidth=1pt]{<-}(2.3,0.3)(2.4,0.3)
\psline[linewidth=1pt]{->}(1.8,0.9)(1.8,1.0)
\psline[linewidth=1pt]{->}(1.2,0.3)(1.1,0.3)
\psline[linewidth=1pt]{<-}(2.5,1.5)(2.4,1.5)
\rput[cc]{0}(0.3,1.5){$\widetilde{N^c_i}$}
\rput[cc]{0}(0.3,0.3){$\widetilde{N^c_j}$}
\rput[cc]{0}(2.1,0.9){$l$}
\rput[cc]{0}(3.3,0.3){$\widetilde{h}$}
\rput[cc]{0}(3.3,1.5){$\widetilde{h}$}
\endpspicture\\[2ex]
\mbox{ }\hspace{\fill}
\pspicture(-2,0)(3.5,1.8)
\rput[rc]{0}(-1,1){$\g_{\scr \sni\snj}^{(2)}$:}
\psline[linewidth=1pt,linestyle=dashed](0.6,1.5)(1.8,1.5)
\psline[linewidth=1pt,linestyle=dashed](1.8,1.5)(3,1.5)
\psline[linewidth=1pt,linestyle=dashed](1.8,1.5)(1.8,0.3)
\psline[linewidth=1pt,linestyle=dashed](0.6,0.3)(1.8,0.3)
\psline[linewidth=1pt,linestyle=dashed](1.8,0.3)(3,0.3)
\psline[linewidth=1pt]{->}(1.2,1.5)(1.3,1.5)
\psline[linewidth=1pt]{<-}(2.5,1.5)(2.4,1.5)
\psline[linewidth=1pt]{->}(1.8,0.9)(1.8,0.8)
\psline[linewidth=1pt]{->}(1.2,0.3)(1.1,0.3)
\psline[linewidth=1pt]{<-}(2.3,0.3)(2.4,0.3)
\rput[cc]{0}(0.3,1.5){$\widetilde{N^c_i}$}
\rput[cc]{0}(0.3,0.3){$\widetilde{N^c_j}$}
\rput[cc]{0}(2.2,0.9){$H_2$}
\rput[cc]{0}(3.3,1.5){$\widetilde{l}$}
\rput[cc]{0}(3.3,0.3){$\widetilde{l}$}
\rput[cc]{0}(4.0,0.9){$+$}
\endpspicture\hspace{5ex}
\pspicture(-0.6,0)(2.5,2)
\psline[linewidth=1pt,linestyle=dashed](0.6,0.3)(1.3,1)
\psline[linewidth=1pt,linestyle=dashed](0.6,1.7)(1.3,1)
\psline[linewidth=1pt,linestyle=dashed](1.3,1)(2,1.7)
\psline[linewidth=1pt,linestyle=dashed](1.3,1)(2,0.3)
\psline[linewidth=1pt]{->}(0.85,1.45)(0.95,1.35)
\psline[linewidth=1pt]{->}(1.72,1.42)(1.82,1.52)
\psline[linewidth=1pt]{->}(0.88,0.58)(0.78,0.48)
\psline[linewidth=1pt]{->}(1.78,0.52)(1.68,0.62)
\rput[cc]{0}(0.3,1.7){$\sni$}
\rput[cc]{0}(0.3,0.3){$\snj$}
\rput[cc]{0}(2.3,0.3){$\wt{l}$}
\rput[cc]{0}(2.3,1.7){$\wt{l}$}
\endpspicture\hspace{\fill}\mbox{ }\\[2ex]
\mbox{ }\hspace{\fill}
\pspicture(-2,0)(3.5,1.8)
\rput[rc]{0}(-1,1){$\g_{\scr \sni\snj}^{(4)}$:}
\psline[linewidth=1pt,linestyle=dashed](0.6,1.5)(1.8,1.5)
\psline[linewidth=1pt,linestyle=dashed](1.8,1.5)(3,1.5)
\psline[linewidth=1pt,linestyle=dashed](1.8,1.5)(1.8,0.3)
\psline[linewidth=1pt,linestyle=dashed](0.6,0.3)(1.8,0.3)
\psline[linewidth=1pt,linestyle=dashed](1.8,0.3)(3,0.3)
\psline[linewidth=1pt]{->}(1.2,1.5)(1.3,1.5)
\psline[linewidth=1pt]{<-}(2.5,1.5)(2.4,1.5)
\psline[linewidth=1pt]{->}(1.8,0.9)(1.8,0.8)
\psline[linewidth=1pt]{->}(1.2,0.3)(1.1,0.3)
\psline[linewidth=1pt]{<-}(2.3,0.3)(2.4,0.3)
\rput[cc]{0}(0.3,1.5){$\widetilde{N^c_i}$}
\rput[cc]{0}(0.3,0.3){$\widetilde{N^c_j}$}
\rput[cc]{0}(2.2,0.9){$\widetilde{l}$}
\rput[cc]{0}(3.3,1.5){$H_2$}
\rput[cc]{0}(3.3,0.3){$H_2$}
\rput[cc]{0}(4.0,0.9){$+$}
\endpspicture\hspace{5ex}
\pspicture(-0.6,0)(2.5,2)
\psline[linewidth=1pt,linestyle=dashed](0.6,0.3)(1.3,1)
\psline[linewidth=1pt,linestyle=dashed](0.6,1.7)(1.3,1)
\psline[linewidth=1pt,linestyle=dashed](1.3,1)(2,1.7)
\psline[linewidth=1pt,linestyle=dashed](1.3,1)(2,0.3)
\psline[linewidth=1pt]{->}(0.85,1.45)(0.95,1.35)
\psline[linewidth=1pt]{->}(1.72,1.42)(1.82,1.52)
\psline[linewidth=1pt]{->}(0.88,0.58)(0.78,0.48)
\psline[linewidth=1pt]{->}(1.78,0.52)(1.68,0.62)
\rput[cc]{0}(0.3,1.7){$\sni$}
\rput[cc]{0}(0.3,0.3){$\snj$}
\rput[cc]{0}(2.3,0.3){$H_2$}
\rput[cc]{0}(2.3,1.7){$H_2$}
\endpspicture\hspace{\fill}\mbox{ }
}
\end{center}
    \caption{\it Sneutrino pair annihilation. \label{NtNt}}
  \end{figure}
  The Yukawa couplings of the right-handed neutrinos also allow lepton
  number conserving processes like neutrino pair creation and
  annihilation. The two superfield diagrams in fig.~\ref{NNsuper}
  yield the following contributions to the $T$-matrix
  \beqa
    iT^{(a)}_{\scr NN}&=&
      -i\left(\l_{\n}\right)_{ki}(\l_{\n}^{\dg})_{jl}
      \int d^4x_1\,d^4x_2\int{d^4k\over(2\p)^4}\,
      {\mbox{e}^{ik(x_1-x_2)}\over k^2+i\ve}\times\\[1ex]
    &&\times\int d^2\q\,d^2\Bar{\q}\,N_i^c(x_1,\q,\Bar{\q}\,)
      \left(L_k(x_1,\q,\Bar{\q}\,)\,\Bar{L}_l(x_2,\q,\Bar{\q}\,)\right)
      \Bar{N_j^c}(x_2,\q,\Bar{\q}\,)\;,\NO\\[2ex]
    iT^{(b)}_{\scr NN}&=&i(\l_{\n}\l_{\n}^{\dg})_{ji}
      \int d^4x_1\,d^4x_2\int{d^4k\over(2\p)^4}\,
      {\mbox{e}^{ik(x_1-x_2)}\over k^2+i\ve}\times\\[1ex]
    &&\times\int d^2\q\,d^2\Bar{\q}\,N_i^c(x_1,\q,\Bar{\q}\,)
      \left(H_2(x_1,\q,\Bar{\q}\,)\,\Bar{H}_2(x_2,\q,\Bar{\q}\,)\right)
      \Bar{N_j^c}(x_2,\q,\Bar{\q}\,)\;.\NO
  \eeqa
  Decomposing the superfield product into component fields with the
  help of eq.~(\ref{phi4prod2}), we get the processes depicted in
  fig.~\ref{NN} for the neutrinos
  \beqx
    \begin{array}{l@{\qquad\qquad}l}
      \g_{\scr N_iN_j} ^{(1)}=\g\Big(N_i+N_j\leftrightarrow\wt{l}+
      {\wt{l}{}}^{\;\dg}\;\Big)\;,&
      \g_{\scr N_iN_j} ^{(2)}=\g\Big(N_i+N_j\leftrightarrow l+\Bar{l}
      \;\Big)\;,\\[2ex]
      \g_{\scr N_iN_j} ^{(3)}=\g\Big(N_i+N_j\leftrightarrow H_2+H_2^{\dg}
      \Big)\;,&
      \g_{\scr N_iN_j} ^{(4)}=\g\Big(N_i+N_j\leftrightarrow\wt{h}
      +\Bar{\wt{h}}\;\Big)\;.
    \end{array}
  \eeqx

  For the scalar neutrinos we have similar diagrams and additional
  contributions from quartic scalar couplings (cf.~fig.~\ref{NtNt}).
  We have the following reaction densities
  \beqx
    \begin{array}{l@{\qquad\qquad}l}
      \g_{\scr\sni\snj} ^{(1)}=\g\Big(\sni+\snj^{\,\dg}
        \leftrightarrow l+\Bar{l}\;\Big)\;,&
        \g_{\scr \sni\snj} ^{(2)}=\g\Big(\sni+\snj^{\,\dg}
        \leftrightarrow\wt{l}+{\wt{l}{}}^{\;\dg}\;\Big)\;,\\[2ex]
      \g_{\scr\sni\snj} ^{(3)}=\g\Big(\sni+\snj^{\,\dg}
        \leftrightarrow\wt{h}+\Bar{\wt{h}}\;\Big)\;,&
        \g_{\scr \sni\snj} ^{(4)}=\g\Big(\sni+\snj^{\,\dg}
        \leftrightarrow H_2+H_2^{\dg}\Big)\;.
    \end{array}
  \eeqx  
  It is interesting to note that the contributions to
  $\g_{\scr\sni\snj}^{(2)}$ and $\g_{\scr \sni\snj}^{(4)}$ from the
  scalar potential are not contained in the superfield diagrams in
  fig.~\ref{NNsuper}. They originate from the contribution of the
  auxiliary field $F_{\scr H_2}$ (cf.~eq.~\ref{aux2}) to the decay
  diagram in fig.~\ref{chap3_fig01a}a.

\begin{figure}[ht]
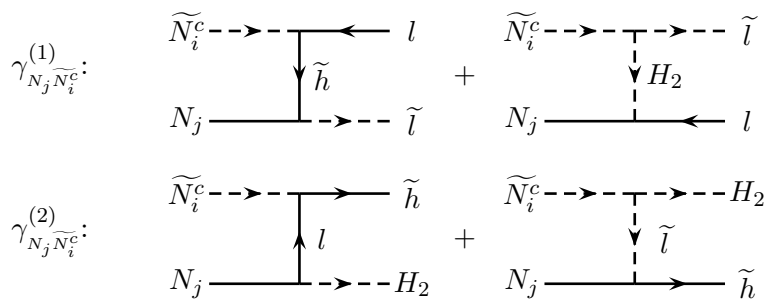

    \begin{center}
\parbox[c]{10.7cm}{
\pspicture(-2,0)(3.5,1.8)
\rput[rc]{0}(-1,1){$\g_{\scr N_j\sni}^{(1)}$:}
\psline[linewidth=1pt,linestyle=dashed](0.6,1.5)(1.8,1.5)
\psline[linewidth=1pt](1.8,1.5)(3,1.5)
\psline[linewidth=1pt](1.8,1.5)(1.8,0.3)
\psline[linewidth=1pt](0.6,0.3)(1.8,0.3)
\psline[linewidth=1pt,linestyle=dashed](1.8,0.3)(3,0.3)
\psline[linewidth=1pt]{->}(1.2,1.5)(1.3,1.5)
\psline[linewidth=1pt]{<-}(2.3,1.5)(2.4,1.5)
\psline[linewidth=1pt]{->}(1.8,0.9)(1.8,0.8)
\psline[linewidth=1pt]{<-}(2.5,0.3)(2.4,0.3)
\rput[cc]{0}(0.3,1.5){$\widetilde{N^c_i}$}
\rput[cc]{0}(0.3,0.3){$N_j$}
\rput[cc]{0}(2.1,0.9){$\widetilde{h}$}
\rput[cc]{0}(3.3,1.5){$l$}
\rput[cc]{0}(3.3,0.3){$\widetilde{l}$}
\rput[cc]{0}(4.0,0.9){$+$}
\endpspicture\hspace{5ex}
\pspicture(0,0)(3.5,1.8)
\psline[linewidth=1pt,linestyle=dashed](0.6,1.5)(1.8,1.5)
\psline[linewidth=1pt,linestyle=dashed](1.8,1.5)(3,1.5)
\psline[linewidth=1pt,linestyle=dashed](1.8,1.5)(1.8,0.3)
\psline[linewidth=1pt](0.6,0.3)(1.8,0.3)
\psline[linewidth=1pt](1.8,0.3)(3,0.3)
\psline[linewidth=1pt]{->}(1.2,1.5)(1.3,1.5)
\psline[linewidth=1pt]{<-}(2.5,1.5)(2.4,1.5)
\psline[linewidth=1pt]{->}(1.8,0.9)(1.8,0.8)
\psline[linewidth=1pt]{<-}(2.4,0.3)(2.5,0.3)
\rput[cc]{0}(0.3,1.5){$\widetilde{N^c_i}$}
\rput[cc]{0}(0.3,0.3){$N_j$}
\rput[cc]{0}(2.2,0.9){$H_2$}
\rput[cc]{0}(3.3,1.5){$\widetilde{l}$}
\rput[cc]{0}(3.3,0.3){$l$}
\endpspicture\\[2ex]
\pspicture(-2,0)(3.5,1.8)
\rput[rc]{0}(-1,1){$\g_{\scr N_j\sni}^{(2)}$:}
\psline[linewidth=1pt,linestyle=dashed](0.6,1.5)(1.8,1.5)
\psline[linewidth=1pt](1.8,1.5)(3,1.5)
\psline[linewidth=1pt](1.8,1.5)(1.8,0.3)
\psline[linewidth=1pt](0.6,0.3)(1.8,0.3)
\psline[linewidth=1pt,linestyle=dashed](1.8,0.3)(3,0.3)
\psline[linewidth=1pt]{->}(1.2,1.5)(1.3,1.5)
\psline[linewidth=1pt]{<-}(2.5,1.5)(2.4,1.5)
\psline[linewidth=1pt]{->}(1.8,0.9)(1.8,1.0)
\psline[linewidth=1pt]{<-}(2.5,0.3)(2.4,0.3)
\rput[cc]{0}(0.3,1.5){$\widetilde{N^c_i}$}
\rput[cc]{0}(0.3,0.3){$N_j$}
\rput[cc]{0}(2.1,0.9){$l$}
\rput[cc]{0}(3.3,1.5){$\widetilde{h}$}
\rput[cc]{0}(3.3,0.3){$H_2$}
\rput[cc]{0}(4.0,0.9){$+$}
\endpspicture\hspace{5ex}
\pspicture(0,0)(3.5,1.8)
\psline[linewidth=1pt,linestyle=dashed](0.6,1.5)(1.8,1.5)
\psline[linewidth=1pt,linestyle=dashed](1.8,1.5)(3,1.5)
\psline[linewidth=1pt,linestyle=dashed](1.8,1.5)(1.8,0.3)
\psline[linewidth=1pt](0.6,0.3)(1.8,0.3)
\psline[linewidth=1pt](1.8,0.3)(3,0.3)
\psline[linewidth=1pt]{->}(1.2,1.5)(1.3,1.5)
\psline[linewidth=1pt]{<-}(2.5,1.5)(2.4,1.5)
\psline[linewidth=1pt]{->}(1.8,0.9)(1.8,0.8)
\psline[linewidth=1pt]{<-}(2.5,0.3)(2.4,0.3)
\rput[cc]{0}(0.3,1.5){$\widetilde{N^c_i}$}
\rput[cc]{0}(0.3,0.3){$N_j$}
\rput[cc]{0}(2.2,0.9){$\widetilde{l}$}
\rput[cc]{0}(3.3,1.5){$H_2$}
\rput[cc]{0}(3.3,0.3){$\widetilde{h}$}
\endpspicture
}
\end{center}
    \caption{\it Neutrino-sneutrino scattering. \label{NNt}}
  \end{figure}

  Finally, there are neutrino-sneutrino scattering processes 
  (cf.~fig.~\ref{NNt}),
  \beqx
    \g_{\scr N_j\sni}^{(1)}=\g\Big(\sni+N_j\leftrightarrow 
      \Bar{l}+\wt{l}\;\Big)\;,\qquad\qquad
    \g_{\scr N_j\sni}^{(2)}=\g\Big(\sni+N_j\leftrightarrow
      \wt{h}+H_2\Big)\;.
  \eeqx 
  Such diagrams also give neutrino-sneutrino transitions like
  $\sni+l\leftrightarrow N_j+\wt{l}$. These processes transform
  neutrinos into sneutrinos and leptons into sleptons, i.e.\ they tend
  to balance out the number densities of the fermions and their
  supersymmetric partners, but they cannot wash out any generated
  asymmetry. As we will see in the next chapter, the number densities
  of the neutrinos and the scalar neutrinos are already equal without
  taking into account these interactions, while the equality of the
  number densities of leptons and sleptons is ensured by
  MSSM-processes, which we are going to discuss in the next section.
  Finally, the dominant contributions to these neutrino-sneutrino
  transitions come from inverse decays, decays and scatterings off a
  (s)top which we have already considered. Hence, we can neglect these
  additional processes.
  \pagebreak
\section{MSSM Processes \label{MSSMsect}}
    \begin{figure}[t]
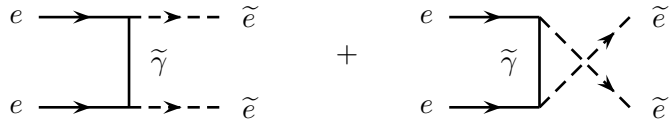

      \begin{center}
\parbox[c]{9.5cm}{
\pspicture(0,0)(4.0,1.8)
\psline[linewidth=1pt](0.6,1.5)(1.8,1.5)
\psline[linewidth=1pt](1.8,1.5)(1.8,0.3)
\psline[linewidth=1pt](0.6,0.3)(1.8,0.3)
\psline[linewidth=1pt,linestyle=dashed](1.8,1.5)(3,1.5)
\psline[linewidth=1pt,linestyle=dashed](1.8,0.3)(3,0.3)
\psline[linewidth=1pt]{->}(1.2,1.5)(1.3,1.5)
\psline[linewidth=1pt]{->}(1.2,0.3)(1.3,0.3)
\psline[linewidth=1pt]{<-}(2.5,1.5)(2.4,1.5)
\psline[linewidth=1pt]{<-}(2.5,0.3)(2.4,0.3)
\rput[cc]{0}(0.3,1.5){$e$}
\rput[cc]{0}(0.3,0.3){$e$}
\rput[cc]{0}(2.2,0.9){$\widetilde{\g}$}
\rput[lc]{0}(3.3,1.5){$\widetilde{e}$}
\rput[lc]{0}(3.3,0.3){$\widetilde{e}$}
\rput[cc]{0}(4.7,1.0){$+$}
\endpspicture\hspace{8ex}
\pspicture(0,0)(3.5,1.8)
\psline[linewidth=1pt](0.6,1.5)(1.8,1.5)
\psline[linewidth=1pt](1.8,1.5)(1.8,0.3)
\psline[linewidth=1pt](0.6,0.3)(1.8,0.3)
\psline[linewidth=1pt,linestyle=dashed](1.8,1.5)(3,0.3)
\psline[linewidth=1pt,linestyle=dashed](1.8,0.3)(3,1.5)
\psline[linewidth=1pt]{->}(1.2,1.5)(1.3,1.5)
\psline[linewidth=1pt]{->}(1.2,0.3)(1.3,0.3)
\psline[linewidth=1pt]{->}(2.7,1.2)(2.8,1.3)
\psline[linewidth=1pt]{->}(2.7,0.6)(2.8,0.5)
\rput[cc]{0}(0.3,1.5){$e$}
\rput[cc]{0}(0.3,0.3){$e$}
\rput[cc]{0}(1.4,0.9){$\widetilde{\g}$}
\rput[lc]{0}(3.3,1.5){$\widetilde{e}$}
\rput[lc]{0}(3.3,0.3){$\widetilde{e}$}
\endpspicture
}
\end{center}
      \caption{\it Example of a $L_f$ and $L_s$ violating MSSM 
        process. \label{MSSMproc}}
    \end{figure}
    In the MSSM the fermionic lepton number $L_f$ and the lepton
    number stored in the scalar leptons $L_s$ are not separately
    conserved. There are processes transforming leptons into scalar
    leptons and vice versa. As an example we have considered the
    process $e+e\leftrightarrow\wt{e}+\wt{e}$
    (cf.~fig.~\ref{MSSMproc}). For large temperatures, i.e.\ $s\gg
    m_{\wt{\g}}^2$, the reduced cross section for this process is
    given by \cite{keung}
    \beq
      \hat{\s}_{\mbox{\tiny MSSM}}\approx 4\p\a^2\left[
      \ln\left(s\over m_{\wt{\g}}^2\right)-4\right]\;.
    \eeq
    This translates into the following reaction density
    \beq
       \g_{\mbox{\tiny MSSM}}\approx {M_1^4\,\a^2\over4\p^3}\,{1\over z^4}
       \left[\ln\left({4\over z^2a_{\wt{\g}}}\right)-2\g_{\scr\rm E}
       -3\right]\;,
    \eeq
    where we have introduced the dimensionless squared photino mass
    \beq
      a_{\wt{\g}} := \left({m_{\wt{\g}}\over M_1}\right)^2\;.
    \eeq
    These processes are in thermal equilibrium if the reaction
    rates are larger than the Hubble parameter $H$. This condition
    gives a very weak upper bound on the photino mass,
    \beq
      m_{\wt{\g}}\;\ltap\;2.5\times10^{9}\;\mbox{GeV}\;
        \left({T\over10^{10}\;\mbox{GeV}}\right)\;
        \exp\left[{-{1\over412}\left({T\over10^{10}\;\mbox{GeV}}
        \right)}\right]\;.
    \eeq
    In the calculations we assume $m_{\wt{\g}}=100\;$GeV.

    \clearpage\chapter{Numerical Results \label{results}}
  Now that we have identified all the relevant processes we can write
  down the network of Boltzmann equations which governs the time
  evolution of the neutrino and sneutrino number densities and of the
  lepton asymmetry\footnote{See app.~\ref{appB} for a short review of
    kinetic theory in an expanding universe.}. In this chapter we work
  out the parameter dependence of the generated baryon asymmetry by
  solving the Boltzmann equations, and we discuss the role of the
  different scattering and decay processes \cite{pluemi2}.

\section{The Boltzmann Equations \label{boltzeq}}
  The evolution of the neutrino number $\Ynj$ as a function of the
  inverse dimensionless temperature $z={M_1/T}$ is given by
  \beqa
    \lefteqn{{\mbox{d}\Ynj\over\mbox{d}z}={-z\over sH(M_1)}\left\{
      \left(\Ynjratio-1\right)\left[\gnj+4\g_{t_j}^{(0)}+4\g_{t_j}^{(1)}
      +4\g_{t_j}^{(2)}+2\g_{t_j}^{(3)}+4\g_{t_j}^{(4)}\right]\right.}
      \\[1ex]
    &&\left.\qquad\qquad+\sum\limits_i\left[\left(\Ynjratio\Yniratio
      -1\right)\sum\limits_{k=1}^4\g_{\scr N_iN_j}^{(k)}
      +\left(\Ynjratio\Ypiratio-2\right)
      \sum\limits_{k=1}^2\g_{\scr N_j\sni}^{(k)}
      \right]\right\}\NO\;.
  \eeqa 
  For the scalar neutrinos and their antiparticles it is convenient to
  use the sum and the difference of the particle numbers per comoving
  volume element as independent variables,
  \beq
    Y_{j\pm} := Y_{\scr \snj}\pm Y_{\scr \snj^{\dg}}\;.
  \eeq
  The Boltzmann equations for these quantities read 
  \beqa
    \lefteqn{{\mbox{d}\Yp\over\mbox{d}z}={-z\over sH(M_1)}\left\{
      \left(\Ypratio-2\right)\left(\gsnj+\gtr+3\g_{22_j}+2\g_{t_j}^{(5)}
      +2\g_{t_j}^{(6)}+2\g_{t_j}^{(7)}+\g_{t_j}^{(8)}+2\g_{t_j}^{(9)}
      \right)\right.}\NO\\[1ex]
    &&\qquad\qquad{}+{1\over2}\Ymratio\YLtratio\left(\g_{22_j}
      -\g_{t_j}^{(8)}\right)+\Ymratio\YLratio\g_{t_j}^{(5)}\\[1ex]
    &&\left.\qquad\qquad+\sum\limits_i
      \left[{1\over2}\left(\Ypratio\Ypiratio-\Ymratio\Ymiratio-4\right)
      \sum\limits_{k=1}^4\g_{\scr \sni\snj}^{(k)}
      +\left(\Ypratio\Yniratio-2\right)\sum\limits_{k=1}^2
      \g_{\scr N_i\snj}^{(k)}\right]
      \right\}\;,\NO\\[3ex]
    \lefteqn{{\mbox{d}\Ym\over\mbox{d}z}={-z\over sH(M_1)}\left\{
      \Ymratio\left(\gsnj+\gtr+3\g_{22_j}+2\g_{t_j}^{(5)}+2\g_{t_j}^{(6)}
      +2\g_{t_j}^{(7)}+\g_{t_j}^{(8)}+2\g_{t_j}^{(9)}\right)
      \right.}\NO\\[1ex]
    &&\qquad\qquad{}+\YLtratio\left[\gtr-{1\over2}\gsnj-2\g_{t_j}^{(9)}
      -{1\over2}\Ypratio\g_{t_j}^{(8)}+\left(2+{1\over2}\Ypratio\right)
      \g_{22_j}\right]\\[1ex]
    &&\qquad\qquad{}+\YLratio\left[{1\over2}\gsnj+2\left(\g_{t_j}^{(6)}
      +\g_{t_j}^{(7)}\right)+\Ypratio\g_{t_j}^{(5)}\right]\NO\\[1ex]
    &&\left.{}+\sum\limits_i\left[{1\over2}\left(\Ymratio\Ypiratio
      -\Ypratio\Ymiratio\right)\sum\limits_{k=1}^4\g_{\scr \sni\snj}^{(k)}
      +\Ymratio\Yniratio\sum\limits_{k=1}^2\g_{\scr N_i\snj}^{(k)}
      +\left(\YLratio-\YLtratio\right)\g_{\scr N_i\snj}^{(1)}\right]
      \right\}\NO\;.
  \eeqa 
  Furthermore, we have to discern the lepton asymmetry stored in the
  standard model particles $\YL$ and the asymmetry $\YLt$ in the
  scalar leptons. Their evolution is governed by
  \beqa
    \lefteqn{{\mbox{d}\YL\over\mbox{d}z}={-z\over sH(M_1)}\left\{
      \sum\limits_j\left[\left({1\over2}\YLratio+\ve_j\right)
      \left({1\over2}\gnj+\gsnj\right)-{1\over2}\ve_j\left(
      \Ynjratio\gnj+\Ypratio\gsnj\right)+{1\over2}\Ymratio\gsnj
      \right]\right.}\NO\\[1ex]
    &&\qquad{}+\YLratio\left(\g_{\scr A}^{\scr \D L}
      +\g_{\scr C}^{\scr \D L}\right)+\YLtratio\left(
      \g_{\scr B}^{\scr \D L}-\g_{\scr C}^{\scr \D L}\right)
      +\left(\YLratio-\YLtratio\right)\g_{\mbox{\tiny MSSM}}\\[1ex]
    &&\qquad{}\!+\sum\limits_j\left[\YLratio\left(\Ynjratio\g_{t_j}^{(3)}
      +\Ypratio\g_{t_j}^{(5)}+2\g_{t_j}^{(4)}+2\g_{t_j}^{(6)}
      +2\g_{t_j}^{(7)}\right)+2\Ymratio\left(\g_{t_j}^{(5)}
      +\g_{t_j}^{(6)}+\g_{t_j}^{(7)}\right)\right]\NO\\[1ex]
    &&\left.\qquad{}+\sum\limits_{i,j}\left(\YLratio-\YLtratio
      +\Ynjratio\Ymiratio\right)\g_{\scr N_j\sni}^{(1)}
      \right\}\;,\NO\\[3ex]
    \lefteqn{{\mbox{d}\YLt\over\mbox{d}z}={-z\over sH(M_1)}\left\{
      \sum\limits_j\left[\left({1\over2}\YLtratio+\ve_j\right)
      \left({1\over2}\gnj+\gsnj\right)-{1\over2}\ve_j\left(
      \Ynjratio\gnj+\Ypratio\gsnj\right)\right.\right.}\NO\\[1ex]
    &&\left.\left.\qquad{}-{1\over2}\Ymratio\gsnj+\left(\YLtratio+\Ymratio
      \right)\gtr+\left({1\over2}\Ypratio\YLtratio+2\YLtratio
      +3\Ymratio\right)\g_{22_j}\right]\right.\NO\\[1ex]
    &&\qquad{}+\YLtratio\left(\g_{\scr A}^{\scr \D L}
      +\g_{\scr D}^{\scr \D L}\right)+\YLratio\left(
      \g_{\scr B}^{\scr \D L}-\g_{\scr C}^{\scr \D L}\right)
      +\left(\YLtratio-\YLratio\right)\g_{\mbox{\tiny MSSM}}\\[1ex]
    &&\qquad{}+\sum\limits_j\left[\YLtratio\left(2\Ynjratio\g_{t_j}^{(0)}
      +{1\over2}\Ypratio\g_{t_j}^{(8)}+2\g_{t_j}^{(1)}+2\g_{t_j}^{(2)}
      +2\g_{t_j}^{(9)}\right)-\Ymratio\left(\g_{t_j}^{(8)}
      +2\g_{t_j}^{(9)}\right)\right]\NO\\[1ex]
    &&\left.\qquad{}+\sum\limits_{i,j}\left(\YLtratio-\YLratio
      -\Ynjratio\Ymiratio\right)\g_{\scr N_j\sni}^{(1)}\right\}\NO\;,
  \eeqa
  where we have introduced the following abbreviations for the
  lepton number violating scatterings mediated by a heavy
  (s)neutrino
  \beqa
    \g_{\scr A}^{\scr \D L}&=&2\g_{\scr N}^{(1)}+\g_{\scr N}^{(3)}
      +\g_{\scr N}^{(4)}+\g_{\scr N}^{(6)}+\g_{\scr N}^{(7)}
      +2\g_{\scr N}^{(12)}+\g_{\scr N}^{(14)}\;,\\[1ex]
    \g_{\scr B}^{\scr \D L}&=&\g_{\scr N}^{(3)}+\g_{\scr N}^{(4)}
      -\g_{\scr N}^{(6)}-\g_{\scr N}^{(7)}+\g_{\scr N}^{(14)}\;,\\[1ex]
    \g_{\scr C}^{\scr \D L}&=&3\g_{\scr N}^{(9)}+\g_{\scr N}^{(17)}
      +\g_{\scr N}^{(18)}+6\g_{\scr N}^{(19)}\;,\\[1ex]
    \g_{\scr D}^{\scr \D L}&=&4\g_{\scr N}^{(5)}+2\g_{\scr N}^{(8)}
      +8\g_{\scr N}^{(10)}+3\g_{\scr N}^{(9)}+4\g_{\scr N}^{(15)}
      +2\g_{\scr N}^{(16)}+\g_{\scr N}^{(17)}+\g_{\scr N}^{(18)}
      +6\g_{\scr N}^{(19)}\;.
  \eeqa  
  The numerical factors in front of the reaction densities arise due
  to the change in quantum numbers in the corresponding scattering,
  e.g.\ processes transforming leptons into sleptons appear with a
  relative minus sign in the Boltzmann equations for $\YL$ and $\YLt$.
  Furthermore, any reaction density is multiplied by the number of
  different processes (cf.~chapter \ref{theory}) contributing 
  independently to the Boltzmann equations.
    
  This set of Boltzmann equations is valid for the most general case
  with arbitrary masses of the right-handed neutrinos. However, if the
  heavy neutrinos are mass degenerate, it is always possible to find a
  basis where the mass matrix $M$ and the Yukawa matrix $\l_{\n}$ are
  diagonal, i.e.\ no asymmetry is generated.  Therefore, one has to
  assume a mass hierarchy for the right-handed neutrinos, which in
  turn implies that the lepton number violating processes induced by
  the lightest right-handed neutrino are in thermal equilibrium as
  long as the temperature is higher than the mass of this neutrino.
  Hence, the lepton asymmetries generated in the decays of the heavier
  right-handed neutrinos are washed out and the asymmetry that we
  observe today must have been generated by the lightest right-handed
  neutrino. We will assume that the first generation neutrino $N_1$ is
  the lightest.
  
  Hence, we will always neglect the heavier right-handed neutrinos as
  free particles. However, they have to be taken into account as
  intermediate states, since they give a substantial contribution to
  the effective lepton number violating processes at low energies.
  
  The fermionic part $\YL$ of the generated lepton asymmetry will be
  transformed into a $(B-L)$ asymmetry by the action of sphalerons.  But
  since MSSM processes like the one in section \ref{MSSMsect} enforce
  the relation
  \beq
    \YL=\YLt\; ,
    \label{equal}
  \eeq
  the total lepton asymmetry $Y_L=\YL+\YLt$ will be proportional to 
  the baryon asymmetry \cite{sphal},
  \beq
    Y_B=-\left({8N_f+4N_H\over22N_f+13N_H}\right)Y_L\; ,
  \eeq
  where $N_f$ is the number of quark-lepton families, and $N_H$
  the number of Higgs doublets. In our model with $N_f=3$ and
  $N_H=2$ we have
  \beq
    Y_B=-{8\over23}Y_L\; .
    \label{sphal}
  \eeq
  {}From the observed baryon asymmetry
  \beq
    Y_B=(0.6-1)\cdot10^{-10}\;,
  \eeq
  and eq.~(\ref{equal}) we can infer the asymmetries that we have to 
  generate,
  \beq
    \YL=\YLt=-(0.9-1.4)\cdot 10^{-10}\;.
    \label{requested}
  \eeq
  The additional anomalous global symmetries in supersymmetric
  theories at high temperatures have no influence on these
  considerations, since they are broken well before the electroweak
  phase transition \cite{ibanez}.

\section{The Generated Lepton Asymmetry \label{gen}}
    \begin{figure}[t]
      \hspace{-0.3cm}
      \begin{minipage}[t]{8.5cm}
        \centerline{\hspace{1cm}(a)}
        \epsfig{file=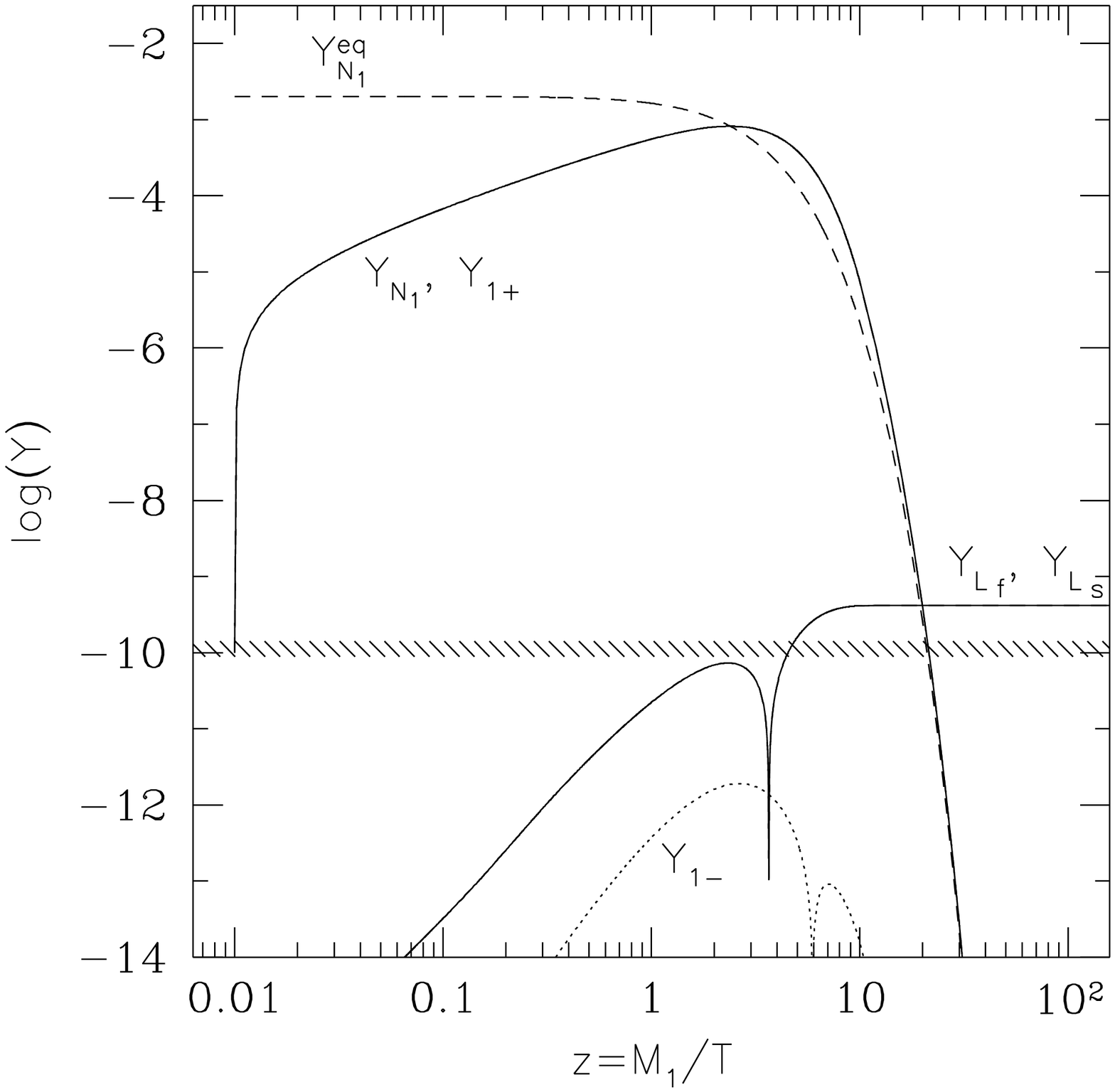,width=8.5cm}
      \end{minipage}
      \hspace{-0.2cm}
      \begin{minipage}[t]{8.5cm}
        \centerline{\hspace{1cm}(b)}
        \epsfig{file=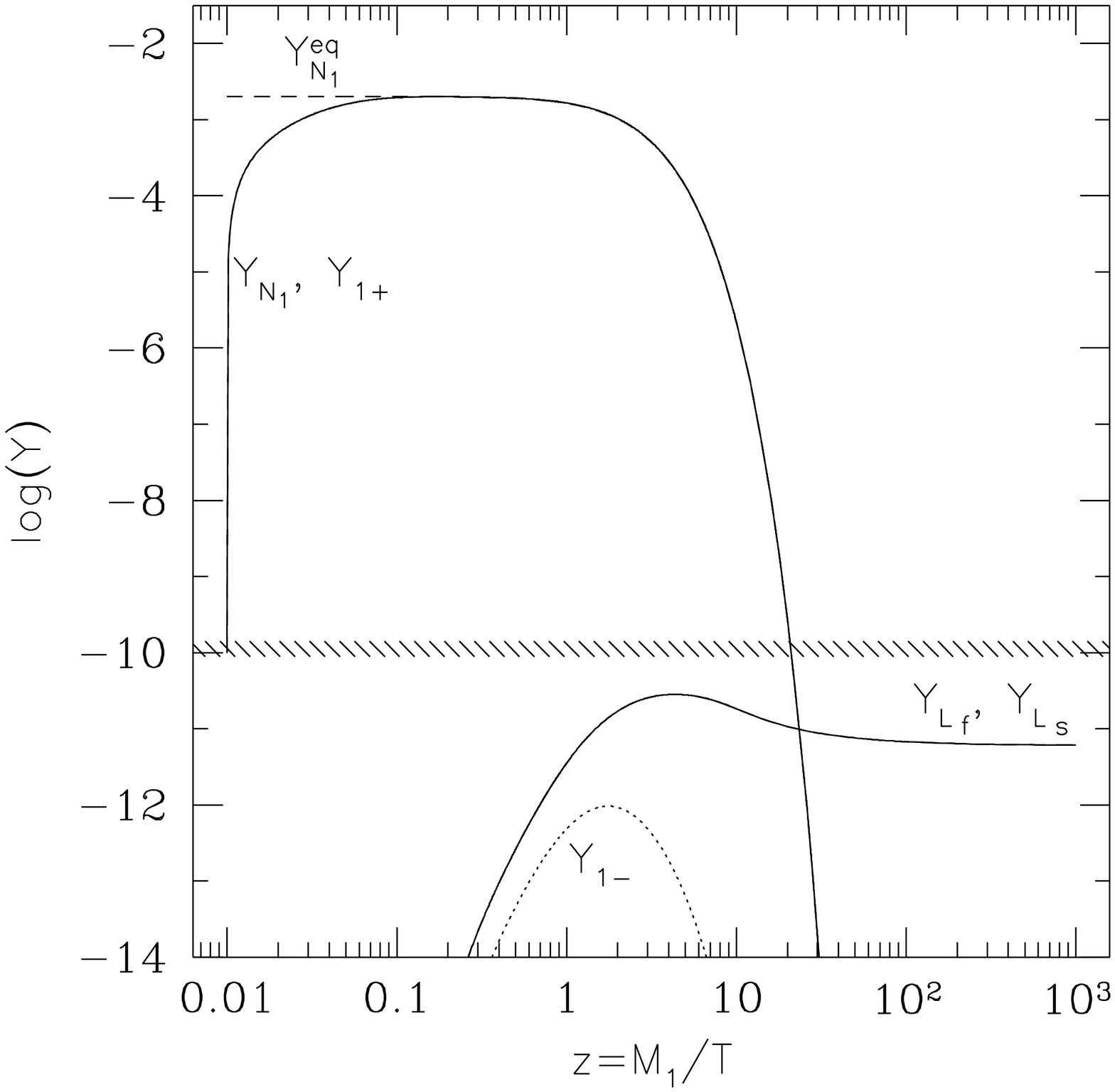,width=8.5cm}
     \end{minipage}  
       \caption{\it Typical solutions of the Boltzmann equations. The
        dashed line represents the equilibrium distribution for the
        neutrinos $N_1$ and the solid lines show the solutions for the
        (s)neutrino number and the absolute values of the asymmetries
        in $L_f$ and $L_s$, while the dotted line represents the
        absolute value of the scalar neutrino asymmetry $Y_{1-}$. The
        lines for $Y_{\scr N_1}$ and $Y_{1+}$  and for the two asymmetries
        $\YL$ and $\YLt$ cannot be distinguished, since they are lying
        one upon another. The hatched area shows the measured value
        (\ref{requested}).\label{sol1fig}}
    \end{figure}
    Typical solutions of the Boltzmann equations are shown in
    fig.~\ref{sol1fig}, where we have assumed a neutrino mass
    $M_1=10^{10}\;$GeV, and a mass hierarchy of the form 
    \beqa
      a_2=10^3\;,&\qquad&(m_{\scr D}^{\dg}m_{\scr D})_{22}
         =a_2\,(m_{\scr D}^{\dg}m_{\scr D})_{11}\;,\\[1ex]
      a_3=10^6\;,&\qquad&(m_{\scr D}^{\dg}m_{\scr D})_{33}
         =a_3\,(m_{\scr D}^{\dg}m_{\scr D})_{11}\;.
    \eeqa
    Furthermore, we have assumed a $CP$ asymmetry $\ve_1=-10^{-6}$. The
    only difference between both figures lies in the choice of 
    $(m_{\scr D}^{\dg}m_{\scr D})_{11}$,
    \beq
      \wt{m}_1:={(m_{\scr D}^{\dg}m_{\scr D})_{11}
        \over M_1}=\left\{
        \begin{array}{rl}
          10^{-4}\;\mbox{eV}&\mbox{for fig.~\ref{sol1fig}a,}\\[1ex]
          10^{-2}\;\mbox{eV}&\mbox{for fig.~\ref{sol1fig}b.}
        \end{array}\right.
    \eeq    
    Finally, as starting condition we have assumed that all the number
    densities vanish at high temperatures $T\gg M_1$, including the
    neutrino numbers $Y_{\scr N_1}$ and $Y_{1+}$. As one can see, the
    Yukawa interactions are strong enough to create a substantial
    number of neutrinos and scalar neutrinos in fig.~\ref{sol1fig}a,
    even if $Y_{\scr N_1}$ and $Y_{1+}$ do not reach their equilibrium
    values as long as $z<1$. However, the generated asymmetries
    \beq
      \YL=\YLt=-4\cdot10^{-10}
      \label{sol1}
    \eeq
    are of the requested magnitude. On the other hand, in
    fig.~\ref{sol1fig}b the Yukawa interactions are much stronger,
    i.e.~the neutrinos are driven into equilibrium rapidly at high
    temperatures. However, the large Yukawa couplings also increase
    the reaction rates for lepton number violating processes which
    can wash out a generated asymmetry, i.e.\ the final asymmetries
    are much smaller than in the previous case,
    \beq
      \YL=\YLt=-6\cdot10^{-12}\;.
      \label{sol2}
    \eeq
    In both cases a small scalar neutrino asymmetry $Y_{1-}$ is
    temporarily generated. However, $Y_{1-}$ is very small and has
    virtually no influence on the generated lepton asymmetries.

    Usually it is assumed that one has a thermal population of
    right-handed neutrinos at high temperatures which decay at very
    low temperatures $T\ll M_1$ where one can neglect lepton number
    violating scatterings. Then the generated lepton asymmetry is
    proportional to the $CP$ asymmetry and the number of decaying
    neutrinos and sneutrinos \cite{kt1},
    \beq
      Y_L\;\approx\;\ve_1\,\left[\;Y_{\scr N_1}^{eq}(T\gg M_1)
        +Y_{1+}^{eq}(T\gg M_1)\;\right]\approx{\ve_1\over250}\;.
    \eeq
    With $\ve_1=-10^{-6}$ this gives
    \beq
      \YL=\YLt\approx-2\cdot10^{-9}\;.
    \eeq
    By comparison with eqs.~(\ref{sol1}) and (\ref{sol2}) one sees
    that by assuming a thermal population of heavy neutrinos at high
    temperatures and neglecting the lepton number violating
    scatterings, one largely overestimates the generated lepton
    asymmetries. 
    
\begin{figure}[t]
       \centerline{\epsfig{file=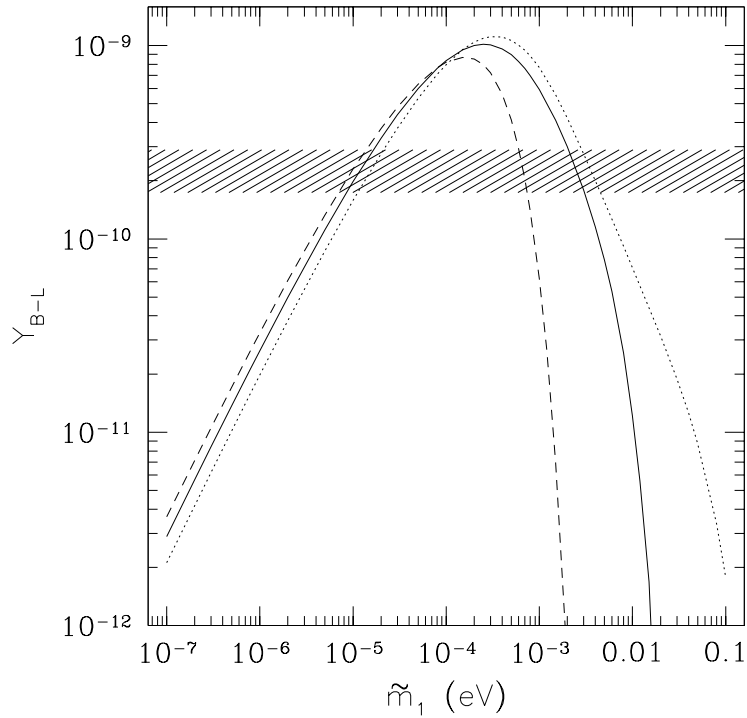,width=10cm}}
       \caption{\it Generated $(B-L)$ asymmetry as a function of 
         $\wt{m}_1$ for $M_1=10^8\;$GeV (dotted line), 
         $M_1=10^{10}\;$GeV (solid line) and $M_1=10^{12}\;$GeV
         (dashed line). The shaded area shows the measured value for
         the asymmetry.\label{fig13}}
    \end{figure}

    A characteristic feature of the non-supersymmetric version of this
    baryogenesis mechanism is that the generated asymmetry does not
    depend on the neutrino mass $M_1$ and $(m_{\scr D}^{\dg} m_{\scr
      D})_{11}$ separately but only on the ratio $\wt{m}_1$
    \cite{pluemi}.  To check if this is also the case in the
    supersymmetric scenario we have varied $\wt{m}_1$ while keeping
    all the other parameters fixed. In fig.~\ref{fig13} we have
    plotted the total lepton asymmetry $Y_L=\YL+\YLt$ as a function of
    $\wt{m}_1$ for the right-handed neutrino masses
    $M_1=10^{12}\;$GeV, $10^{10}\;$GeV and $10^8\;$GeV and the $CP$
    asymmetry $\ve_1=-10^{-6}$.
    
    The main difference between the supersymmetric and
    non-supersymmetric scenarios concerns the necessary production of
    the neutrinos at high temperatures. In the non-supersymmetric
    scenario the Yukawa interactions are too weak to account for this,
    i.e.\ additional interactions of the right-handed neutrinos have
    to be introduced. This is no longer the case here. The
    supersymmetric Yukawa interactions are much more important, and
    can produce a thermal population of right-handed neutrinos, i.e.\ 
    the same vertices which are responsible for the generation of the
    asymmetry can also bring the neutrinos into thermal equilibrium at
    high temperatures. However, these lepton number violating
    processes will also erase a part of the generated asymmetry,
    hereby giving rise to the $\wt{m}_1$ dependence of the generated
    asymmetry which we shall discuss in detail.
    
    First one sees that in the whole parameter range the generated
    asymmetry is much smaller than the naively expected value
    $4\cdot10^{-9}$. For low $\wt{m}_1$ the reason being that the
    Yukawa interactions are too weak to bring the neutrinos into
    equilibrium at high temperatures, like in fig.~\ref{sol1fig}a. For
    high $\wt{m}_1$ on the other hand, the lepton number violating
    scatterings wash out a large part of the generated asymmetry at
    temperatures $T<M_1$, like in fig.~\ref{sol1fig}b. Hence, the
    requested asymmetry can only be generated if $\wt{m}_1$ is larger
    than $\sim10^{-5}\;$eV and smaller than $\sim5\cdot10^{-3}\;$eV,
    depending on the heavy neutrino mass $M_1$.
    
    The asymmetry in fig.~\ref{fig13} depends almost only on
    $\wt{m}_1$ for small $\wt{m}_1\;\ltap\;10^{-4}\;$eV, since in this
    region of parameter space the asymmetry depends mostly on the
    number of neutrinos generated at high temperatures, i.e.\ on the
    strength of the processes in which a right-handed neutrino can be
    generated or annihilated. The dominant reactions are decays,
    inverse decays and scatterings with a (s)top, which all give
    contributions proportional to $\wt{m}_1$ to the Boltzmann
    equations at high temperatures,
    \beqa
       &&{-z\over sH(M_1)}\;\gnone\propto\wt{m}_1\;,\qquad\qquad
         {-z\over sH(M_1)}\;\gsnone\propto\wt{m}_1\;,\qquad\qquad
         {-z\over sH(M_1)}\;\gtrone\propto\wt{m}_1\;,\NO\\[1ex]
       &&{-z\over sH(M_1)}\;\g_{22_1}\propto\wt{m}_1\;,\qquad\qquad
         {-z\over sH(M_1)}\;\g_{t_1}^{(i)}(T\gg M_1)\propto\wt{m}_1\;.
         \label{propor}
    \eeqa
    For large $\wt{m}_1\;\gtap\;10^{-4}\;$eV on the other hand, the
    neutrinos reach thermal equilibrium at high temperatures, i.e.\
    the generated asymmetry depends mostly on the influence of the lepton
    number violating scatterings at temperatures $T\ltap M_1$. In
    contrast to the relations of eq.~(\ref{propor}) the lepton number
    violating processes mediated by a heavy neutrino behave like
    \beq
      {-z\over sH(M_1)}\;\g_i^{\scr \D L}\propto
      M_1\sum\limits_j\wt{m}_j^2\;,
      \qquad i=A,\ldots,D
      \label{propor2}
    \eeq
    at low temperatures. Hence, one expects that the generated
    asymmetry becomes smaller for growing neutrino mass $M_1$ and this
    is exactly what one observes in fig.~\ref{fig13}.

    Eq.~(\ref{propor2}) can also explain the small dependence of the
    asymmetry on the heavy neutrino mass $M_1$ for
    $\wt{m}_1\,\ltap\,10^{-4}\;$eV. The inverse decay processes which
    take part in producing the neutrinos at high temperatures are
    $CP$ violating, i.e.\ they generate a lepton asymmetry at high
    temperatures.  Due to the interplay of inverse decay processes and
    lepton number violating $2\to2$ scatterings this asymmetry has a
    different sign compared to the one generated in neutrino
    decays at low temperatures, i.e.\ the asymmetries will
    partially cancel each other, as one can see in the change of sign
    of the asymmetry in fig.~\ref{sol1fig}a.  This cancellation can
    only be avoided if the asymmetry generated at high temperatures is
    washed out before the neutrinos decay. At high temperatures the
    lepton number violating scatterings behave like
    \beq
      {-z\over sH(M_1)}\;\g_i^{\scr \D L}\propto
      M_1\sum\limits_ja_j\wt{m}_j^2\;,
      \qquad i=A,\ldots,D\;.
    \eeq 
    Hence, the wash-out processes are more efficient for
    larger neutrino masses, i.e.\ the final asymmetry should grow with
    the neutrino mass $M_1$. The finally generated asymmetry is not
    affected by the stronger wash-out processes, since for small
    $\wt{m}_1$ the neutrinos decay late, where one can neglect the
    lepton number violating scatterings.

    This change of sign in the asymmetry is not observed in
    fig.~\ref{sol1fig}b. Due to the larger $\wt{m}_1$ value the
    neutrinos are brought into equilibrium at much higher
    temperatures, where decays and inverse decays are suppressed by a
    time dilatation factor, i.e.~the (s)neutrinos are produced in CP
    invariant scatterings off a (s)top.

    \clearpage\chapter{SO(10) Unification and Neutrino Mixing \label{Yuk}}
  In order to study the implications of leptogenesis for low-energy
  neutrino physics and leptonic flavour mixing we will assume a
  similar pattern of masses and mixings for the leptons and the quarks
  in this chapter \cite{bp,pluemi2}.

\section{Neutrino Masses and Mixings}
  If we choose a basis where the Majorana mass matrix $M$ and the
  Dirac mass matrix $m_l$ for the charged leptons are diagonal with
  real and positive eigenvalues,
  \beq
    m_l=\left(\begin{array}{ccc} m_e&0&0\\0&m_{\m}&0\\0&0&m_{\t}
        \end{array}\right)\;,\qquad\qquad
    M=\left(\begin{array}{ccc} M_1&0&0\\0&M_2&0\\0&0&M_3
        \end{array}\right)\;,
  \eeq
  the Dirac mass matrix of the neutrinos can be written in the form
  \beq
    m_{\scr D}=V\,\left(\begin{array}{ccc}
    m_1&0&0\\0&m_2&0\\0&0&m_3\end{array}\right)\,U^{\dag}\;,
    \label{u_def}
  \eeq
  where $V$ and $U$ are unitary matrices and the $m_i$ are real
  and positive.
  
  Since the Majorana masses $M$ are assumed to be much larger than the
  Dirac masses $m_{\scr D}$, we have 6 Majorana neutrinos as mass
  eigenstates \cite{seesaw}. In the weak eigenstate basis the mass
  matrix of the light neutrinos reads \cite{wyler}
  \beq
    m'_{\n}=-m_{\scr D}{1\over M}m_{\scr D}^T
    +\co\left({1\over M^3}\right)\;.\label{mweak}
  \eeq
  It can be diagonalized by a unitary matrix $K$, i.e.\ the light mass 
  eigenstates
  \beq
    \n_i\simeq(K^{\dg}\n_{\mbox{\tiny L}})_i
    +(\n^{\mbox{\tiny C}}_{\mbox{\tiny L}}K)_i\;,\qquad i=e,\m,\t\;,
  \eeq
  have masses
  \beq
    m_{\n}=-K^{\dg}m_{\scr D}{1\over M}m_{\scr D}^TK^*
    +\co\left({1\over M^3}\right)\equiv\left(
    \begin{array}{ccc} m_{\n_e}&0&0\\0&m_{\n_{\m}}&0\\0&0&m_{\n_{\t}}
        \end{array}\right)\;,
  \eeq
  whereas the heavy neutrino mass eigenstates
  \beq
    N_i\simeq\n_{i,\mbox{\tiny R}}
    +\n^{\mbox{\tiny C}}_{i,\mbox{\tiny R}}\;,\qquad i=1,2,3\;,
  \eeq
  have masses
  \beq
    m_{\scr N}= M+\co\left({1\over M}\right)\;.
  \eeq
  
  As we have seen in chapter \ref{theory}, all the quantities relevant
  for baryogenesis, i.e.\ the decay widths, $CP$ asymmetries and
  scattering cross sections, depend only on the product 
  $m_{\scr D}^{\dg}m_{\scr D}$, where the mixing matrix $V$ drops
  out. On the other hand, the mixing matrix $K$ in the leptonic
  charged current depends on the parameters of both unitary matrices
  $U$ and $V$. Hence, leptonic mixing and $CP$ violation at high and
  low energies are to leading order independent, i.e.\ the $CP$
  violation needed for baryogenesis does not allow to infer on $CP$
  violating interactions of light leptons.
  
  The mixing matrix $U$ can be parametrized by three mixing angles and
  six phases. Five of these phases can be factored out with the
  Gell-Mann matrices $\l_i$,
  \beq
    U=\mbox{e}^{i\g}\,\mbox{e}^{i\l_3\a}\,\mbox{e}^{i\l_8\b}\,U_1\,
    \mbox{e}^{i\l_3\s}\,\mbox{e}^{i\l_8\t}\;.
  \eeq  
  In analogy to the Cabibbo-Kobayashi-Maskawa (CKM) matrix for quarks
  the remaining matrix $U_1$ depends on three mixing angles and one
  phase. In unified theories based on SO(10) it is natural to assume
  a similar pattern of masses and mixings for leptons and quarks. This
  suggests the Wolfenstein parametrization \cite{wolfenstein} as an 
  ansatz for $U_1$,
  \beq
    U_1=\left(\begin{array}{ccc}
    1-{\l^2\over2}  &      \l        & A\l^3(\r-i\h) \\[1ex]
        -\l         & 1-{\l^2\over2} & A\l^2 \\[1ex]
    A\l^3(1-\r-i\h) &    -A\l^2      &  1
    \end{array}\right)\;,
    \label{mm}
  \eeq  
  where $A$ and $|\r+i\h|$ are of order one, while the mixing
  parameter $\l$ is assumed to be small. For the Dirac masses $m_i$,
  SO(10) unification motivates a hierarchy like for the up-type
  quarks,
  \beq
    m_1=b\l^4m_3\qquad m_2=c\l^2m_3\qquad b,c=\co(1)\;.
    \label{DMass}
  \eeq  
  We have mentioned in section \ref{boltzeq} that we also need a
  hierarchy in the Majorana masses $M_i$ to get a non-vanishing lepton
  asymmetry. We choose a similar hierarchy as in
  eq.~(\ref{DMass}),
  \beq
    M_1=B\l^4M_3\qquad M_2=C\l^2M_3\qquad B,C=\co(1)\;.
    \label{Majomass}
  \eeq
  Later on we will vary the parameters $B$ and $C$ to
  investigate different hierarchies for the right-handed neutrinos.

  Diagonalizing the neutrino mass matrix (\ref{mweak}) in powers of
  $\l$ yields the light neutrino masses
  \beqa
     m_{\n_e}&=&{b^2\over\left|C+\mbox{e}^{4i\a}\;B\right|}\;\l^4
           \;m_{\n_{\t}}+\co\left(\l^6\right)\label{mne}\;,\\[1ex]
     m_{\n_{\m}}&=&{c^2\left|C+\mbox{e}^{4i\a}\;B\right|\over BC}
           \;\l^2\;m_{\n_{\t}}+\co\left(\l^4\right)\label{mnm}\;,\\[1ex]
     m_{\n_{\t}}&=&{m_3^2\over M_3}+\co\left(\l^4\right)\;.\label{mnt}
  \eeqa
  We will not discuss the masses of the light scalar neutrinos here,
  since they depend on unknown soft breaking terms.

  In section \ref{gen} we have seen that the lepton asymmetry is
  largely determined by the mass parameter $\wt{m}_1$, which is
  given by
  \beq
    \wt{m}_1={c^2+A^2|\r+i\h|^2\over B}\;\l^2\;m_{\n_{\t}}=
    {C(c^2+A^2|\r+i\h|^2)\over c^2\left|C+\mbox{e}^{4i\a}\;B\right|}
    \;m_{\n_{\m}}\;,\label{mt}
  \eeq
  i.e.\ $\wt{m}_1$ is of the same order as the $\n_{\m}$ mass. 
  According to eq.~(\ref{SUSY_CP}) the $CP$ asymmetry in the decay of
  the lightest right-handed neutrino reads
  \beq
    \ve_1={3\over8\p}\;{B\;A^2\over c^2+A^2\;|\r+i\h|^2}\;\l^4\;
    {m_3^2\over v_2^2}\;\mbox{Im}\left[(\r-i\h)^2
    \mbox{e}^{i2(\a+\sqrt{3}\b)}\right]
    \;+\;\co\left(\l^6\right)\;.
  \eeq
  In the next section we will always assume maximal phases, i.e.\ 
  we will set
  \beq
    \ve_1= \;-{3\over8\p}\;{B\;A^2\;|\r+i\h|^2\over 
     c^2+A^2\;|\r+i\h|^2}\;\l^4\;
    {m_3^2\over v_2^2}\;+\;\co\left(\l^6\right)\;.
    \label{cpa}
  \eeq
  Hence, the lepton asymmetries that we are going to calculate may
  be viewed as upper bounds on the attainable asymmetries. 

  Like in the non-supersymmetric scenario a large value of the
  Yukawa-coupling $m_3/v_2$ will be preferred by this baryogenesis
  mechanism, since $\ve_1\propto m_3^2/v_2^2$. This holds irrespective
  of our ansatz for neutrino mixings.

\section{Numerical Results}

    The neutrino masses (\ref{mne})-(\ref{mnt}) can be used to
    constrain the free parameters of our ansatz. The strongest hint
    for a non-vanishing neutrino mass being the solar neutrino
    deficit\footnote{For a review and references, see \cite{haxt}.}, 
    we will fix the $\n_{\m}$ mass to the value preferred by the
    Mikheyev-Smirnov-Wolfenstein (MSW) solution \cite{msw},
    \beq
      m_{\n_{\m}}\simeq3\cdot10^{-3}\,\mbox{eV}\;.
      \label{numass}
    \eeq
    Hence the parameter $\wt{m}_1$, which is of the same order as
    $m_{\n_{\m}}$ according to eq.~(\ref{mt}), will be in the
    interval allowed by fig.~\ref{fig13}. 

\begin{figure}[t]
      \begin{minipage}[t]{8.5cm}
        \centerline{\hspace{1cm}(a)}
        \epsfig{file=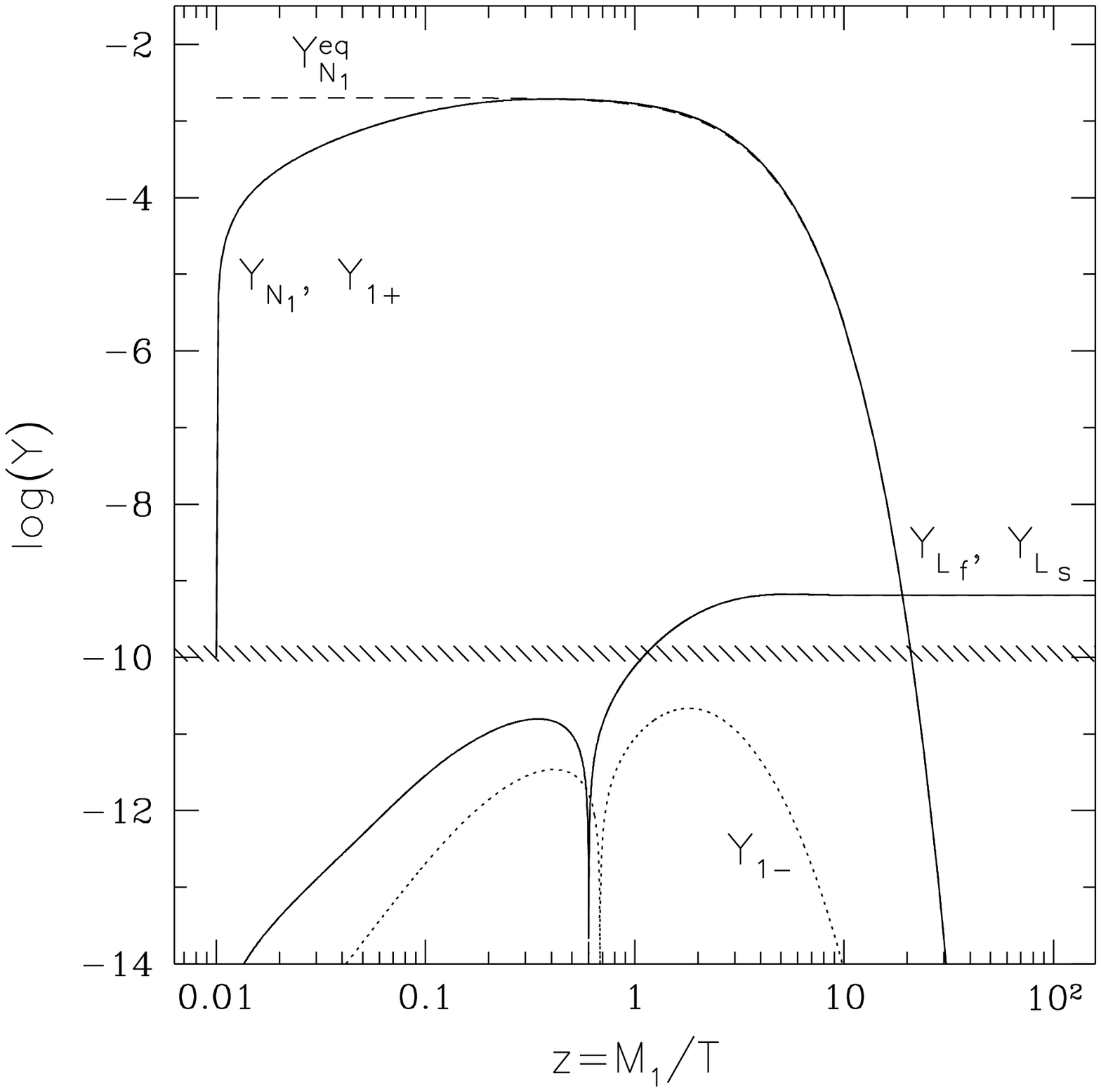,width=8.5cm}
      \end{minipage}
      \hspace{-0.5cm}
      \begin{minipage}[t]{8.5cm}
        \centerline{\hspace{1cm}(b)}
        \epsfig{file=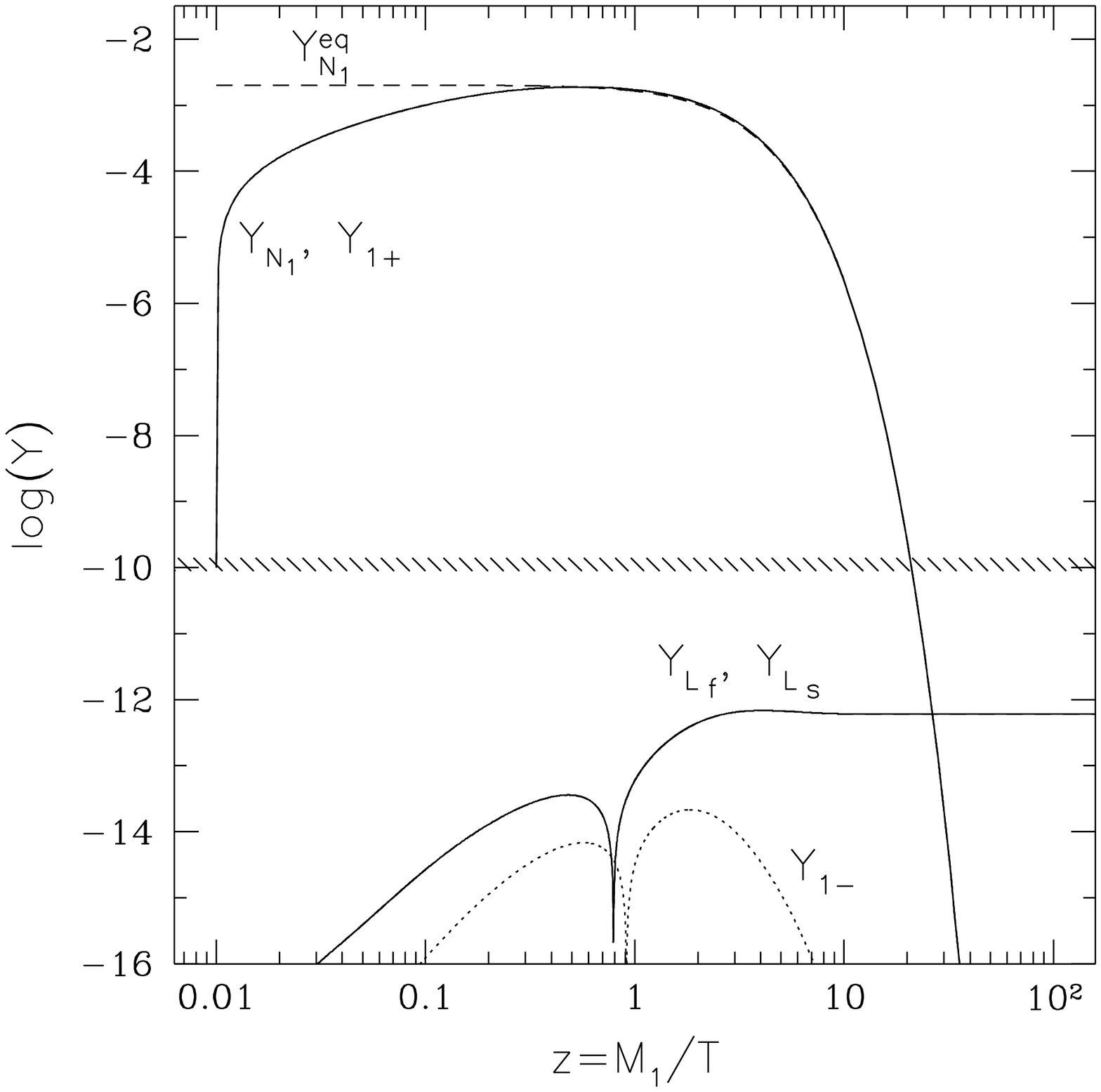,width=8.5cm}
     \end{minipage}  
       \caption{\it Generated asymmetry if one assumes a similar
        pattern of masses and mixings for the leptons and the
        quarks. In both figures we have $\l=0.1$ and $m_3=m_t$ (a) and
        $m_3=m_b$ (b).
       \label{sol2fig}}
    \end{figure}

    The most obvious parameter choice is to take all $\co(1)$
    parameters equal to one and to fix $\l$ to a similar
    value as the $\l$ parameter of the quark mixing matrix,
    \beqa
      &A=B=C=b=c=|\r+i\h|\simeq 1\;,&\label{p1a}\\[1ex]
      &\qquad \l\simeq 0.1\;.&\label{p1b}
    \eeqa
    The $\n_{\m}$ mass in eqs.~(\ref{mnm}) and (\ref{numass}) then fixes
    the $\n_e$ and $\n_{\t}$ masses,
    \beq
      m_{\n_e}\simeq 8\cdot10^{-6}\;\mbox{eV}\; , \quad
      m_{\n_{\t}}\simeq 0.15\;\mbox{eV}\;,
      \label{m1}
    \eeq
    and $\wt{m}_1$ reads
    \beq
      \wt{m}_1 \simeq 3\cdot10^{-3}\;\mbox{eV}\;.
    \eeq
    SO(10) unification suggests that the Dirac neutrino mass
    $m_3$ is equal to the top-quark mass,
    \beq
      m_3=m_t\simeq 174\;\mbox{GeV}\;.
      \label{3t}
    \eeq
    This leads to a large Majorana mass scale for the right-handed
    neutrinos,
    \beq
      M_3 \simeq 2\cdot10^{14}\;\mbox{GeV}\quad\Rightarrow\quad
      M_1\simeq 2\cdot10^{10}\;\mbox{GeV}\mbox{ and }
      M_2\simeq 2\cdot10^{12}\;\mbox{GeV}\;,
      \label{M3}
    \eeq
    and eq.~(\ref{cpa}) gives  the $CP$ asymmetry $\ve_1 \simeq 
    -6\cdot10^{-6}$. Integration of the Boltzmann equations yields 
    the $(B-L)$ asymmetry (cf.~fig.~\ref{sol2fig}a)
    \beq 
       Y_{B-L} \simeq 1\cdot10^{-9}\; , 
    \eeq 
    which is of the correct order of magnitude. It is interesting
    to note that in the non-supersymmetric scenario one has $Y_{B-L} 
    \simeq 9\cdot10^{-10}$ for the same choice of parameters. 
    \begin{figure}[t]
        \centerline{\epsfig{file=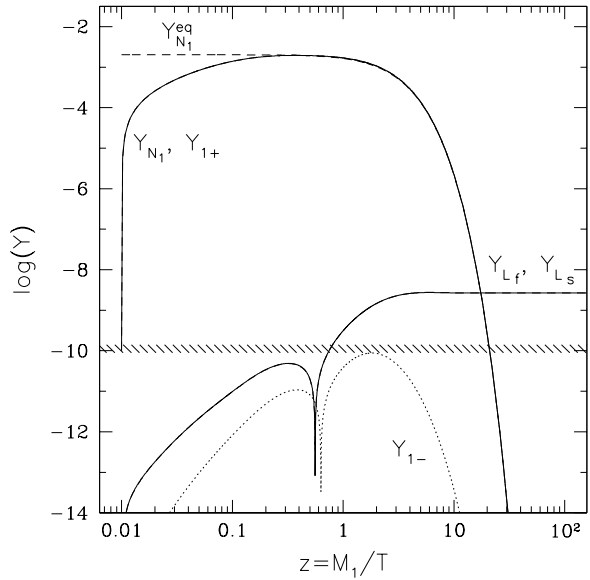,width=8.5cm}}
        \caption{\it Solution of the Boltzmann equations suggested by
          the atmospheric neutrino problem. \label{atm}}
    \end{figure}
    \begin{figure}[t]
        \centerline{\epsfig{file=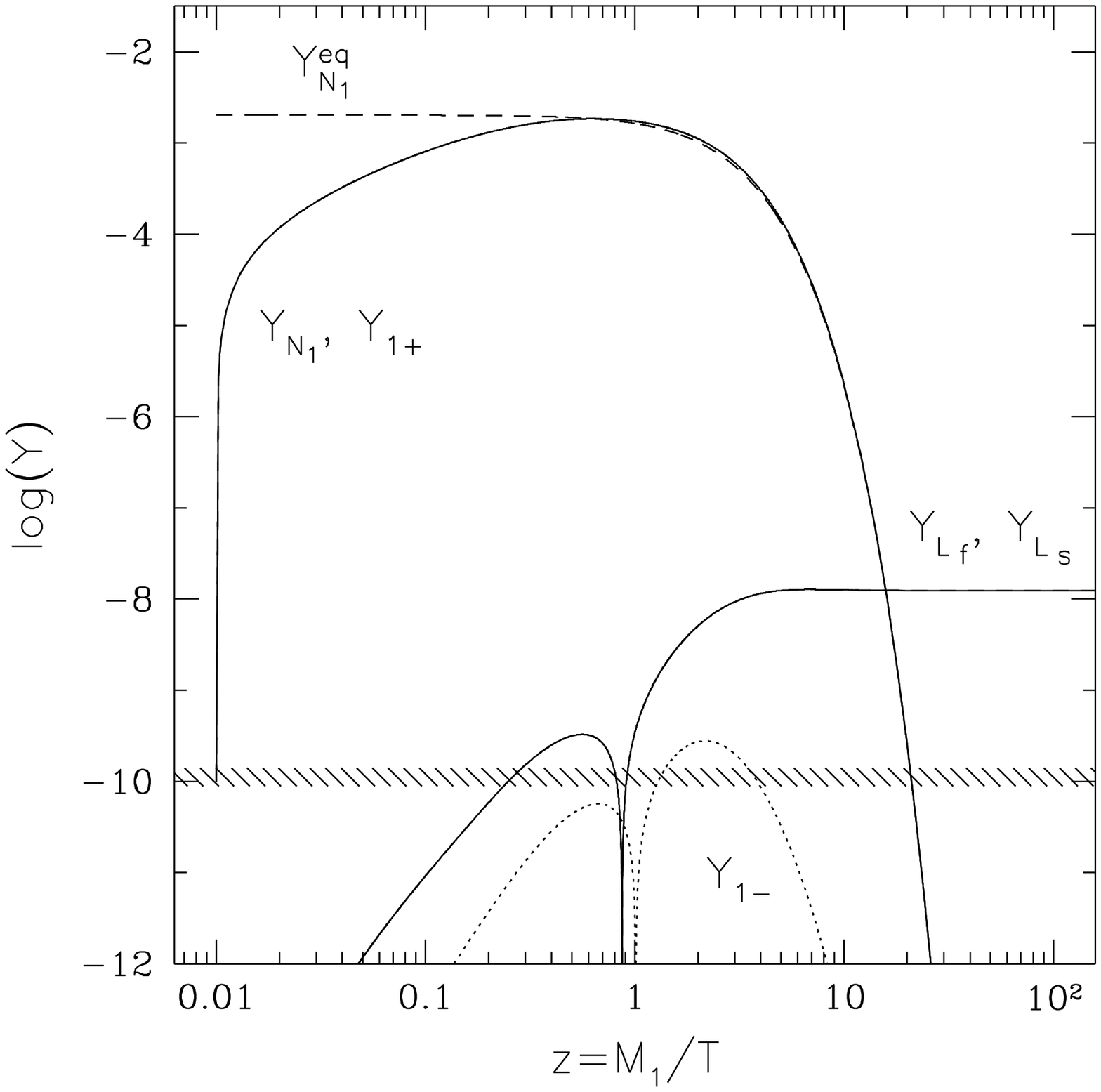,width=8.5cm}}
        \caption{\it Generated lepton asymmetry if one assumes a
        similar mass hierarchy for the right-handed neutrinos and the
        down-type quarks.
        \label{hierarch}}
    \end{figure}

    Our assumption (\ref{3t}), $m_3 \simeq m_t$ led to a large
    Majorana mass scale $M_3$ in eq.~(\ref{M3}). To check the
    sensitivity of our result for the baryon asymmetry on this choice,
    we have envisaged a lower Dirac mass scale
    \beq 
      m_3 = m_b \simeq 4.5\; \mbox{GeV}\; , 
    \eeq 
    while keeping all other parameters in eqs.~(\ref{p1a}) and
    (\ref{p1b}) fixed. The assumed $\n_{\m}$ mass (\ref{numass}) then
    yields a much lower value for the Majorana mass scale,
    \beq
      M_3\simeq1\cdot10^{11}\;\mbox{GeV}\quad\Rightarrow\quad
      M_1\simeq1\cdot10^{7}\;\mbox{GeV}\mbox{ and }
      M_2\simeq1\cdot10^{9}\;\mbox{GeV}\;,
    \eeq
    while the light neutrino masses (\ref{m1}) remain unchanged. The
    $CP$ asymmetry $\ve_1\simeq-4\cdot10^{-9}$ becomes very small.
    Consequently, the generated baryon asymmetry
    (cf.~fig.~\ref{sol2fig}b)
    \beq 
      Y_{B-L} \simeq 1\cdot10^{-12}\;, 
    \eeq    
    is too small by two orders of magnitude. We can conclude that high
    values for both masses $m_3$ and $M_3$ are preferred.  This
    suggests that $(B-L)$ is already broken at the unification scale
    $\L_{\mbox{\tiny GUT}} \sim 10^{16}\;$GeV, without any intermediate
    scale of symmetry breaking, which is natural in SO(10)
    unification. Alternatively, a Majorana mass scale of the order of
    $10^{12}$ to $10^{14}\;$GeV can also be generated radiatively if
    SO(10) is broken into SU(5) at some high scale between
    $10^{16}\;$GeV and the Planck-scale, and SU(5) is subsequently
    broken into the MSSM gauge group at the usual GUT scale $\sim
    10^{16}\;$GeV \cite{su5}.
    
    Such a Majorana mass scale naturally leads to a baryogenesis scale
    $M_1\sim10^{10}\;$GeV. As discussed in section \ref{gen} and as
    one can see in fig.~\ref{sol2fig}, the neutrinos can be brought
    into equilibrium at temperatures slightly above their mass, i.e.\ 
    this scenario requires a reheating temperature $\sim10^{10}\;$GeV
    at the end of inflation. This is well compatible with the
    constraints on the reheating temperature from the gravitino
    problem \cite{gravitino}.

    It is interesting to note that the $\n_{\t}$ mass in eq.~(\ref{m1})
    has got almost the value suggested by the atmospheric
    neutrino problem \cite{atmospheric}
    \beq
      m_{\n_{\t}}\approx 7\cdot10^{-2}\;\mbox{eV}\;,\label{nutau_atm}
    \eeq
    if one assumes that the anomalous $\m$ to $e$ ratio produced by
    atmospheric neutrinos is due to oscillations from $\n_{\m}$ to
    $\n_{\t}$, and when the $\n_{\m}$ mass is again given by the MSW
    value in eq.~(\ref{numass}). If we then use eq.~(\ref{nutau_atm})
    to fix the $\n_{\t}$ mass,the Dirac mass scale (\ref{3t})
    determines the Majorana mass scale
    \beq
      M_3\simeq 4\cdot10^{14}\;\mbox{GeV}\;.
    \eeq
    The ratio of the $\n_{\t}$ and $\n_{\m}$ masses then yields the
    mixing parameter 
    \beq
      \l=0.15\;,
    \eeq
    if the $\co(1)$ parameters are given by eq.~(\ref{p1a}). The
    remaining neutrino masses read
    \beq
      M_1\simeq2\cdot10^{11}\;\mbox{GeV}\;,\qquad
      M_2\simeq9\cdot10^{12}\;\mbox{GeV}\;,\qquad
      m_{\n_e}\simeq2\cdot10^{-5}\;\mbox{eV}\;.
    \eeq
    Consequently, we again get a large $CP$ asymmetry $\ve_1\simeq
    -3\cdot10^{-5}$, and a large $(B-L)$ asymmetry (cf.~fig.~\ref{atm})
    \beq
      Y_{B-L}\simeq5\cdot10^{-9}\;.
    \eeq
    This shows that the parameters required to explain the anomalous
    results of neutrino experiments by neutrino oscillations also
    predict a baryon asymmetry of the correct order of magnitude,
    although the large mixing angle which seems to be required to
    solve the atmospheric neutrino problem is difficult to accommodate
    within our small-mixing ansatz.
    
    Up to now we have always assumed a mass hierarchy for the heavy
    Majorana neutrinos like for the up-type quarks.  Alternatively,
    one can assume a weaker hierarchy, like for the down-type quarks
    by choosing
    \beq
      B=10 \qquad \mbox{and} \qquad C=3\;.
    \eeq  
    Keeping the other parameters in eqs.~(\ref{p1a}) and (\ref{p1b}) 
    unchanged fixes the $\n_e$ and $\n_{\t}$ masses,
    \beq
      m_{\n_e} \simeq 5\cdot10^{-6}\;\mbox{eV}\;,\qquad\qquad 
        m_{\n_{\t}} \simeq 0.7\;\mbox{eV}\;,
    \eeq
    and the mass parameter $\wt{m}_1$,
    \beq
      \wt{m}_1 \simeq 1\cdot10^{-3}\;\mbox{eV}\;.
    \eeq
    Choosing the Dirac mass scale (\ref{3t}) we get a large Majorana 
    mass scale 
    \beq
      M_3 \simeq 4\cdot10^{13}\;\mbox{GeV}\quad\Rightarrow\quad
      M_1 \simeq 4\cdot10^{10}\;\mbox{GeV}\mbox{ and }
      M_2 \simeq 10^{12}\;\mbox{GeV}\;.
    \eeq
    {}From eq.~(\ref{cpa}) one obtains the $CP$ asymmetry 
    $\ve_1\simeq-6\cdot10^{-5}$. The corresponding solutions of the
    Boltzmann equations are shown in fig.~\ref{hierarch}. The final
    $(B-L)$ asymmetry,
    \beq
      Y_{B-L} \simeq 2\cdot10^{-8}\;,
    \eeq    
    is much larger than requested, but this value can always be
    lowered by adjusting the unknown phases. Hence, the possibility to
    generate a lepton asymmetry does not depend on the special form of
    the mass hierarchy assumed for the right-handed neutrinos.

    In the non-supersymmetric scenario one finds for the same parameter
    choice
    \beq
      Y_{B-L} \simeq 2\cdot10^{-8}\;.
    \eeq
    Hence, when comparing the supersymmetric and the
    non-supersymmetric scenario, one sees that the larger $CP$
    asymmetry in the former and the additional contributions from the
    sneutrino decays are compensated by the wash-out processes which
    are stronger than in the latter.

    \clearpage\chapter*{Conclusions}
\addcontentsline{toc}{chapter}{Conclusions}
\markboth{CONCLUSIONS}{CONCLUSIONS}
  We have analysed in detail the generation of a cosmological baryon
  asymmetry by out-of-equilibrium decays of heavy right-handed Majorana
  neutrinos and their scalar partners in a supersymmetric extension of
  the Standard Model. By developing a resummed perturbative expansion
  in flavour non-diagonal self-energies, we could show how self-energy
  contributions to $CP$ asymmetries in heavy neutrino decays are
  consistently taken into account.
  
  We have discussed all the decays and scattering processes relevant
  for leptogenesis, and by solving the Boltzmann equations we have
  shown that, in order to be consistent, one has to pay attention to
  two phenomena which can hamper the generation of a lepton asymmetry.
  
  First, one has to take into consideration lepton number violating
  scatterings mediated by a heavy neutrino or its scalar partner.
  These processes, which are usually neglected, can wash out a large
  part of the asymmetry if the Yukawa couplings of the right-handed
  neutrinos become too large.
  
  On the other hand, the neutrinos have to be brought into thermal
  equilibrium at high temperatures. We could show that for this
  purpose it is not necessary to assume additional interactions of the
  right-handed neutrinos in our theory, since the Yukawa interactions
  can be sufficiently strong to produce a thermal population of heavy
  neutrinos at high temperatures, while still being weak enough to
  prevent the final asymmetry from being washed out.
  
  The observed baryon asymmetry can be obtained without any fine
  tuning of parameters if one assumes a similar pattern of mixings and
  Dirac masses for the neutrinos and the up-type quarks. Then the
  generated asymmetry is related to the $\n_{\m}$ mass, and fixing this
  mass to the value preferred by the MSW-solution to the solar
  neutrino problem leads to a baryon asymmetry of the requested order,
  provided $(B-L)$ is broken at the unification scale, as suggested by
  supersymmetric SO(10) unification. The baryon asymmetry is generated
  at a scale of approximately $10^{10}\;$GeV, which looks promising
  with respect to the gravitino problem.
  
  \pagebreak   
  In supersymmetric theories there are further possible sources of a
  $(B-L)$ asymmetry, e.g.\ it may be possible to combine inflation
  with leptogenesis by using a right-handed scalar neutrino as the
  inflaton (cf.~refs.~\cite{infl}). In this connection, possible
  constraints on the neutrino masses and on the reheating temperature
  from lepton number violating processes at low temperatures require
  further studies.
  
  Furthermore, it should be studied to which extent a ``primordial''
  baryon asymmetry, i.e.\ an asymmetry generated during or shortly
  after reheating, is affected if right-handed neutrinos come into
  equilibrium after reheating. This may yield interesting constraints
  on Yukawa interactions of first generation leptons, since an
  asymmetry in right-handed electrons might be protected from being
  washed-out if right-handed electrons decouple from the thermal
  plasma \cite{kimmo}.
  
  Finally, as discussed in the first chapter, one has to separate
  ``on-shell'' and ``off-shell'' contributions to lepton number
  violating scatterings, in order to be able to describe the
  generation of a baryon asymmetry in terms of Boltzmann equations. To
  avoid this separation and treat these contributions simultaneously,
  one has to go beyond the semi-classical approximation realized in
  the Boltzmann equations by constructing a complete nonequilibrium
  quantum kinetic theory. Although such a theory has not been realized
  up to now, quantum corrections to the Boltzmann equations have
  recently been investigated in refs.~\cite{beyond}.

  \begin{appendix}
    \clearpage\chapter{One-Loop Integrals \label{integrals}}

  We summarize some standard formulae for dimensionally
  regularized one-loop integrals in Minkowski space. We follow the 
  notation of refs.~\cite{passarino}, although we use a metric 
  $g_{\m\n}=(+,-,-,-)$ (cf.~e.g.~ref.~\cite{denner}).
 
\section{One-Point Function}
  In $n=4-2\e$ dimensions the scalar one-point function is defined by
  \beq
    A(m_1)={\m^{4-n}\over i\p^2}\int d^nk\,
    {1\over k^2-m_1^2+i\ve}\;.
  \eeq
  In the limit $\e\to0$, $A(m_1)$ is given by
  \beq
    A(m_1)=m_1^2\left(\D-\ln\left({m_1^2\over\m^2}\right)+1\right)\;,
  \eeq
  where the UV-divergence is contained in
  \beq
    \D={1\over\e}-C+\ln(4\p)\;,
  \eeq
  and $C=0.577216$ is Euler's constant. Note that the massless tadpole
  $A(0)$ vanishes in dimensional regularization, and that $A(m_1)$ has
  no absorptive contribution
  \beq
    \mbox{Im}\left[A(m_1)\right]=0\;.\label{ImA}
  \eeq

\section{Two-Point Functions}
  Three different two-point integrals can occur
  \beqa
    B_0(p_1^2,m_1,m_2)&=&{\m^{4-n}\over i\p^2}\int d^nk\,
      {1\over (k^2-m_1^2+i\ve)[(k+p_1)^2-m_2^2+i\ve] }\label{B0def}\\[1ex]
    B_{\m}(p_1^2,m_1,m_2)&=&{\m^{4-n}\over i\p^2}\int d^nk\,
      {k_{\m}\over (k^2-m_1^2+i\ve)[(k+p_1)^2-m_2^2+i\ve]}
      \label{Bmdef}\\[1ex]
    B_{\m\n}(p_1^2,m_1,m_2)&=&{\m^{4-n}\over i\p^2}\int d^nk\,
      {k_{\m}k_{\n}\over (k^2-m_1^2+i\ve)[(k+p_1)^2-m_2^2+i\ve]}
      \label{Bmndef}
  \eeqa
  Lorentz covariance of the integrals allows to decompose the tensor
  integrals into tensors constructed from the external momentum $p_1$,
  and the metric tensor $g_{\m\n}$
  \beqa
    B_{\m}(p_1^2,m_1,m_2)&=&p_{1,\m}B_1(p_1^2,m_1,m_2)\;,
      \label{B1def}\\[1ex]
    B_{\m\n}(p_1^2,m_1,m_2)&=&p_{1,\m}p_{1,\n}B_{21}(p_1^2,m_1,m_2)
      +g_{\m\n}B_{22}(p_1^2,m_1,m_2)\;.\label{B21def}
  \eeqa
  Using the Feynman parametrization, one can derive an integral
  representation for $B_0$,
  \beq
    B_0(p_1^2,m_1,m_2)=\D-\int\limits_0^1dx\,\ln\left({x^2p_1^2-
    x(p_1^2+m_1^2-m_2^2)+m_1^2-i\ve\over\m^2}\right)+\co(n-4)\;,
  \eeq
  which yields the following useful identities in the limit $n\to4$
  \beqa
    B_0(p_1^2,0,0)&=&\D-\ln\left({|p_1^2|\over\m^2}\right)+2
      +i\p\q(p_1^2)\;,\label{B0finite}\\[1ex]
    B_0(0,0,m)&=&B_0(0,m,0)=\D-\ln\left({m^2\over\m^2}\right)+1
      ={1\over m^2}A(m^2)\;.
  \eeqa
  Contracting eqs.~(\ref{Bmdef})-(\ref{B21def}) with $p_{1,\m}$ and 
  $g_{\m\n}$ yields a set of coupled linear equations, which determine
  the scalar coefficients $B_1$, $B_{21}$ and $B_{22}$,
  \beqa
    B_1(p_1^2,m_1,m_2)&=&{1\over2p_1^2}\left[A(m_1)-A(m_2)
      +(m_2^2-m_1^2-p_1^2)B_0(p_1^2,m_1,m_2)\right]\\
    &=&-{1\over2\e}+\mbox{ UV-finite parts,}\\[1ex]
    B_{21}(p_1^2,m_1,m_2)&=&{1\over3p_1^2}\left[A(m_2)
      -m_1^2B_0(p_1^2,m_1,m_2)\right.\NO\\
    &&\left.-2(p_1^2+m_1^2-m_2^2)B_1(p_1^2,m_1,m_2)
      -{1\over2}(m_1^2+m_2^2-{1\over3}p_1^2)\right]\\
    &=&{1\over3\e}+\mbox{ UV-finite parts,}\\[1ex]
    B_{22}(p_1^2,m_1,m_2)&=&{1\over6}\left[A(m_2)
      +2m_1^2B_0(p_1^2,m_1,m_2)\right.\NO\\
    &&\left.+(p_1^2+m_1^2-m_2^2)B_1(p_1^2,m_1,m_2)
      +m_1^2+m_2^2-{1\over3}p_1^2\right]\\
    &=&-{1\over12\e}\left(p_1^2-3m_1^2-3m_2^2\right)
    +\mbox{ UV-finite parts.}
  \eeqa
  In the case of equal or vanishing masses one has
  \beqa
    B_1(p_1^2,m,m)&=&-{1\over2}B_0(p_1^2,m,m)\;,\\[1ex]
    B_{21}(p_1^2,0,0)&=&
      {1\over3}\left[B_0(p_1^2,0,0)+{1\over6}\right]\;,\\[1ex]
    B_{22}(p_1^2,0,0)&=&
      {-p_1^2\over12}\left[B_0(p_1^2,0,0)+{2\over3}\right]\;.
  \eeqa

\section{Three-Point Functions \label{A3}}
  In general one has four different three-point functions
  \beqa
    C_0(p_1^2,p_2^2,m_1,m_2,m_3)&=&{\m^{4-n}\over i\p^2}
      \int d^nk\,{1\over D_{m_1}D_{m_2}(p_1)D_{m_3}(p_1,p_2)}
      \;,\label{defC0}\\[1ex]
    C_{\m}(p_1^2,p_2^2,m_1,m_2,m_3)&=&{\m^{4-n}\over i\p^2}
      \int d^nk\,{k_{\m}\over D_{m_1}D_{m_2}(p_1)D_{m_3}(p_1,p_2)}
      \;,\label{defC1}\\[1ex]
    C_{\m\n}(p_1^2,p_2^2,m_1,m_2,m_3)&=&{\m^{4-n}\over i\p^2}
      \int d^nk\,{k_{\m}k_{\n}\over D_{m_1}D_{m_2}(p_1)D_{m_3}(p_1,p_2)}
      \;,\label{defC2}\\[1ex]
    C_{\m\n\r}(p_1^2,p_2^2,m_1,m_2,m_3)&=&{\m^{4-n}\over i\p^2}
      \int d^nk\,{k_{\m}k_{\n}k_{\r}\over D_{m_1}D_{m_2}(p_1)
      D_{m_3}(p_1,p_2)}\;,\label{defC3}
  \eeqa
  where we have introduced the following abbreviations for inverse
  propagators 
  \beqa
    D_{m_1}&=&k^2-m_1^2+i\ve\;,\\[1ex]
    D_{m_2}(p_1)&=&(k+p_1)^2-m_2^2+i\ve\;,\\[1ex]
    D_{m_3}(p_1,p_2)&=&(k+p_1+p_2)^2-m_3^2+i\ve\;.
  \eeqa
  Like for the two point-functions, Lorentz covariance of the
  integrals suggests the following tensor decomposition of the tensor
  three-point functions
  \beqa
    C_{\m}(p_1^2,p_2^2,m_1,m_2,m_3)&=&p_{1,\m}C_{11}+p_{2,\m}C_{12}
      \;,\label{defC4}\\[1ex]
    C_{\m\n}(p_1^2,p_2^2,m_1,m_2,m_3)&=&p_{1,\m}p_{1,\n}C_{21}+
      p_{2,\m}p_{2,\n}C_{22}\NO\\
    &&+(p_1p_2)_{(\m\n)}C_{23}+g_{\m\n}C_{24}\;,\label{defC5}\\[1ex]    
    C_{\m\n\r}(p_1^2,p_2^2,m_1,m_2,m_3)&=&p_{1,\m}p_{1,\n}p_{1,\r}C_{31}+
      p_{2,\m}p_{2,\n}p_{2,\r}C_{32}\NO\\
    &&+(p_2p_1p_1)_{(\m\n\r)}C_{33}+(p_1p_2p_2)_{(\m\n\r)}C_{34}\NO\\
    &&+(p_1g)_{(\m\n\r)}C_{35}+(p_2g)_{(\m\n\r)}C_{36}\;,\label{defC6}
  \eeqa
  where we have used the following abbreviations for index
  symmetrizations 
  \beqa
    (p_1p_2)_{(\m\n)}&=&p_{1,\m}p_{2,\n}+p_{1,\n}p_{2,\m}\;,\\[1ex]
    (p_1p_2p_2)_{(\m\n\r)}&=&p_{1,\m}p_{2,\n}p_{2,\r}+
      p_{2,\m}p_{1,\n}p_{2,\r}+p_{2,\m}p_{2,\n}p_{1,\r}\;,\\[1ex]
    (p_1g)_{(\m\n\r)}&=&p_{1,\m}g_{\n\r}+p_{1,\n}g_{\m\r}+
      p_{1,\r}g_{\m\n}\;.
  \eeqa

  The form factors $C_{ij}$ can be related to the scalar functions
  $A$, $B_0$ and $C_0$ by contracting the definitions
  (\ref{defC1})-(\ref{defC3}) and (\ref{defC4})-(\ref{defC6}) with
  external momenta $p_1$, $p_2$ and the metric $g^{\m\n}$.

  In our calculation we only need three-point functions with two
  vanishing masses ($m_2=m_3=0$), and two light-like momenta $p_1^2=0$
  and $(p_1+p_2)^2=0$. Then $C_{12}$ and $C_{11}$ read
  \beqa
    C_{12}(p_1,p_2,m,0,0)&=&{1\over2p_1\cdot p_2}\left[B_0(0,m,0)
      -B_0(p_2^2,0,0)-m^2C_0(p_1,p_2,m,0,0)\right]\;,\\[1ex]
    C_{11}(p_1,p_2,m,0,0)&=&2C_{12}(p_1,p_2,m,0,0)\;,
  \eeqa
  and the imaginary parts of $C_0$ and $C_{12}$ are given by
  \beqa
    \mbox{Im}\left[C_0(p_1,p_2,m,0,0)\right]&=&
      -{\p\q(p_2^2)\over p_2^2}\ln\left(1+{p_2^2\over m^2}\right)\;,\\[1ex]
    \mbox{Im}\left[C_{12}(p_1,p_2,m,0,0)\right]&=&
      {\p\q(p_2^2)\over p_2^2}\left[1-{m^2\over p_2^2}
      \ln\left(1+{p_2^2\over m^2}\right)\right]\;.
  \eeqa

    \clearpage\chapter{Spinor Notation and Conventions\label{SpinorConv}}
  We will use the conventions of ref.~\cite{wb} with flat space-time
  metric $g_{\m\n}=(+,-,-,-).$ Greek indices $\a$,
  $\b$, $\dt{\a}$ and $\dt{\b}$ run from one to two and denote
  two-component Weyl spinors, while all other Greek letters denote
  Lorentz-indices. 

\section{Weyl Spinors}
  Two-component spinors $\j$ and $\Bar{\j}$ transform under the
  $({1\over2},0)$ and $(0,{1\over2})$ representations of the Lorentz
  group SO$(1,3)$. Matrix representations are given by the universal
  covering group SL$(2,\mathbb{C})$ of SO$(1,3)$, i.e.\ under a
  Lorentz transformation $M\in\mbox{SL}(2,\mathbb{C})$ the Weyl
  spinors transform like
  \beqa
    \j_{\a}'={M_{\a}}^{\b}\j_{\b}\;,&\qquad&\Bar{\j'}_{\dt{\a}}=
    {{M^*}_{\dt{\a}}}^{\dt{\b}}\,\Bar{\j}_{\dt{\b}}\;,\\[1ex]
    {\j'}^{\,\a}={{M^{-1}}_{\b}}^{\a}\j^{\b}\;,&\qquad&
    \Bar{\j'}^{\,\dt{\a}}={{(M^*)^{-1}}_{\dt{\b}}}^{\dt{\a}}\,
    \Bar{\j}^{\,\dt{\b}}\;.
  \eeqa
  The Pauli matrices ${\s_{\a\dt{\a}}}^{\m}$ form a basis of 
  SL$(2,\mathbb{C})$,
  \beqa
    \s^0=\left(\begin{array}{cc}1&0\\0&1\end{array}\right)\;,&\qquad&
    \s^1=\left(\begin{array}{cc}0&1\\1&0\end{array}\right)\;,\\[1ex]
    \s^2=\left(\begin{array}{cc}0&-i\\i&0\end{array}\right)\;,&\qquad&
    \s^3=\left(\begin{array}{cc}1&0\\0&-1\end{array}\right)\;.
  \eeqa
  Spinor indices can be raised and lowered using the antisymmetric
  $\ve$-tensors
   \beqa
    \ve_{\a\b}=\left(\begin{array}{cc}0&-1\\1&0\end{array}\right)\;,
    &\qquad&
    \ve^{\a\b}=\left(\begin{array}{cc}0&1\\-1&0\end{array}\right)\;,\\[1ex]
    \ve_{\dt{\a}\dt{\b}}=\left(\begin{array}{cc}0&-1\\1&0\end{array}
    \right)\;,&\qquad&
    \ve^{\dt{\a}\dt{\b}}=\left(\begin{array}{cc}0&1\\-1&0
    \end{array}\right)\;,
  \eeqa
  i.e.\ one has
  \beqa
    \j^{\a}=\ve^{\a\b}\j_{\b}\;,&\qquad&\j_{\a}=\ve_{\a\b}\j^{\b}\;,\\
    \j^{\dt{\a}}=\ve^{\dt{\a}\dt{\b}}\j_{\dt{\b}}\;,&\qquad&
    \j_{\dt{\a}}=\ve_{\dt{\a}\dt{\b}}\j^{\dt{\b}}\;.
  \eeqa
  Correspondingly, spinor indices of Pauli matrices can be raised and
  lowered,
  \beq
    \Bar{\s}^{\,\m\dt{\a}\a}
    =\ve^{\dt{\a}\dt{\b}}\ve^{\a\b}\s^{\m}_{\b\dt{\b}}\;,
  \eeq
  i.e.\ the matrices $\Bar{\s}^{\,\m}$ are given by
  \beq
    \Bar{\s}^{\,0}=\s^0\;,\qquad\Bar{\s}^{\,1,2,3}=-\s^{1,2,3}\;.
  \eeq
  The Pauli matrices fulfil
  \beqa
    {\left(\s^{\m}\Bar{\s}^{\,\n}+\s^{\n}\Bar{\s}^{\,\m}\right)_{\a}}^{\b}
    &=&2\,g^{\m\n}{\d_{\a}}^{\b}\;,\\[1ex]
    {\left(\Bar{\s}^{\,\m}\s^{\n}+\Bar{\s}^{\,\n}\s^{\m}\right)
    ^{\dt{\a}}}_{\dt{\b}}&=&2\,g^{\m\n}{\d^{\dt{\a}}}_{\dt{\b}}\;,\\[1ex]
    \mbox{Tr}\;\s^{\m}\Bar{\s}^{\,\n}&=&2\,g^{\m\n}\;,\\[1ex]
    {\s_{\a\dt{\a}}}^{\m}\,{\Bar{\s}_{\m}}^{\dt{\b}\b}
    &=&2\,{\d_{\a}}^{\b}{\d_{\dt{\a}}}^{\dt{\b}}\;.
  \eeqa
  Products of two-component spinors are defined such that
  \beqa
    &\j\c\equiv\j^{\a}\c_{\raisebox{-0.5ex}{$\scriptstyle \a$}}=
      -\j_{\a}\c^{\a}=\c^{\a}\j_{\a}=\c\j\;,&\\[1ex]
    &\Bar{\j}\,\Bar{\c}\equiv\Bar{\j}_{\,\dt{\a}}\,\Bar{\c}^{\,\dt{\a}}=
      -\Bar{\j}^{\,\dt{\a}}\,
      \Bar{\c}_{\raisebox{-0.5ex}{$\scriptstyle\dt{\a}$}}=
      \Bar{\c}_{\raisebox{-0.5ex}{$\scriptstyle\dt{\a}$}}\,
      \Bar{\j}^{\,\dt{\a}}=\Bar{\c}\,\Bar{\j}\;,&\\[1ex]
    &(\c\j)^{\dg}=\left(\c^{\a}\j_{\a}\right)^{\dg}=\Bar{\j}_{\dt{\a}}\,
    \Bar{\c}^{\,\dt{\a}}=\Bar{\j}\,\Bar{\c}\;.&
  \eeqa
  Products of Weyl-spinors involving Pauli matrices read
  \beqa
    \c\s^{\m}\,\Bar{\j}&=&-\Bar{\j}\,\Bar{\s}^{\,\m}\c\;,\\[1ex]
    \left(\c\s^{\m}\,\Bar{\j}\,\right)^{\dg}&=&
      \j\,\s^{\m}\,\Bar{\c}\;,\\[1ex]
    \c\s^{\m}\Bar{\s}^{\,\n}\j&=&
      \j\,\s^{\n}\Bar{\s}^{\,\m}\c\;,\\[1ex]
    \left(\c\s^{\m}\,\Bar{\s}^{\,\n}\j\right)^{\dg}&=&
      \Bar{\j}\,\Bar{\s}^{\,\n}\s^{\m}\,\Bar{\c}\;,\\[1ex]
    (\j\l)\Bar{\c}_{\raisebox{-0.5ex}{$\scriptstyle\dt{\a}$}}&=&
      {1\over2}\left(\l\,\s^{\m}\,\Bar{\c}\right)
      \left(\j\,\s_{\m}\right)_{\dt{\a}}\;.
  \eeqa
  Furthermore, when computing superfield products
  (cf.~app.~\ref{sfproducts}), one can take advantage of the
  relations
  \beqa
    \q^{\a}\q^{\b}&=&-{1\over2}\,\ve^{\a\b}\,\q^2\;,\\[1ex]
    \q_{\a}\q_{\b}&=&{1\over2}\,\ve_{\a\b}\,\q^2\;,\\[1ex]
    \Bar{\q}^{\,\dt{\a}}\Bar{\q}^{\,\dt{\b}}&=&
      {1\over2}\,\ve^{\dt{\a}\dt{\b}}\,\Bar{\q}^{\,2}\;,\\[1ex]
    \Bar{\q}_{\dt{\a}}\Bar{\q}_{\dt{\b}}&=&
      -{1\over2}\,\ve_{\dt{\a}\dt{\b}}\,\Bar{\q}^{\,2}\;,\\[1ex]
    \q\s^{\m}\Bar{\q}\,\q\s^{\n}\Bar{\q}&=&
      {1\over2}\,\q^2\,\Bar{\q}^{\,2}\,g^{\m\n}\;.
  \eeqa

\section{Four-Component Spinors}
  We define the Dirac $\g$-matrices as
  \beq
    \g^{\m}=\left(\begin{array}{cc}0&\s^{\m}\\{\Bar{\s}}^{\,\m}&0
    \end{array}\right)\;,\qquad\g^{5}=i\g^0\g^1\g^2\g^3=\left(
    \begin{array}{cc}-\openone&0\\0&\openone\end{array}\right)\;,
  \eeq
  obeying
  \beqa
    &&\{\g^{\m},\g^{\n}\}=2g^{\m\n}\;,\\[1ex]
    &&\{\g^5,\g^{\m}\}=0\;,\\[1ex]
    &&(\g^5)^2=1\;.
  \eeqa
  As usual, $\g^0$ intertwines the $\g^{\m}$ representation of the
  Dirac algebra with the equivalent hermitian conjugated
  representation ${\g^{\m}}^{\dg}$
  \beqa
    \g^0\,\g^{\m}\,\g^0&=&{\g^{\m}}^{\dg}\;,\\[1ex]
    \g^0\,\g^5\,\g^0&=&-{\g^5}^{\dg}=-\g^5\;.
  \eeqa
  This representation of the $\g$-matrices can be used to relate Weyl
  spinors to the more familiar four-component spinors. A Dirac spinor
  $\J_{\mbox{\tiny D}}$ consists of two Weyl spinors
  \beq
    \J_{\mbox{\tiny D}}={\c_{\raisebox{-0.5ex}{$\scriptstyle \a$}}
    \choose\,\Bar{\l}^{\,\dt{\a}}}\;,\label{dirac}
  \eeq
  i.e.\ its hermitian conjugate reads
  \beq
    \Bar{\J}_{\mbox{\tiny D}}\equiv\J^{\dg}_{\mbox{\tiny D}}\g^0=
    \left(\l^{\a},\;\Bar{\c}_{\raisebox{-0.5ex}{$\scriptstyle\dt{\a}$}}
    \,\right)\;.
  \eeq
  The chiral projectors read
  \beq
    P_{\mbox{\tiny R}}={1\over2}\left(1+\g^5\right)=
    \left(\begin{array}{cc}0&0\\0&\openone\end{array}\right)\;,\qquad
    P_{\mbox{\tiny L}}={1\over2}\left(1-\g^5\right)=
    \left(\begin{array}{cc}\openone&0\\0&0\end{array}\right)\;,
  \eeq
  i.e.\ for the Dirac spinor (\ref{dirac}) one has
  \beqa
    P_{\mbox{\tiny L}}\J_{\mbox{\tiny D}}=
      \c_{\raisebox{-0.5ex}{$\scriptstyle \a$}}\;,&\qquad&
    P_{\mbox{\tiny R}}\J_{\mbox{\tiny D}}=\Bar{\l}^{\,\dt{\a}}\;,\\[1ex]
    \Bar{\J}_{\mbox{\tiny D}}P_{\mbox{\tiny L}}=\l^{\a}\;,&\qquad&
      \Bar{\J}_{\mbox{\tiny D}}P_{\mbox{\tiny R}}=
      \Bar{\c}_{\raisebox{-0.5ex}{$\scriptstyle\dt{\a}$}}\;.
  \eeqa

  The charge conjugation matrix $C$ is defined by
  \beq
    C=-i\g^2\g^0=\left(\begin{array}{cc}\ve_{\a\b}&0\\[1ex]0&
    \ve^{\dt{\a}\dt{\b}}\end{array}\right)\;,
  \eeq
  and it fulfils the following useful identities
  \beq
    C^T=C^{\dg}=C^{-1}=-C\;,\qquad C^2=-1\;.
  \eeq
  $C$ intertwines the $\g^{\m}$ and $-\g_{\m}^T$ representations of
  the Dirac algebra
  \beqa
    &&C\g_{\m}C^{-1}=-\g_{\m}^T\;,\\[1ex]
    &&C\g_5C^{-1}=\g_5^T\;,
  \eeqa
  The charge conjugated Dirac spinor (\ref{dirac}) then reads
  \beq
    \J_{\mbox{\tiny D}}^{\mbox{\tiny C}}\equiv 
    C\Bar{\J}_{\mbox{\tiny D}}^T=
    {\l_{\a}\choose\,\Bar{\c}^{\,\dt{\a}}}\;,\label{cc}
  \eeq
  i.e.\ one has
  \beq
    \Bar{\J}_{\mbox{\tiny D}}=
      {\J_{\mbox{\tiny D}}^{\mbox{\tiny C}}}^T\,C\;,\qquad
    \J_{\mbox{\tiny D}}=C\,\Bar{\J_{\mbox{\tiny D}}^{\mbox{\tiny C}}}^T\;,
    \qquad\Bar{\J_{\mbox{\tiny D}}^{\mbox{\tiny C}}}=
    \J_{\mbox{\tiny D}}^T\;C\;.
  \eeq

  On the other hand, a Majorana spinor $\J_{\mbox{\tiny M}}$ contains
  only one Weyl spinor
  \beq
    \J_{\mbox{\tiny M}}={\c_{\raisebox{-0.5ex}{$\scriptstyle \a$}}
    \choose\,\Bar{\c}^{\,\dt{\a}}}\;,
  \eeq
  i.e.\ eq.~(\ref{cc}) immediately implies
  \beq
    \J_{\mbox{\tiny M}}^{\mbox{\tiny C}}=\J_{\mbox{\tiny M}}\;.
  \eeq

  In Lagrange densities we can switch from Weyl spinors to
  four-component spinors and vice versa by means of the following 
  relations
  \beqa
    \Bar{\J_1}\,P_{\mbox{\tiny L}}\,\J_2&=&
      \l_1\c_{\raisebox{-0.5ex}{$\scriptstyle 2$}}\;,\\[1ex]
    \Bar{\J_1}\,P_{\mbox{\tiny R}}\,\J_2&=&
      \Bar{\c_{\raisebox{-0.5ex}{$\scriptstyle 1$}}}\,\Bar{\l_2}\;,\\[1ex]
    \Bar{\J_1}\,P_{\mbox{\tiny L}}\,\g^{\m}\,\J_2&=&
      \l_1\,\s^{\m}\,\Bar{\l_2}=-\Bar{\l_2}\,\Bar{\s}^{\,\m}\l_1
      \;,\\[1ex]
    \Bar{\J_1}\,P_{\mbox{\tiny R}}\,\g^{\m}\,\J_2&=&
      \Bar{\c_{\raisebox{-0.5ex}{$\scriptstyle 1$}}}\;\Bar{\s}^{\,\m}
      \c_{\raisebox{-0.5ex}{$\scriptstyle 2$}}=
      -\c_{\raisebox{-0.5ex}{$\scriptstyle 2$}}\,\s^{\m}\,
      \Bar{\c_{\raisebox{-0.5ex}{$\scriptstyle 1$}}}
      \;,\\[1ex]
    \Bar{\J_1}\,P_{\mbox{\tiny L}}\,\g^{\m}\,\g^{\n}\,\J_2&=&
      \l_1\,\s^{\m}\,\Bar{\s}^{\,\n}\,
      \c_{\raisebox{-0.5ex}{$\scriptstyle 2$}}=
      \c_{\raisebox{-0.5ex}{$\scriptstyle 2$}}\,\s^{\n}\,
      \Bar{\s}^{\,\m}\,\l_1\;,\\[1ex]
    \Bar{\J_1}\,P_{\mbox{\tiny R}}\,\g^{\m}\,\g^{\n}\,\J_2&=&
      \Bar{\c_{\raisebox{-0.5ex}{$\scriptstyle 1$}}}\;
      \Bar{\s}^{\,\m}\,\s^{\n}\,\Bar{\l_2}=
      \Bar{\l_2}\;\Bar{\s}^{\,\n}\,\s^{\m}\,
      \Bar{\c_{\raisebox{-0.5ex}{$\scriptstyle 1$}}}\;,
  \eeqa
  where $\J_i$ $(i=1,2)$ is a generic Dirac spinor consisting of two
  Weyl spinors $\c_{\raisebox{-0.5ex}{$\scriptstyle i$}}$ and
  $\Bar{\l_i}$ (cf.~eq.~(\ref{dirac})).

\section{Superfield Products \label{sfproducts}}
  Since products of chiral superfields are again chiral, these
  products can be conveniently computed in the $y$-basis
  (\ref{chiral1}). For the products of two or three chiral superfields
  $\F_i$ one finds
  \beqa
    \lefteqn{\F_i(y)\F_j(y)=A_i(y)A_j(y)
      +\sqrt{2}\q\Big[\j_i(y)A_j(y)+A_i(y)\j_j(y)\Big]}
      \label{chiralprod1}\\[1ex]
    &&\qquad+\q^2\Big[A_i(y)F_j(y)+A_j(y)F_i(y)-\j_i(y)\j_j(y)\Big]
      \NO\\[2ex]
    \lefteqn{\F_i(y)\F_j(y)\F_k(y)=A_i(y)A_j(y)A_k(y)}
      \label{chiralprod2}\\[1ex]
    &&\qquad+\sqrt{2}\q\Big[\j_i(y)A_j(y)A_k(y)+\j_j(y)A_k(y)A_i(y)
      +\j_k(y)A_i(y)A_j(y)\Big]\NO\\[1ex]
    &&\qquad+\q^2\Big[F_i(y)A_j(y)A_k(y)+F_j(y)A_k(y)A_i(y)
      +F_k(y)A_i(y)A_j(y)\NO\\[1ex]
    &&\qquad-\j_i(y)\j_j(y)A_k(y)-\j_j(y)\j_k(y)A_i(y)
      -\j_k(y)\j_i(y)A_j(y)\Big]\NO\;.
  \eeqa
  On the other hand, the product of a chiral superfield $\F_j$ and an
  antichiral superfield $\F_i^{\dg}$ is neither chiral nor antichiral.
  In terms of the variables $x^{\m}$, $\q$ and $\Bar{\q}$ it reads
  \beqa
    \lefteqn{\Bar{\F}_i(x)\F_j(x)=A_i^*(x)A_j(x)
      +\sqrt{2}(\q\j_j(x))A_i^*(x)
      +\sqrt{2}(\Bar{\q}\,\Bar{\j_i}(x))A_j(x)}
      \label{phidagphi}\\[1ex]
    &&\qquad+\q^2 A_i^*(x)F_j(x)+\Bar{\q}^{\,2}A_j(x)F_i^*(x)
      \NO\\[1ex]
    &&\qquad+\q^{\a}\Bar{\q}^{\,\dt{\a}}\Big[-i{\s_{\a\dt{\a}}}^{\m}
      \left(A_i^*(x)\partial_{\m}A_j(x)-A_j(x)\partial_{\m}A_i^*(x)
      \right)-2\Bar{\j_i}_{\dt{\a}}(x){\j_j}_{\a}\Big]\NO\\[1ex]
    &&\qquad+\q^2\,\Bar{\q}^{\,\dt{\a}}\left[{-i\over\sqrt{2}}
      {\s_{\a\dt{\a}}}^{\m}\left(A_i^*(x)\partial_{\m}\j_j^{\a}(x)
      -\j_j^{\a}(x)\partial_{\m}A_i^*(x)\right)-\sqrt{2}F_j(x)
      \Bar{\j_i}_{\dt{\a}}(x)\right]\NO\\[1ex]
    &&\qquad+\Bar{\q}^{\,2}\,\q^{\a}\left[{i\over\sqrt{2}}
      {\s_{\a\dt{\a}}}^{\m}\left(\Bar{\j_i}^{\,\dt{\a}}(x)
      \partial_{\m}A_j(x)-A_j(x)\partial_{\m}\Bar{\j_i}^{\,\dt{\a}}(x)
      \right)+\sqrt{2}F_i^*(x){\j_j}_{\a}(x)\right]\NO\\[1ex]
    &&\qquad+\q^2\,\Bar{\q}^{\,2}\,\left[F_i^*(x)F_j(x)
      -{1\over4}A_i^*(x)\bo A_j(x)-{1\over4}A_j(x)\bo A_i^*(x)
      \right.\NO\\[1ex]
    &&\left.\qquad\qquad+{1\over2}(\partial_{\mu}A_i^*(x))
      (\partial^{\m}A_j(x))-{i\over2}\partial_{\m}\Bar{\j_i}(x)
      \Bar{\s}^{\,\m}\j_j(x)+{i\over2}\Bar{\j_i}(x)\Bar{\s}^{\,\m}
      \partial_{\m}\j_j(x)\right]\;.\NO
  \eeqa
  Superspace integration discussed in Chapter \ref{SuperPert} will
  project out $F$-terms, i.e.\ terms proportional to $\q^2$ in chiral
  superfields and the $D$-term (proportional to $\q^2\,\Bar{\q}^{\,2}$)
  in eq.~(\ref{phidagphi}). These are the terms which can be used to
  construct SUSY-invariant actions since they transform into a
  spacetime derivative under SUSY transformations.

  When computing $2\to2$ scatterings in superspace one has to evaluate
  products of four chiral or anti-chiral superfields, with or without
  covariant derivatives acting on them. In our calculations we need
  the following two products, which can be computed by successively
  using eqs.~(\ref{chiralprod1}) and (\ref{phidagphi})
  \beqa
    &&\int d^2{\q}\,d^2\Bar{\q}\,\F_r(x_1,\q,\Bar{\q}\,)\,
      \F_s(x_1,\q,\Bar{\q}\,)\,{1\over4}{D_2^2\over\bo_2}\,
      \F_i(x_2,\q,\Bar{\q}\,)\,\F_j(x_2,\q,\Bar{\q}\,)=
      \qquad\qquad\qquad\qquad\qquad\qquad\mbox{ }
      \label{phi4prod}\\[1ex]
    &&\qquad\qquad\qquad=\;\Big[A_r(x_1)F_s(x_1)+A_s(x_1)F_r(x_1)
      -\j_r(x_1)\j_s(x_1)\Big]A_i(x_2)A_j(x_2)\NO\\[1ex]
    &&\qquad\qquad\qquad\quad+\Big[A_i(x_2)F_j(x_2)+A_j(x_2)F_i(x_2)
      -\j_i(x_2)\j_j(x_2)\Big]A_r(x_1)A_s(x_1)\NO\\[1ex]
    &&\qquad\qquad\qquad\quad+\Big[\j_s(x_1)A_r(x_1)+\j_r(x_1)A_s(x_1)
      \Big]\Big[\j_i(x_2)A_j(x_2)+\j_j(x_2)A_i(x_2)\Big]\;.\NO
  \eeqa
  \beqa
    \lefteqn{\int d^2{\q}\,d^2\Bar{\q}\,\F_r(x_1,\q,\Bar{\q}\,)\,
      \F_s(x_1,\q,\Bar{\q}\,)\,\Bar{\F}_i(x_2,\q,\Bar{\q}\,)\,
      \Bar{\F}_j(x_2,\q,\Bar{\q}\,)=}\label{phi4prod2}\\[1ex]
    &\qquad\qquad=&-A_i^{\dg}(x_2)A_j^{\dg}(x_2)\bo_1
      A_r(x_1)A_s(x_1)\NO\\[1ex]
    &&+\;i\left[\Bar{\j_i}(x_2)A_j^{\dg}(x_2)
      +\Bar{\j_j}(x_2)A_i^{\dg}(x_2)\right]\Bar{\s}^{\,\m}\partial_{1,\m}
      \Big[\j_s(x_1)A_r(x_1)+\j_r(x_1)A_s(x_1)\Big]\NO\\[1ex]
    &&+\Big[A_r(x_1)F_s(x_1)+A_s(x_1)F_r(x_1)-\j_r(x_1)\j_s(x_1)\Big]
      \times\NO\\[1ex]
    &&\mbox{ }\times\left[A_j^{\dg}(x_2)F_i^{\dg}(x_2)
      +A_i^{\dg}(x_2)F_j^{\dg}(x_2)
      -\Bar{\j_j}(x_2)\Bar{\j_i}(x_2)\right]\;.\NO
  \eeqa
  Here we have partially integrated derivatives and dropped total
  derivatives which do not contribute to the action.

    \clearpage\chapter{Feynman Rules \label{AppFeynm}}
  In this appendix we present the component field Feynman rules that
  we have used in the calculations of chapters \ref{decaychapter} and
  \ref{theory}.  Feynman rules for Majorana fermions which yield the
  correct relative minus signs between different diagrams contributing
  to a process without explicit recourse to Wick's theorem are proposed in
  ref.~\cite{denner2}. The basic idea is to introduce a continuous
  fermion flow, i.e.\ an arbitrary orientation of each fermion line.
  Then one can form chains of Dirac matrices by proceeding in a
  direction opposite to the chosen fermion flow. Relative signs of
  interfering diagrams are determined like for Dirac fermions, i.e.\ 
  any permutation of two external fermion lines gives a minus sign.
  One only needs the Dirac propagator for all fermions. However, one
  has to introduce two analytical expressions for each vertex
  involving fermions, corresponding to the two different choices of
  the fermion flow.

  As an example, consider the coupling of a right-handed neutrino to
  a SM Higgs and lepton doublet. This coupling can be written in two 
  equivalent ways
  \beq
    \Bar{N}\,\l_{\n}^{T}\,P_{\mbox{\tiny L}}\,\Big(l\,\e\,H_2\Big)=
    -\,\Big(H_2\,\e\,\Bar{l^c}\Big)\,P_{\mbox{\tiny L}}\,\l_{\n}\,\Bar{N}\;,
  \eeq
  corresponding to the two possible choices for the fermion flow. This
  gives rise to the following equivalent vertices, where the thin
  arrow denotes the chosen fermion flow\\[2ex]
  \pspicture(0,0)(6.5,3.0)
    \psline[linewidth=1pt](0.5,0.3)(1.5,1.3)
    \psline[linewidth=1pt](1.5,1.3)(2.5,0.3)
    \psline[linewidth=1pt,linestyle=dashed](1.5,1.3)(1.5,2.3)
    \psline[linewidth=1pt]{->}(1.5,1.9)(1.5,1.8)
    \psline[linewidth=1pt]{->}(1.0,0.8)(1.1,0.9)
    \psarc[linewidth=0.1pt](1.5,-0.2){0.8}{45}{135}
    \psline[linewidth=1pt]{->}(1.5,0.6)(1.6,0.6)
    \rput[cc]{0}(0.1,0.3){$l^b_{i,\a}$}
    \rput[cc]{0}(2.9,0.3){$N_{j,\b}$}
    \rput[cc]{0}(1.5,2.7){$H_2^a$}
    \rput[lc]{0}(3.3,1.3){
      $-(i\l_{\n})_{ij}\,\e^{ab}(P_{\mbox{\tiny L}})_{\b\a}$}
  \endpspicture
  \hspace{2cm}
  \pspicture(0,0)(6.5,3.0)
    \psline[linewidth=1pt](0.5,0.3)(1.5,1.3)
    \psline[linewidth=1pt](1.5,1.3)(2.5,0.3)
    \psline[linewidth=1pt,linestyle=dashed](1.5,1.3)(1.5,2.3)
    \psline[linewidth=1pt]{->}(1.5,1.9)(1.5,1.8)
    \psline[linewidth=1pt]{->}(1.0,0.8)(1.1,0.9)
    \psarc[linewidth=0.1pt](1.5,-0.2){0.8}{45}{135}
    \psline[linewidth=1pt]{<-}(1.4,0.6)(1.5,0.6)
    \rput[cc]{0}(0.1,0.3){$l^b_{i,\a}$}
    \rput[cc]{0}(2.9,0.3){$N_{j,\b}$}
    \rput[cc]{0}(1.5,2.7){$H_2^a$}
    \rput[lc]{0}(3.3,1.3){
      $-(i\l_{\n})_{ij}\,\e^{ab}(P_{\mbox{\tiny L}})_{\a\b}$}
  \endpspicture\\[2ex]
  In general, one has spinor structures of the kind
  \beq
    \Bar{\j_1}\,\G\j_2\;,\label{struct1}
  \eeq
  where the four-component spinors $\j_1$ and $\j_2$ are either Dirac
  or Majorana fermions, and $\G$ is a product of Dirac $\g$-matrices,
  \beq
    \G=1,\g^{\m},\g^5,\g^5\g^{\m},\s^{\m\n}\;.
  \eeq
  Reverting the fermion flow corresponds to replacing particles by
  antiparticles, i.e.\ (\ref{struct1}) is rewritten in the equivalent
  form
  \beq
    \Bar{\j_1}\,\G\j_2=\Bar{\j_2^c}\,\G'\j_1^c\;,\label{struct2}
  \eeq
  where
  \beq
    \G'=C\,\G^T\,C^{-1}=\left\{\begin{array}{rl}
      \G & \mbox{for } \G=1,\g^5,\g^5\g^{\m}\\[1ex]
      -\G & \mbox{for } \G=\g^{\m},\s^{\m\n}
    \end{array}\right.\;.\label{minus}
  \eeq
  Hence, when stating Feynman rules in the following we can restrict
  ourselves to one fermion flow. Changing the fermion flow just
  amounts to replacing $\G$ by $\pm\G$, according to eq.~(\ref{minus}). 

  Decomposing the superfield products in the superpotential
  (\ref{spotential}) into component fields, we get the Yukawa
  interactions of a right-handed Majorana neutrino\\[2ex]
  \pspicture(-0.3,0)(6.5,3.0)
    \psline[linewidth=1pt](0.5,0.3)(1.5,1.3)
    \psline[linewidth=1pt](1.5,1.3)(2.5,0.3)
    \psline[linewidth=1pt,linestyle=dashed](1.5,1.3)(1.5,2.3)
    \psline[linewidth=1pt]{->}(2.0,0.8)(2.1,0.7)
    \psline[linewidth=1pt]{->}(1.5,1.9)(1.5,1.8)
    \psarc[linewidth=0.1pt](1.5,-0.2){0.8}{45}{135}
    \psline[linewidth=1pt]{->}(1.5,0.6)(1.6,0.6)
    \rput[cc]{0}(0.1,0.3){$N_{j,\b}$}
    \rput[cc]{0}(2.8,0.3){${\wt{h}{ }}^a_{\a}$}
    \rput[cc]{0}(1.5,2.7){$\wt{l_i}^b$}
    \rput[lc]{0}(3.3,1.3){
      $(i\l_{\n})_{ij}\,\d^{ab}(P_{\mbox{\tiny L}})_{\a\b}$}
  \endpspicture
  \hspace{2cm}
  \pspicture(0,0)(6.5,3.0)
    \psline[linewidth=1pt](0.5,0.3)(1.5,1.3)
    \psline[linewidth=1pt](1.5,1.3)(2.5,0.3)
    \psline[linewidth=1pt,linestyle=dashed](1.5,1.3)(1.5,2.3)
    \psline[linewidth=1pt]{->}(1.0,0.8)(1.1,0.9)
    \psline[linewidth=1pt]{->}(1.5,1.95)(1.5,2.05)
    \psarc[linewidth=0.1pt](1.5,-0.2){0.8}{45}{135}
    \psline[linewidth=1pt]{->}(1.5,0.6)(1.6,0.6)
    \rput[cc]{0}(0.1,0.3){${\wt{h}{ }}_{\a}^a$}
    \rput[cc]{0}(2.9,0.3){$N_{j,\b}$}
    \rput[cc]{0}(1.5,2.7){$\wt{l_i}^b$}
    \rput[lc]{0}(3.3,1.3){
      $(i\l_{\n}^{\dg})_{ji}\,\d^{ab}(P_{\mbox{\tiny R}})_{\b\a}$}
  \endpspicture\\[2ex]
  \pspicture(-0.3,0)(6.5,3.0)
    \psline[linewidth=1pt](0.5,0.3)(1.5,1.3)
    \psline[linewidth=1pt](1.5,1.3)(2.5,0.3)
    \psline[linewidth=1pt,linestyle=dashed](1.5,1.3)(1.5,2.3)
    \psline[linewidth=1pt]{->}(1.5,1.9)(1.5,1.8)
    \psline[linewidth=1pt]{->}(1.0,0.8)(1.1,0.9)
    \psarc[linewidth=0.1pt](1.5,-0.2){0.8}{45}{135}
    \psline[linewidth=1pt]{->}(1.5,0.6)(1.6,0.6)
    \rput[cc]{0}(0.1,0.3){$l^b_{i,\a}$}
    \rput[cc]{0}(2.9,0.3){$N_{j,\b}$}
    \rput[cc]{0}(1.5,2.7){$H_2^a$}
    \rput[lc]{0}(3.3,1.3){
      $-(i\l_{\n})_{ij}\,\e^{ab}(P_{\mbox{\tiny L}})_{\b\a}$}
  \endpspicture
  \hspace{2cm}
  \pspicture(0,0)(6.5,3.0)
    \psline[linewidth=1pt](0.5,0.3)(1.5,1.3)
    \psline[linewidth=1pt](1.5,1.3)(2.5,0.3)
    \psline[linewidth=1pt,linestyle=dashed](1.5,1.3)(1.5,2.3)
    \psline[linewidth=1pt]{->}(1.5,1.95)(1.5,2.05)
    \psline[linewidth=1pt]{->}(2.0,0.8)(2.1,0.7)
    \psarc[linewidth=0.1pt](1.5,-0.2){0.8}{45}{135}
    \psline[linewidth=1pt]{->}(1.5,0.6)(1.6,0.6)
    \rput[cc]{0}(0.1,0.3){$N_{j,\b}$}
    \rput[cc]{0}(2.9,0.3){$l^b_{i,\a}$}
    \rput[cc]{0}(1.5,2.7){$H_2^a$}
    \rput[lc]{0}(3.3,1.3){
      $-(i\l_{\n}^{\dg})_{ji}\,\e^{ab}(P_{\mbox{\tiny R}})_{\a\b}$}
  \endpspicture\\[2ex]
  Correspondingly, the interactions of a scalar neutrino are given 
  by\\[2ex]
  \pspicture(-0.3,0)(6.5,3.0)
    \psline[linewidth=1pt](0.5,0.3)(1.5,1.3)
    \psline[linewidth=1pt](1.5,1.3)(2.5,0.3)
    \psline[linewidth=1pt,linestyle=dashed](1.5,1.3)(1.5,2.3)
    \psline[linewidth=1pt]{->}(1.0,0.8)(1.1,0.9)
    \psline[linewidth=1pt]{->}(2.0,0.8)(2.1,0.7)
    \psline[linewidth=1pt]{->}(1.5,1.9)(1.5,1.8)
    \rput[cc]{0}(0.1,0.3){$l^b_{i,\b}$}
    \rput[cc]{0}(2.9,0.3){${\wt{h}{ }}_{\a}^a$}
    \rput[cc]{0}(1.5,2.7){$\snj$}
    \rput[lc]{0}(3.3,1.3){
      $(i\l_{\n})_{ij}\,\d^{ab}(P_{\mbox{\tiny L}})_{\a\b}$}
  \endpspicture
  \hspace{2cm}
  \pspicture(0,0)(6.5,3.0)
    \psline[linewidth=1pt](0.5,0.3)(1.5,1.3)
    \psline[linewidth=1pt](1.5,1.3)(2.5,0.3)
    \psline[linewidth=1pt,linestyle=dashed](1.5,1.3)(1.5,2.3)
    \psline[linewidth=1pt]{->}(1.0,0.8)(1.1,0.9)
    \psline[linewidth=1pt]{->}(2.0,0.8)(2.1,0.7)
    \psline[linewidth=1pt]{->}(1.5,1.95)(1.5,2.05)
    \rput[cc]{0}(0.1,0.3){${\wt{h}{ }}_{\a}^a$}
    \rput[cc]{0}(2.8,0.3){$l^b_{i,\b}$}
    \rput[cc]{0}(1.5,2.7){$\snj$}
    \rput[lc]{0}(3.3,1.3){
      $(i\l_{\n}^{\dg})_{ji}\,\d^{ab}(P_{\mbox{\tiny R}})_{\b\a}$}
  \endpspicture\\[2ex]
  Here we have not specified an explicit fermion flow, since these
  diagrams have a natural orientation of the fermion lines.\\
  The mass term in the auxiliary neutrino field $F_{\scr N^c_i}$
  (cf.~eq.~(\ref{auxN})) yields trilinear scalar couplings\\[2ex]
  \pspicture(-0.3,0)(6.5,3.0)
    \psline[linewidth=1pt,linestyle=dashed](0.5,0.3)(1.5,1.3)
    \psline[linewidth=1pt,linestyle=dashed](1.5,1.3)(2.5,0.3)
    \psline[linewidth=1pt,linestyle=dashed](1.5,1.3)(1.5,2.3)
    \psline[linewidth=1pt]{->}(1.0,0.8)(1.1,0.9)
    \psline[linewidth=1pt]{->}(2.0,0.8)(2.1,0.7)
    \psline[linewidth=1pt]{->}(1.5,1.9)(1.5,1.8)
    \rput[cc]{0}(0.1,0.3){$H_2^a$}
    \rput[cc]{0}(2.9,0.3){$\snj$}
    \rput[cc]{0}(1.5,2.7){$\wt{l_i}^b$}
    \rput[lc]{0}(3.3,1.3){$-(i\l_{\n}M)_{ij}\,\e^{ab}$}
  \endpspicture
  \hspace{2cm}
  \pspicture(0,0)(6.5,3.0)
    \psline[linewidth=1pt,linestyle=dashed](0.5,0.3)(1.5,1.3)
    \psline[linewidth=1pt,linestyle=dashed](1.5,1.3)(2.5,0.3)
    \psline[linewidth=1pt,linestyle=dashed](1.5,1.3)(1.5,2.3)
    \psline[linewidth=1pt]{->}(1.0,0.8)(1.1,0.9)
    \psline[linewidth=1pt]{->}(2.0,0.8)(2.1,0.7)
    \psline[linewidth=1pt]{->}(1.5,1.95)(1.5,2.05)
    \rput[cc]{0}(0.1,0.3){$\snj$}
    \rput[cc]{0}(2.8,0.3){$H_2^a$}
    \rput[cc]{0}(1.5,2.7){$\wt{l_i}^b$}
    \rput[lc]{0}(3.3,1.3){$-(iM\l_{\n}^{\dg})_{ji}\,\e^{ab}$}
  \endpspicture\\
  Furthermore, we take into account the following Yukawa couplings of
  the (s)top\\[2ex]
  \pspicture(-0.3,0)(6.5,3.0)
    \psline[linewidth=1pt](0.5,0.3)(1.5,1.3)
    \psline[linewidth=1pt](1.5,1.3)(2.5,0.3)
    \psline[linewidth=1pt,linestyle=dashed](1.5,1.3)(1.5,2.3)
    \psline[linewidth=1pt]{->}(1.0,0.8)(1.1,0.9)
    \psline[linewidth=1pt]{->}(2.0,0.8)(2.1,0.7)
    \psline[linewidth=1pt]{->}(1.5,1.9)(1.5,1.8)
    \rput[cc]{0}(0.1,0.3){$q^b_{i,\b}$}
    \rput[cc]{0}(2.9,0.3){${\wt{h}{ }}_{\a}^a$}
    \rput[cc]{0}(1.5,2.7){$\surj$}
    \rput[lc]{0}(3.3,1.3){
      $(i\l_u)_{ij}\,\d^{ab}(P_{\mbox{\tiny L}})_{\a\b}$}
  \endpspicture
  \hspace{2cm}
  \pspicture(0,0)(6.5,3.0)
    \psline[linewidth=1pt](0.5,0.3)(1.5,1.3)
    \psline[linewidth=1pt](1.5,1.3)(2.5,0.3)
    \psline[linewidth=1pt,linestyle=dashed](1.5,1.3)(1.5,2.3)
    \psline[linewidth=1pt]{->}(1.0,0.8)(1.1,0.9)
    \psline[linewidth=1pt]{->}(2.0,0.8)(2.1,0.7)
    \psline[linewidth=1pt]{->}(1.5,1.95)(1.5,2.05)
    \rput[cc]{0}(0.1,0.3){${\wt{h}{ }}_{\a}^a$}
    \rput[cc]{0}(2.8,0.3){$q^b_{i,\b}$}
    \rput[cc]{0}(1.5,2.7){$\surj$}
    \rput[lc]{0}(3.3,1.3){
      $(i\l_u^{\dg})_{ji}\,\d^{ab}(P_{\mbox{\tiny R}})_{\b\a}$}
  \endpspicture\\[1.5ex]
  \pspicture(-0.3,0)(6.5,3.0)
    \psline[linewidth=1pt](0.5,0.3)(1.5,1.3)
    \psline[linewidth=1pt](1.5,1.3)(2.5,0.3)
    \psline[linewidth=1pt,linestyle=dashed](1.5,1.3)(1.5,2.3)
    \psline[linewidth=1pt]{->}(1.0,0.8)(0.9,0.7)
    \psline[linewidth=1pt]{->}(2.0,0.8)(2.1,0.7)
    \psline[linewidth=1pt]{->}(1.5,1.9)(1.5,1.8)
    \psarc[linewidth=0.1pt](1.5,-0.2){0.8}{45}{135}
    \psline[linewidth=1pt]{->}(1.5,0.6)(1.6,0.6)
    \rput[cc]{0}(0.1,0.3){${\wt{h}{ }}_{\a}^a$}
    \rput[cc]{0}(2.9,0.3){$u_{j,\b}$}
    \rput[cc]{0}(1.5,2.7){$\wt{q_i}^b$}
    \rput[lc]{0}(3.3,1.3){
      $(i\l_u)_{ij}\,\d^{ab}(P_{\mbox{\tiny L}})_{\b\a}$}
  \endpspicture
  \hspace{2cm}
  \pspicture(0,0)(6.5,3.0)
    \psline[linewidth=1pt](0.5,0.3)(1.5,1.3)
    \psline[linewidth=1pt](1.5,1.3)(2.5,0.3)
    \psline[linewidth=1pt,linestyle=dashed](1.5,1.3)(1.5,2.3)
    \psline[linewidth=1pt]{->}(1.0,0.8)(1.1,0.9)
    \psline[linewidth=1pt]{->}(2.0,0.8)(1.9,0.9)
    \psline[linewidth=1pt]{->}(1.5,1.95)(1.5,2.05)
    \psarc[linewidth=0.1pt](1.5,-0.2){0.8}{45}{135}
    \psline[linewidth=1pt]{->}(1.5,0.6)(1.6,0.6)
    \rput[cc]{0}(0.1,0.3){${\wt{h}{ }}_{\a}^a$}
    \rput[cc]{0}(2.9,0.3){$u_{j,\b}$}
    \rput[cc]{0}(1.5,2.7){$\wt{q_i}^b$}
    \rput[lc]{0}(3.3,1.3){
      $(i\l_u^{\dg})_{ji}\,\d^{ab}(P_{\mbox{\tiny R}})_{\b\a}$}
  \endpspicture\\[1.5ex]
  \pspicture(-0.3,0)(6.5,3.0)
    \psline[linewidth=1pt](0.5,0.3)(1.5,1.3)
    \psline[linewidth=1pt](1.5,1.3)(2.5,0.3)
    \psline[linewidth=1pt,linestyle=dashed](1.5,1.3)(1.5,2.3)
    \psline[linewidth=1pt]{->}(1.0,0.8)(1.1,0.9)
    \psline[linewidth=1pt]{->}(2.0,0.8)(2.1,0.7)
    \psline[linewidth=1pt]{->}(1.5,1.9)(1.5,1.8)
    \rput[cc]{0}(0.1,0.3){$q^b_{i,\b}$}
    \rput[cc]{0}(2.9,0.3){$u_{j,\a}$}
    \rput[cc]{0}(1.5,2.7){$H_2^a$}
    \rput[lc]{0}(3.3,1.3){
      $-(i\l_u)_{ij}\,\e^{ab}(P_{\mbox{\tiny L}})_{\a\b}$}
  \endpspicture
  \hspace{2cm}
  \pspicture(0,0)(6.5,3.0)
    \psline[linewidth=1pt](0.5,0.3)(1.5,1.3)
    \psline[linewidth=1pt](1.5,1.3)(2.5,0.3)
    \psline[linewidth=1pt,linestyle=dashed](1.5,1.3)(1.5,2.3)
    \psline[linewidth=1pt]{->}(1.0,0.8)(1.1,0.9)
    \psline[linewidth=1pt]{->}(2.0,0.8)(2.1,0.7)
    \psline[linewidth=1pt]{->}(1.5,1.95)(1.5,2.05)
    \rput[cc]{0}(0.1,0.3){$u_{j,\a}$}
    \rput[cc]{0}(2.9,0.3){$q^b_{i,\b}$}
    \rput[cc]{0}(1.5,2.7){$H_2^a$}
    \rput[lc]{0}(3.3,1.3){
      $-(i\l_u^{\dg})_{ji}\,\e^{ab}(P_{\mbox{\tiny R}})_{\b\a}$}
  \endpspicture\\
  Finally, the scalar potential
  \beq
    \cv=\sum\limits_iF_i^{\dg}F_i\;,\qquad
    i=H_1,H_2,Q_i,\ldots,N_i^c\;,
  \eeq
  yields quartic scalar couplings involving one or two scalar
  neutrinos\\[2ex]
  \pspicture(-0.3,0)(6.5,2.6)
    \psline[linewidth=1pt,linestyle=dashed](0.5,0.3)(2.5,2.3)
    \psline[linewidth=1pt,linestyle=dashed](0.5,2.3)(2.5,0.3)
    \psline[linewidth=1pt]{->}(0.98,0.78)(1.08,0.88)
    \psline[linewidth=1pt]{->}(0.98,1.82)(1.08,1.72)
    \psline[linewidth=1pt]{->}(2.0,0.8)(2.1,0.7)
    \psline[linewidth=1pt]{->}(2.0,1.8)(2.1,1.9)
    \rput[cc]{0}(0.1,0.3){$\snj$}
    \rput[cc]{0}(0.1,2.3){$\wt{l_i}^b$}
    \rput[cc]{0}(2.9,0.3){$\wt{q_s}^a$}
    \rput[cc]{0}(2.9,2.3){$\surr$}
    \rput[lc]{0}(3.3,1.3){
      $-i(\l_{\n})_{ij}(\l_u^{\dg})_{rs}\,\d^{ab}$}
  \endpspicture
  \hspace{2cm}
  \pspicture(0,0)(6.5,2.6)
    \psline[linewidth=1pt,linestyle=dashed](0.5,0.3)(2.5,2.3)
    \psline[linewidth=1pt,linestyle=dashed](0.5,2.3)(2.5,0.3)
    \psline[linewidth=1pt]{->}(0.98,0.78)(1.08,0.88)
    \psline[linewidth=1pt]{->}(0.98,1.82)(1.08,1.72)
    \psline[linewidth=1pt]{->}(2.0,0.8)(2.1,0.7)
    \psline[linewidth=1pt]{->}(2.0,1.8)(2.1,1.9)
    \rput[cc]{0}(0.1,0.3){$\wt{q_s}^a$}
    \rput[cc]{0}(0.1,2.3){$\surr$}
    \rput[cc]{0}(2.9,0.3){$\snj$}
    \rput[cc]{0}(2.9,2.3){$\wt{l_i}^b$}
    \rput[lc]{0}(3.3,1.3){
      $-i(\l_{\n}^{\dg})_{ji}(\l_u)_{sr}\,\d^{ab}$}
  \endpspicture\\[2ex]
  \pspicture(-0.3,0)(6.5,2.6)
    \psline[linewidth=1pt,linestyle=dashed](0.5,0.3)(2.5,2.3)
    \psline[linewidth=1pt,linestyle=dashed](0.5,2.3)(2.5,0.3)
    \psline[linewidth=1pt]{->}(0.98,0.78)(1.08,0.88)
    \psline[linewidth=1pt]{->}(0.98,1.82)(1.08,1.72)
    \psline[linewidth=1pt]{->}(2.0,0.8)(2.1,0.7)
    \psline[linewidth=1pt]{->}(2.0,1.8)(2.1,1.9)
    \rput[cc]{0}(0.1,0.3){$\snj$}
    \rput[cc]{0}(0.1,2.3){$H_2^a$}
    \rput[cc]{0}(2.9,0.3){$H_1^b$}
    \rput[cc]{0}(2.9,2.3){$\wt{E_i^c}$}
    \rput[lc]{0}(3.3,1.3){$-i(\l_l^{\dg}\l_{\n})_{ij}\,\d^{ab}$}
  \endpspicture
  \hspace{2cm}
  \pspicture(0,0)(6.5,2.6)
    \psline[linewidth=1pt,linestyle=dashed](0.5,0.3)(2.5,2.3)
    \psline[linewidth=1pt,linestyle=dashed](0.5,2.3)(2.5,0.3)
    \psline[linewidth=1pt]{->}(0.98,0.78)(1.08,0.88)
    \psline[linewidth=1pt]{->}(0.98,1.82)(1.08,1.72)
    \psline[linewidth=1pt]{->}(2.0,0.8)(2.1,0.7)
    \psline[linewidth=1pt]{->}(2.0,1.8)(2.1,1.9)
    \rput[cc]{0}(0.1,0.3){$\wt{E_i^c}$}
    \rput[cc]{0}(0.1,2.3){$H_1^b$}
    \rput[cc]{0}(2.9,0.3){$H_2^a$}
    \rput[cc]{0}(2.9,2.3){$\snj$}
    \rput[lc]{0}(3.3,1.3){$-i(\l_{\n}^{\dg}\l_l)_{ji}\,\d^{ab}$}
  \endpspicture\\[2ex]
  \pspicture(-0.3,0)(6.5,2.6)
    \psline[linewidth=1pt,linestyle=dashed](0.5,0.3)(2.5,2.3)
    \psline[linewidth=1pt,linestyle=dashed](0.5,2.3)(2.5,0.3)
    \psline[linewidth=1pt]{->}(0.98,0.78)(1.08,0.88)
    \psline[linewidth=1pt]{->}(0.98,1.82)(1.08,1.72)
    \psline[linewidth=1pt]{->}(2.0,0.8)(2.1,0.7)
    \psline[linewidth=1pt]{->}(2.0,1.8)(2.1,1.9)
    \rput[cc]{0}(0.1,0.3){$\snj$}
    \rput[cc]{0}(0.1,2.3){$H_2^a$}
    \rput[cc]{0}(2.9,0.3){$\sni$}
    \rput[cc]{0}(2.9,2.3){$H_2^b$}
    \rput[lc]{0}(3.3,1.3){
      $-i(\l_{\n}^{\dg}\l_{\n})_{ij}\,\d^{ab}$}
  \endpspicture
  \hspace{2cm}
  \pspicture(0,0)(6.5,2.6)
    \psline[linewidth=1pt,linestyle=dashed](0.5,0.3)(2.5,2.3)
    \psline[linewidth=1pt,linestyle=dashed](0.5,2.3)(2.5,0.3)
    \psline[linewidth=1pt]{->}(0.98,0.78)(1.08,0.88)
    \psline[linewidth=1pt]{->}(0.98,1.82)(1.08,1.72)
    \psline[linewidth=1pt]{->}(2.0,0.8)(2.1,0.7)
    \psline[linewidth=1pt]{->}(2.0,1.8)(2.1,1.9)
    \rput[cc]{0}(0.1,0.3){$\snj$}
    \rput[cc]{0}(0.1,2.3){$\wt{l_r}^a$}
    \rput[cc]{0}(2.9,0.3){$\sni$}
    \rput[cc]{0}(2.9,2.3){$\wt{l_s}^b$}
    \rput[lc]{0}(3.3,1.3){
      $-i(\l_{\n})_{rj}(\l_{\n}^{\dg})_{is}\,\d^{ab}$}
  \endpspicture\\[2ex]
  Internal Dirac or Majorana fermion lines are all represented by the
  usual Dirac propagator\\[1ex]
  \centerline{
  \pspicture[0.5](0,0.1)(6.5,1.1)
    \psline[linewidth=1pt]{*-*}(0.3,0.6)(1.9,0.6)
    \psline[linewidth=0.3pt]{->}(0.6,0.4)(1.6,0.4)
    \psline[linewidth=0.3pt]{->}(0.9,0.75)(1.3,0.75)
    \rput[cc]{0}(1.1,1.0){$p$}
    \rput[cc]{0}(0.0,0.6){$\a$}
    \rput[cc]{0}(2.2,0.6){$\b$}
    \rput[cc]{0}(5.0,0.6){$\displaystyle
      \left({i\over\slash{p}-m+i\e}\right)_{\b\a}$}
  \endpspicture$\;.$}\\[2ex]
  Correspondingly, we assign spinors to external fermion lines with
  orientation\\[1ex]
  \centerline{
  \pspicture[0.5](0,0.4)(11.0,0.7)
    \psline[linewidth=1pt]{*-}(0.3,0.6)(1.9,0.6)
    \psline[linewidth=1pt]{->}(1.1,0.6)(1.2,0.6)
    \psline[linewidth=0.3pt]{->}(0.6,0.4)(1.6,0.4)
    \psline[linewidth=1pt]{*-}(3.3,0.6)(4.9,0.6)
    \psline[linewidth=1pt]{<-}(3.9,0.6)(4.0,0.6)
    \psline[linewidth=0.3pt]{->}(3.6,0.4)(4.6,0.4)
    \psline[linewidth=1pt]{*-}(6.3,0.6)(7.9,0.6)
    \psline[linewidth=0.3pt]{->}(6.6,0.4)(7.6,0.4)
    \rput[cc]{0}(10.0,0.6){$\displaystyle\Bar{u}(p,s)\;,$}
  \endpspicture}\\[2ex]
  \centerline{
  \pspicture[0.5](0,0.4)(11.0,0.7)
    \psline[linewidth=1pt]{*-}(0.3,0.6)(1.9,0.6)
    \psline[linewidth=1pt]{->}(1.1,0.6)(1.2,0.6)
    \psline[linewidth=0.3pt]{<-}(0.6,0.4)(1.6,0.4)
    \psline[linewidth=1pt]{*-}(3.3,0.6)(4.9,0.6)
    \psline[linewidth=1pt]{<-}(3.9,0.6)(4.0,0.6)
    \psline[linewidth=0.3pt]{<-}(3.6,0.4)(4.6,0.4)
    \psline[linewidth=1pt]{*-}(6.3,0.6)(7.9,0.6)
    \psline[linewidth=0.3pt]{<-}(6.6,0.4)(7.6,0.4)
    \rput[cc]{0}(10.0,0.6){$\displaystyle v(p,s)\;,$}
  \endpspicture}\\[2ex]
  \centerline{
  \pspicture[0.5](0,0.4)(11.0,0.7)
    \psline[linewidth=1pt]{-*}(0.3,0.6)(1.9,0.6)
    \psline[linewidth=1pt]{->}(1.1,0.6)(1.2,0.6)
    \psline[linewidth=0.3pt]{->}(0.6,0.4)(1.6,0.4)
    \psline[linewidth=1pt]{-*}(3.3,0.6)(4.9,0.6)
    \psline[linewidth=1pt]{<-}(3.9,0.6)(4.0,0.6)
    \psline[linewidth=0.3pt]{->}(3.6,0.4)(4.6,0.4)
    \psline[linewidth=1pt]{-*}(6.3,0.6)(7.9,0.6)
    \psline[linewidth=0.3pt]{->}(6.6,0.4)(7.6,0.4)
    \rput[cc]{0}(10.0,0.6){$\displaystyle u(p,s)\;,$}
  \endpspicture}\\[2ex]
  \centerline{
  \pspicture[0.5](0,0.4)(11.0,0.7)
    \psline[linewidth=1pt]{-*}(0.3,0.6)(1.9,0.6)
    \psline[linewidth=1pt]{->}(1.1,0.6)(1.2,0.6)
    \psline[linewidth=0.3pt]{<-}(0.6,0.4)(1.6,0.4)
    \psline[linewidth=1pt]{-*}(3.3,0.6)(4.9,0.6)
    \psline[linewidth=1pt]{<-}(3.9,0.6)(4.0,0.6)
    \psline[linewidth=0.3pt]{<-}(3.6,0.4)(4.6,0.4)
    \psline[linewidth=1pt]{-*}(6.3,0.6)(7.9,0.6)
    \psline[linewidth=0.3pt]{<-}(6.6,0.4)(7.6,0.4)
    \rput[cc]{0}(10.0,0.6){$\displaystyle\Bar{v}(p,s)\;,$}
  \endpspicture}\\[2ex]
  where the momentum $p$ always flows from left to right.
  The spinors $v$ and $u$ are related by charge conjugation
  \beqa
    v(p,s)=C\,\Bar{u}^T(p,s)&\quad\Leftrightarrow\quad&
      \Bar{u}(p,s)=v^T(p,s)\,C\;,\\[1ex]
    u(p,s)=C\,\Bar{v}^T(p,s)&\quad\Leftrightarrow\quad&
      \Bar{v}(p,s)=u^T(p,s)\,C\;.
  \eeqa

    \clearpage\chapter{Kinetic Theory \label{appB}}
  The microscopic evolution of particle densities and asymmetries is
  governed by a network of Boltzmann equations. In the following we
  will compile some basic formulae to introduce our notation \cite{kt1}.

\section{Thermodynamics in the Expanding Universe}
    The early universe can be assumed to be spatially homogeneous and
    isotropic. Hence, it is described by a Robertson-Walker metric
    \beq
      ds^2=dt^2-R(t)^2\left\{{dr^2\over1-kr^2}+r^2d\q^2
        +r^2\sin^2\q d\f^2\right\}\;,
    \eeq
    where $(t,r,\q,\f)$ are comoving coordinates and $k=\pm1,0$
    describes the spatial curvature of spacetime. The scale factor
    $R(t)$, which describes the expansion of the universe, is given by
    the Friedmann equation
    \beq
      \dt{R}^2+k={8\p G\over3}\r R^2\;,\label{Friedmann}
    \eeq
    where $\r$ is the energy density of the universe, and 
    \beq
      G={1\over m_{\scr Pl}^2}
    \eeq
    denotes Newton's constant in units where $\hbar=c=1$, and
    $m_{\scr Pl}=1.2211\cdot10^{19}\;$GeV is the Planck mass. 
    Neglecting the curvature term $k$ in the Friedmann equation 
    (\ref{Friedmann}), which is a good approximation in the early 
    universe, we get an equation for the Hubble parameter $H$
    \beq
      H\equiv{\dt{R}\over R}={1\over m_{\scr Pl}}
      \sqrt{{8\p\r\over3}}\;.
    \eeq
    In a radiation dominated universe the energy density reads
    \beq
      \r=g_*{\p^2\over30}T^4\;,
    \eeq
    where $g_*$ is the number of effectively massless degrees of
    freedom
    \beq
      g_*=\sum\limits_{\scr i=\mbox{\tiny Bosons}}g_i
        +{7\over8}\sum\limits_{\scr i=\mbox{\tiny Fermions}}g_i\;,
    \eeq
    and $g_i$ is the number of internal degrees of freeedom of the
    corresponding particle. At temperatures far above the electroweak
    scale one has $g_*=106.75$ in the standard model, and $g_*=228.75$
    in the MSSM.

    Hence, the Hubble parameter in a radiation dominated universe reads
    \beq
      H=\sqrt{{4\p^3g*\over45}}{T^2\over m_{\scr Pl}}\;.
    \eeq

\section{Boltzmann Equations\label{boltzmann}}
    It is usually a good approximation to assume Maxwell-Boltzmann
    statistics, so that the equilibrium number density of a particle
    $i$ is given by
    \beq
     n_i^{\rm eq}(T)={g_i\over(2\p)^3}\int\dd^3p_i\,f_i^{\rm eq}
     \qquad\mbox{with}\qquad
     f^{\rm eq}_i\left(E_i,T\right)=\mbox{e}^{-E_i/T}\;.
    \eeq
    For a massive particle one finds
    \beq
      n_i^{\rm eq}(T)={g_iTm_i^2\over2\p^2}
        \mbox{K}_2\left({m_i\over T}\right)\;,
    \eeq
    whereas for a massless particle one gets
    \beq
      n_i^{\rm eq}(T)={g_iT^3\over\p^2}\;.
    \eeq
    
    Particle densities can be changed by interactions and by the
    expansion of the universe. Since we are only interested in the
    effect of interactions, it is useful to scale out the expansion.
    This is done by taking the number of particles per comoving
    volume element, i.e.\ the ratio of the particle density $n_i$
    to the entropy density $s$,
    \beq
      Y_i={n_i\over s}\;,
    \eeq
    as independent variable instead of the number density. In a
    radiation dominated universe the entropy density reads
    \beq
      s=g_*{2\p^2\over45}T^3\;.
    \eeq

    In our case elastic scatterings, which can only change the phase
    space distributions but not the particle densities, occur at a
    much higher rate than inelastic processes. Therefore, we can
    assume kinetic equilibrium, so that the phase space densities are
    given by
    \beq
     f_i(E_i,T)={n_i\over n_i^{\rm eq}}\mbox{e}^{-E_i/T}\;.
    \eeq
    In this framework the Boltzmann equation describing the evolution
    of a particle number $Y_{\j}$ in an isentropically expanding 
    universe reads \cite{kw,luty}
    \beqa
    {\mbox{d}Y_{\j}\over\mbox{d}z}&=&-{z\over sH\left(m_{\j}\right)}
    \sum\limits_{a,i,j,\ldots}\left[{Y_{\j}Y_a\ldots\over
     Y_{\j}^{\rm eq}Y_a^{\rm eq}\ldots}\,\g^{\rm eq}\left(\j+a+\ldots\to 
     i+j+\ldots\right)\right.\NO\\[1ex]
     &&\qquad\qquad\qquad\left.{}-{Y_iY_j\ldots\over 
     Y_i^{\rm eq}Y_j^{\rm eq}\ldots}
     \,\g^{\rm eq}\left(i+j+\ldots\to\j+a+\ldots\right)\right]\;,
    \label{7}
    \eeqa
    where $z=m_{\j}/T$ and $H\left(m_{\j}\right)$ is the Hubble
    parameter at $T=m_{\j}$. The $\g^{\rm eq}$ are space time
    densities of scatterings for the different processes. For
    a decay one finds \cite{luty}
    \beq
     \g_D:=\g^{\rm eq}(\j\to i+j+\ldots)=
     n^{\rm eq}_{\j}{\mbox{K}_1(z)\over\mbox{K}_2(z)}\,\G\;,
     \label{decay}
    \eeq
    where K$_1$ and K$_2$ are modified Bessel functions and
    $\G$ is the usual decay width in the rest system of
    the decaying particle. Neglecting a possible $CP$ violation, one
    finds the same reaction density for the inverse decay.

    The reaction density for a two body scattering reads
    \beq
     \g^{\rm eq}({\j}+a\leftrightarrow i+j+\ldots)=
     {T\over64\p^4}\int\limits_{\left(m_{\j}+m_a\right)^2}^{\infty}
     \hspace{-0.5cm}\dd s\,\hat{\s}(s)\,\sqrt{s}\,
     \mbox{K}_1\left({\sqrt{s}\over T}\right)\;,
     \label{22scatt}
    \eeq 
    where $s$ is the squared centre of mass energy and the reduced
    cross section $\hat{\s}(s)$ for the process ${\j}+a\to i+j+\ldots$
    is related to the usual total cross section $\s(s)$ by
    \beq
      \hat{\s}(s)={2\l(s,m^2_{\j},m^2_a\,)\over s}\,\s(s)\;,
    \eeq
    where $\l$ is the usual kinematical function
    \beq
      \l(s,m^2_{\j},m^2_a\,)\equiv\left[s-(m_{\j}+m_a)^2\right]
      \left[s-(m_{\j}-m_a)^2\right]\;.
    \eeq

    \clearpage\chapter{Reduced Cross Sections \label{appC}}
    In this section we will collect the reduced cross sections for all
    the $2\leftrightarrow2$ and $2\leftrightarrow3$ processes that we
    had discussed in chapter \ref{theory}. The corresponding reaction
    densities, which can be calculated analytically in some interesting
    limiting cases, will be discussed in the next appendix.
\section[Lepton Number Violating Scatterings]{Lepton Number Violating
         Processes Mediated by Right-Handed Neutrinos}
    We have mentioned in the main text that we have to subtract the
    contributions coming from on-shell (s)neutrinos, i.e.\ we have to
    replace the usual propagators by off-shell propagators
    \beq
      {1\over D_j(x)} := {x-a_j\over(x-a_j)^2+a_jc_j}\quad
      \mbox{and}\quad
      {1\over \wt{D}_j(x)} := {x-a_j\over(x-a_j)^2+a_j\wt{c_j}}\;.
    \eeq
    To begin with, let us specify the reduced cross sections 
    for the reactions depicted in fig.~\ref{chap2_fig03}.
    For the processes $\wt{l}+\Bar{\wt{h}}\leftrightarrow
    {\wt{l}{ }}^{\,\dg}+\wt{h}$ and $l+H_2\leftrightarrow\Bar{l}
    +H_2^{\dg}$ one has
    \beqa
      \lefteqn{\hat{\s}_{\scr N}^{(1)}(x) = \hat{\s}_{\scr N}^{(2)}(x) = 
        {1\over2\pi}\Bigg\{\sum\limits_j\lljj^2\;{a_j\over x}\left[
        {x\over a_j}+{x\over D_j(x)}+{x^2\over 2D_j^2(x)}
        -\left(1+{x+a_j\over D_j(x)}\right)\lnaj\right]}
        \NO\\[1ex]
      &&{}+\sum\limits_{n,j\atop j<n}\mbox{Re}\left[\llnj^2\right]
        {\sqrt{a_na_j}\over x}\left[{x\over D_j(x)}+{x\over D_n(x)}
        +{x^2\over D_j(x)D_n(x)}\right.\\[1ex]
      &&\left.{}+\left(x+a_j\right)\left({2\over a_n-a_j}
        -{1\over D_n(x)}\right)\lnaj+\left(x+a_n\right)
        \left({2\over a_j-a_n}-{1\over D_j(x)}\right)\lnan
        \right]\Bigg\}\;,\NO
    \eeqa
    where $n$ and $j$ are the flavour indices of the neutrinos in the 
    intermediate state. The interference terms with $n\ne j$ are
    always very small and can safely be neglected.\\
    The reduced cross section for the process 
    $\wt{l}+\Bar{\wt{h}}\leftrightarrow\Bar{l}+H_2^{\dg}$ reads 
    \beqa
      \lefteqn{\hat{\s}_{\scr N}^{(3)}(x) = {1\over2\pi}\Bigg\{
        \sum\limits_j\lljj^2\;{a_j\over x}\left[{-x\over x+a_j}
        +{x\over D_j(x)}+{x^2\over 2D_j^2(x)}
        +\left(1-{a_j\over D_j(x)}\right)\lnaj\right]}\NO\\[1ex]
      &&\qquad{}+\sum\limits_{n,j\atop j<n}\mbox{Re}\left[\llnj^2\right]
        {\sqrt{a_na_j}\over x}\left[{x\over D_j(x)}+{x\over D_n(x)}
        +{x^2\over D_j(x)D_n(x)}\right.\\[1ex]
      &&\left.\qquad
        {}-\;a_j\left({2\over a_n-a_j}+{1\over D_n(x)}\right)\lnaj
        -a_n\left({2\over a_j-a_n}+{1\over D_j(x)}\right)\lnan
        \right]\Bigg\}\;.\NO
    \eeqa
    The same result is valid for the $CP$ conjugated process.\\
    For the process $l+\Bar{\wt{h}}\leftrightarrow{\wt{l}{ }}^{\,\dg}
    +H_2^{\dg}$ one finds
    \beqa
      \hat{\s}_{\scr N}^{(4)}(x) &=& {1\over2\pi}\Bigg\{
        \sum\limits_j\lljj^2\;{a_j\over x}\left[{x^2\over a_j(x+a_j)}
        +{x^2\over \wt{D_j}^2(x)}+{x\over\wt{D_j}(x)}
        \lnaj\right]\NO\\[1ex]
      &&{}+\sum\limits_{n,j\atop j<n}\mbox{Re}\left[\llnj^2\right]
        {\sqrt{a_na_j}\over x}\left[{2x^2\over\wt{D_j}(x)
        \wt{D_n}(x)}+x\left({2\over a_n-a_j}+
        {1\over\wt{D_n}(x)}\right)\lnaj\right.\NO\\[1ex]
      &&\left.\hspace{2cm}{}+\;x\left({2\over a_j-a_n}
        +{1\over\wt{D_j}(x)}\right)\lnan\right]\Bigg\}\;.
    \eeqa
    For the scattering $\wt{l}+H_2\rightarrow{\wt{l}{ }}^{\,\dg}
    +\sur+\wt{q}$ and the corresponding $CP$ transformed process we
    have
    \beqa
      \lefteqn{\hat{\s}_{\scr N}^{(5)}(x) = {3\,\a_u\over8\pi^2}\Bigg\{
        \sum\limits_j\lljj^2\;{a_j\over x}\left[{x\over a_j}
        +{x\over \wt{D_j}(x)}+{x^2\over \wt{D_j}^2(x)}
        -\left(1+{x+a_j\over\wt{D_j}(x)}\right)
        \lnaj\right]}\NO\\[1ex]
      &&{}+\sum\limits_{n,j\atop j<n}\mbox{Re}\left[\llnj^2\right]
        {\sqrt{a_na_j}\over x}\left[{x\over\wt{D_j}(x)}+
        {x\over\wt{D_n}(x)}+{x^2\over\wt{D_j}(x)
        \wt{D_n}(x)}\right.\\[1ex]
      &&\left.\!{}+\;\left(x+a_j\right)\left({2\over a_n-a_j}
        -{1\over\wt{D_n}(x)}\right)\lnaj
        +\left(x+a_n\right)\left({2\over a_j-a_n}
        -{1\over\wt{D_j}(x)}\right)\lnan\right]\Bigg\}\;.\NO
    \eeqa
    Finally, we have two processes which do not violate lepton number
    but merely transform leptons into scalar leptons and vice versa.
    We have the $2\to2$ scattering $l+H_2\leftrightarrow\wt{l}
    +\Bar{\wt{h}}$,
    \beq
      \hat{\s}_{\scr N}^{(6)}(x) = {1\over4\pi}
      \sum\limits_{j,n}\left|\llnj\right|^2\;{x^2\over D_j(x)D_n(x)}\;,
    \eeq
    and the $2\to3$ process $l+\Bar{\wt{h}}\leftrightarrow\wt{l}
    +\wt{q}^{\dg}+\sur^{\dg}$,
    \beq
      \hat{\s}_{\scr N}^{(7)}(x) = {3\,\a_u\over16\pi^2}
        \sum\limits_{j,n}\left|\llnj\right|^2\;
        {x^2\over\wt{D_j}(x)\wt{D_n}(x)}\;.
    \eeq

    Let us now come to the $2\to3$ processes shown in
    fig.~\ref{chap2_fig04}.\\
    For the transition $\wt{q}+\sur\rightarrow\wt{l}+\wt{l}+H_2$ the
    reduced cross section reads\\
    \beqa
      \lefteqn{\hat{\s}_{\scr N}^{(8)}(x) = {3\,\a_u\over16\pi^2}\Bigg\{
        \sum\limits_j\lljj^2\;{a_j\over x}\left[
        -{x\over a_j+\wt{c_j}}+{x-a_j\over\sqrt{a_j\wt{c_j}}}\atnj
        \right.}\NO\\[1ex]
      &&\left.\hspace{2cm}{}-\lnpropj
        +{1\over2}\int\limits_0^x\;\mbox{d}x_1\;
        {1\over\wt{D_j}(x_1)}\ln\left({(x-x_1-a_j)^2
        +a_j\wt{c_j}\over a_j^2+a_j\wt{c_j}}\right)\right]\NO\\[1ex]
      \lefteqn{{}+2\sum\limits_{n,j\atop j<n}\mbox{Re}
        \left[\llnj^2\right]{\sqrt{a_na_j}\over x}\left[
         {1\over2}\int\limits_0^x\;\mbox{d}x_1\;
         {1\over\wt{D_n}(x_1)}\ln\left({(x-x_1-a_j)^2+a_j\wt{c_j}
         \over a_j^2+a_j\wt{c_j}}\right)
        \right.}\\[1ex]
      &&{}+2\sqrt{a_j\wt{c_j}}{x-a_n\over\left(a_j-a_n\right)^2}\atnj
        +{x-a_j\over a_j-a_n}\lnpropj\NO\\[1ex]
      &&\left.
        {}+2\sqrt{a_n\wt{c_n}}{x-a_j\over\left(a_n-a_j\right)^2}\atnn
        +{x-a_n\over a_n-a_j}\lnpropn\right]\Bigg\}\NO\;.
    \eeqa
    The remaining integral cannot be solved exactly. However,
    it can be neglected for $x>a_j,a_n$ and for $x<a_j,a_n$ it can be
    approximated by
    \beqa
      \lefteqn{{1\over2}\int\limits_0^x\;\mbox{d}x_1\;{1\over\wt{D_n}(x_1)}
        \ln\left({(x-x_1-a_j)^2+a_j\wt{c_j}
        \over a_j^2+a_j\wt{c_j}}\right)}\\[1ex]
      &&\qquad\approx\ln\left({a_j+a_n-x\over a_j}\right)
        \ln\left({a_n-x\over a_n}\right)
        +\mbox{Sp}\left({a_n\over a_n+a_j-x}\right)-
        \mbox{Sp}\left({a_n-x\over a_n+a_j-x}\right)\;,\NO
    \eeqa
    where $\mbox{Sp}(x)$ is the Spence function or dilogarithm
    \beq
      \mbox{Sp}(x)=\mbox{Li}_2(x)=
        -\int\limits_0^x dy\,{\ln(1-y)\over y}\;.
    \eeq
    For the scatterings $\wt{q}+\sur\rightarrow\wt{l}+\Bar{l}+\wt{h}$, 
    ${\wt{l}{ }}^{\,\dg}+\wt{q}\rightarrow\Bar{l}+\sur^{\dg}+\wt{h}$
    and ${\wt{l}{ }}^{\,\dg}+\sur\rightarrow\Bar{l}+\wt{q}^{\dg}+\wt{h}$
    the reduced cross sections are equal,
    \beqa
      \lefteqn{\hat{\s}_{\scr N}^{(9)}(x) = \hat{\s}_{\scr N}^{(11)}(x) = 
        {3\,\a_u\over8\pi^2x}\Bigg\{\sum\limits_j\lljj^2
        \left[-{3\over2}x+{1\over2}(x-2a_j)\lnpropj
        \right.}\NO\\[1ex]
      &&\hspace{2cm}\left.{}+{1\over2}\sqrt{a_j\over\wt{c_j}}
        \left(x-a_j+3\wt{c_j}\right)\atnj\right]\\[1ex]
      &&\hspace{-21pt}{}+2\sum\limits_{n,j\atop j<n}\left|\llnj\right|^2
        \left[-2x+a_j{x-a_j\over a_j-a_n}\lnpropj
      +a_n{x-a_n\over a_n-a_j}\lnpropn\right.\NO\\[1ex]
      &&\hspace{2cm}{}+2\sqrt{a_j\wt{c_j}}\;
        {xa_n-2a_na_j+a_j^2\over(a_j-a_n)^2}\atnj\NO\\[1ex]
      &&\left.\hspace{2cm} 
        {}+2\sqrt{a_n\wt{c_n}}\;{xa_j-2a_na_j+a_n^2\over(a_n-a_j)^2}
        \atnn\right]\Bigg\}\;.\NO
    \eeqa
    For the process ${\wt{l}{ }}^{\,\dg}+\wt{q}\rightarrow\wt{l}
    +\sur^{\dg}+H_2$ and similar reactions one gets
    \beqa
      \lefteqn{\hat{\s}_{\scr N}^{(10)}(x) = {3\,\a_u\over16\pi^2}
        \left\{
        \sum\limits_j\lljj^2\;{a_j\over x}^{\mbox{}}\left[
        {x\over a_j}-2\ln\left({x+a_j\over a_j}\right)
        -\lnpropj\right.\right.}\NO\\[1ex]
      &&\left.\!{}+{x-a_j\over\sqrt{a_j\wt{c_j}}}\atnj
        +2\int\limits_0^x\;\mbox{d}x_1\;
        {1\over\wt{D_j}(x_1)}\left[\mbox{Sp}\left(-{x\over a_j}
        \right)-\mbox{Sp}\left(-{x_1\over a_j}\right)\right]\right]
        \NO\\[1ex]
      \lefteqn{{}+2\sum\limits_{n,j\atop j<n}\mbox{Re}
        \left[\llnj^2\right]{\sqrt{a_na_j}\over x}\left[
        2{x+a_j\over a_n-a_j}\lnaj+2{x+a_n\over a_j-a_n}\lnan
        \right.}\\[1ex]
      &&{}+{x-a_j\over a_j-a_n}\lnpropj+2\sqrt{a_j\wt{c_j}}
        {x-a_n\over(a_j-a_n)^2}\atnj\NO\\[1ex]
      &&{}+{x-a_n\over a_n-a_j}\lnpropn+2\sqrt{a_n\wt{c_n}}
        {x-a_j\over(a_n-a_j)^2}\atnn\NO\\[1ex]
      &&\left.\left.{}+\int\limits_0^x\;\mbox{d}x_1\;\left[
        {1\over\wt{D_j}(x_1)}\left(\mbox{Sp}\left(-{x\over a_n}
        \right)-\mbox{Sp}\left(-{x_1\over a_n}\right)\right)
        +{1\over\wt{D_n}(x_1)}\left(\mbox{Sp}\left(-{x\over a_j}
        \right)-\mbox{Sp}\left(-{x_1\over a_j}\right)\right)\right]
        \right]\right\}\;.\NO
    \eeqa
    The remaining integral can again not be solved exactly. However,
    it can be approximated by
    \beqa
      \lefteqn{\int\limits_0^x\;\mbox{d}x_1\;
        {1\over\wt{D_j}(x_1)}\left[\mbox{Sp}\left(-{x\over a_n}
        \right)-\mbox{Sp}\left(-{x_1\over a_n}\right)\right]}\\[1ex]
      &&\approx{x\over a_n}-{\sqrt{a_j\wt{c_j}}\over a_n}\atnj
        -{x-a_j\over2a_n}\lnpropj\NO
    \eeqa
    for $x<a_n$ and for $x>a_n$ it can be neglected.

    Finally, we have to compute the $t$- and $u$-channel processes
    in fig.~\ref{chap2_fig05}, which give simple contributions.\\     
    For the processes $\wt{l}+\wt{l}\leftrightarrow\wt{h}+\wt{h}$ and 
    $l+l\leftrightarrow H_2^{\dg}+H_2^{\dg}$ we get
    \beqa
      \hat{\s}_{\scr N}^{(12)}(x) &=& \hat{\s}_N^{(13)}(x)
        ={1\over2\pi}\left\{\sum\limits_j\lljj^2\;
        \left[{x\over x+a_j}+{a_j\over x+2a_j}
        \ln\left({x+a_j\over a_j}\right)\right]\right.\NO\\[1ex]
      &&{}+\sum\limits_{n,j\atop j<n}\mbox{Re}\left[\llnj^2\right]
        \sqrt{a_na_j}\left[\left({1\over x+a_n+a_j}
        +{2\over a_n-a_j}\right)\ln\left({x+a_j\over a_j}\right)
        \right.\NO\\[1ex]
      &&\left.\hspace{2cm}{}+\left({1\over x+a_n+a_j}
        +{2\over a_j-a_n}\right)\ln\left({x+a_n\over a_n}\right)
        \right]\Bigg\}\;.
    \eeqa
    In this order of perturbation theory the same result is valid for
    the $CP$ transformed processes.\\
    For the scattering $\wt{l}+l\leftrightarrow\wt{h}+H_2^{\dg}$ one
    has 
    \beqa
      \hat{\s}_{\scr N}^{(14)}(x) &=& {1\over2\pi}\left\{
        \sum\limits_j\lljj^2\;\left[{x\over x+a_j}
        -{a_j\over x+2a_j}\ln\left({x+a_j\over a_j}\right)\right]
        \right.\NO\\[1ex]
      &&{}+2\sum\limits_{n,j\atop j<n}\mbox{Re}\left[\llnj^2\right]
        \sqrt{a_na_j}\left[\left({1\over x+a_n+a_j}
        +{1\over a_n-a_j}\right)\ln\left({x+a_j\over a_j}\right)
        \right.\NO\\[1ex]
      &&\left.\hspace{2cm}{}+\left({1\over x+a_n+a_j}
        +{1\over a_j-a_n}\right)\ln\left({x+a_n\over a_n}\right)
        \right]\Bigg\}\;.
    \eeqa      
    The $2\to3$ process $H_2+\wt{q}^{\,\dg}\leftrightarrow
    {\wt{l}{}}^{\;\dg}+{\wt{l}{}}^{\;\dg}+\sur$ gives
    \beqa
      \lefteqn{\hat{\s}_{\scr N}^{(15)}(x) = {3\a_u\over8\pi^2}\left\{
        \sum\limits_j\lljj^2{a_j\over x}\;\left[{x\over a_j}
        -\left(1-{1\over2}\ln\left({x+2a_j\over a_j}\right)\right)
        \ln\left({x+a_j\over a_j}\right)\right.\right.}\NO\\[1ex]
      &&\left.\hspace{4cm}{}+{1\over2}\mbox{Sp}\left({a_j\over x+2a_j}\right)
        -{1\over2}\mbox{Sp}\left({x+a_j\over x+2a_j}\right)
        \right]\NO\\[1ex]
      &&{}+\sum\limits_{n,j\atop j<n}\mbox{Re}\left[\llnj^2\right]
        {\sqrt{a_na_j}\over x}\left[\left(2{x+a_j\over a_n-a_j}
        +\ln\left({x+a_n+a_j\over a_n}\right)\right)\ln\left(
        {x+a_j\over a_j}\right)\right.\NO\\[1ex]
      &&\hspace{3cm}{}+\left(2{x+a_n\over a_j-a_n}
        +\ln\left({x+a_n+a_j\over a_j}\right)
        \right)\ln\left({x+a_n\over a_n}\right)\\[1ex]
      &&\left.{}+\mbox{Sp}\left({aj\over x+a_n+a_j}\right)
        -\mbox{Sp}\left({x+aj\over x+a_n+a_j}\right)
        +\mbox{Sp}\left({an\over x+a_n+a_j}\right)
        -\mbox{Sp}\left({x+an\over x+a_n+a_j}\right)\right]\Bigg\}\;.\NO
    \eeqa
    For the related transition $\wt{l}+\wt{l}\leftrightarrow
    \sur+\wt{q}+H_2^{\dg}$ we have
    \beqa
      \lefteqn{\hat{\s}_{\scr N}^{(16)}(x) = {3\a_u\over16\pi^2}\left\{
        \sum\limits_j\lljj^2{a_j\over x}\;\left[{x\over a_j}
        +2\,\mbox{Sp}\left(-{x+a_j\over a_j}\right)
        +{\p^2\over6}\right.\right.}\NO\\[1ex]
      &&\left.\hspace{4cm}{}-\left(1-2\ln\left({x+2a_j\over a_j}
        \right)\right)\ln\left({x+a_j\over a_j}\right)\right]\\[1ex]
      &&\hspace{-20pt}{}+2\sum\limits_{n,j\atop j<n}
        \mbox{Re}\left[\llnj^2\right]{\sqrt{a_na_j}\over x}
        \left[\left({x+a_j\over a_n-a_j}+\ln\left({x+a_n+a_j\over 
        a_n}\right)\right)\ln\left({x+a_j\over a_j}\right)
        +\mbox{Sp}\left(-{x+a_j\over a_n}\right)\right.\NO\\[1ex]
      &&\hspace{\fill}\left.{}+\left({x+a_n\over a_j-a_n}
        +\ln\left({x+a_n+a_j\over a_j}\right)\right)
        \ln\left({x+a_n\over a_n}\right)
        +\mbox{Sp}\left(-{x+a_n\over a_j}\right)
        +{\p^2\over6}+{1\over2}\ln^2\left({a_n\over a_j}\right)
        \right]\Bigg\}\;.\NO
    \eeqa
    There are some $2\to2$ processes left which do not violate lepton
    number but simply transform leptons into scalar leptons, like in 
    the process $\wt{l}+\Bar{l}\leftrightarrow\wt{h}+H_2$
    \beqa
      &&\hspace{-2cm}\hat{\s}_{\scr N}^{(17)}(x)={1\over2\pi}\left\{
        \sum\limits_j\lljj^2\;\left[{-x\over x+a_j}
        +\ln\left({x+a_j\over a_j}\right)\right]\right.\NO\\[1ex]
      &&\left.\hspace{-1cm}{}+2\sum\limits_{n,j\atop j<n}
        \left|\llnj\right|^2
        \left[{a_j\over a_j-a_n}\ln\left({x+a_j\over a_j}\right)
        +{a_n\over a_n-a_j}\ln\left({x+a_n\over a_n}\right)
        \right]\right\}\;,
    \eeqa
    or in the similar process $\wt{l}+H_2^{\dg}\leftrightarrow\wt{h}+l$
    \beqa
      &&\hspace{-2cm}\hat{\s}_{\scr N}^{(18)}(x) ={1\over2\pi}\left\{
        \sum\limits_j\lljj^2\;\left[-2+{x+2a_j\over x}
        \ln\left({x+a_j\over a_j}\right)\right]\right.\NO\\[1ex]
      &&\left.\hspace{-1cm}{}+2\sum\limits_{n,j\atop j<n}
        \left|\llnj\right|^2\left[-1+{a_j\over x}{x+a_j\over a_j-a_n}
        \ln\left({x+a_j\over a_j}\right)
        +{a_n\over x}{x+a_n\over a_n-a_j}\ln\left({x+a_n\over a_n}\right)
        \right]\right\}\;.
    \eeqa
    Finally, the last process $l+{\wt{l}{}}^{\;\dg}\leftrightarrow
    \wt{h}+\wt{q}^{\,\dg}+\sur^{\dg}$ gives
    \beqa
      &&\hspace{-2cm}\hat{\s}_{\scr N}^{(19)}(x) ={3\a_u\over8\pi^2}
        \left\{\sum\limits_j\lljj^2\;\left[-2+{x+2a_j\over x}
        \ln\left({x+a_j\over a_j}\right)\right]\right.\NO\\[1ex]
      &&\left.\hspace{-1cm}{}+2\sum\limits_{n,j\atop j<n}
        \left|\llnj\right|^2\left[-1+{a_j\over x}{x+a_j\over a_j-a_n}
        \ln\left({x+a_j\over a_j}\right)
        +{a_n\over x}{x+a_n\over a_n-a_j}\ln\left({x+a_n\over a_n}\right)
        \right]\right\}\;.
    \eeqa

\section{Scattering off a Top or a Stop}
    For the processes specified in fig.~\ref{Ntop} the reduced cross
    sections read
    \beqa
      \hat{\s}_{t_j}^{(0)}&=&{3\,\a_u\over2}\lljj{x^2-a_j^2\over x^2}
        \;,\\[1ex]
      \hat{\s}_{t_j}^{(1)}&=&3\,\a_u\,\lljj{x-a_j\over x}\left[
        -{2x-a_j+2a_h\over x-a_j+a_h}+{x+2a_h\over x-a_j}
        \ln\left({x-a_j+a_h\over a_h}\right)\right]\;, \\[1ex]
      \hat{\s}_{t_j}^{(2)}&=&3\,\a_u\,\lljj{x-a_j\over x}
        \left[-{x-a_j\over x-a_j+2a_h}+\ln\left({x-a_j+a_h\over a_h}
        \right)\right]\;,\\[1ex]
      \hat{\s}_{t_j}^{(3)}&=&3\,\a_u\,\lljj\left({x-a_j\over x}
        \right)^2\;,\\[1ex]
      \hat{\s}_{t_j}^{(4)}&=&3\,\a_u\,\lljj{x-a_j\over x}\left[
        {x-2a_j+2a_h\over x-a_j+a_h}+{a_j-2a_h\over x-a_j}
        \ln\left({x-a_j+a_h\over a_h}\right)\right]\;.
    \eeqa
    To regularize an infrared divergence in the $t$-channel diagrams
    we had to introduce a Higgs-mass
    \beq
       a_h:=\left({\m\over M_1}\right)^2\;.
    \eeq
    In the calculations we have used the value $\m=800\;$GeV.

    The analogous processes involving a scalar neutrino
    (cf.~fig.~\ref{Nttop}) give similar contributions
    \beqa
      \hat{\s}_{t_j}^{(5)}&=&{3\,\a_u\over2}\lljj
        \left({x-a_j\over x}\right)^2\;, \\[1ex]
      \hat{\s}_{t_j}^{(6)}&=&3\,\a_u\,\lljj{x-a_j\over x}\left[
        -2+{x-a_j+2a_h\over x-a_j}\ln\left({x-a_j+a_h\over
        a_h}\right)\right]\;, \\[1ex]
      \hat{\s}_{t_j}^{(7)}&=&3\,\a_u\,\lljj\left[-{x-a_j\over 
        x-a_j+2a_h}+\ln\left({x-a_j+a_h\over a_h}\right)\right] 
        \;,\\[1ex]
      \hat{\s}_{t_j}^{(8)}&=&3\,\a_u\,\lljj{x-a_j\over x}
        \;{a_j\over x}\;,\\[1ex]
      \hat{\s}_{t_j}^{(9)}&=&3\,\a_u\,\lljj{a_j\over x}\left[-{x-a_j\over 
        x-a_j+a_h}+\ln\left({x-a_j+a_h\over a_h}\right)\right]\;.
    \eeqa

\section{Neutrino Pair Creation and Annihilation}
    With the abbreviations
    \beqa
      \l_{ij}&=&\l\left(x,a_i,a_j\right)=
        \left[x-\left(\sqrt{a_i}+\sqrt{a_j}\right)^2\right]\,
        \left[x-\left(\sqrt{a_i}-\sqrt{a_j}\right)^2\right]\;,\\[1ex]
      \mbox{L}_{ij}&=&\ln\left({x-a_i-a_j+\sqrt{\lkin}
        \over x-a_i-a_j-\sqrt{\lkin}}\right)\;,
    \eeqa
    the reduced cross sections for the right-handed neutrino pair
    creation read
    \beqa
      &&\hat{\s}_{\scr N_iN_j} ^{(1)}={1\over4\p}
        \left\{\lljj\llii\left[-{2\over x}\sqrt{\lkin}+\lnNN\right]
        -2\,\mbox{Re}\left[\llji^2\right]{\sqrt{a_ia_j}
        \left(a_i+a_j\right)\over x\left(x-a_i-a_j\right)}\lnNN
        \right\}\;,\NO\\
      &&\\[1ex]
      &&\hat{\s}_{\scr N_iN_j} ^{(2)}={1\over4\p}
        \left\{\lljj\llii\left[{2\over x}\sqrt{\lkin}+
        {a_i+a_j\over x}\lnNN\right]-2\,\mbox{Re}\left[\llji^2\right]
        {\sqrt{a_ia_j}\over x-a_i-a_j}\lnNN\right\}\;,\NO\\
      &&\\[1ex]
      &&\hat{\s}_{\scr N_iN_j} ^{(3)}={1\over4\p}
        \left\{\left|\llji\right|^2\left[-{2\over x}\sqrt{\lkin}
        +\lnNN\right]-2\,\mbox{Re}\left[\llji^2\right]{\sqrt{a_ia_j}
        \left(a_i+a_j\right)\over x\left(x-a_i-a_j\right)}\lnNN
        \right\}\;,\\[1ex]
      &&\hat{\s}_{\scr N_iN_j} ^{(4)}={1\over4\p}
        \left\{\left|\llji\right|^2\left[{2\over x}\sqrt{\lkin}+
        {a_i+a_j\over x}\lnNN\right]-2\,\mbox{Re}\left[\llji^2\right]
        {\sqrt{a_ia_j}\over x-a_i-a_j}\lnNN\right\}\;.
    \eeqa
    For the scalar neutrinos one has similarly
    \beqa
      \hat{\s}_{\scr \sni\snj} ^{(1)}&=&{1\over4\p}
        \lljj\llii\left[-{2\over x}\sqrt{\lkin}
        +{x-a_i-a_j\over x}\lnNN\right]\;,\\[1ex]
      \hat{\s}_{\scr \sni\snj} ^{(2)}&=&{1\over4\p}
        \left\{\lljj\llii{2\over x}\sqrt{\lkin}
        -2\,\mbox{Re}\left[\llji^2\right]
        {\sqrt{a_ia_j}\over x}\lnNN\right\}\;,\\[1ex]
      \hat{\s}_{\scr \sni\snj} ^{(3)}&=&{1\over4\p}
        \left|\llji\right|^2\left[-{2\over x}\sqrt{\lkin}
        +{x-a_i-a_j\over x}\lnNN\right]\;,\\[1ex]
      \hat{\s}_{\scr \sni\snj} ^{(4)}&=&{1\over4\p}
        \left\{\left|\llji\right|^2{2\over x}\sqrt{\lkin}
        -2\,\mbox{Re}\left[\llji^2\right]
        {\sqrt{a_ia_j}\over x}\lnNN\right\}\;.
    \eeqa
    For the diagrams involving one neutrino and one sneutrino
    (cf.~fig.~\ref{NNt}) one finally has
    \beqa
      &&\hat{\s}_{\scr N_j\sni}^{(1)}={1\over4\p}
        \left\{\lljj\llii{x+a_i-a_j\over x}\lnNN
        -2\,\mbox{Re}\left[\llji^2\right]{\sqrt{a_ia_j}\over x}
        {x+a_i-a_j\over x-a_i-a_j}\lnNN\right\}\;,\NO\\
      &&\\[1ex]
      &&\hat{\s}_{\scr N_j\sni} ^{(2)}={1\over4\p}
        \left\{\left|\llji\right|^2{x+a_i-a_j\over x}\lnNN
        -2\,\mbox{Re}\left[\llji^2\right]{\sqrt{a_ia_j}\over x}
        {x+a_i-a_j\over x-a_i-a_j}\lnNN\right\}\;.
    \eeqa

    \clearpage\chapter{Reaction Densities \label{appD}}
  In general the reaction densities corresponding to the reduced
  cross sections discussed in appendix~\ref{appC} have to be calculated
  numerically. However, there exist some interesting limiting cases
  where one can calculate them analytically. Since thermal averaging
  of reduced cross sections via eq.~(\ref{22scatt}) involves modified
  Bessel functions, we start by summarizing a few useful formulae for
  Bessel functions before discussing the reaction densities. 

\section{Bessel functions}
  Modified Bessel functions with different indices are related via
  recursion relations \cite{gr},
  \beqa
    &&x\mbox{K}_{\nu-1}(x)+x\mbox{K}_{\nu+1}(x)=2\nu\mbox{K}_{\nu}(x)
      \;,\\[1ex]
    &&\mbox{K}_{\nu-1}(x)-\mbox{K}_{\nu+1}(x)=
      2{d\over dx}\mbox{K}_{\nu}(x)\;.
  \eeqa
  For integer index Bessel functions have the following series 
  representation 
  \beqa
    \mbox{K}_n(x)&=&{1\over2}\,\sum\limits_{k=0}^{n-1}\,(-1)^k\,
      {(n-k-1)!\over k!\,\left(\displaystyle {z\over 2}\right)^{n-2k}}
      \,+\\[1ex]
    &&+\,(-1)^{n+1}\sum\limits_{k=0}^{\infty}
      {\left(\displaystyle {z\over 2}\right)^{n+2k}\over k!(n+k)!}
      \left[\ln\left({x\over2}\right)-{1\over2}\psi(k+1)
      -{1\over2}\psi(n+k+1)\right]\;,\NO
  \eeqa
  where $\psi$ denotes the derivative of the logarithm of the Gamma
  function
  \beq
    \psi(x)={d\over dx}\ln\Gamma(x)\;.
  \eeq
  For integer argument it reads
  \beq
    \psi(n)=-\g_{\scr\rm E}+\sum\limits_{k=1}^{n-1}{1\over k}\;,
  \eeq
  where $\g_{\scr\rm E}=0.577216$ is Euler's constant. Hence, the
  leading terms of the series are given by
  \beqa
    \mbox{K}_0(x)&=&\ln\left({2\over x}\right)-\g_{\scr\rm E}+\ldots
      \;,\label{seriesBK0}\\[1ex]
    \mbox{K}_n(x)&=&{(n-1)!\over2}\left({2\over x}\right)^n+\ldots
      \;,\qquad\mbox{for }n\geq1\;.\label{seriesBKn}
  \eeqa
  The asymptotic expansion of modified Bessel functions reads 
  \beq
    \mbox{K}_{\nu}(x)=\sqrt{{\pi\over2x}\,}\,\mbox{e}^{-x}
    \sum\limits_{k=0}^\infty
    {1\over k!(2x)^k}\,{\Gamma\left(\nu+k+{1\over2}\right)\over
    \Gamma\left(\nu-k+{1\over2}\right)}\;,
  \eeq
  i.e.\ to leading order all Bessel functions have the same asymptotic
  behaviour,
  \beq
    \mbox{K}_{\nu}(x)=\sqrt{{\pi\over2x}\,}\,\mbox{e}^{-x}+\ldots\;.
  \eeq

  Furthermore, when evaluating reaction densities according to
  eq.~(\ref{22scatt}), one has to compute integrals involving Bessel
  functions. In the following we compile the integrals that we have
  used in our calculations.\\
  \noindent{\bf Bessel functions and powers}
  \beqa
    &&\int\limits_0^{\infty}dx\,x^{\m}\mbox{K}_{\n}(ax)
      =2^{\m-1}a^{-\m-1}\Gamma\left({1+\m+\n\over2}\right)
       \Gamma\left({1+\m-\n\over2}\right)\\
    &&\qquad\qquad\qquad\qquad\mbox{for Re}(1+\m\pm\n)>0 
      \mbox{ and Re}(a)>0\;,\NO\\[2ex]
    &&\int\limits_0^1dx\,x^{\n+1}\mbox{K}_{\n}(ax)
      =2^{\n}a^{-\n-2}\Gamma(\n+1)-{1\over a}\mbox{K}_{\n+1}(a)
      \qquad\mbox{for Re}(\n)>-1\;,\\[2ex]
    &&\int\limits_0^1dx\,x^{1-\n}\mbox{K}_{\n}(ax)
      =2^{-\n}a^{\n-2}\Gamma(1-\n)-{1\over a}\mbox{K}_{\n-1}(a)
      \qquad\mbox{for Re}(\n)<1\;,\\[2ex]
    &&\int\limits_a^{\infty}dx\,\mbox{K}_1(zx)=
      {1\over z}\mbox{K}_0(za)\;,\\[2ex]
    &&\int\limits_a^{\infty}dx\,x^2\,\mbox{K}_1(zx)=
      {a^2\over z}\mbox{K}_2(za)\;,\\[2ex]
    &&\int\limits_0^{\infty}dx\,x^2\,\mbox{K}_1(zx)=
      {2\over z^3}\;,\\[2ex]
    &&\int\limits_a^{\infty}dx\,x\,\mbox{K}_0(zx)=
      {a\over z}\mbox{K}_1(za)\;,\\[2ex]
    &&\int\limits_0^{\infty}dx\,x\,\mbox{K}_0(zx)={1\over z^2}\;,\\[1ex]
    &&\int\limits_a^{\infty}dx\,\sqrt{x}\,\mbox{K}_1(z\sqrt{x}\,)=
      2{a\over z}\mbox{K}_2(z\sqrt{a})\;,\\[2ex]
    &&\int\limits_0^{\infty}dx\,\sqrt{x}\,\mbox{K}_1(z\sqrt{x}\,)=
      {4\over z^3}\;,\\[2ex]
    &&\int\limits_a^{\infty}dx\,{1\over\sqrt{x}}\,\mbox{K}_1(z\sqrt{x}\,)=
      {2\over z}\mbox{K}_0(z\sqrt{a})\;.
  \eeqa

\noindent{\bf Bessel functions and logarithms}
  \beqa
    &&\int\limits_a^{\infty}dx\,x\ln(x)\mbox{K}_0(zx)=
      {1\over z^2}\mbox{K}_0(za)+a\ln(a){1\over z}\mbox{K}_1(za)
      \;,\qquad\qquad\qquad\qquad\mbox{ }\\[2ex]
    &&\int\limits_0^{\infty}dx\,x\ln(x)\mbox{K}_0(zx)=
      {1\over z^2}\left[-\g_{\scr\rm E}+\ln\left({2\over z}\right)\right]
      \;,\\[2ex]
    &&\int\limits_a^{\infty}dx\,\sqrt{x}\ln\left({x\over b}\right)
      \mbox{K}_1(z\sqrt{x}\,)=\left[{8\over z^3}+{2a\over z}
      \ln\left({a\over b}\right)\right]\mbox{K}_0(z\sqrt{a})\NO\\
    &&\hspace{5cm}
      +4{\sqrt{a}\over z^2}\left[1+\ln\left({a\over b}\right)\right]
      \mbox{K}_1(z\sqrt{a})\;,\\[2ex]
    &&\int\limits_0^{\infty}dx\,\sqrt{x}\ln\left({x\over b}\right)
      \mbox{K}_1(z\sqrt{x}\,)={4\over z^3}\left[1-2\g_{\scr\rm E}
      +\ln\left({4\over bz^2}\right)\right]\;.
  \eeqa
  In all these integrals we have always assumed that $a$, $b$ and $z$
  are real and positive.

\section{Lepton Number Violating Scatterings}
  In the Boltzmann equations we do not need every reaction density
  $\g_N^{(i)}$, $i=1,\ldots,19$ separately (cf.~sect.~\ref{boltzeq}).
  We only have to consider the combined reaction densities
  \beqa
    \g_{\scr A}^{\scr \D L}&=&2\g_{\scr N}^{(1)}+\g_{\scr N}^{(3)}
      +\g_{\scr N}^{(4)}+\g_{\scr N}^{(6)}+\g_{\scr N}^{(7)}
      +2\g_{\scr N}^{(12)}+\g_{\scr N}^{(14)}\;,\\[1ex]
    \g_{\scr B}^{\scr \D L}&=&\g_{\scr N}^{(3)}+\g_{\scr N}^{(4)}
      -\g_{\scr N}^{(6)}-\g_{\scr N}^{(7)}+\g_{\scr N}^{(14)}\;,\\[1ex]
    \g_{\scr C}^{\scr \D L}&=&3\g_{\scr N}^{(9)}+\g_{\scr N}^{(17)}
      +\g_{\scr N}^{(18)}+6\g_{\scr N}^{(19)}\;,\\[1ex]
    \g_{\scr D}^{\scr \D L}&=&4\g_{\scr N}^{(5)}+2\g_{\scr N}^{(8)}
      +8\g_{\scr N}^{(10)}+3\g_{\scr N}^{(9)}+4\g_{\scr N}^{(15)}
      +2\g_{\scr N}^{(16)}+\g_{\scr N}^{(17)}+\g_{\scr N}^{(18)}
      +6\g_{\scr N}^{(19)}\;.
  \eeqa  
  For low temperatures, i.e.~$z\gg1/\sqrt{a_j}\;$, the dominant
  contribution to the integrand of the reaction densities comes from
  small centre of mass energies, i.e.~$x\ll a_j$. In this limit the
  reduced cross sections $\hat{\s}_{\scr N}^{(i)}$ for the $(L+\wt{L})$
  violating or conserving processes behave differently. For the
  $(L+\wt{L})$ violating scatterings ($i=1,\ldots,5,8,10,12,\ldots,16$)
  one finds
  \beq
    \hat{\s}_{\scr N}^{(i)}\propto x\quad\mbox{for }x\ll a_j\;,
  \eeq
  while one has
  \beq
    \hat{\s}_{\scr N}^{(i)}\propto x^2\quad\mbox{for }x\ll a_j
  \eeq  
  for the $(L+\wt{L})$ conserving processes ($i=6,7,9,11,17,18,19$).
  In diagrams with an intermediate neutrino this different behaviour
  is due to the different chiral parts of the fermionic propagator
  contributing to the scatterings. $(L+\wt{L})$ violating processes
  contain the chirality violating propagators $\sqrt{a_j}/(x-a_j)$,
  whereas $(L+\wt{L})$ conserving processes depend on the chirality
  conserving propagator $\sqrt{x}/(x-a_j)$. For diagrams with an
  intermediate scalar neutrino the corresponding kinematical factors
  originate in the couplings of sneutrinos to different initial and
  final states.

  Hence, the reaction densities can be calculated analytically in
  this limit and one finds
  \beqa
    &&\hspace{-1cm}\g_{\scr A}^{\scr \D L}={M_1^4\over\p^5}
      {1\over z^6}\left\{\sum\limits_j\lljj^2{2\over a_j}+
      \sum\limits_{n,j\atop j<n}\mbox{Re}\left[\llnj^2\right]
      {19\over4\sqrt{a_na_j}}\right\}\;,\\[1ex]
    &&\hspace{-1cm}\g_{\scr B}^{\scr \D L}={M_1^4\over\p^5}
      {1\over z^6}\left\{\sum\limits_j\lljj^2{1\over2a_j}+
      \sum\limits_{n,j\atop j<n}\mbox{Re}\left[\llnj^2\right]
      {7\over4\sqrt{a_na_j}}\right\}\;,\\[1ex]
    &&\hspace{-1cm}\g_{\scr C}^{\scr \D L}={M_1^4\over\p^5}
      {1\over z^8}\left\{\sum\limits_j\lljj^2{1\over a_j^2}
      \left(4+{27\a_u\over4\p}\right)
      +\sum\limits_{n,j\atop j<n}\left|\llnj\right|^2
      {1\over a_ja_n}\left(8+{18\a_u\over\p}\right)\right\}\;,\\[1ex]
    &&\hspace{-1cm}\g_{\scr D}^{\scr \D L}={M_1^4\over\p^5}{\a_u\over\p}
      {1\over z^6}\left\{\sum\limits_j\lljj^2{153\over32a_j}+
      \sum\limits_{n,j\atop j<n}\mbox{Re}\left[\llnj^2\right]
      {147\over16\sqrt{a_na_j}}\right\}\;.
  \eeqa
  For high temperatures, i.e.~$z\ll1/\sqrt{a_j}\;$, we can use
  the asymptotic expansions of the reduced cross sections to compute
  the reactions densities and we get
  \beqa
    \g_{\scr A}^{\scr \D L}&=&{M_1^4\over64\p^5}{1\over z^4}\left\{
      \left(13+{3\a_u\over4\p}\right)\sum\limits_j\lljj^2+
      \sum\limits_{n,j\atop j<n}\mbox{Re}\left[\llnj^2\right]
      {24\sqrt{a_na_j}\over a_n-a_j}\ln\left(a_n\over a_j\right)
      \right.\NO\\[1ex]
    &&\left.\hspace{3cm}{}+\left(2+{3\a_u\over2\p}\right)
      \sum\limits_{n,j\atop j<n}\left|\llnj\right|^2\right\}\;,\\[2ex]
    \g_{\scr B}^{\scr \D L}&=&{M_1^4\over64\p^5}{1\over z^4}\left\{
      \left(3-{3\a_u\over4\p}\right)\sum\limits_j\lljj^2+
      \sum\limits_{n,j\atop j<n}\mbox{Re}\left[\llnj^2\right]
      {8\sqrt{a_na_j}\over a_n-a_j}\ln\left(a_n\over a_j\right)
      \right.\NO\\[1ex]
    &&\left.\hspace{3cm}{}-\left(2+{3\a_u\over2\p}\right)
      \sum\limits_{n,j\atop j<n}\left|\llnj\right|^2\right\}\;,\\[2ex]
    \g_{\scr C}^{\scr \D L}&=&{M_1^4\over32\p^5}{1\over z^4}\left\{
      \sum\limits_j\lljj^2\left[-1-{45\a_u\over8\p}
      +{9\a_u\over8}\sqrt{a_j\over\wt{c_j}}
      +\left(4+{27\a_u\over2\p}\right)
      \left(\ln\left({2\over z\sqrt{a_j}}\right)-\g_{\scr\rm E}\right)\right]
      \right.\NO\\[1ex]
    &&{}+\sum\limits_{n,j\atop j<n}\left|\llnj\right|^2
      \left[2+{9\a_u\over\left(a_n-a_j\right)^2}
      \left(a_n\sqrt{a_j\wt{c_j}}+a_j\sqrt{a_n\wt{c_n}}\right)
      \right.\\[1ex]
    &&\left.\hspace{2cm}{}+\left(8+{36\a_u\over\p}\right)
      \left({a_n\over a_n-a_j}\ln\left({2\over\sqrt{a_n}z}\right)
      +{a_j\over a_j-a_n}\ln\left({2\over\sqrt{a_j}z}\right)
      -\g_{\scr\rm E}\right)\right]\Bigg\}\;,\NO\\[2ex]
    \g_{\scr D}^{\scr \D L}&=&{M_1^4\over32\p^5}{1\over z^4}\left\{
      \sum\limits_j\lljj^2\left[-1+{27\a_u\over8\p}
      +{39\a_u\over8}\sqrt{a_j\over\wt{c_j}}
      +\left(4+{27\a_u\over2\p}\right)
      \left(\ln\left({2\over z\sqrt{a_j}}\right)
      -\g_{\scr\rm E}\right)\right]\right.\NO\\[1ex]
    &&{}+\sum\limits_{n,j\atop j<n}\left|\llnj\right|^2
      \left[4-8\g_{\scr\rm E}
      +{8\over a_n-a_j}\left(a_n\ln\left({2\over\sqrt{a_n}z}\right)
      -a_j\ln\left({2\over\sqrt{a_j}z}\right)\right)\right]\\[1ex]
    &&\left.{}+{3\a_u\over\p}\sum\limits_{n,j\atop j<n}\mbox{Re}
      \left[\llnj^2\right]\sqrt{a_na_j}\left[
      {7\over2}{1\over a_n-a_j}\ln\left({a_n\over a_j}\right)
      +{5\p\over\left(a_n-a_j\right)^2}\left(\sqrt{a_j\wt{c_j}}
      +\sqrt{a_n\wt{c_n}}\right)\right]\right\}\NO\;,
  \eeqa
  i.e.\ these reaction densities are proportional to $T^4$ at high
  temperatures, as expected on purely dimensional grounds.
  
  For intermediate temperatures $z\sim 1/\sqrt{a_j}$ the reaction
  densities have to be computed numerically. This becomes increasingly
  difficult in the narrow width limit, where $1/ D_j(x)$ has two very
  sharp peaks. However, in the limit $c_j\to0$ the two peaks in
  $1/D_j(x)$ cancel each other, since they have a different sign,
  while the peaks in $1/D^2_j(x)$ add up.  Therefore, the terms
  proportional to $1/D_j(x)$ or $1/D_j(x)D_n(x)$ with $n\ne j$ can be
  neglected in the narrow width limit, while $1/D^2_j(x)$ can be
  approximated by a $\d$-function
  \beq 
    {1\over D^2_j(x)}\approx{\p\over2\sqrt{a_jc_j}}\,
      \d\left(x-a_j\right)\;.
  \eeq
  An analogous relation holds for $1/\wt{D_j}^2(x)$.

  These relations allow to calculate the contributions from the
  $s$-channel diagrams to the reaction densities analytically in the
  limit $c_j\to0$, while the contributions from the $t$-channel
  diagrams can easily be evaluated numerically. 

\section{Stop Neutrino Scatterings}
  The reaction densities $\g_{t_j}^{(i)}$ for the interaction of a
  (s)neutrino with a top or a stop can also be calculated
  analytically in the limit of high temperatures
  $z\ll1/\sqrt{a_j}\;$. For the $s$-channel processes one finds
  \beqa
    \g_{t_j}^{(0)}&=&{3\a_uM_1^4\over64\p^4}
      \lljj a_j{\mbox{K}_2\left(z\sqrt{a_j}\right)\over z^2}\;,\\[1ex]
    \g_{t_j}^{(3)}&=&2\g_{t_j}^{(0)}\;,\qquad\qquad
      \g_{t_j}^{(5)}=\g_{t_j}^{(0)}\;.
  \eeqa
  For the $t$-channel reaction densities one has analogously
  \beqa
    &&\hspace{-1.8cm}\g_{t_j}^{(1)}={3\a_uM_1^4\over8\p^4}
      \lljj{1\over z^4}\left[\left(1-{z^2a_j\over4}\right)
      \mbox{K}_0\left(z\sqrt{a_j}\right)
      +{z^2a_j\over4}\left(\ln\left({a_j\over a_h}\right)-1
      \right)\mbox{K}_2\left(z\sqrt{a_j}\right)\right]\;,\\[1ex]
    &&\hspace{-1.8cm}\g_{t_j}^{(2)}={3\a_uM_1^4\over8\p^4}
      \lljj{1\over z^4}\left[
      \left(1-{z^2a_j\over4}\right)\mbox{K}_0\left(z\sqrt{a_j}\right)
      +{z^2a_j\over4}\ln\left({a_j\over a_h}\right)
      \mbox{K}_2\left(z\sqrt{a_j}\right)\right]\;,\\[1ex]
    &&\g_{t_j}^{(4)}=2\g_{t_j}^{(0)}\;,\qquad\qquad
      \g_{t_j}^{(6)}=\g_{t_j}^{(1)}\;,\qquad\qquad
      \g_{t_j}^{(7)}=\g_{t_j}^{(2)}\;.
  \eeqa
  $\g_{t_j}^{(8)}$ and $\g_{t_j}^{(9)}$ are several orders of
  magnitude smaller than the other $\g_{t_j}^{(i)}$ for small $z$
  and can therefore be neglected at high temperatures. 
  
  By using the series expansions (\ref{seriesBK0}) and
  (\ref{seriesBKn}), one sees that the processes with a higgsino in
  the $t$-channel, i.e.\ $\g_{t_j}^{(1)}$, $\g_{t_j}^{(2)}$,
  $\g_{t_j}^{(6)}$ and $\g_{t_j}^{(7)}$, behave like 
  $T^4\ln\left(T/M_j\right)$ at high temperatures, whereas the
  other reaction densities are proportional to $T^4$.

\section{Pair Creation and Annihilation of Neutrinos}
  In the Boltzmann equations we only need certain combinations of
  reaction densities which can easily be evaluated for high
  temperatures
  \beqa
    \sum\limits_{k=1}^2\g_{\scr N_iN_j}^{(k)}&=&
      \sum\limits_{k=1}^2\g_{\scr \sni\snj}^{(k)}=
      \g_{\scr N_j\sni}^{(1)}=\\[1ex]
     &=&{M_1^4\over16\p^5}{1\over z^4}\lljj\llii\left\{
      \left[1-{z^2\over4}\left(\sqrt{a_i}+\sqrt{a_j}\right)^2\right]
      \mbox{K}_0\left(z\left(\sqrt{a_i}+\sqrt{a_j}\right)\right)
      \right.\NO\\[1ex]
   &&\left.{}+{z^2\over4}\left(\sqrt{a_i}+\sqrt{a_j}\right)^2\left[
      1+\ln\left(2+{a_i+a_j\over\sqrt{a_ia_j}}\right)\right]
      \mbox{K}_2\left(z\left(\sqrt{a_i}+\sqrt{a_j}\right)\right)
      \right\}\;,\NO\\[1ex]
   \sum\limits_{k=3}^4\g_{\scr N_iN_j}^{(k)}&=&
      \sum\limits_{k=3}^4\g_{\scr \sni\snj}^{(k)}=
      \g_{\scr N_j\sni}^{(2)}=\\[1ex]
   &=&{M_1^4\over16\p^5}{1\over z^4}
      \left|\llji\right|^2
      \left\{\left[1-{z^2\over4}\left(\sqrt{a_i}+\sqrt{a_j}\right)^2\right]
      \mbox{K}_0\left(z\left(\sqrt{a_i}+\sqrt{a_j}\right)\right)
      \right.\NO\\[1ex]
   &&\left.{}+{z^2\over4}\left(\sqrt{a_i}+\sqrt{a_j}\right)^2\left[
      1+\ln\left(2+{a_i+a_j\over\sqrt{a_ia_j}}\right)\right]
      \mbox{K}_2\left(z\left(\sqrt{a_i}+\sqrt{a_j}\right)\right)
      \right\}\;,\NO
  \eeqa
  i.e.\ these reaction densities are proportional to 
  $T^4\ln\left(T/(M_i+M_j)\right)$ at high temperatures.

  \end{appendix}

\newpage\noindent
\markboth{}{}
{\bf\LARGE Acknowledgments}\\[1ex]

\noindent
I would like to thank my advisor, W.~Buchm\"uller, who suggested this
investigation, for continuous support and a close and fruitful
collaboration.

I would also like to acknowledge highly instructive discussions with
W.~Beenakker, J.~Hein, G.~Hiller, A.~Jakov\'ac, S.~Moch, T.~Plehn and
U.~Nierste.

Finally, I would like to thank all my colleagues and friends at DESY
for the enjoyable working atmosphere.
\end{document}